%% file: MEG_Upgrade_Proposal.tex
\documentclass[11pt,nofootinbib,floatfix]{revtex4}
\pdfoutput=1
\usepackage{graphicx}
\usepackage{epstopdf}   
\usepackage{import}
\usepackage{amssymb}
\usepackage{txfonts}
\usepackage{epsfig}
\usepackage{natbib}
\usepackage{listings}
\usepackage{mdwlist}
\usepackage{subfigure}
\usepackage{pennames}
\usepackage{hyperref}

\usepackage{pgfgantt}

\usepackage{pdflscape}

\usepackage{float}
\floatstyle{plaintop}
\newfloat{fgantt}{thbp}{fgantt}
\floatname{fgantt}{Gantt chart}

\usepackage{multirow}

\usepackage{eurosym}

\newcommand*{\INFNPi}{\affiliation{INFN Sezione di Pisa$^{a}$; Dipartimento di Fisica$^{b}$ dell'Universit\`a, Largo B.~Pontecorvo~3, 56127 Pisa, Italy\vspace{-2.2mm}}}
\newcommand*{\INFNGe}{\affiliation{INFN Sezione di Genova$^{a}$; Dipartimento di Fisica$^{b}$ dell'Universit\`a, Via Dodecaneso 33, 16146 Genova, Italy\vspace{-2.2mm}}}
\newcommand*{\INFNPv}{\affiliation{INFN Sezione di Pavia$^{a}$; Dipartimento di Fisica Nucleare e Teorica$^{b}$ dell'Universit\`a, Via Bassi 6, 27100 Pavia, Italy\vspace{-2.2mm}}}
\newcommand*{\INFNRm}{\affiliation{INFN Sezione di Roma$^{a}$; Dipartimento di Fisica$^{b}$ dell'Universit\`a ``Sapienza'', Piazzale A.~Moro, 00185 Roma, Italy\vspace{-2.2mm}}}
\newcommand*{\INFNLe}{\affiliation{INFN Sezione di Lecce$^{a}$; Dipartimento di Matematica e  Fisica$^{b}$ dell'Universit\`a; 
Dipartimento di Ingegneria dell'innovazione$^{c}$ dell'Universit\`a, Via per Arnesano, 73100 Lecce, Italy; Universit\`a ``G.~Marconi''$^{d}$, Via Plinio 44, 00193 Roma, Italy\vspace{-2.2mm}}}
\newcommand*{\ICEPP} {\affiliation{ICEPP, The University of Tokyo 7-3-1 Hongo, Bunkyo-ku, Tokyo 113-0033, Japan \vspace{-2.2mm}}}
\newcommand*{\UCI}   {\affiliation{University of California, Irvine, CA 92697, USA\vspace{-2.2mm}}}
\newcommand*{\KEK}   {\affiliation{KEK, High Energy Accelerator Research Organization 1-1 Oho, Tsukuba, Ibaraki 305-0801, Japan\vspace{-2.2mm}}}

\newcommand*{\PSI}   {\affiliation{Paul Scherrer Institut PSI, CH-5232 Villigen, Switzerland\vspace{-2.2mm}}}

\newcommand*{\BINP}   {\affiliation{Budker Institute of Nuclear Physics, 630090 Novosibirsk, Russia\vspace{-2.2mm}}}
\newcommand*{\JINR}   {\affiliation{Joint Institute for Nuclear Research, 141980, Dubna, Russia\vspace{-2.2mm}}}

\newcommand*{\meg}{\mu\to\rm{e}\gamma}
\newcommand*{\megc}{\ifmmode\mu^+ \to \mathrm{e}^+ \gamma\else$\mu^+ \to \mathrm{e}^+ \gamma$\fi}
\newcommand*{\conv}{\mu \to \rm{e}}

\newcommand*{\tmug}{\tau \to  \mu \gamma}

\newcommand*{\BR}     { {\cal B} }

\newcommand*{\michelsign}        {\mu^+ \to \mathrm{e}^+ \nu\bar{\nu}}

\newcommand*{\egamma}         {E_{\gamma}}
\newcommand*{\epositron}      {E_\mathrm{e}}
\newcommand*{\ppositron}      {P_\mathrm{e}}
\newcommand*{\tegamma}        {t_{\mathrm{e}\gamma}}
\newcommand*{\Thetaegamma}    {\Theta_{\mathrm{e}\gamma}}

\newcommand*{\thetae}    {\theta_\mathrm{e}}
\newcommand*{\phie}      {\phi_\mathrm{e}}

\begin{document}


\title{MEG Upgrade Proposal}
\author{{ A.~M.~Baldini}$^{a}$}\thanks{Spokespersons}   \INFNPi
\author{D.~Bagliani$^{ab}$}        \INFNGe
\author{E.~Baracchini}         \ICEPP
\author{G.~Boca$^{ab}$}        \INFNPv
\author{P.~W.~Cattaneo$^{a}$}  \INFNPv
\author{M.~Cascella$^{ab}$}      \INFNLe
\author{G.~Cavoto$^{a}$}       \INFNRm
\author{F.~Cei$^{ab}$}         \INFNPi
\author{C.~Cerri$^{a}$}        \INFNPi
\author{A.~de~Bari$^{a}$}      \INFNPv
\author{M.~De~Gerone$^{a}$}   \INFNGe
\author{S.~Dussoni$^{a}$}     \INFNPi
\author{Y.~Fujii}              \ICEPP
\author{L.~Galli$^{a}$}        \INFNPi
\author{F.~Gatti$^{ab}$}        \INFNGe
\author{F.~Grancagnolo$^{a}$}      \INFNLe
\author{M.~Grassi$^{a}$}        \INFNPi
\author{A.~Graziosi$^{a}$}       \INFNRm
\author{D.~N.~Grigoriev}        \BINP
\author{T.~Haruyama}            \KEK
\author{M.~Hildebrandt}         \PSI
\author{F.~Ignatov}             \BINP
\author{T.~Iwamoto}             \ICEPP
\author{T.~I.~Kang}              \UCI
\author{D.~Kaneko}             \ICEPP
\author{P.-R.~Kettle}           \PSI
\author{B.~I.~Khazin}           \BINP
\author{N.~Khomutov}         \JINR
\author{A.~Korenchenko}         \JINR
\author{N.~Kravchuk}            \JINR
\author{N.~Kuchinksy}         \JINR
\author{A.~L'Erario$^{ac}$}      \INFNLe
\author{G.~Lim}              \UCI
\author{A.~Maffezzoli$^{ac}$}      \INFNLe
\author{A. Miccoli$^{a}$}            \INFNLe
\author{S.~Mihara}              \KEK
\author{W.~Molzon}              \UCI
\author{{ T.~Mori}}  \thanks{Spokespersons}              \ICEPP
\author{R.~Nard\'o$^{a}$}      \INFNPv
\author{D.~Nicol\`o$^{ab}$}     \INFNPi
\author{H.~Nishiguchi}          \KEK
\author{M.~Nishimura}           \ICEPP
\author{G.~Onorato$^{ad}$}      \INFNLe
\author{W.~Ootani}              \ICEPP
\author{G.~Palam\'a$^{ab}$}      \INFNLe
\author{M.~Panareo$^{ab}$}      \INFNLe
\author{A.~Papa}                \PSI
\author{A.~Pepino$^{a}$}      \INFNLe
\author{G.~Piredda$^{a}$}       \INFNRm
\author{A.~Popov}               \BINP
\author{F.~Raffaelli$^{a}$}     \INFNPi
\author{S.~Rella$^{ac}$}      \INFNLe
\author{F.~Renga}        \PSI
\author{E.~Ripiccini$^{ab}$}       \INFNRm
\author{S.~Ritt}                \PSI
\author{M.~Rossella$^{a}$}      \INFNPv
\author{R.~Sawada}              \ICEPP
\author{F.~Sergiampietri$^{a}$}\INFNPi
\author{G.~Signorelli$^{a}$} \INFNPi
\author{A.~Stoykov}            \PSI
\author{G.F.~Tassielli$^{ad}$}      \INFNLe
\author{F.~Tenchini$^{ab}$}     \INFNPi
\author{Y.~Uchiyama}            \ICEPP
\author{C.~Voena$^{a}$}         \INFNRm
\author{A.~Yamamoto}            \KEK
\author{Z.~You}                \UCI
\author{Yu.~V.~Yudin}           \BINP
\author{G.~Zavarise$^{ac}$}      \INFNLe



\maketitle
\newpage

\newpage
\tableofcontents

\newpage
\subimport{./00_Executive_Summary/}{Executive_Summary.tex}
\clearpage

\newpage
\subimport{./01_MEG_Status/}{MEG_Status.tex}
\clearpage

\newpage
\subimport{./02_Scientific_Merits/}{Scientific_Merits.tex}

\clearpage

\newpage
\subimport{./03_Upgrade_Overview/}{Upgrade_Overview.tex}

\clearpage

\newpage
\subimport{./04_Beam_Line_and_Target/}{Beam_Line_and_Target.tex}
\clearpage

\newpage
\subimport{./05_Positron_Tracker/}{Positron_Tracker.tex}

\clearpage


\newpage
\subimport{./07_Photon_Calorimeter/}{Photon_Calorimeter.tex}

\clearpage

\newpage
\subimport{./08_Trigger_and_DAQ/}{Trigger_and_DAQ.tex}

\clearpage

\newpage
\subimport{./09_Final_Sensitivity/}{Final_Sensitivity.tex}
\clearpage

\newpage
\subimport{./10_Budget_and_Responsibilities/}{Budget_and_Responsibilities.tex}

\clearpage

\newpage
\subimport{./11_Time_Schedule/}{Time_Schedule.tex}

\clearpage

\newpage
\subimport{./12_Summary/}{Summary.tex}

\clearpage

\newpage
\subimport{./13_Appendix/}{Appendix.tex}

\clearpage

\newpage
\bibliographystyle{unsrt}
\bibliography{MEG_Upgrade_Proposal}
\end{document}

%% file: 00_Executive_Summary/Executive_Summary.tex
\section{Executive Summary}
%
%

We propose the continuation of the MEG experiment to search for the charged lepton flavour violating decay (cLFV) $\meg$, based on an upgrade of the experiment, which aims for a sensitivity enhancement of one order of magnitude compared to the final MEG result. The current MEG experiment can be considered, apart from its discovery potential, as a benchmark for next-generation cLFV decay experiments. This, not only by having imposed one of the most stringent constraints to date on models predicting large LFV-enhancements through "New Physics" beyond the Standard Model (BSM), but also by setting the tightest upper limit on the decay itself, that of $2.4\times10^{-12}$ (90\% C.L.). On the experimental side, a benchmark has also been set by both having designed and run the experiment close to the intensity frontier, made possible by the PSI high-intensity proton accelerator facility (HIPA).

 The planned sensitivity enhancement for an upgraded MEG experiment would, together with the planned next generation $\conv$ conversion experiments COMET at J-PARC and Mu2e at Fermilab and the next-generation $\mu\rightarrow \rm{eee}$ experiment at PSI, which seek to probe the other two "Golden" cLFV muon channels, and in addition to efforts at future B-factories to measure the charged LFV tau-decays, test BSM models with unprecedented sensitivity, in a complementary way, to the direct searches at the energy frontier of high-energy colliders. Furthermore, the cLFV experiments also allow access to mass scales for "New Physics" well beyond the reach of the direct searches at colliders.
 
  The MEG experiment is currently close to finishing its 2012 run on schedule, having been able to double its statistics of 2011 and having achieved more than a factor of three more statistics than used for the previously published best limit . The experiment is expected to continue data-taking until mid 2013, when a sensitivity of about $\sim 6 \cdot 10^{-13}$ is expected to be reached, beyond which only limited improvement is expected due to the expected increasing dominance of background events in the signal region. Therefore, in order to significantly improve the sensitivity reach with a goal of being able to detect the $\meg$ decay at a level of about one order of magnitude lower, a new upgraded MEG experiment is required.
  
  The key features of this new MEG upgrade, aimed at significantly improving the experimental sensitivity, are to increase the rate capability of all detectors to enable running at the intensity frontier, while also improving the energy and angular and timing resolutions, for both the positron and photon arms of the detector. This is especially valid on the positron-side, where a new low-mass, single volume, high granularity tracker is under development. This, in combination with a thinner stopping target and hence a reduction in the multiple scattering of the positrons, will lead to the spatial, angular and energy requirements being met on the positron side. A new highly segmented, fast timing counter array will replace the old system, so allowing improved timing resolution capabilities in order to minimize the number of background events entering the signal timing window. The photon-arm, with the largest liquid xenon (LXe) detector in the world, totalling 900 l, will also be improved by increasing the granularity at the incident face, by replacing the current photomultiplier tubes (PMTs) with a larger number of smaller photosensors and optimizing the photosensor layout also on the lateral faces. This should also lead to improved energy and spatial resolutions for the LXe detector. Finally, in order to meet the stringent requirements of an increased number of readout channels and to cope with the necessary bandwidth required by such a system, a new DAQ scheme involving the implementation of a new combined readout board capable of integrating the various functions of digitization, trigger capability and splitter functionality into one condensed unit, is also under development.
  
  During the R$\&$D that has been on-going since 2011, various complementary and auxiliary devices and technologies have been  studied in order to reach the baseline solution outlined in this proposal. Some of these devices have been developed to a significant level, such that once prototypes have been rigorously tested under realistic beam conditions they could then be introduced into the running experiment, so allowing for further improvements.
  
\begin{figure}[hbct]
\begin{center}
\includegraphics[width=0.8\linewidth]{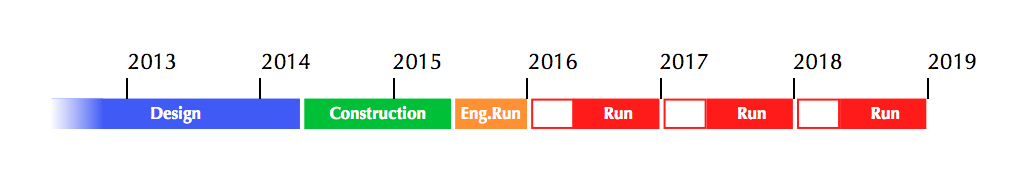}
\caption{\label{fig:timeline} 
Planned Schedule for the MEG Upgrade, showing the periods for R$\&$D, construction, and implementation, as well as  an Engineering Run followed by three years of data-taking.
}
\end{center}
\end{figure}

  The overall planned schedule for the upgrade and its implementation is shown in the timeline in Figure~\ref{fig:timeline}. An initial period of design and development, with the planned end of construction date of around mid 2015, is to be followed by an engineering run in the latter half of 2015 and, providing the performance is as expected, data-taking could start in 2016. The present sensitivity estimate is based on a muon stopping rate of $7\cdot10^{7}$ muons/s for a running time of 3 years, assuming 180 DAQ days per year.
  
In the following sections of the proposal, a detailed description of the current status of the MEG experiment, the scientific merits of an upgrade and the detailed overview of the key features involved will be addressed. In conclusion to this, the sensitivity reach of such an upgrade will be given, as well as the necessary collaboration infrastructure, costs, manpower and a detailed time schedules presented. In the appendix, an overview of the complementary and auxiliary devices and technologies studied will be given.
  
  Finally, the MEG collaboration is confident that the goals outlined of such a MEG upgrade will lead to the new MEG experiment being a further benchmark for future LFV experiments.

%% file: 01_MEG_Status/MEG_Status.tex
\section{Status of the MEG experiment in the framework  of experimental charged Lepton Flavour Violation (cLFV) searches }
%
%

\begin{figure}[hbct]
\begin{center}
\includegraphics[width=10cm]{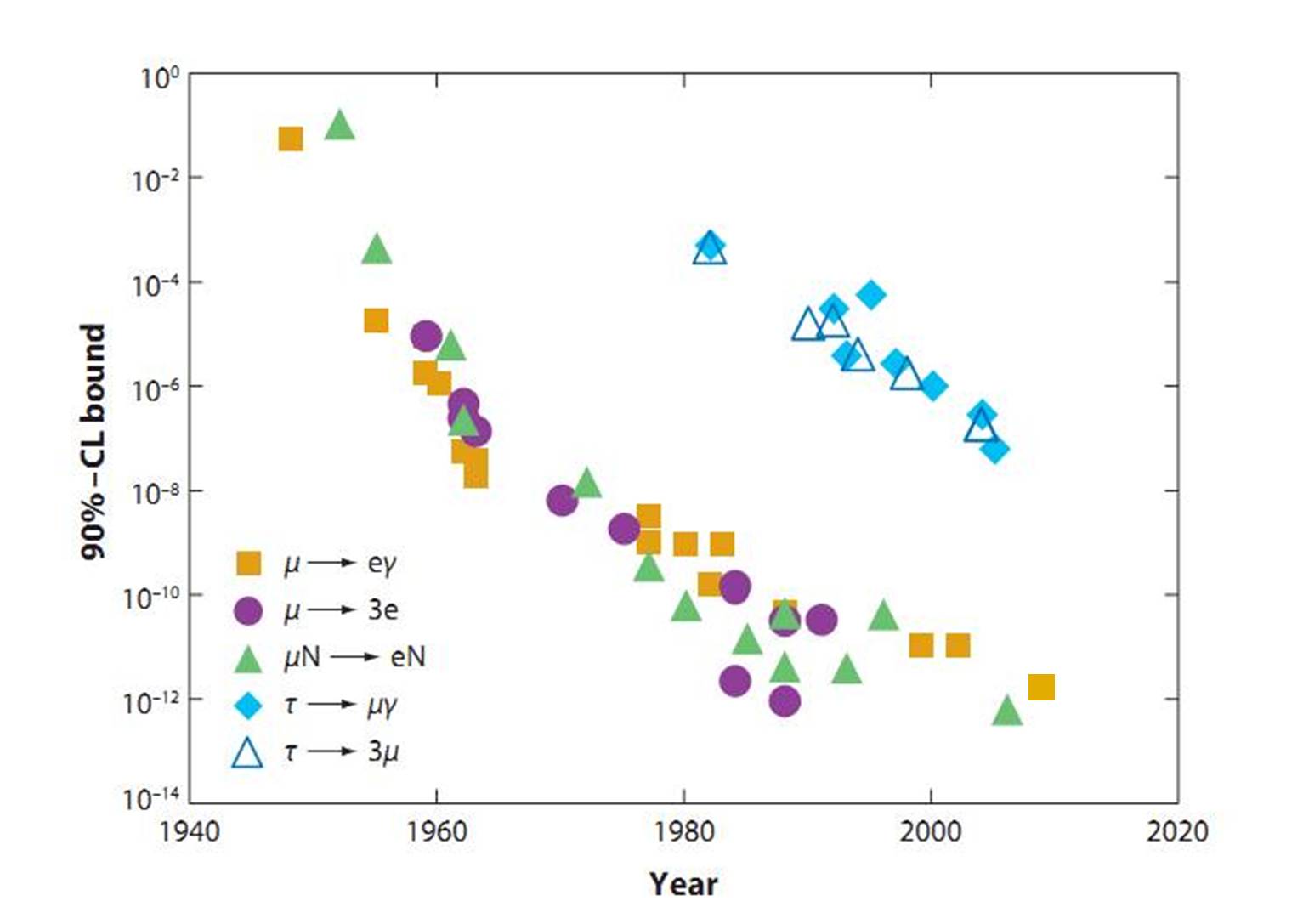}
\caption{\label{fig:LFVlimits} Upper limits on cLFV processes as a function of the year}
\end{center}
\end{figure}

The experimental upper limits established in searching for cLFV processes including the $\meg$ decay 
are shown in Fig.\,\ref{fig:LFVlimits} as a function of the year. 
Historically, the negative results of these experiments led to 
the formulation of the Standard Model (SM) of elementary particles interactions, 
in which lepton flavour conservation is put directly in from the beginning. 
During the past 35 years the experimental sensitivity to the $\meg$ decay has improved by almost three orders of magnitude,
thanks to improvements in detector and beam technologies. 
In particular
‘surface’ muon beams (i.e. beams of muons originating in the decay of $\pi^+$’s that stopped in the pion production target)
with a 
momentum of $\sim29\,\mathrm{MeV}/c$, 
offer the highest muon stop densities obtainable at present, 
allowing for the low-mass experimental targets 
that are required to reach the ultimate resolution in positron momentum and emission angle 
and to suppress the generation of the unwanted $\gamma$-background.

The MEG experiment at the Paul Scherrer Institute (PSI, Zurich, Switzerland) uses the world's most intense 
(higher than $10^8\,\mu$/s) continuous ‘surface’ muon beam 
but, for reasons explained in the following, the stopping intensity is limited to
$3 \times 10^7\,\mu$/s. The signal of  the possible two-body $\meg$  decay at rest 
is distinguished from the background by measuring the photon energy $\egamma$, 
the positron momentum $\ppositron$, their relative angle $\Thetaegamma$  
and timing $\tegamma$ with the best possible resolutions
\footnote{In the following we will indicate the (1$\sigma$) resolution on a variable 
with a $\Delta$ in front of that variable}.

\begin{figure}[hbct]
\begin{center}
\includegraphics[width=15cm]{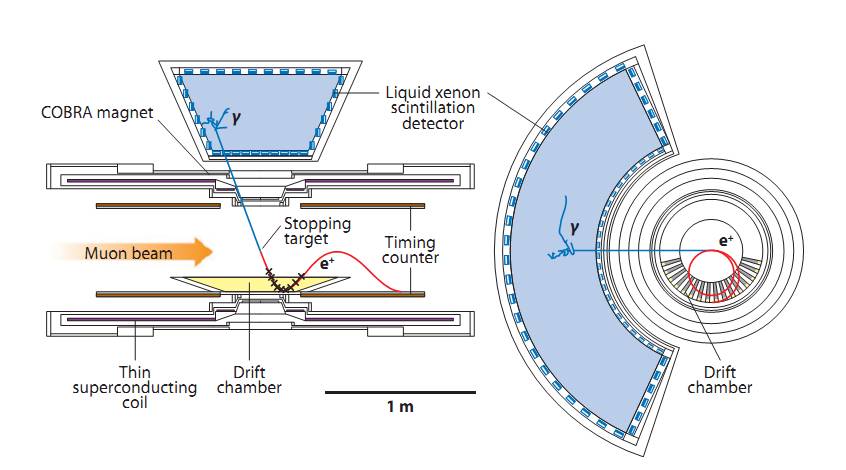}
\caption{\label{fig:MEG} 
A sketch of the MEG experiment
}
\end{center}
\end{figure}

Muons are stopped in a thin (205 $\mu$m) polyethylene target, 
placed at the centre of the experimental set-up which includes a positron spectrometer and a photon detector, 
as sketched in Fig.\,\ref{fig:MEG}. 
The positron spectrometer consists of a set of drift chambers 
and scintillating timing counters located inside a superconducting solenoid 
with a gradient magnetic field along the beam axis, 
ranging from 1.27 Tesla at the centre to 0.49 Tesla at either end. 
The photon detector, located outside of the solenoid, is a homogeneous volume ($900\,\ell$) of liquid xenon (LXe)
viewed by 846 UV-sensitive photomultipliers tubes (PMTs) submerged in the liquid. 
The spectrometer measures the positron momentum vector and timing, 
while the LXe detector is used to reconstruct the $\gamma$-ray energy 
as well as the position and time of its interaction in LXe. 
All the signals are individually digitized by in-house designed waveform digitizers (DRS) \cite{Ritt2010486}.

The background comes either from radiative muon decays $\michelsign \gamma$ (RMD) 
in which the neutrinos carry away little energy or from an accidental coincidence of an energetic positron
from a normal Michel decay with a photon coming from RMD, bremsstrahlung or positron annihilation-in-flight.  

The number of accidental coincidences ($N_\mathrm{acc}$), for given selection criteria, 
depends on the experimental resolutions with which the four relevant quantities 
($\egamma$,  $\ppositron$, $\Thetaegamma$,  $\tegamma$) are measured. 
By integrating the RMD photon and Michel positron spectra over respectively the photon energy
and positron momentum resolution it can be shown that:

\begin{eqnarray}
\label{eq:bacc}
N_\mathrm{acc} \propto R_\mu^2 \times {\Delta\egamma}^2 \times {\Delta\ppositron} \times  \Delta\Thetaegamma^2 \times \Delta\tegamma \times T
\end{eqnarray}

where $R_\mu$ is the rate of stopping muons and T is the measurement time.
The number of expected signal events for a given branching ratio ($\BR$) 
is instead related to the solid angle $\Omega$ subtended by the photon and positron detectors, 
the total acquisition time (T), the efficiencies of these detectors ($\epsilon_\gamma, \epsilon_\mathrm{e}$) 
and the efficiency of the selection criteria ($\epsilon_\mathrm{s}$):
\footnote{A usual selection criteria is to choose 90\% efficient cuts on each of the variables 
($\egamma$,  $\ppositron$, $\Thetaegamma$,  $\tegamma$) 
around the values expected for the signal: 
this criterion defines the selection efficiency to be $\epsilon_\mathrm{s} = (0.9)^4$. 
This kind of analysis in which one counts the number of events within some selection cuts 
and compares the number found with predictions for the background is named \lq\lq box analysis\rq\rq. 
The MEG experiments usually adopt much more refined analyses 
which take into account the different distribution of 
($\egamma$,  $\ppositron$, $\Thetaegamma$,  $\tegamma$) for background and signal by using maximum likelihood methods. }

\begin{eqnarray}
\label{eq:signal}
N_\mathrm{sig} = R_\mu \times T \times \Omega \times \BR \times \epsilon_\gamma \times \epsilon_\mathrm{e} \times \epsilon_\mathrm{s}
\end{eqnarray}

The Single Event Sensitivity (SES) is defined as the $\BR$ for which the experiment would see one event. 
In principle the lowest SES, and therefore the largest possible $R_\mu$, 
is experimentally desirable in order to be sensitive to the lowest possible $\BR$. 
However, due to the quadratic dependence on the muon stop rate, 
the accidental coincidences are largely dominant over the background coming from RMD 
(which is linearly dependent on $R_{\mu}$). 
It is then clear from Eq.\,(\ref{eq:bacc}) and  (\ref{eq:signal}) that, 
for fixed experimental resolutions, the muon stop rate cannot be increased too much 
but it must be chosen in order to keep a reasonable signal over background ratio.

\begin{figure}[hbct]
\begin{center}
\begin{tabular}{cc}
\includegraphics[width=0.46\columnwidth]{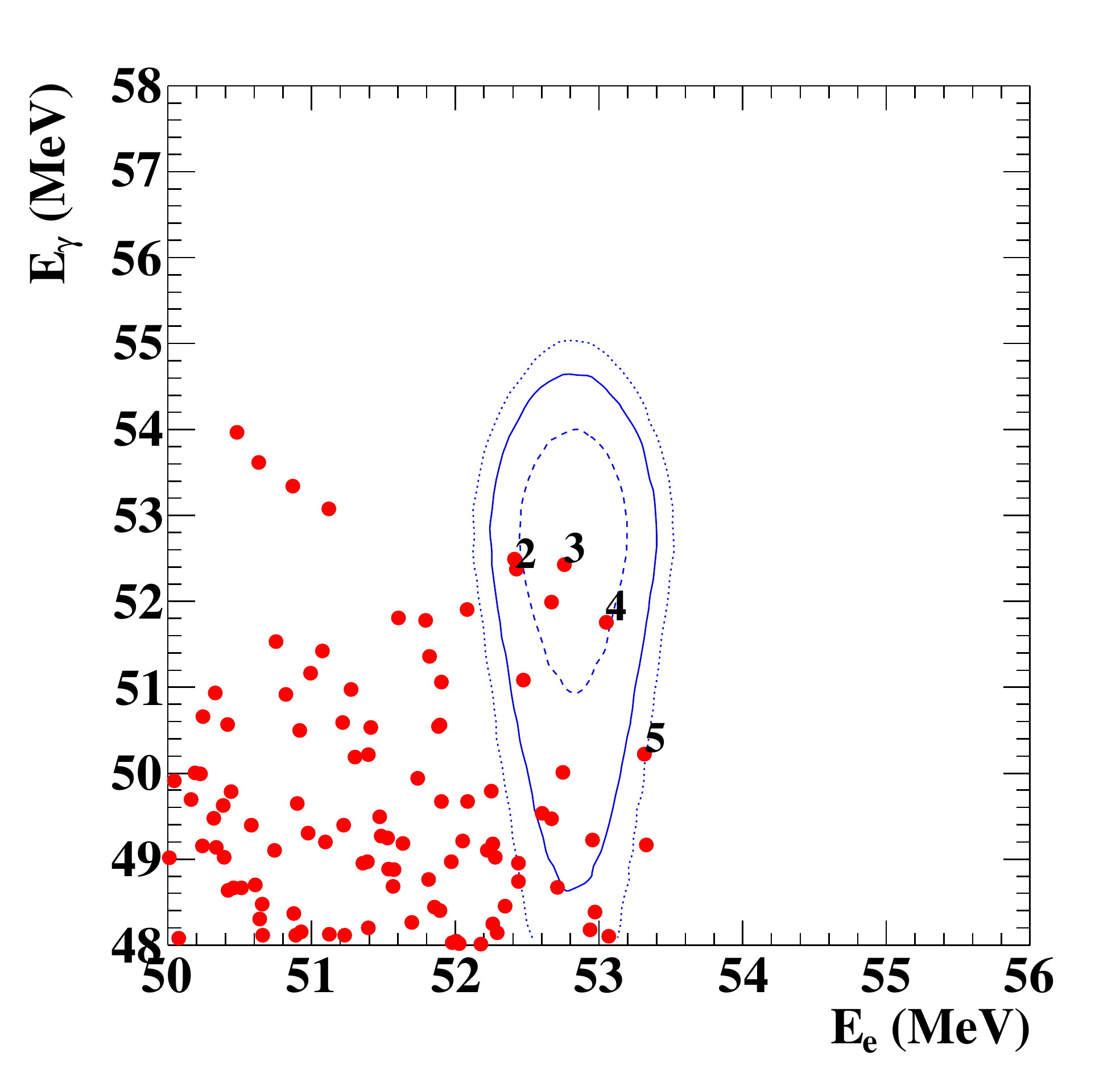} &
\includegraphics[width=0.46\columnwidth]{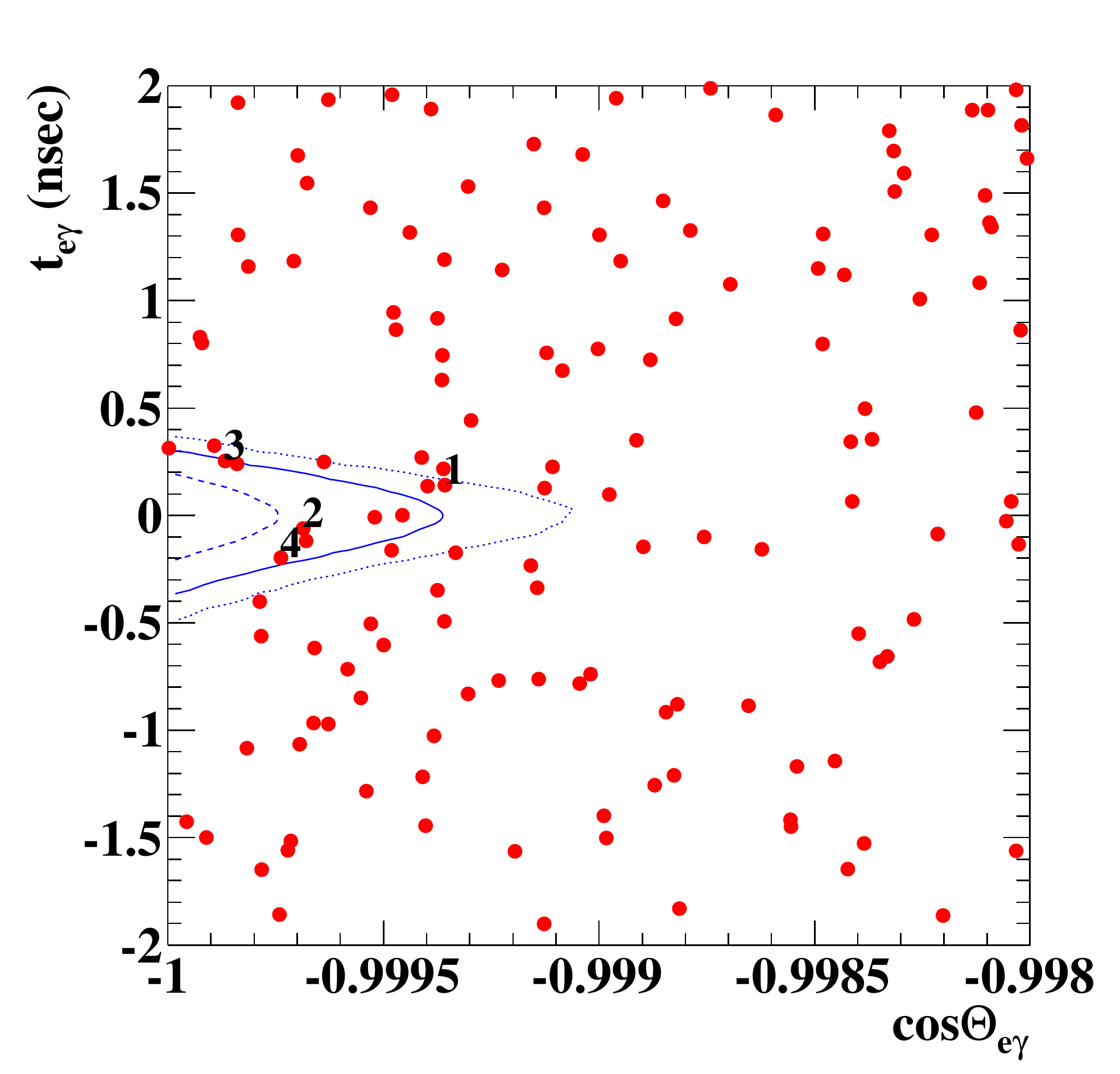}
\end{tabular}
\caption{\label{fig:MEGevents} 
Event distribution in $\epositron$ vs $\egamma$ and $\tegamma$ vs $\cos\Thetaegamma$ planes. 
The curves shown are the 68, 90 and 95\% probability contour levels for signal events. 
Events numbered are the ones having a largest probability of being signal rather than background
}
\end{center}
\end{figure}

In Fig.\,\ref{fig:MEGevents} the event distribution 
in the $\epositron$ vs $\egamma$ and $\tegamma$ vs $\cos\Thetaegamma$ plane 
for data acquired in year 2009 and 2010 are shown in a region where the possible signal is expected. 
The (accidental) background extends into the signal region. 
This implies that the sensitivity of the experiment will not increase linearly with the statistics. 
The upper limit (90\% C.L.) of $2.4\times10^{-12}$ on $\megc$ published by MEG in year 2011 
is the present best experimental result on this dacay. 
Based on this result we estimate to reach a sensitivity $\sim 6 \times 10^{-13}$ 
in the middle of year 2013 and then stop data taking due to the limited statistical significance of a further continuation.

There are three other cLFV channels, in competition with $\meg$, which are currently being actively considered: 
$\conv$ conversion, $\mu \rightarrow \rm{eee}$  and $\tmug$.  

In $\conv$ conversion experiments negative muons are stopped in a thin target and form muonic atoms. 
The conversion of the muon into an electron in the field of the nucleus results 
in the emission of a monochromatic electron of $\simeq100\,\mathrm{MeV}/c$. 
Here the backgrounds to be rejected are totally different from the $\meg$ case. 
The dominant ones are those correlated with the presence of beam
 impurities, mostly pions. In order to reduce these backgrounds the experiments planned at Fermilab (Mu2e)\cite{Car2008}
 and JPARC (COMET)\cite{Bry2007} will use accelerators with bunched proton beams to produce muons. 
 Since muonic atoms have lifetimes of the order of hundreds 
of nanoseconds conversion electrons will be searched for at times different from those of the bunches. 
These experiments can in principle
 reach sensitivities in the BR $\conv$ conversion below $10^{-17}$. 
The Mu2e experiment is now foreseen to start in year 2020 with a first phase 
goal of $7 \times 10^{-17}$ sensitivity. 

In a recent letter of intent presented to PSI\cite{mu3eLoI} 
it is planned to perform an experiment to search for the $\mu \rightarrow
\rm{eee}$ decay down to a sensitivity of $10^{-15}$ 
(three orders of magnitude improvement with respect to the present best
experimental limit) by utilizing  the same beam line of MEG. 
In the second phase a new muon beam line with an increased muon stopping rate in the target ($10^9\,\mu$/s 
instead of the $10^8$ presently possible with the MEG muon beam line) 
could bring the experiment to reach a sensitivity of $10^{-16}$.

$\tmug$ will be explored at future high intensities B factories \cite{Ole2010}\cite{PhysicsAtSuperKEKB}
where sensitivities of the order
 of $10^{-9}$ to the BR for this decay will be possible.

A comparison between the sensitivity  planned for the MEG upgrade and that envisaged for the other above mentioned cLFV processes will be discussed in the next section after a very short introduction on cLFV predictions  in theories beyond the standard model.

%% file: 02_Scientific_Merits/Scientific_Merits.tex
\section{Scientific merits of the MEG upgrade}
\begin{figure}[!hb]
\begin{center}
\includegraphics[width=0.85\columnwidth]{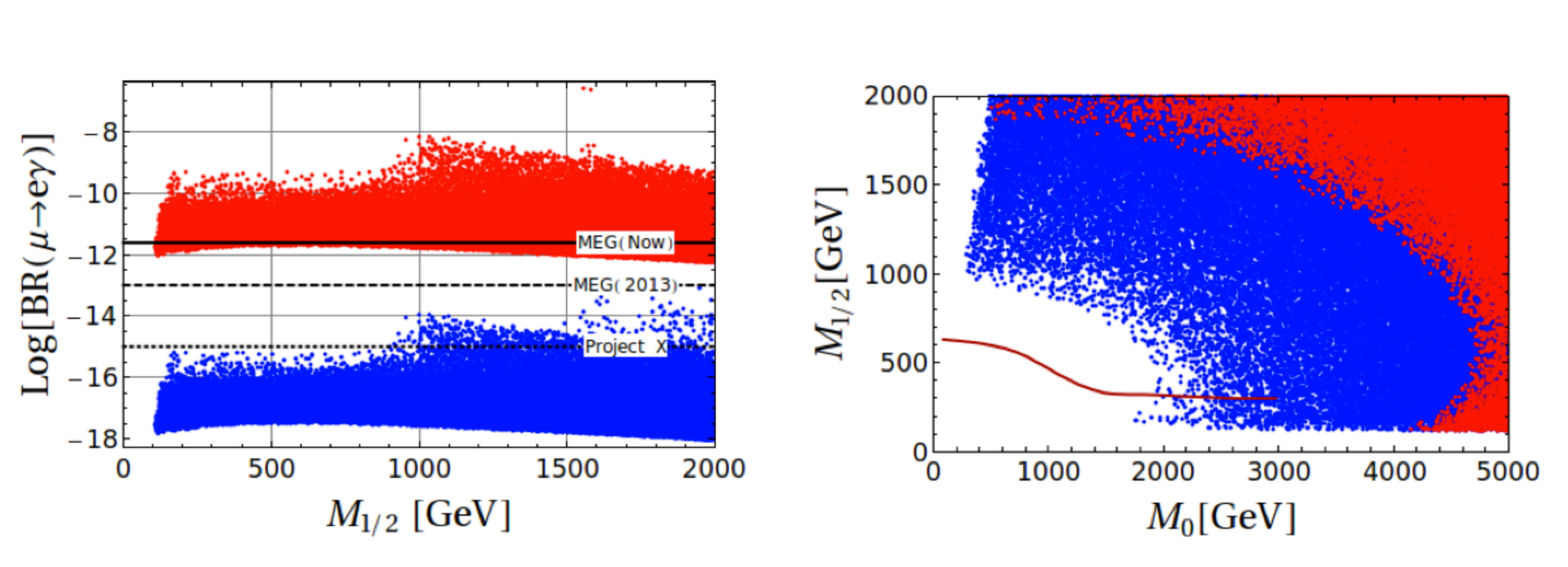}
\caption{\label{fig:Calibbi_LHC}Predictions of the $\meg$ decay rate obtained by scanning the mSUGRA parameter space, for $\tan\beta = 10$ and
$U_{e3} = 0.11$. The red points correspond to PMNS-like mixing, the blue ones to CKM-like. The figure on the right shows the allowed space
in the $m_0-m_{1/2}$ plane which satisfies the current MEG bound. The region below the red line is excluded by the current LHC searches%
~\cite{Calibbi:2012gr}.}
\end{center}
\end{figure}

%
%
The Standard Model (SM) practically forbids Lepton Flavour Violation in the charged lepton sector (cLFV). In fact even introducing massive
neutrinos in the model, in order to account for the experimentally measured phenomenon of neutrino oscillations, the SM predicts
a branching ratio ($\BR$) for $\meg$ below $10^{-50}$, which cannot be experimentally observed. cLFV processes are therefore clean channels 
to look for possible new physics beyond the SM. 
Although no experiment has until now observed any discrepancy from its predictions, the SM model is widely considered to be a low 
energy approximation of a more complete and general theory. 
Several candidates for such a theory, among which Supersymmetric Grand-Unified Theories (SUSY-GUT), predict cLFV with rates close to the present 
$\meg$ experimental upper limit. 
%
%
According to SUSY several low mass states should have been observed. However, these have not 
been discovered yet in energy frontier experiments and this represents one of the aims, 
unfortunately not yet reached, of the LHC program. Experiments looking for cLFV processes 
can provide another approach to clarify these theories since they are very sensitive to
SUSY and particularly to SUSY-GUT models. In this sense, cLFV experiments are complementary 
to LHC in testing these theories.
%
%
(see, {\em e.g.} Figure~\ref{fig:Calibbi_LHC}, where the sensitivity of $\meg$ to the high-energy scale is compared
to direct LHC searches.)

In such theories the  mass matrix of the supersymmetric partners of the  leptons (sleptons) is considered to be diagonal 
in flavour space at the Planck mass scale but radiative corrections 
in the renormalization group evolution from the GUT  to the weak scale generate relevant non diagonal terms, giving rise to cLFV.
So far two terms have been discussed .  The
 first term considered comes from  the fact that at the Planck scale leptons and quarks belong to the same group representation 
(Grand Unification); radiative corrections induce large non diagonal terms in the slepton mass matrix owing to the heavy top quark mass
 \cite{Barb1994}. 
The second term, which is independent from the previous one and adds up to  it in contributing to the slepton mass mixing, therefore giving rise to cLFV,
 is linked to neutrino oscillations. The introduction of see-saw mechanisms to explain the neutrino mass pattern, with the addition
 of large mass right-handed neutrinos, in SUSY models   gives rise to these non diagonal mass terms \cite{hisano-1999},\cite{Cali2006}.

\begin{figure}[hbct]
\begin{center}
\includegraphics[width=0.5\columnwidth]{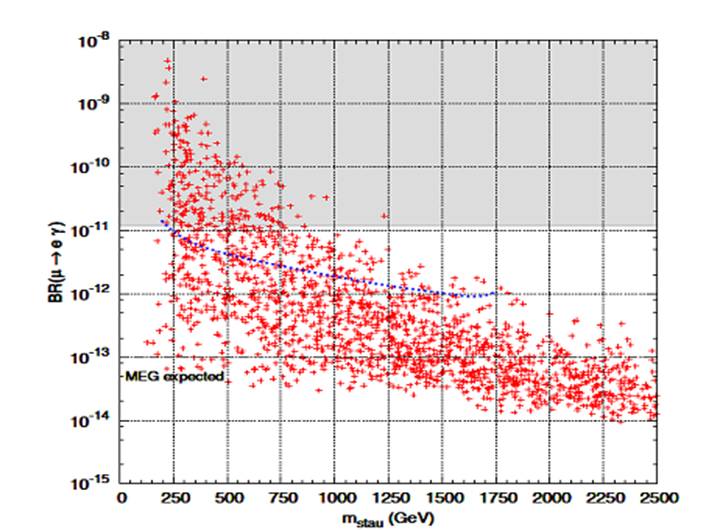}
\caption{\label{fig:SUSY-GUT SO(10)}
SUSY-GUT SO(10) predictions for the $\meg$ decay~\cite{Calibbi:2009wk}.}
\end{center}
\end{figure}

Predictions of the $\meg$  branching ratio depend on the particular SUSY-GUT model taken into consideration and 
on the several other parameters of the theory such as the masses of the (yet unobserved) supersymmetric particles 
and the vacuum expectation values of the Higgs particles.
However the requirement of constructing a stable theory without the need of fine tuning of the parameters implies that the new SUSY particles must have 
masses not much higher than 1 TeV.  In this case 
most models predict that the sum of the two cLFV terms described above gives a predicted $\BR$ for $\meg$ larger than $10^{-13}$.
This is shown for instance in Fig. \ref{fig:SUSY-GUT SO(10)} where the predicted $\meg$ branching ratio is shown as a function of the mass
of the stau particle for values around 1 TeV in a SUSY-GUT model based on SO(10).


\begin{figure}[hbct]
\begin{center}
\includegraphics[width=13cm]{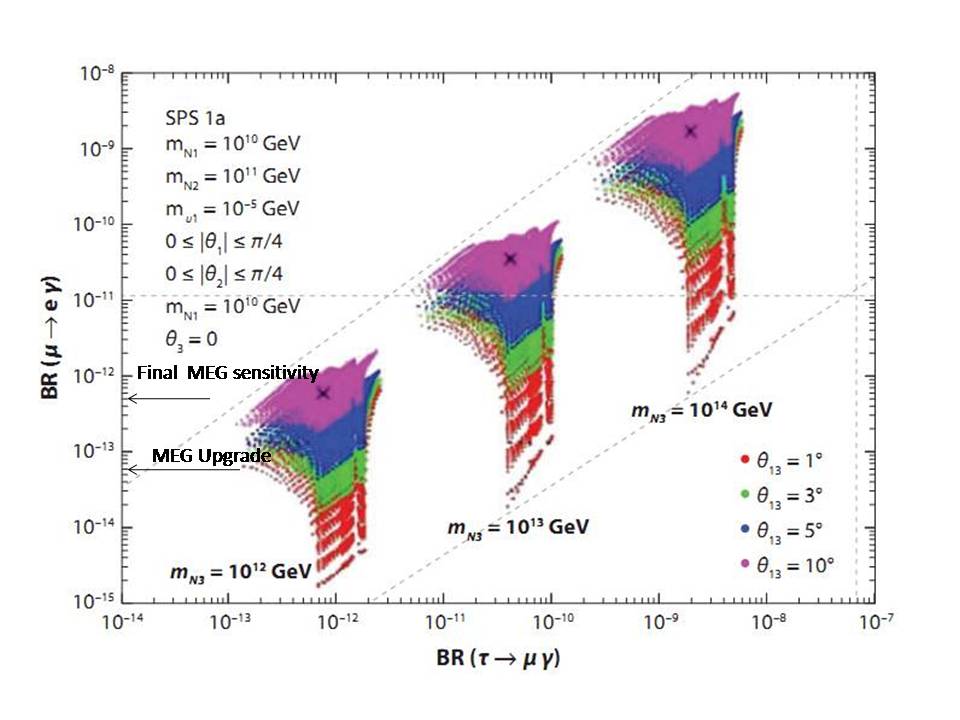}
\caption{\label{fig:Nu Susy} 
SUSY see-saw model predictions for $\meg$  vs $\tau \rightarrow \mu \gamma$: see the text for an explanation
}
\end{center}
\end{figure}

In non-GUT SUSY models cLFV predictions are more dependent on parameters choices but the recent observations \cite{DayaBay},\cite{Reno},\cite{Abe2011},\cite{Abe2012}
of a non vanishing value for $\theta_{13}$ ( the mixing angle between the first and the third neutrino mass eigenstates, 
measured  to be about 8.5 degrees) bring again models to predict large
branching ratios for $\meg$. This can be seen for instance in Fig. \ref{fig:Nu Susy} where we 
show the predictions for $\meg$  vs  for $\tmug$ in a supersymmetric 
see-saw model \cite{Ant2009} as a function of the largest of the masses of
the right-handed neutrinos introduced and of $\theta_{13}$. 
This can be understood by the noting that $\theta_{13}$ represents a mixing between the first and the other generations of neutral leptons,
which is mapped to a large mixing, in a model-dependent way, in the charged sector.

\begin{figure}[hbct]
\begin{center}
\includegraphics[width=10cm]{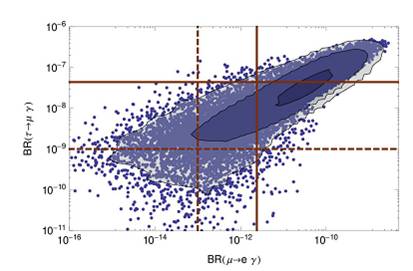}
\caption{\label{fig:Blanck} 
SUSY see-saw model predictions for $\meg$  vs $\tau \rightarrow \mu \gamma$: see the text for an explanation
}
\end{center}
\end{figure}

Another example of the predictions of a non GUT SUSY model\cite{Blank2012}  is shown Fig. \ref{fig:Blanck}, 
where again a plot of $\tmug$ vs $\meg$ branching ratio is given. In this model only one
of the squarks families is considered to have a mass in the TeV scale while the other two can have higher mass values  as suggested
by the most recent LHC measurements. The regions with different colour intensities correspond to points densities differing by one
order of magnitude.

The sensitivity of the $\meg$ decay with respect to $\tmug$ roughly ranges from 500 to $10^4$  in SUSY-GUT models. This is not true in general
for non-GUT SUSY models, as  shown in Fig. \ref{fig:Nu Susy}, from which it seems that the sensitivity in the search for cLFV at future B factories,
in some of these models, cannot compete with the sensitivity of the present proposal. 

\begin{figure}[hbct]
\begin{center}
\includegraphics[width=11cm,clip=TRUE]{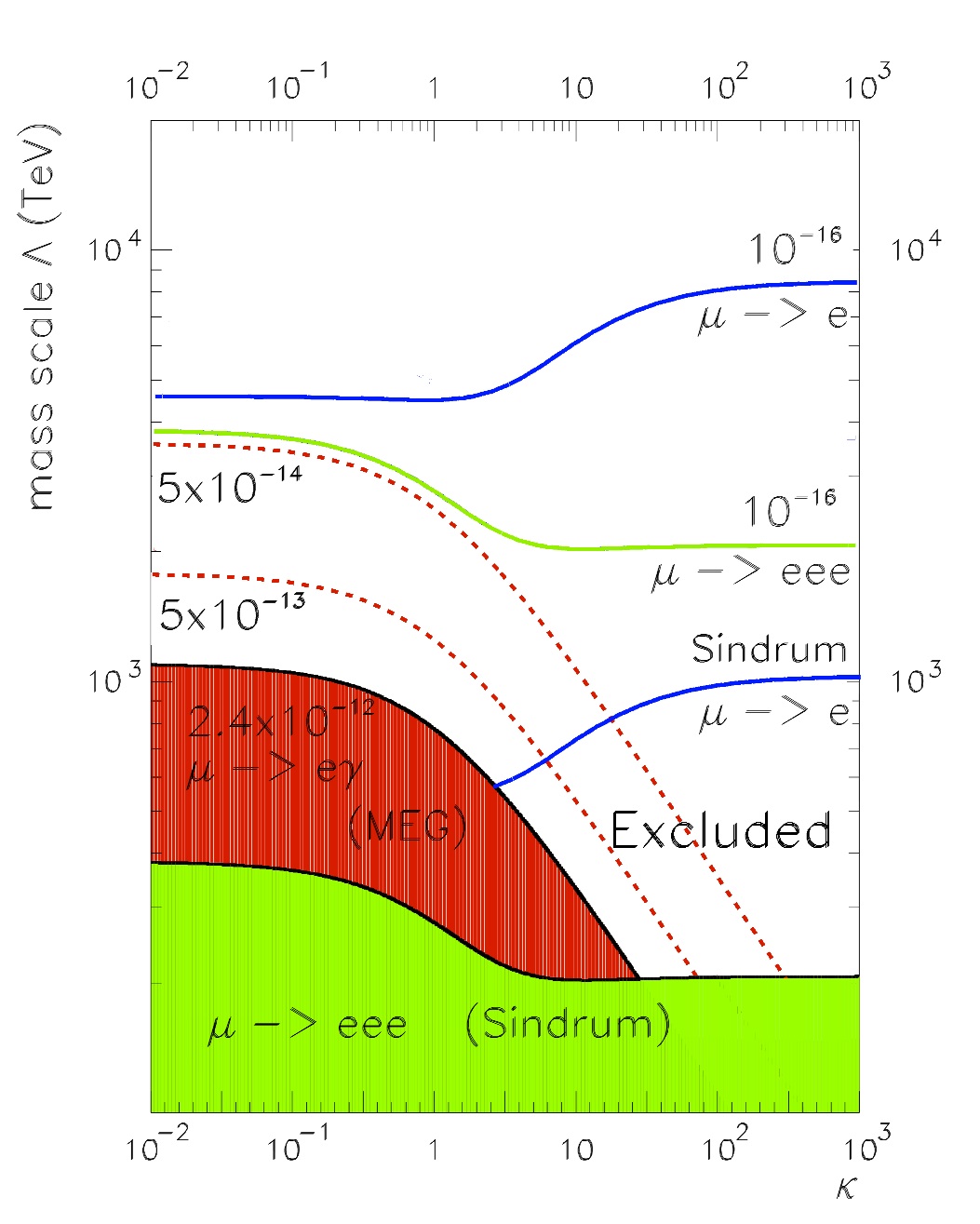}
\caption{\label{fig:klsensi} 
the range of parameters in the ($\Lambda$ , $\kappa$ ) plane that are  
explored by $\meg$, $\mu \rightarrow \rm{3 e}$     and $\conv$  conversion experiments (adapted from~\cite{Car2008}). 
}
\end{center}
\end{figure}

The comparison between $\meg$ versus $\conv$ conversion  and $\mu \rightarrow \rm{3 e}$  is usually done in a model independent way by using the 
effective lagrangian\cite{AdG2010}
\begin{eqnarray}
{\cal L}_{CLFV} =   \frac {m_\mu}{\left(\kappa+1\right)\Lambda^2} \bar{\mu}_R \sigma_{\mu \nu} e_L F^{\mu \nu} + 
                                \frac {\kappa}{\left(\kappa+1\right)\Lambda^2} \bar{\mu}_R \gamma_{\mu} e_L  \bar{f}\gamma^{\mu }f
\end{eqnarray}
which contains two possible terms contributing to cLFV. In the second one $f$ stands for the appropriate fermion field: the electron in
the $\mu \rightarrow \rm{3 e}$ case or the relevant quarks in the $\conv$ conversion case. While $\meg$ 
proceeds only via the first term, corresponding to $\kappa = 0$, the 
$\conv$ conversion process and the $\mu \rightarrow \rm{3 e}$ decay may proceed also through the other one (large $\kappa$ values).
Fig. \ref{fig:klsensi}  shows the range of parameters in the ($\kappa$ , $\Lambda$) 
plane which can be explored for the sensitivities that can be reached by $\conv$  conversion,  $\mu \rightarrow \rm{3 e}$ or $\meg$ experiments. All the relevant SUSY - GUT models 
privilege the $\kappa = 0$ term for which one can see from the figure that MEG is not only competitive with the second phase
of the $\mu \rightarrow \rm{3 e}$ experiment but also, with a much shorter timescale and a far lower budget, with the first phase of the Mu2e project. 

 We finally care to note that the upgraded MEG will represent the best effort to address the search of the  $\meg$ rare decay with the
 available detector technology coupled with the most intense dc muon beam in the world. We know that to achieve any significant improvement in this field several years are needed (one decade was necessary to pass from MEGA to MEG), and therefore we feel committed to push the sensitivity to the ultimate limits.

%% file: 03_Upgrade_Overview/Upgrade_Overview.tex
\section{Upgrade Overview}
\label{sec:overview}
\subsection{Key elements to a MEG upgrade}
\begin{figure}[!h]
\begin{center}
\includegraphics[width=1.0\columnwidth]{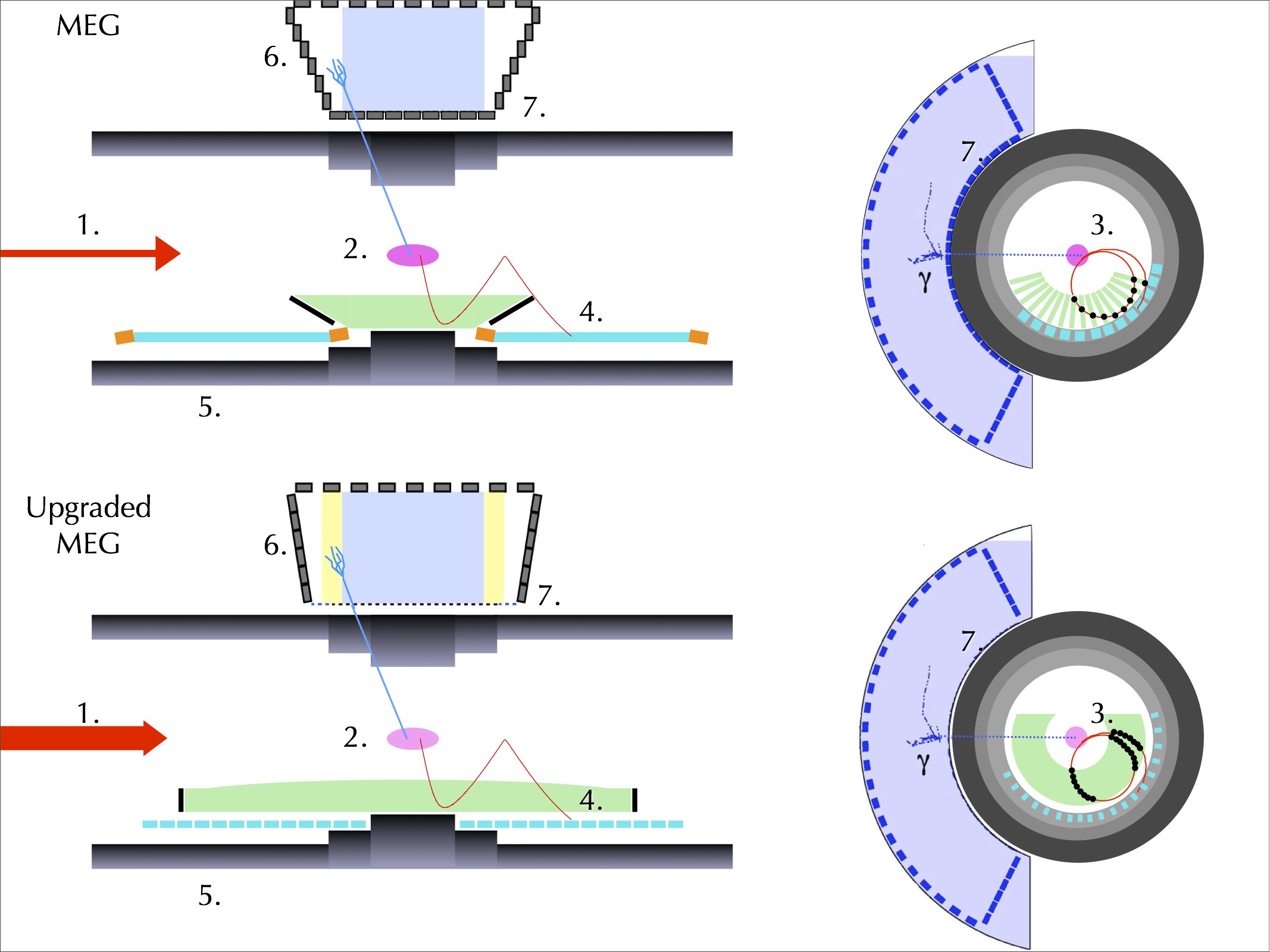}
\caption{\label{fig:upgrade} An overview of the present MEG experiment versus the proposed upgrade. The 
numbers refer to the items listed in the text.}
\end{center}
\end{figure}

The MEG upgrade relies on the following improvements compared with the present MEG experiment, shown 
schematically in Figure~\ref{fig:upgrade} and discussed below:
\begin{enumerate}
\item Increasing the number of stopping muons on target;
\item Reducing the target thickness to minimize the material traversed by photons and positrons on their trajectories
towards the detector;
\item Replacing the positron tracker,  reducing its radiation length and improving its granularity and
resolutions;
\item Improving the positron tracking and timing integration, by measuring the $e^+$ trajectory  to the TC interface;
\item Improving the timing counter granularity for better timing and reconstruction;
\item Extending the $\gamma-$ray detector acceptance;
\item Improving the $\gamma-$ray energy, position and timing resolution for shallow events;
\item Integrating splitter, trigger and DAQ while maintaining a high bandwidth.
\end{enumerate}
\subsection{Discussion}
A major improvement in the sensitivity to the $\meg$ decay requires a higher muon stopping rate and 
improved detectors efficiencies, to achieve a better single event sensitivity. 
This requires improvements to the experimental resolutions to proceed in parallel, to keep the accidental 
background low.
The MEG measured resolutions and efficiencies are compared in Tab. \ref{tab:resolutions} with the original
MEG proposal.
\begin{table}[hbct]
\begin{center}
\begin{tabular}{lrr}
\\{\bf Variable} \hspace{2cm} & Foreseen& \hspace{1cm} Obtained  \\[1mm] 
\hline
\hline\\[-3mm]
$\Delta\egamma$ (\%)          & 1.2                     & 1.7  \\
$\Delta t_\gamma$ (psec)      & 43                      & 67     \\
$\gamma$ position (mm)        & 4(u,v),6(w)             & 5(u,v),6(w)\\
$\gamma$ efficiency (\%)      & $>40$                   & 63 \\
$\Delta\ppositron$ (KeV)      & $ 200$                  & 306   \\
$\rm{e}^+$ angle (mrad)       & 5($\phie$),5($\thetae$) & 8.7($\phie$),9.4($\thetae$)  \\
$\Delta t_{\rm{e}^+ }$ (psec) & 50                      & 107 \\
$\rm{e}^+$  efficiency (\%)   & 90                      & 40 \\
$\Delta\tegamma$ (ps)         & 65                      & 122   \\
\hline \hline
\end{tabular}
\end{center}
\caption{\label{tab:resolutions}Foreseen and measured resolutions for the MEG detector. 
In this Table $\thetae$ is the angle w.r.t. the $z$-axis (beam direction) and $\phie$ is the angle in the 
transverse plane; $u$,$v$ and $w$ are local Cartesian coordinates that refer the photon position 
reconstruction to the liquid xenon detector ($w$ represents the reconstructed photon conversion depth).
}
\end{table}

While the photon detector and the timing counter almost met their requirements, 
the resolutions of the positron spectrometer are significantly worse than the design values,
with consequences also on the relative e-$\gamma$ timing.
$\tegamma$ in fact contains the length of the positron track from the target to the timing counter, 
as measured by the positron tracker.

The photon detector showed somewhat degraded reconstruction capabilities for photons converting at the
edge of its acceptance. Close to the entrance face the size of the 
2'' PMTs introduces a strong non-uniformity, while close to the lateral faces the 
PMTs introduce shadows in the acceptance.
As explained in section~\ref{sec:photoncalorimeter} 
a different solution is now envisaged for the front and lateral
faces, to recover resolutions and efficiencies.

Furthermore there is also room for improving the tracker efficiency. 
The main part of the MEG tracking inefficiency is 
mainly due to the DC front-end electronic boards and mechanical support which intercept 
a large fraction of positrons on their path to the timing counters. 
The use of segmented cathode foils (Vernier pads) to reconstruct the $z-$coordinate was partially 
limited by
the low amplitude of the induced signals on the cathodes, making the $z-$measurement more sensitive
to the noise.
The chamber operation presented some instabilities: their use in a high radiation environment led to 
ageing related problems, with discharges preventing their usage. This  implied  the impossibility of operating some of the 
chamber planes during part of the MEG runs.

We propose to build a new tracking chamber, designed to overcome the previous limitations,
with improved efficiency, momentum and angular resolutions and capable of steady operation at 
high rates.  
The planned resolutions for the proposed tracker, together with a thinner stopping target (reported  
in section \ref{sec:positrontracker}) lead to a substantial improvement in the determination 
of the positron kinematic variables. 

In section \ref{sec:beamandtarget} we show the prospects from upgrading the beam-target combination,
with studies performed under different 
configurations of beam (surface/sub-surface) and target thicknesses, showing that a $140~\mu$m target 
at an angle of 15$^\circ$ with respect to the incoming surface muon beam is 
considered as the baseline solution for a three-years run.

We further propose to upgrade the liquid xenon detector to improve the photon acceptance, and its 
energy and position resolutions, by using a larger number of photo sensors with 
smaller dimensions, as described in section \ref{sec:photoncalorimeter}.

A new pixelated timing counter is described in section \ref{sec:timingcounter}, that 
can withstand the increased positron rate, with improved resolution in the $\tegamma$ measurement, 

The increased number of channels will be handled by  a new mixed trigger/digitizer DAQ board 
(described in section \ref{sec:Trigger and DAQ})  maintaining the high 
bandwidth of the DRS analogue front-end.

\subsection{Auxiliary devices}
The detector described so far represents the baseline design for the MEG upgrade. In parallel other
R\&D studies are being performed within our collaboration to study possible sensitivity improvements by
complementing the measurements of positron and photon kinematic variables.

To further improve the positron angular and momentum resolutions 
an Active TARget (ATAR) and a Silicon Vertex Tracker (SVT) options are under study.
A scintillation fibre-based active target (already independently financed) aims at tagging the incoming
muons and the decay electrons with excellent position and timing resolutions.
The silicon vertex tracker relies on an extremely innovative technology (high-voltage 
monolythic active pixel sensors, HV-MAPS)
that is not yet mature.

A radiative muon decay counter (RDC) could also be used to tag low energy positrons 
associated with the high energy photons 
in the signal region. The existence of a companion positron would exclude a tagged
 photon from our signal candidate sample.

Those auxiliary devices, when beam tested and having demonstrated their usefulness
without having significant impact on the backgroun of the experiment,
could then be accommodated  in the experiment during the running phase. 

These complementary and auxiliary devices  will be described in detail 
in the appendix of the present document together with
a tracker based on a different technology (a Time Projection Chamber). 
The TPC option was considered as a possible positron tracker at the early stages of this proposal, 
but its R\&D turned out to be incompatible with the present proposal time scale.

Needless to say, all the calibration and monitoring techniques developed for the MEG experiment will be carried 
on also in the upgraded detector~\cite{calibration_cw, baldini:2006}. 
Furthermore we plan to implement auxiliary calibration and monitoring techniques 
for the new tracker already at its design stage.

%% file: 04_Beam_Line_and_Target/Beam_Line_and_Target.tex
\section{Beam Line and Target}
%
\label{sec:beamandtarget}
\subsection{The MEG beam line and muon target}
\begin{figure}[hbct]
\centering
\includegraphics[width=1.\linewidth]{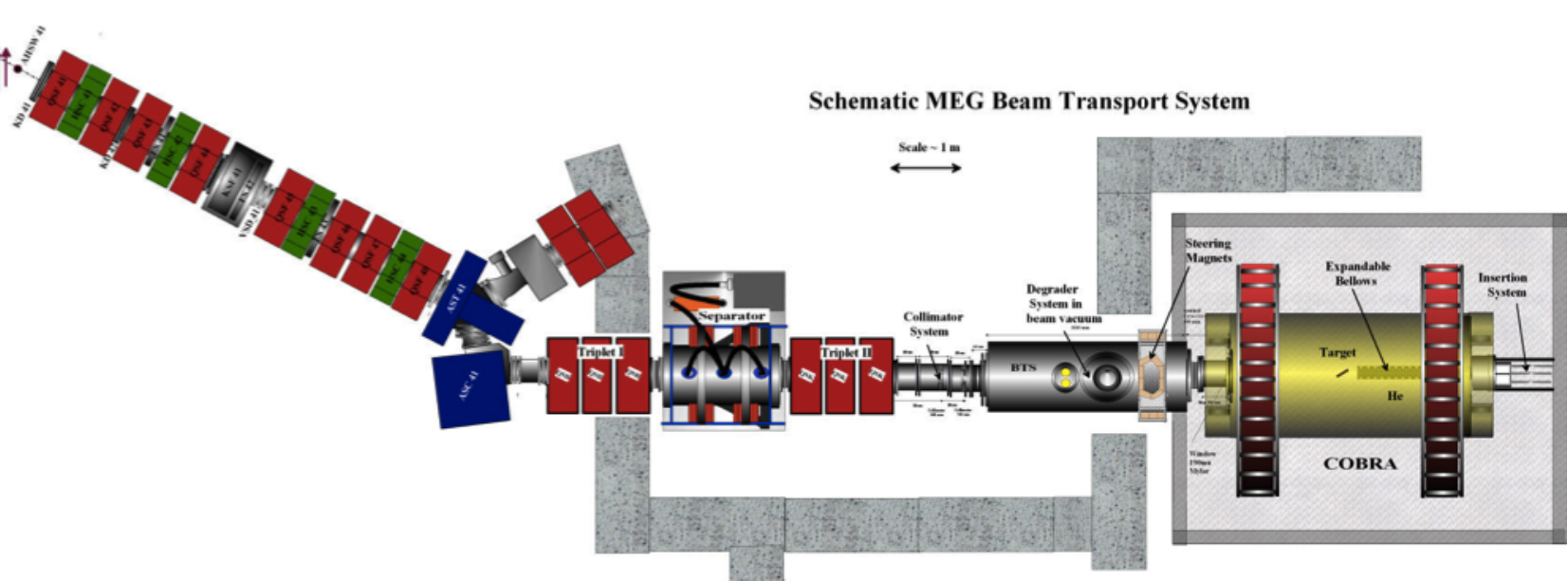}
\caption{(Left-part) shows the $\pi E5$ channel, connecting the production target $E$ to the $\pi E5$ area. The MEG beam line starts from the extraction element Triplet I exiting the wall, followed by a Wien-filter, Triplet II and a collimator system, used to eliminate the beam contamination. The final range adjustment and focusing is performed by a superconducting solenoid BTS, before the muons are stopped in an ultra-thin target placed at the centre of the COBRA positron spectrometer.}
\label{fig:beamline}
\end{figure}

A schematic of the MEG beam line and the $\pi E5$ channel is show in Fig.~\ref{fig:beamline}. Driven by the world's most intense DC proton machines at the Paul Scherrer Institut's high-intensity proton accelerator complex HIPA, it constitutes the intensity frontier in continuous muon beams around the world (c.f. Table~\ref{tab:beamline}) and as such, is capable of delivering more than $10^8~\mu^+/s$ at $28~{\rm MeV}/c$ to the MEG experiment. The surface muon beam has distinct advantages over a conventional 2-step pion decay-channel.

The main beam characteristics are:
\begin{itemize}
\item High intensity
\item Small beam emittance
\item High-transmission optics for \lq\lq~surface muons~\rq\rq ($28~{\rm MeV}/c$)
\item Small, adjustable momentum-byte (less than $10\%$ FWHM), implying a \lq\lq~monochromatic\rq\rq~beam with a high muon stopping density
\item Use of an ultra-thin, slanted muon target and minimization of material along the beam line (vacuum and He-environments) to reduce multiple scattering for both muons and Michel decay positrons as well as to reduce the probability of background photon production, such as annihilation-in-flight or Bremsstrahlung
\item Minimization and separation of beam-related background, such as beam $e^+$ from $\pi^0$-decay in the production target or decay particles along the beam line
\end{itemize}

\begin{table}
{\small
\hfill{}
\caption{\label{tab:beamline}Shows the currently running muon beam facilities around the world that are used for particle 
              physics experiments and materials science $\mu$SR investigations. Also shown are the planned   
              next-generation facilities designed for cLFV experiments, together with an estimate of their 
              starting date. The PSI HiMB solution is currently only under study and is included purely for   
              completeness.}
\begin{center}
\resizebox{0.7\textwidth}{!}{
\begin{tabular}{llll}
\hline
\hline
\textbf{Laboratory}/  & \textbf{Energy}/ & \textbf{Present Surface} &  \textbf{Future estimated}  \\
\textbf{Beam line}    & \textbf{Power} &  \textbf{$\mu^+$ rate (Hz)} & \textbf{$\mu^+ /\mu^-$ rate (Hz)} \\
\hline
\textbf{PSI (CH)} &  (590 MeV, 1.3 MW, DC) &  & \\
LEMS & " & $4\cdot10^8 $ & \\
$\pi E5$ & " & $1.6\cdot10^8 $ & \\ 
HiMB & (590 MeV, 1 MW, DC)  &  & $4\cdot10^{10} (\mu^+)$\\
\hline
\textbf{J-PARC (JP)} &  (3 GeV, 1 MW, Pulsed) &  & \\
                                    &  currently 210 KW           &  & \\
MUSE D-line          &              "                             &$3\cdot10^7$ & \\                                   
MUSE U-line          &              "                             & & $4\cdot10^8 (\mu^+)$  (2012)\\                                                           
COMET                      & (8 GeV, 56 kW, Pulsed) & & $10^{11} (\mu^-)$   (2019/20)\\
PRIME/PRISM          & (8 GeV, 300 kW, Pulsed) & & $10^{11-12} (\mu^-)$   ($> 2020$)\\
\hline
\textbf{FNAL (USA)} & & & \\
Mu2e  &  (8 GeV, 25 kW, Pulsed) &  & $5\cdot10^{10} (\mu^-)$  (2019/20) \\
Project X Mu2e  &  (3 GeV, 750 kW, Pulsed) &  & $2\cdot10^{12} (\mu^-)$  ($>2022$) \\
\hline
\textbf{TRIUMF (CA)} & (500 MeV, 75 kW, DC)  & & \\
M20 & " & $2\cdot10^{6}$ &  \\
\hline
\textbf{KEK (JP)} & (500 MeV, 2.5 kW, Pulsed)  & & \\
Dai Omega & " & $4\cdot10^5$ & \\
\hline
\textbf{RAL -ISIS (UK)} & (800 MeV, 160 kW, Pulsed)  & & \\
RIKEN-RAL & & $1.5\cdot10^6$ & \\
\hline
\textbf{RCNP Osaka Univ. (JP)} & (400 MeV, 400 W, Pulsed)  & & \\
MUSIC & currently max 4W & & $10^8 (\mu^+)$  (2012)\\
              & & & means $> 10^{11}$ per MW\\
\hline
\textbf{DUBNA (RU)} & (660 MeV, 1.65 kW, Pulsed)  & & \\
Phasatron Ch:I-III & & $3\cdot10^4$ & \\              
\hline
\hline
\end{tabular}
}
\hfill{}
\end{center}
}
\end{table}

\begin{figure}[hbct]
\centering
\includegraphics[width=0.6\linewidth]{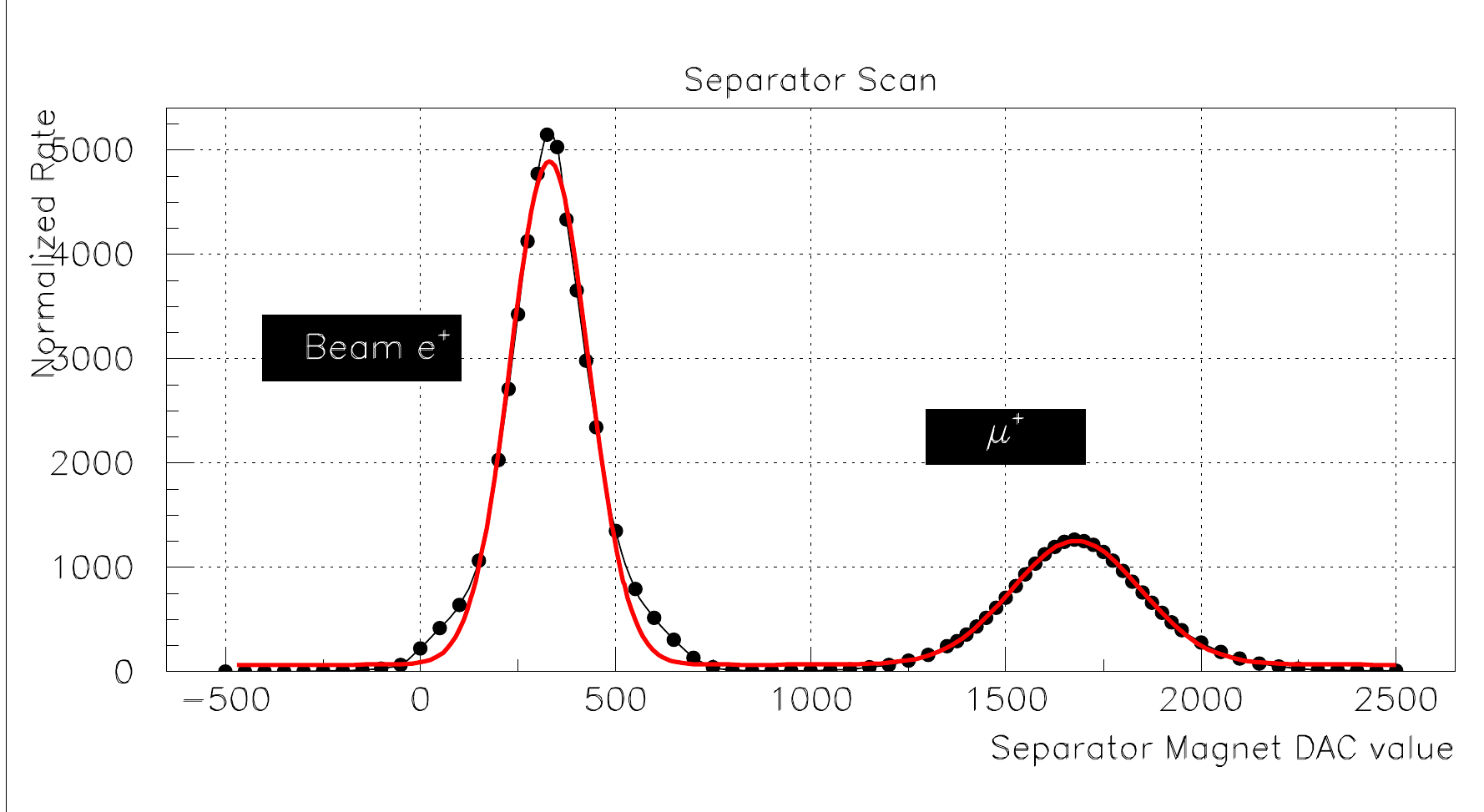}
\caption{Separator scan plot, measured post collimator system. The black dots represent on-axis
intensity measurements at a given separator magnetic field value during the scan, for a fixed electric field
value of -195 KV. The red curve represents a double Gaussian fit to the data points, with a constant
background. A separation of 8.1 muon beam $\sigma$s is found, corresponding to more than 12 cm separation of
beam-spot centres at the collimator. The raw muon peak, measured at low threshold, also contains a contribution from Michel $e^{+}$ due to decays in the detector, these can be eliminated by also measuring at high threshold}
\label{fig:sepmu_pos}
\end{figure}

The beam line layout (Fig.~\ref{fig:beamline}) consists of the $\pi E5$ channel, a high-acceptance $165 ^{\circ}$ backward directed quadrupole and sextupole channel connecting the main target $E$ station to the $\pi E5$ area. Coupled to this is the MEG beam line starting with the set of extraction quadrupole magnets, Triplet I, exiting the shielding wall and allowing for an optimal high transmission through the Wien-filter ($E\times B$ field separator). This together with Triplet II and a collimator system placed after the second triplet, separates the muons from the factor eight-times higher beam positron contamination coming from the target. A separation quality between muons and beam positrons of 8.1 $\sigma_{\mu}$ is achieved, corresponding to a 12 cm physical separation at the collimator system as shown in Fig.~\ref{fig:sepmu_pos}. This allows an almost pure muon beam to propagate to the superconducting transport solenoid BTS, which has a degrader system placed at its intermediate focus to minimize the multiple scattering contribution to the beam and to adjust the muon range for a maximum stop-density at the centre of muon target, placed inside a helium atmosphere, at the centre of the positron spectrometer. 

The 205 $\mu m$ thick muon stopping target, an elliptical shaped sheet of polyethylene foil, sandwiched between a Rohacell foam-structured frame which maintains it in a vertical plane, has a major axis length of 200.5 mm and minor axis length of 79.8 mm and is placed at an angle of $20.5^{\circ}$ to the beam axis. This allows, together with the $300~\mu m$ thick degrader, $190~\mu m$ vacuum window and the $1475~mm$ of helium gas, a maximal stopping efficiency of about $82\%$. A series of 10 mm diameter holes are also punched into the foil to allow an independent check of the planarity and target plane position from reconstructed tracks.

\subsection{Beam Line and Target Upgrade Potential}

There also exists the potential of enhancing the sensitivity of the experiment by possible improvements to the beam line and target system. As already explained in section I, the optimal signal sensitivity for the current experiment is achieved for a muon stopping rate of $3\cdot10^7$ Hz, approximately a factor of 3 lower than the maximal rate achievable. However, an increased muon rate must be accompanied by increased detector efficiencies, while at the same time minimizing accidental backgrounds by improving the experimental resolutions. 

Several scenarios combining a higher beam intensity with various target thicknesses, ranging from a 250 $\mu m$ thick active scintillating fibre target  down to a thin 100 $\mu m$ passive target version have been investigated. The former version not only has the potential of improving the positron momentum resolution but also improving the angular resolutions associated with the positron, mainly due to overcoming the multiple scattering restriction of the vertex determination on the target plane when back-projecting the track. A detailed status of the Active-Target study ATAR is given later. Also, the use of a lower momentum \lq\lq sub-surface\rq\rq muon beam in combination with the various target types was investigated. For a thicker target, this would allow the possibility of tuning the muon straggling distribution to minimizing the multiple scattering of the out-going positrons, while in the case of thinner targets a higher stopping quality (target stops versus ranging-out particles) could be achieved.
\subsection{Use of a Sub-surface Muon Beam}
Surface muons~\cite{Pifer:1976ia}, or low-energy muons derived from stopped pion decay close to the surface of the production target, have a unique momentum of 29.79 MeV/c due to the 2-body decay kinematics. However, due to the finite momentum-byte of the channel and the requirement to maximize the muon intensity, the central momentum is chosen to be approximately 28 MeV/c, thus selecting muons originating from a target surface layer of thickness equivalent to a few hundred microns. The range of such muons is of the order of 120 mg cm$^{-2}$ and so must be degraded in order to stop them in an ultra-thin target.

The target thickness is governed by the requirements to have a maximal stopping density, i.e. maximal stopping rate in the thinnest possible target, while minimizing the multiple scattering and photon background production possibilities of the out-going positrons on the one hand and on the other, minimizing the amount of material seen by the photon travelling in the opposite direction. Hence these requirements are directly related to the amount of range-straggling of the muon beam. The total range-straggling is comprised of two components, one from energy-loss straggling of the intervening material, mainly the target, degrader, vacuum window and He-gas and the other component being due to the momentum-byte of the beam. At these muon momenta ($\sim$ 30 MeV/c) the energy-loss component amounts to a constant contribution of about $9\%$ of the range~\cite{Pifer:1976ia}, while the range varies strongly with momentum P and is proportional to $a\cdot P^{3.5}$, where the factor \lq\lq a\rq\rq\ is a material constant. Hence the total range straggling is given by:
\begin{equation}
\Delta R_{TOT} = a \cdot \sqrt{\big( (0.09)^2+(3.5 \Delta P/P)^2 \big)}\cdot P^{3.5}
\label{eq:range}
\end{equation}

\begin{figure}[h!]
\centering
\includegraphics[width=0.6\linewidth]{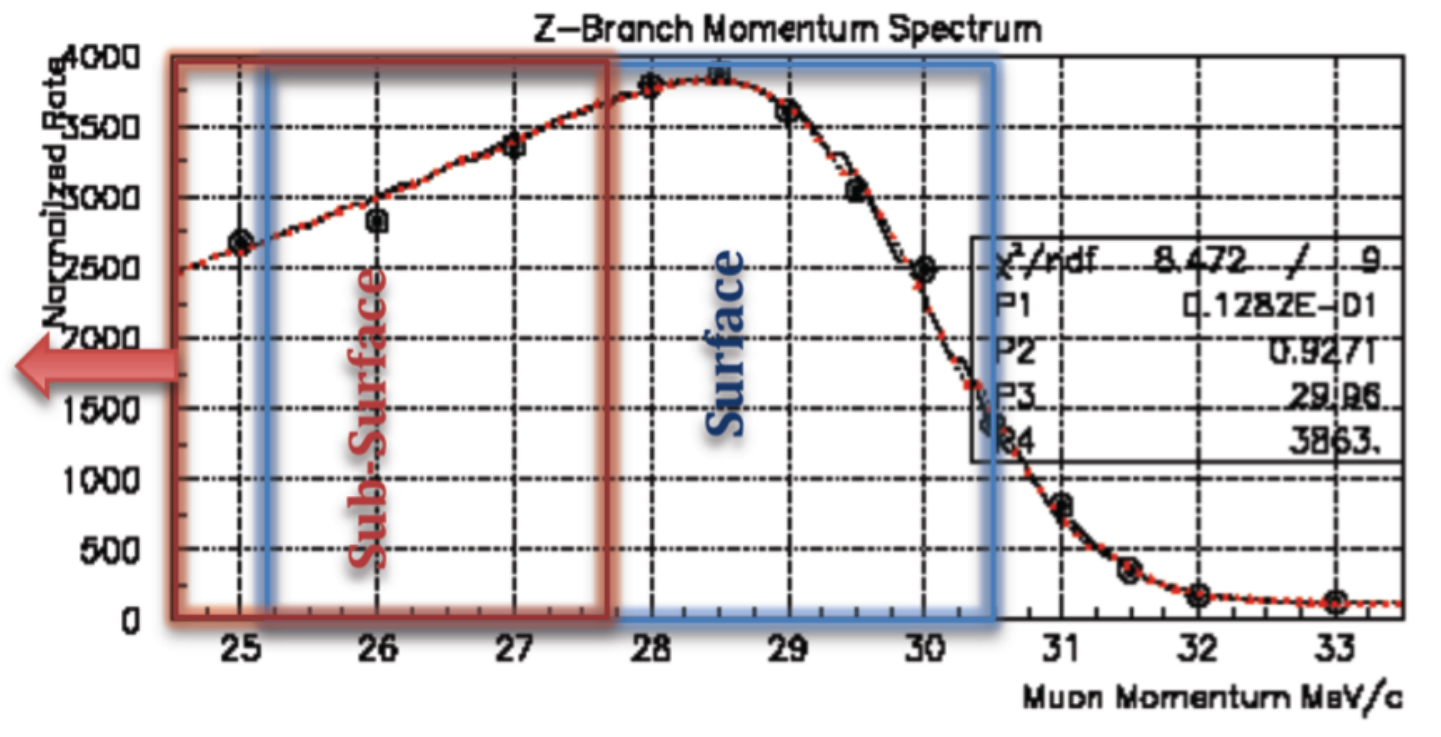}
\caption{Shows the measured muon momentum spectrum from the $\pi E5$ channel with the momentum slits fully \lq\lq open\rq\rq. The red curve is a fit to the data of a theoretical $P^{3.5}$-distribution folded with a Gaussian resolution function corresponding to the momentum-byte plus a constant cloud-muon background. The blue and red boxes show the full momentum-byte for surface and sub-surface muons respectively.}
\label{fig:spect}
\end{figure}

From Eq.~\ref{eq:range} it can be seen that the most efficient way of reducing the straggling and hence the target thickness, is by reducing the beam momentum, rather than making the momentum-byte smaller.
By reducing the beam momentum to 25 MeV/c, which constitutes a so-called \lq\lq sub-surface\rq\rq\  muon beam, since the muon acceptance layer in the production target now lies below the surface, as demonstrated by the measured muon momentum spectrum shown in Fig.~\ref{fig:spect}. 

This was obtained during commissioning by tuning the entire beam line to match the central momentum of each point. Here the momentum slits are fully \lq\lq open\rq\rq. The kinematic edge, smeared by the momentum resolution and defined by the momentum-byte can clearly be seen together with the characteristic $P^{3.5}$ empirical behaviour of the spectrum towards lower momenta. The blue and the red (truncated) boxes show the full $\pm 3\sigma$ momentum-byte for surface and sub-surface muons, respectively $\pm 2.7$~MeV/c and $\pm 2.5$~MeV/c. The spectrum fit shows that the muon rate also falls-off with $P^{3.5}$ consistent with a more recent set of measurements made in $\pi$E5 and summarized in Tab.~\ref{tab:slitopening}.

In order to take advantage of either reduced range-straggling and therefore a potentially thinner target, or the ability to optimize the position of the straggling distribution within a thicker target, a careful balance of rate, sensitivity and background reduction are mandatory. The basic scenarios studied for a MEG upgrade strategy are summarized in Tab.~\ref{tab:beamtarscenario} and involved the use of either a surface or sub-surface muon beam with target thicknesses ranging from 100 - 250 $\mu m$. The muon stopping numbers are from Monte-Carlo simulations based on real phase space measurements, while the target stopping rates are scaled from measured muon intensities.

\begin{table}[h!]
\caption{\label{tab:slitopening}Shows a series of beam measurements taken at the intermediate collimator position and at the centre of COBRA (CC), for a sub-surface muon beam of 25 MeV/c. No Degrader was used and the rates at CC are those of muons arriving at the centre, for a proton beam intensity of 2.0 mA.}
\begin{center}
\resizebox{0.7\textwidth}{!}{
\begin{tabular}{ccccccc}
\hline
\hline
\textbf{Slit opening}  & & \textbf{Collimator position} & &  & \textbf{COBRA center} &\\
\hline
 & $R_{\mu}$ (Hz) at 2mA & $\sigma_{x}$ (mm) &  $\sigma_{y}$ (mm) &$R_{\mu}$ (Hz) at 2mA & $\sigma_{x}$ (mm) &  $\sigma_{y}$ (mm)\\
\hline
250/280 & $9\cdot10^7$ & 21.8 & 18.6 & $7\cdot10^7$ & 9.6 & 10.1 \\
115/115 & $3.5\cdot10^7$ & 21.4 & 15.5 & $2.9\cdot10^7$ & 8.9 & 8.8 \\
70/70 & $6.5\cdot10^6$ & 20.4 & 15.8 & $5.8\cdot10^6$ & 8.4 & 8.3 \\
\hline
\hline
\end{tabular}
}
\end{center}
\end{table}

\begin{table}[h!]
\caption{\label{tab:beamtarscenario}Monte-Carlo results for a Surface and Sub-surface muon beam + various target combinations based on a proton beam intensity of 2.3 mA.}
\begin{center}
\resizebox{0.6\textwidth}{!}{
\begin{tabular}{cccccccccccc}
\hline
\hline
\textbf{Beam}  & \textbf{Target}       & \textbf{Target} & \textbf{US}         &  \textbf{Tg}   &\textbf{DS}       &\textbf{Stop Rate 2.3mA}&\textbf{Stopping}&\textbf{Stopping}&\textbf{Measuring}  \\
                          & \textbf{Tickness}   & \textbf{Angle}     &                                 &                                &                          &\textbf{Whole Target}                   &\textbf{Efficiency}         &\textbf{Quality} &\textbf{Time} \\
                          & (\textbf{$\mu$ m}) & (\textbf{deg})   & (\textbf{$\%$})  &   (\textbf{$\%$})    &(\textbf{$\%$})&\textbf{x$10^7$Hz}          &(\textbf{\%})          &\textbf{Factor SQF} &\textbf{yrs} \\
\hline
Surface & 250 & 20.5 & 8.4 & 75.3 & 16.2 & 9.6 & 82.3 & 3.0 & 2.2 \\
Surface & 205 & 20.5 & 7.2 & 65.9 & 26.8 & 8.4 & 71.1 & 1.7 & 2.5 \\
Surface & 180 & 20.5 & 7.3 & 61.6 & 31.0 & 7.8 & 66.5 & 1.4 & 2.7 \\
Surface & 160 & 20.5 & 9.3 & 57.5 & 33.2 & 7.3 & 63.4 & 1.2 & 2.9 \\
Surface & 140 & 20.5 & 13.7 & 53.4 & 32.8 & 6.8 & 62.0 & 1.0 & 3.1 \\
Surface & 100 & 20.5 & 23.6 & 41.8 & 34.5 & 5.3 & 54.8 & 0.6 & 4.0 \\
Surface & 180 & 15.0 & 5.7 & 64.9  & 29.3 & 8.2 & 68.9 & 1.5 & 2.6\\
Surface & 160 & 15.0 & 7.6 & 62.3  & 29.9 & 7.9 & 67.6 & 1.3 & 2.7 \\
Surface & 140 & 15.0 & 7.5 & 59.4  & 33.0 & 7.5 & 64.3 & 1.2 & 2.8 \\
Surface & 120 & 16.0 & 9.7 & 52.8  & 37.4 & 6.7 & 58.6 & 0.9 & 3.1 \\ 
\hline
Sub-Surf & 250 & 20.5 & 5.8 & 78.4 & 15.7 & 8.2 & 83.4 & 3.5 & 2.6 \\
Sub-Surf & 205 & 20.5 & 5.3 & 70.2 & 24.3 & 7.3 & 74.3 & 2.1 & 2.9 \\
Sub-Surf & 140 & 20.5 & 17.3 & 60.7 & 22.0 & 6.3 & 73.4 & 1.4 & 3.3 \\
Sub-Surf & 100 & 20.5 & 32.5 & 47.8 & 19.7 & 5.0 & 70.8 & 1.1 & 4.2 \\  
Sub-Surf & 180 & 15.0 & 4.8 & 69.6 & 25.6 & 7.2 & 73.1 & 1.9 & 2.9 \\
Sub-Surf & 160 & 15.0 & 5.5 & 66.6 & 27.8 & 6.9 & 70.6 & 1.6 & 3.0 \\
Sub-Surf & 140 & 15.0 & 7.2 & 64.8 & 27.8 & 6.7 & 69.6 & 1.4 & 3.1 \\
Sub-Surf & 120 & 16.0 & 9.7 & 59.1 & 31.0 & 6.1 & 65.6 & 1.1 & 3.4 \\  
\hline
\hline
\end{tabular}
}
\end{center}
\end{table}

\begin{figure}[h!]
\centering
\includegraphics[width=0.6\linewidth]{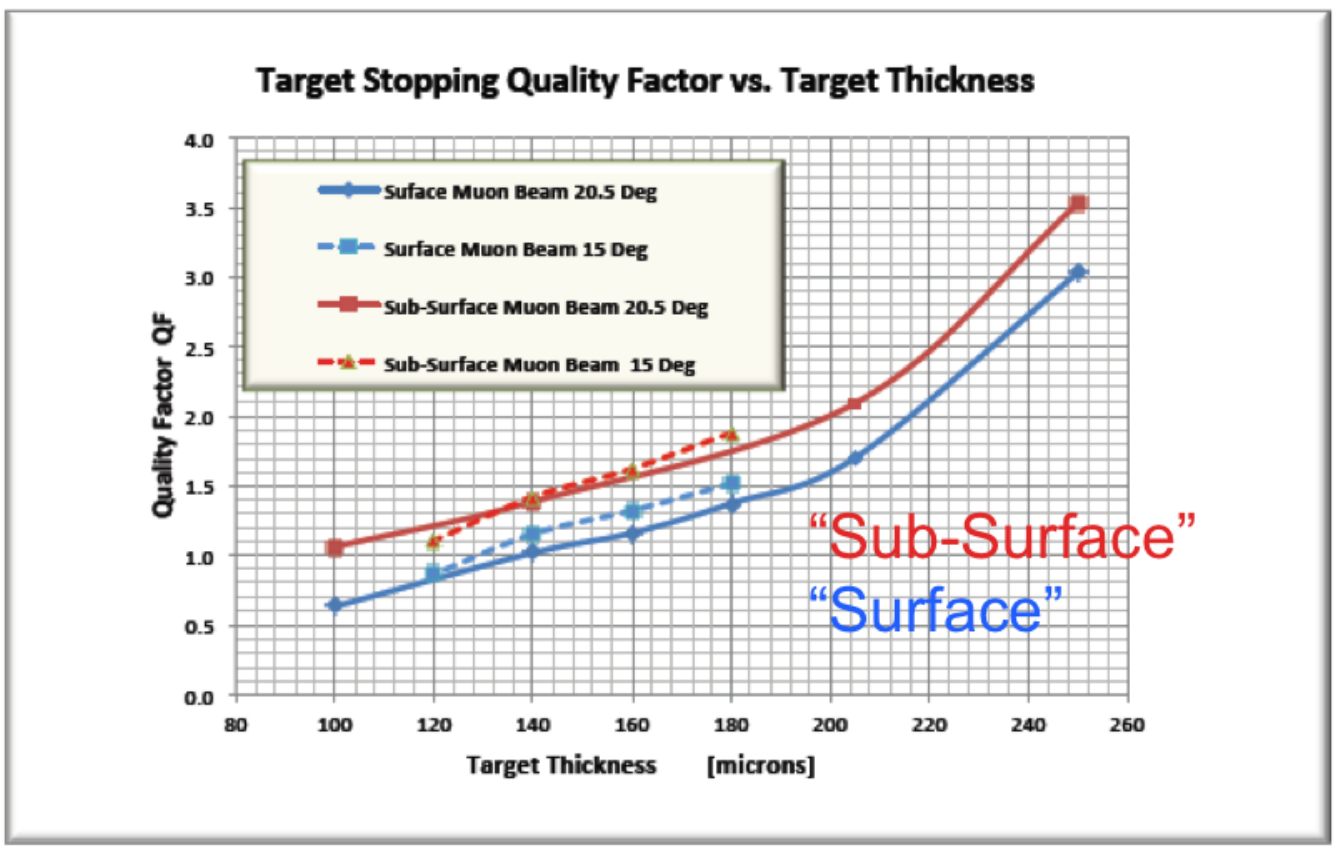}
\caption{Shows the stopping quality, a factor based on the maximization of the muon stops in the target and the minimization of potential background, caused by muons stopping elsewhere, for the case of  a surface and sub-surface beam.}
\label{fig:sqf}
\end{figure}

In summary, the results show that both surface and sub-surface beams yield solutions within a reasonable measuring time span of  3 years. The estimated stopping rate at a proton beam intensity of 2.3 mA is also shown, as is the estimated running time for each scenario, relative to the baseline sensitivity goal solution of 7$\cdot10^{7}$ muons/s and three years of running. The stopping quality (SQF-factor) , as defined in the table, is the ratio of target stops to stops elsewhere. As expected, a better stopping quality, in the case of a sub-surface beam is demonstrated  in Fig.~\ref{fig:sqf}, which shows the relative virtue of each beam-target combination for muons stopping in the target compared to muons stopping elsewhere. As one goes to smaller target angle orientations ($15^\circ$), shown by the dashed lines in the figure, a higher gain is achieved in the surface case compared to the sub-surface case. 

This can be understood since the sub-surface straggling width is closer to being optimal than in the case of the surface beam, such that the effect of going to smaller angles is larger for a surface beam. A further asset of the sub-surface muon beam is that the separation between muons and beam positrons in the Wien filter, as shown in Fig.~\ref{fig:sepmu_pos}, should also be enhanced due to the larger velocity difference. 

 In order to select a beam/target solution that will provide the optimal sensitivity many factors must be considered such as stopping-rate, which will influence the statistics, the central momentum and the momentum-byte, which determine the stopping distribution and optimal target size. This in-turn, dictates the amount of material encountered by both the in-coming muon and the out-going positron and photon. Restrictions of minimal multiple scattering and hence better positron tracking resolutions on the one hand and minimal probability for background photon processes such as annihilation-in-flight or Bremsstrahlung on the other, dominate the choice for a baseline solution for the MEG upgrade. 
This basically allows two possible solutions, achievable within a time of 3 years, namely:
\begin{itemize}
\item Surface Muon Beam, $140~\mu m$ thick Target, placed at $15^\circ$
\item Sub-surface Muon Beam, $160~\mu m$ thick Target, placed at $15^\circ$
\end{itemize}

Until the full background simulations of the two scenarios have been compared in detail, the former surface muon beam version with the thinner target is taken as the baseline solution, for the reasons already mentioned.

%% file: 05_Positron_Tracker/Positron_Tracker.tex
\section{Positron detector}
\begin{figure}[!hb]
\begin{center}
\includegraphics[width=0.75\columnwidth]{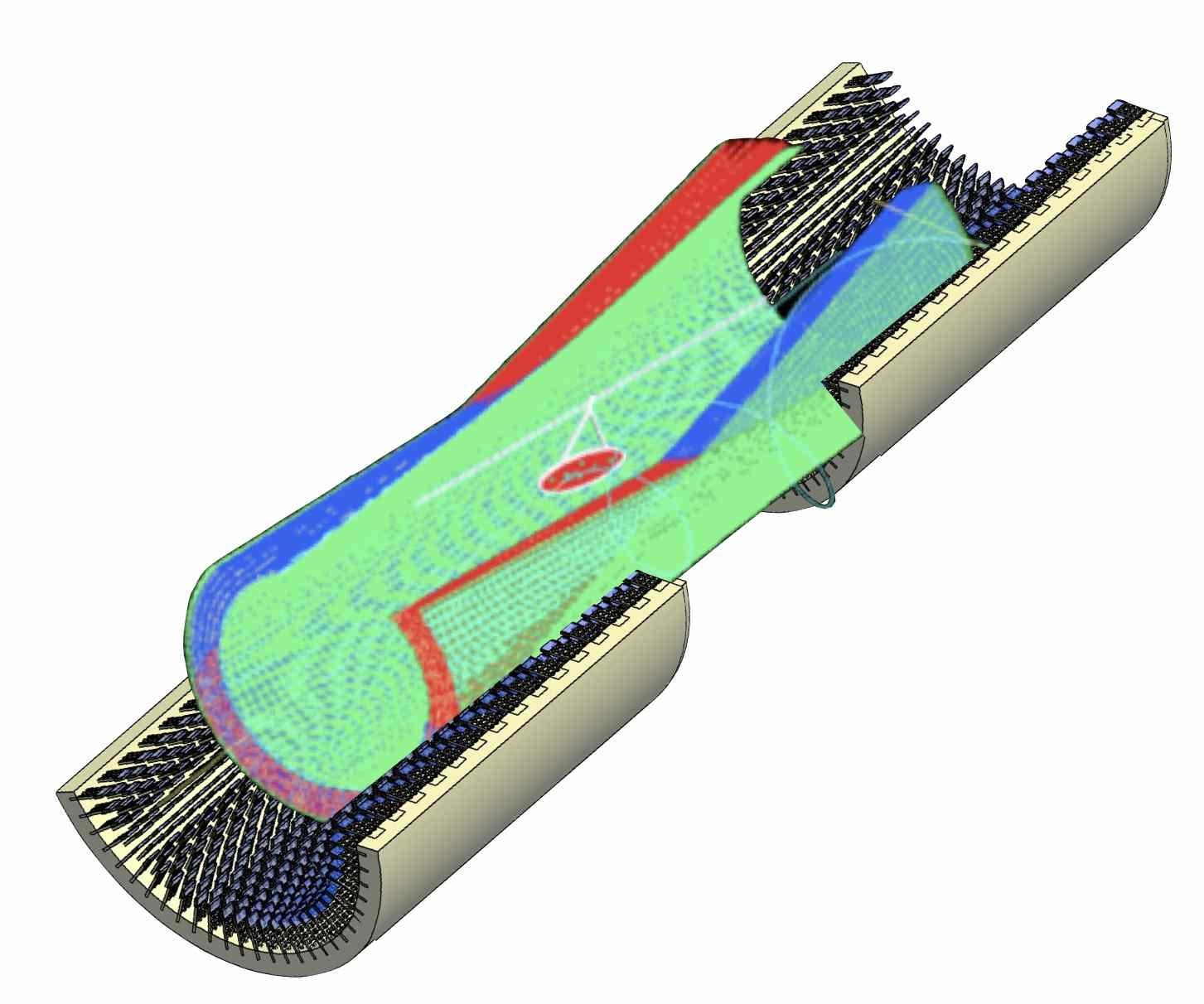}
\caption{\label{fig:pos-det}Schematics design of the positron spectrometer}
\end{center}
\end{figure}

%
\label{sec:positrontracker}
The positron detector, shown schematically in Figure~(\ref{fig:pos-det}), consists of a low mass stereo drift
chamber (DC) followed by a multi-tile scintillation timing counter (TC) for a precise determination of the
particle momentum and production time.

Both detectors are placed inside COBRA, the gradient field magnet specifically designed for the MEG experiment.
As in the MEG experiment the positron tracker is located at a large radius ($r > 18$~cm) so low energy positrons
are swept out of the magnet by the magnetic field without crossing the sensitive volume; positrons with momentum
larger than $\sim 45$~MeV/c, on the other hand, are tracked until they reach the TC tiles, with  minimum presence
of passive material.
\subsection{The positron tracker}
The positron tracker is a unique volume, cylindrical wire drift chamber, with the axis parallel to the muon beam, inspired by
the one used in the KLOE experiment \cite{Adi2002}. 
The external radius of the chamber is constrained by the available room inside COBRA, while its length is dictated by the necessity of
tracking positron trajectories until they hit the TC. This minimizes  
the contribution of the track length measurement to the positron timing resolution
and increases the positron reconstruction efficiency
avoiding any material along the positrons path to the TC.
The requirements are satisfied for a chamber of  $\sim$180 cm length. 

\begin{figure}
\begin{center}
\begin{tabular}{cc}
\includegraphics[width=0.45\columnwidth]{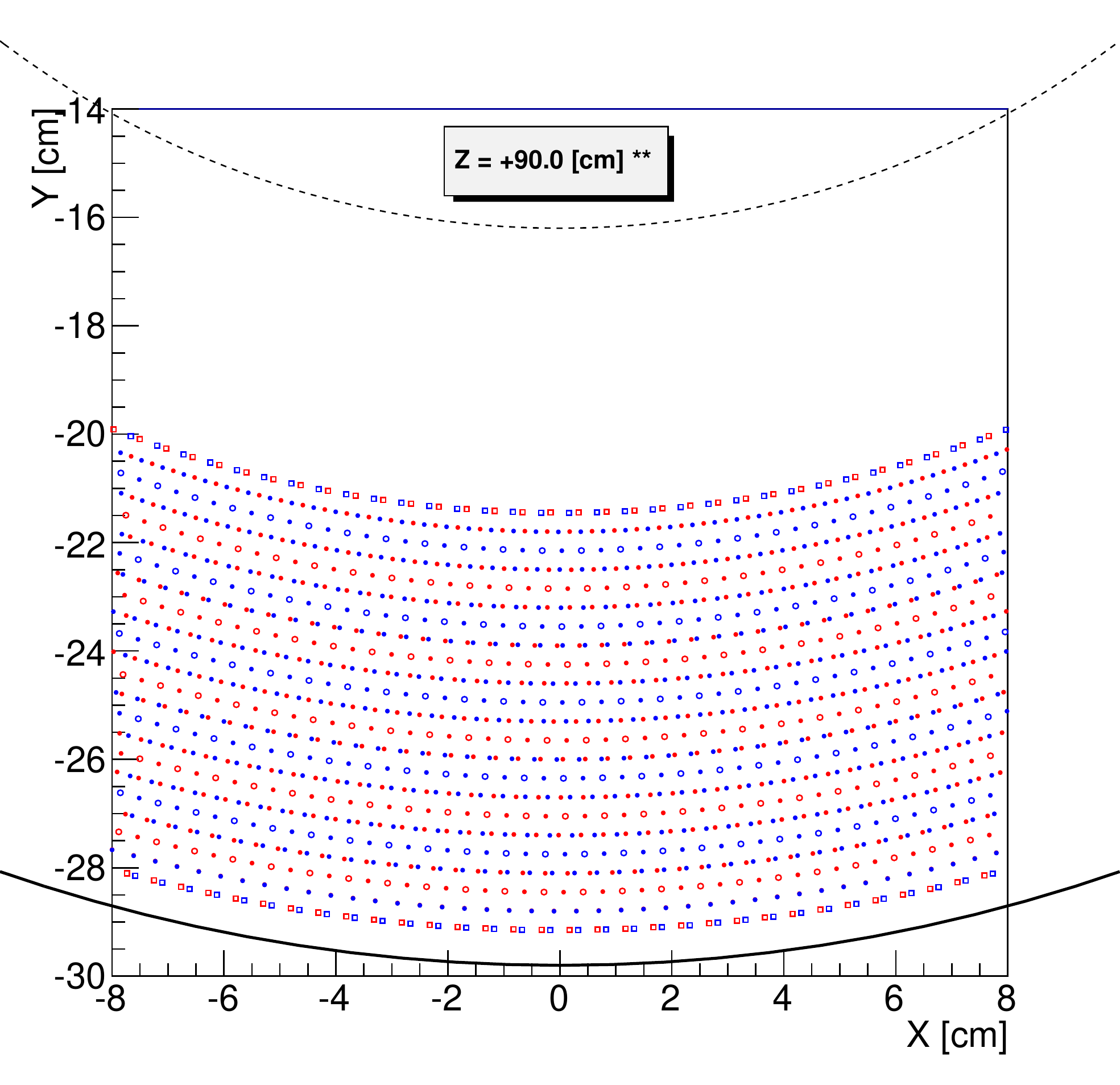} &
\includegraphics[width=0.45\columnwidth]{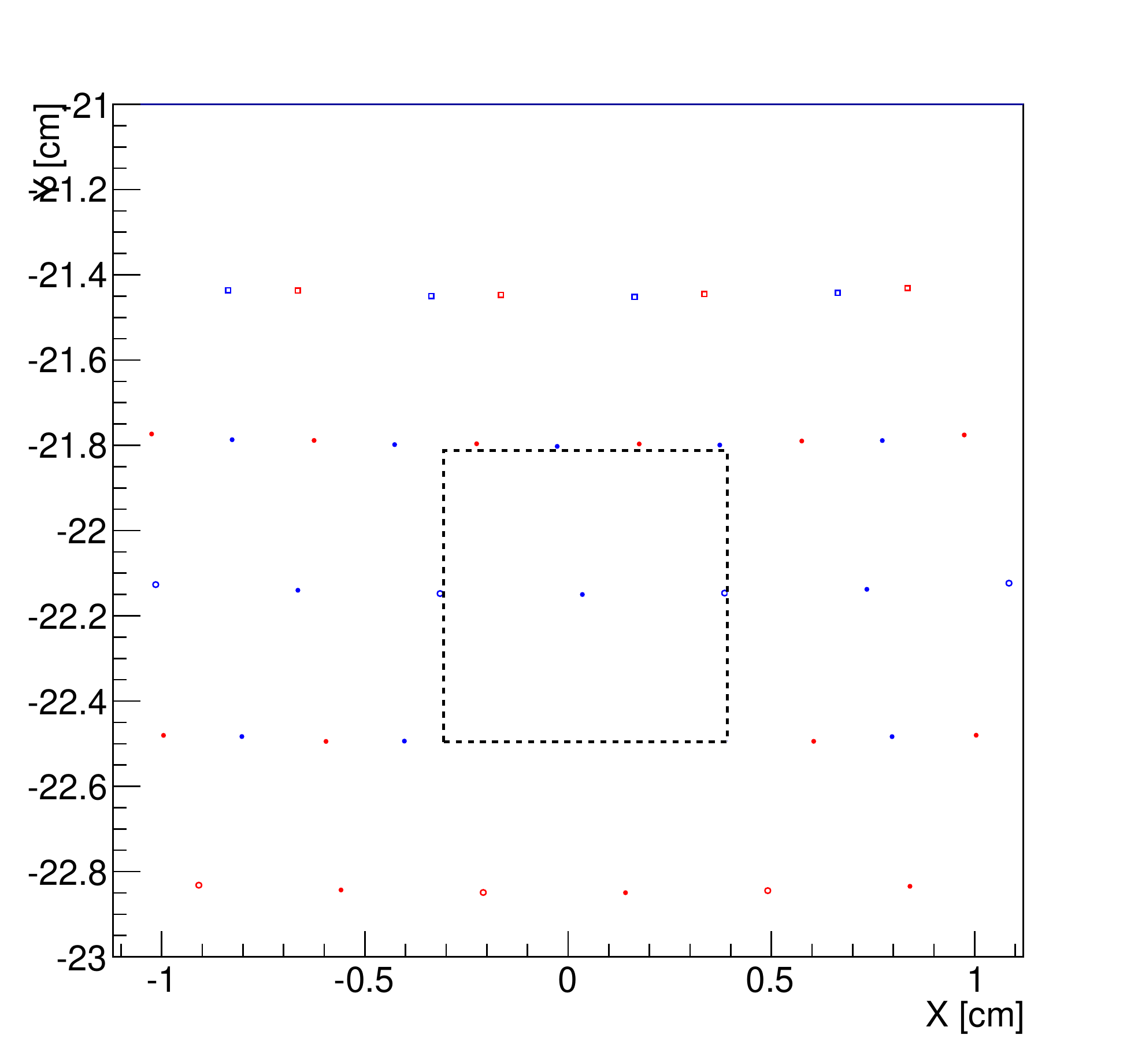} \\
$(a)$ & $(b)$ \\
\includegraphics[width=0.45\columnwidth]{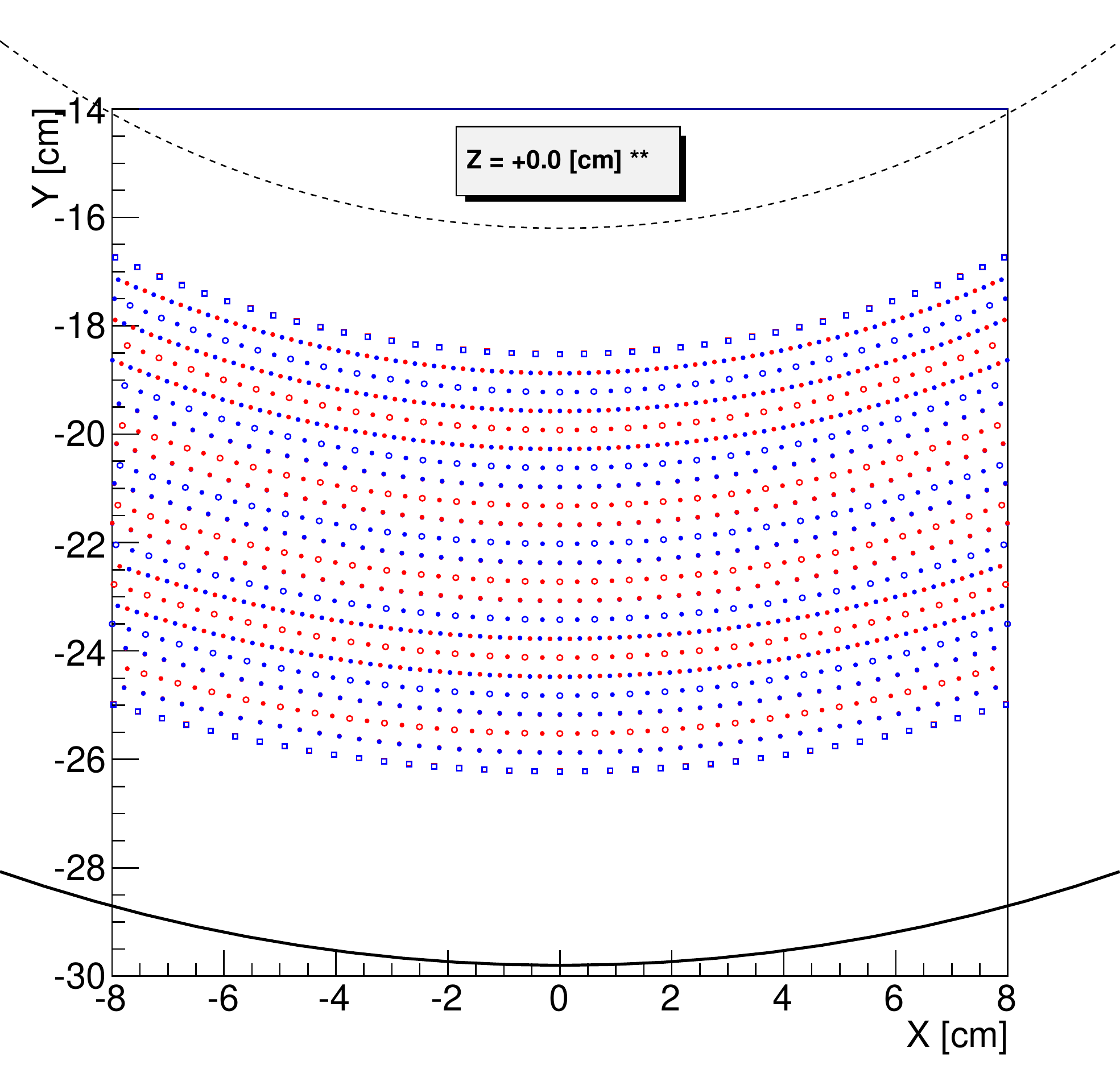} &
\includegraphics[width=0.45\columnwidth]{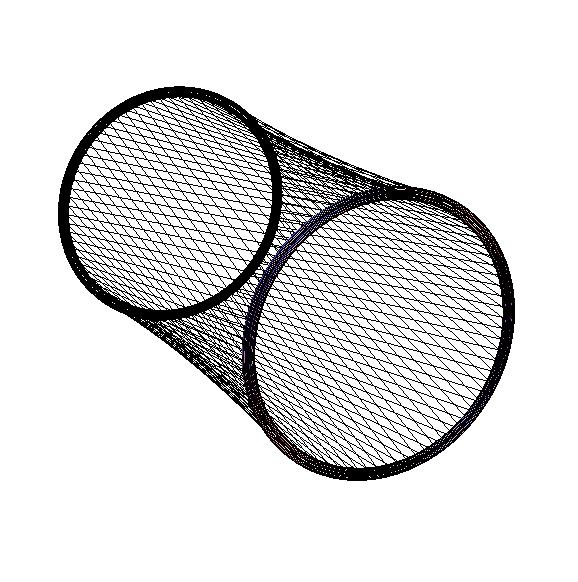} \\
$(c)$ & $(d)$ \\
\end{tabular}
\caption{\label{fig:DRAGOsketch} Schematic distribution of field and anode wires in the proposed DC. Blue and red 
colors correspond to $\pm$stereo angles. Sense wires are drawn an open circles, while closed dots are field wires.
Guard fields are depicted as square markers.
$(a)$ at the end-plate anchor point, $(b)$ a zoomed version where a single $7\times 7$~mm$^2$ cell is outlined.
$(c)$ the wire configuration at the centre of the COBRA magnet ($z = 0$). $(d)$ is a schematic representation
of one of the hyperbolic mesh ground planes.}
\end{center}
\end{figure}

The DC is composed of 10 criss-crossing sense wire planes with wires extending along
the beam axis with alternating stereo angles in order to reconstruct the coordinate along the axis of the chamber by combining the information of adjacent layers. The stereo angle varies from $8^{\circ}$ in the outermost layers to $7^{\circ}$ in the innermost ones.
There are alternate field planes and planes containing sense wires. The field planes
are common to the two $\pm$stereo views and are stringed in both directions, creating
a ground mesh between sense planes of alternating stereo view.
In this way the chamber takes the shape of a rotation hyperboloyd, whose surface 
is given by the envelope of the wire planes (see Fig.~\ref{fig:DRAGOsketch}).
Drift cells have an almost square shape (eight field wires surrounding the central anodic wire).
The side of each cell is 7 mm in order to guarantee a tolerable occupancy of the innermost wires, which are placed at roughly 18 cm from the beam axis where the rate is $\sim$1 MHz for a stopping rate of $7 \times 10^7 \mu$/s. 
With a maximum drift time of $\sim$150 ns, {\em e.g.} with the same gas mixture of 
KLOE (He/Isobutane 90:10), this corresponds to a $\sim$15\% occupancy of the innermost wires.

\begin{figure}[hbct]
\centering
\includegraphics[width = 0.6\columnwidth]{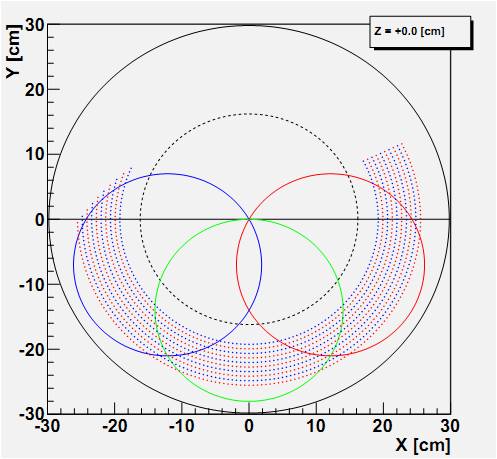}
\caption{\label{fig:DRAGOtracks} The pattern of anodic wires with three tracks originating from the central target superimposed}
\end{figure}

The distribution of cells inside the magnet volume is dictated by the angular coverage of the calorimeter. Fig.\ref{fig:DRAGOtracks} shows the pattern of anodic wires for three 52.8 MeV/c tracks originating from the target and with no momentum component out of the plane. The number of anodic wires is $\sim$1200 while the cathode wires are $\sim$6400. 

To minimize multiple scattering the DC runs with a low $Z$ gas mixture. A (90:10)
helium-isobutane (He:iC$_4$H$_{10}$) is presently foreseen, with the possibility
to increase the isobutane fraction to meet the best compromise between track resolution
and multiple scattering.
In Tab. \ref{tab:DRAGOmaterials} we show the average contribution to the multiple scattering of the various materials for tracks contained in a plane perpendicular to the beam axis. 
The corresponding total number of radiation lengths for the new chamber is smaller to that of the current MEG drift chamber 
($\approx 1.7 \times 10^{-3}$ $X_0$). 
This is crucial to keep the multiple scattering contribution to the momentum and angular positron resolutions under control, and 
also the rate of background photons in the electromagnetic calorimeter
generated by positron annihilation in the chamber.
\begin{table}[hbct]
\begin{center}
\begin{tabular}{lcr}
\\{\bf Item} \hspace{2cm} & Description& \hspace{1cm} Thickness  \\
&&($10^{-3}X_0$)\\
\hline
\hline 
Target & (140 $\mu$m Polyethilene) & 0.21  \\
Sense wires & (25 $\mu$m Ni/Cr) & 0.16 \\
Field wires & (40 $\mu$m Al) & 0.38 \\
Protective foil & (20 $\mu$m Kapton) & 0.14 \\
Inner gas & (Pure He) & 0.06 \\
Tracker gas & He/iBut. 85:15 (90:10) & 0.50 (0.36)\\
Total & One full turn w/o target & 1.24 (1.10)\\
\hline \hline
\end{tabular}
\end{center}
\caption{\label{tab:DRAGOmaterials}Material budget of the new MEG drift chamber along one track. Note that the decay positron 
crosses only part of the polyethilene target. Two options for the gas composition are shown.}
\end{table}
\subsubsection{Mechanics}
\begin{figure}[hbct]
\begin{center}
\includegraphics[width=10cm]{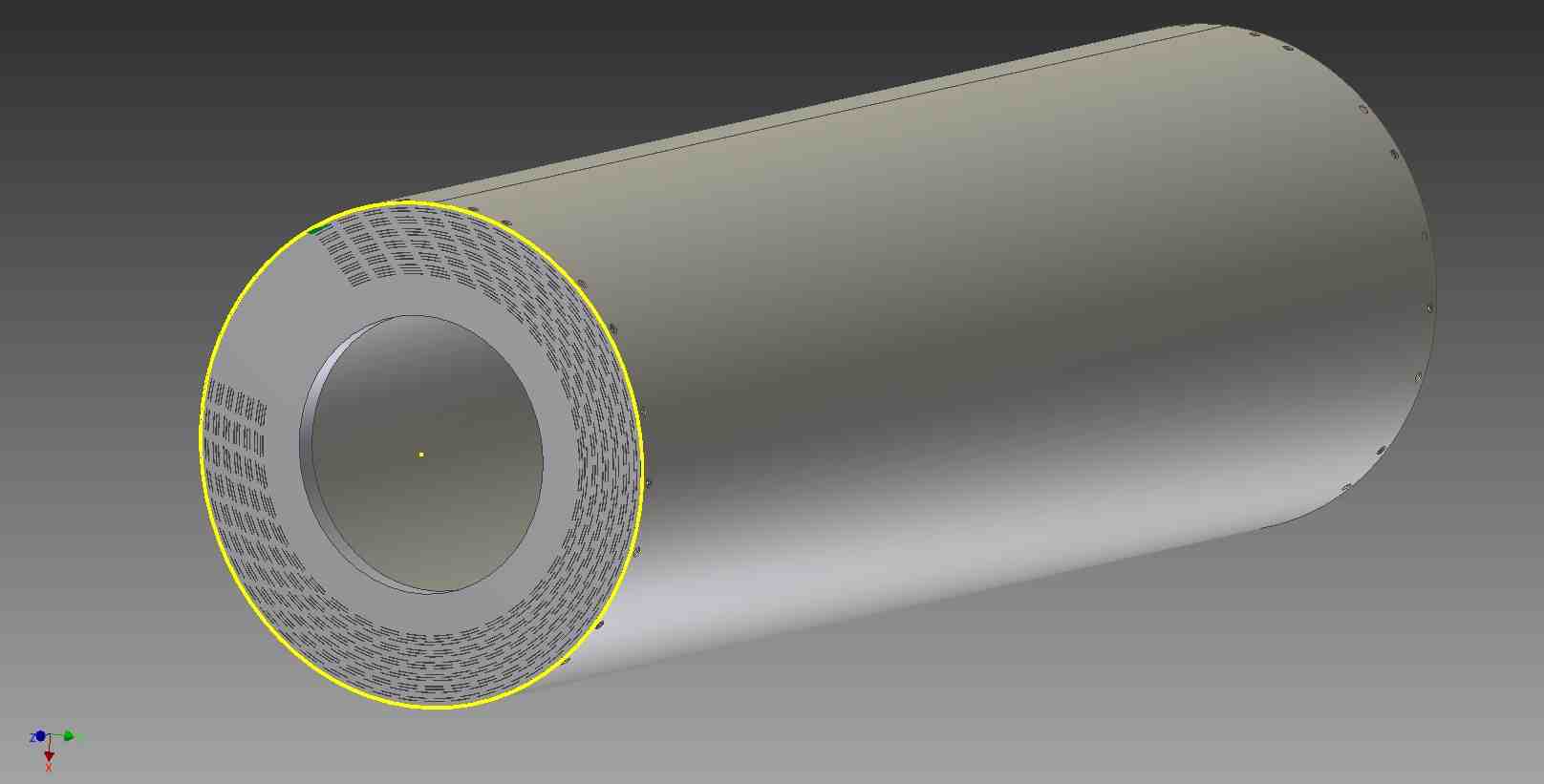}
\caption{\label{fig:MODELsketch} 
Model of the new drift chamber used in FEA simulation
}
\end{center}
\end{figure}

\begin{figure}[hbct]
\centering
\begin{minipage}[t]{9 cm}
\centering
\includegraphics[width= 7 cm]{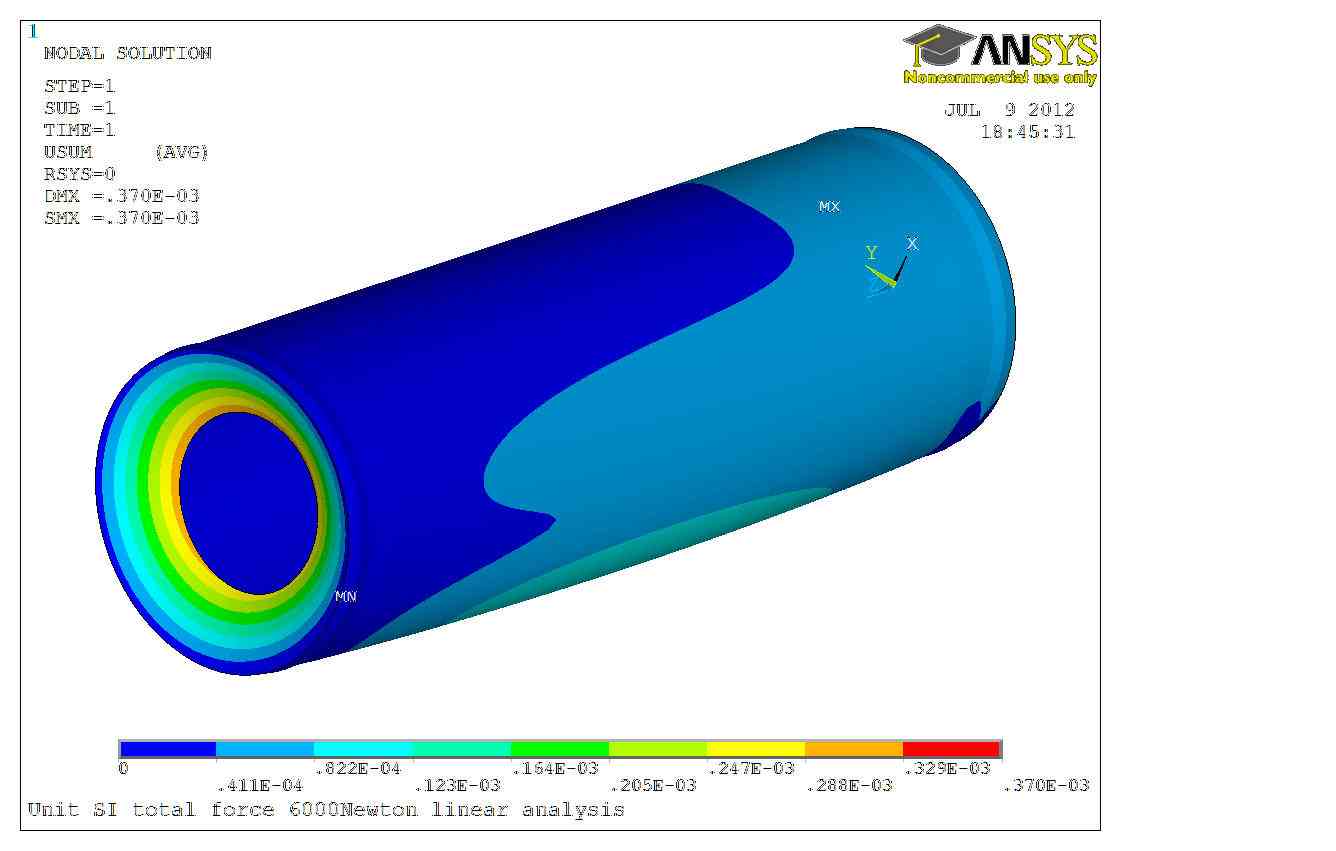}
\caption{\label{fig:FEAplate} Simulated end-plates deflection} 
\end{minipage}
\begin{minipage}[t]{7 cm}
\centering
\includegraphics[width = 7 cm]{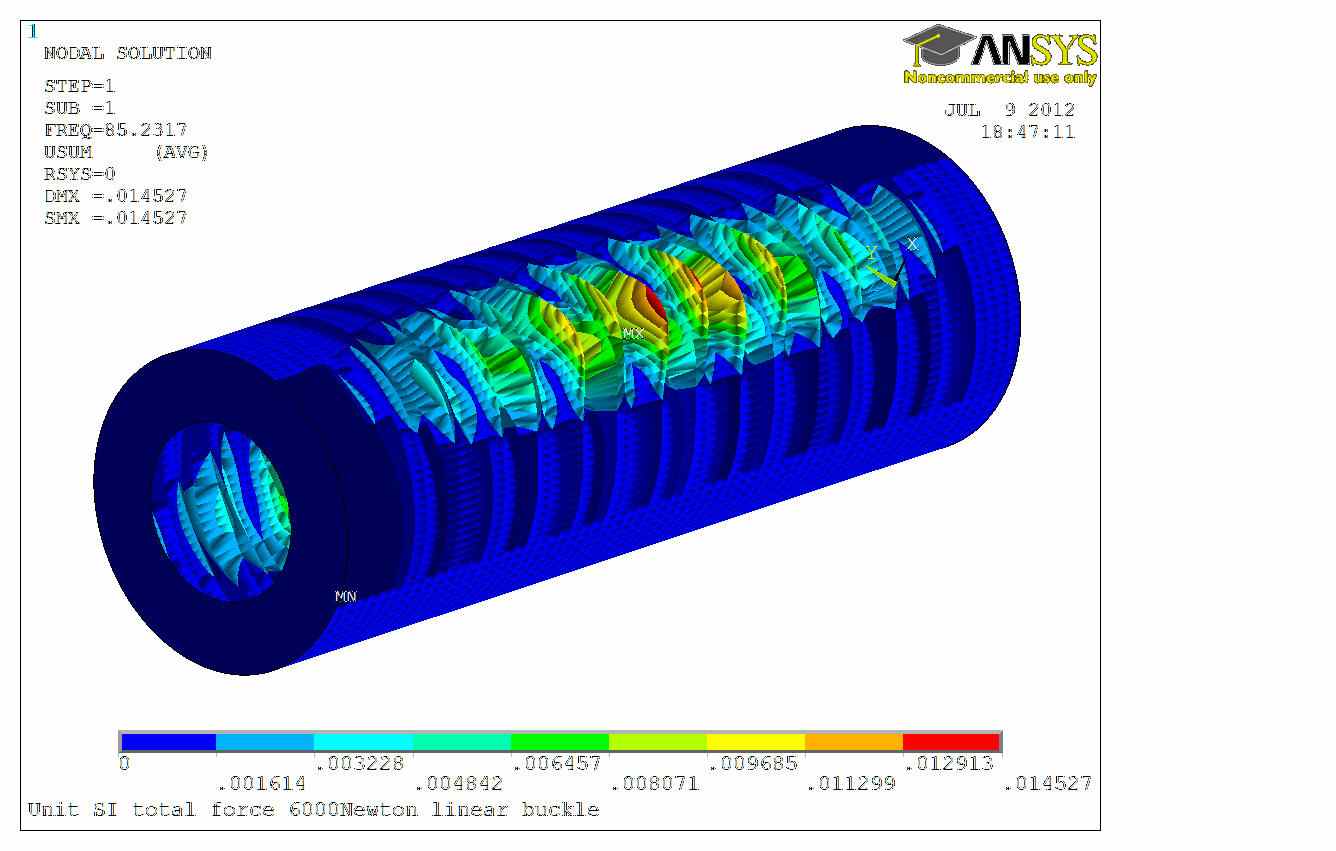}
\caption{\label{fig:FEAcylinder} Simulated cylinder linear buckling}
\end{minipage}
\end{figure}

A first evaluation of the mechanical feasibility of the DC was done simulating, with a commercial
Finite Elements Analysis program (FEA)~\cite{ansys}, a model composed of two circular aluminum end-plates 
with an equivalent thickness of 20 mm (corresponding to a 30 mm end-plate with slots for PCB cards where wires are soldered) kept
in position by a 180~cm long, 2 mm thick, external carbon-fiber cylinder, made of 16 intermediate high-module (E460-MJ46) pre-preg layers, as in Fig.~\ref{fig:MODELsketch}. 
Fig. \ref{fig:FEAplate} shows the simulation result when end-plates
are loaded with a total wires pressure of 6000 N, uniformly distributed over 300$^\circ$ end-plate sectors. The maximum deflection is 0.37 mm which
is tolerable given the stretching of the wires at the proposed mechanical tension. In Fig. \ref{fig:FEAcylinder} it is also shown that the 
linear buckling of the carbon-fiber cylinder happens at about 85 times the nominal tension, indicating that 
distorsions are kept within good safety margins.

%
\begin{figure}[htbc]
\begin{center}
\includegraphics[width=0.8\columnwidth]{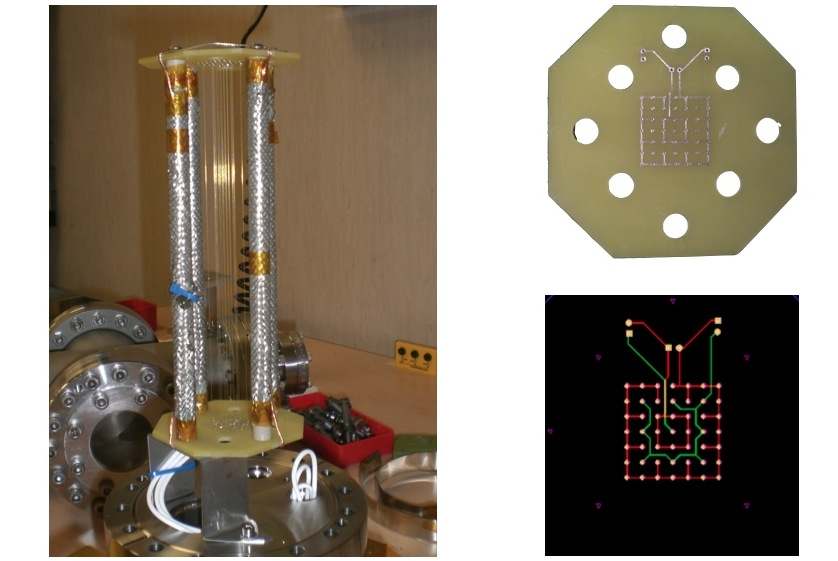}
\caption{\label{fig:DRAGOage} 
DC prototype used for ageing test. The PCB and its layout are also shown.
}
\end{center}
\end{figure}
%
%
\subsubsection{Ageing tests}
During the entire acquisition time of the upgraded MEG experiment the total 
charge collected by the innermost cells of the new DC was evaluated to be $\leq
0.4$~C/cm for a gain of $1 \times 10^{5}$ and a muon rate of $7 \times 10^{7} \mu^+/$sec (see Table~\ref{tab:ageing}).
\begin{table}
\begin{center}
\begin{tabular}{rclrclrcl}
\hline \hline
$\mu^+$ rate & = & $7\times10^{7} \mu^+/$sec & 
& \quad & &
Maximum positron rate & = & $45~$kHz/cm$^2$ at 10$^8 \mu^+/$sec \\ 

Innermost Radius & = & 17.6~cm & 
& \quad & &
Cell size & = & 7 mm$^2$ \\

DC gain & = & $1\times 10^5$ &
& \quad & &
Years of run & = & 3 \\

DAQ days/year & = & 210 &
& \quad & &
electrons/MIP & $\simeq$ & 20/cm \\

\hline \hline
\end{tabular}
\caption{\label{tab:ageing}Operating conditions for the DC of the positron tracker,
used to evaluate the possible detector ageing.}
\end{center}
\end{table}
This represents a huge amount of charge collected on anode and cathode and therefore
a study of the DC ageing is mandatory. DC ageing induces gain loss, excessive
chamber current, self-sustained discharges, sparking, high voltage instability,
and it is thought to be predominantly caused by free radical polymerization~\cite{niebhur}.
In test set-ups ageing is usually accelerated by irradiating a DC prototype with
high intensity sources, and the chamber gain is monitored as a function of the total
integrated charge.

A DC prototype of $20$~cm length was prepared, which implements a single $7 \times 7$~mm$^2$ cell 
surrounded by field shaping wires that mimic the presence of all other cells. All wires are gold-coated tungsten wires; the central sense is $25~\mu$m diameter, the field and shaping wires are $80~\mu$m diameter (see Figure~\ref{fig:DRAGOage}).
An $^{241}$Am $\alpha-$source of $\sim 1$~Bq activity is placed on the protoype supporting 
structure to provide large ionization signals visible in the very light He:iC$_4$H$_{10}$ mixture. 

The DC prototype is placed in a stainless-steel chamber of $3500$~cc volume, made
of standard CF100 high vacuum components, equipped with HV and signal feed-throughs as well as two $150$~$\mu$m thin windows in order to let
ionizing radiation through.
The wires are soldered on a FR4 printed circuit board (PCB) and
all internal connections other than the ones on the PCB are made with PTFE-coated coaxial cables. 
We work in a shifted-potential configuration, {\em i.e.} negative HV is applied 
to the field wires while the central (signal) wire is grounded through a Keithley~2635 picoammeter in order to read the DC current. The central signal can be also used to monitor the pulses seen by the DC. The measuring set-up is placed inside a lead/alimunum box for radiation safety.

A commercial MKS gas system was assembled to deliver the gas from bottles containing pre-mixed (90:10) helium-isobutane into the test chamber. 
The gas can be sampled by a residual gas analyzer (RGA) to monitor the gas composition at the percent level. 
Gas is flushed at 50~cc/min to provide a complete volume exchange in one hour.

\begin{figure}[!hb]
\begin{center}
\begin{tabular}{cc}
\includegraphics[width=0.55\columnwidth]{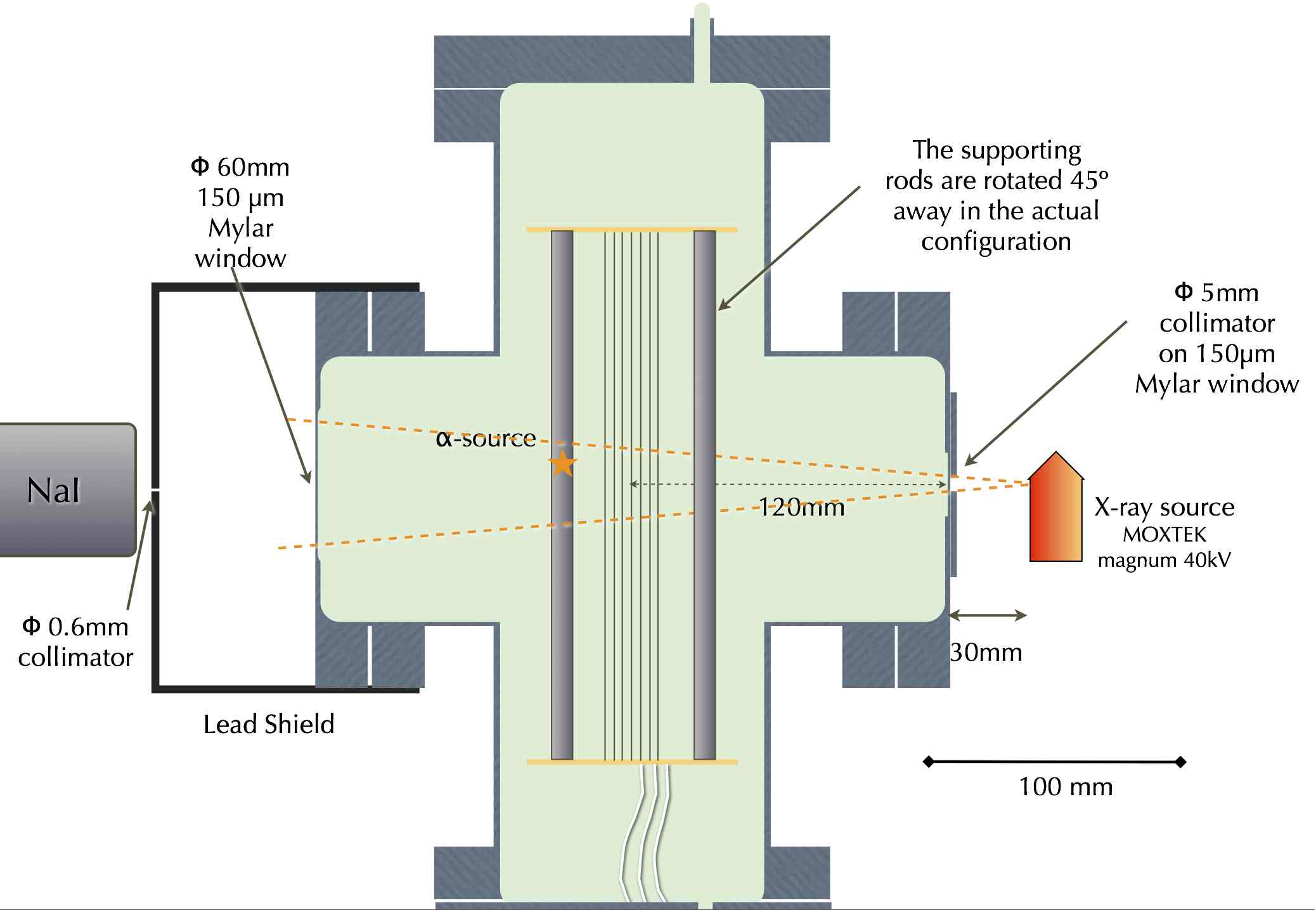} &
\includegraphics[width=0.5\columnwidth]{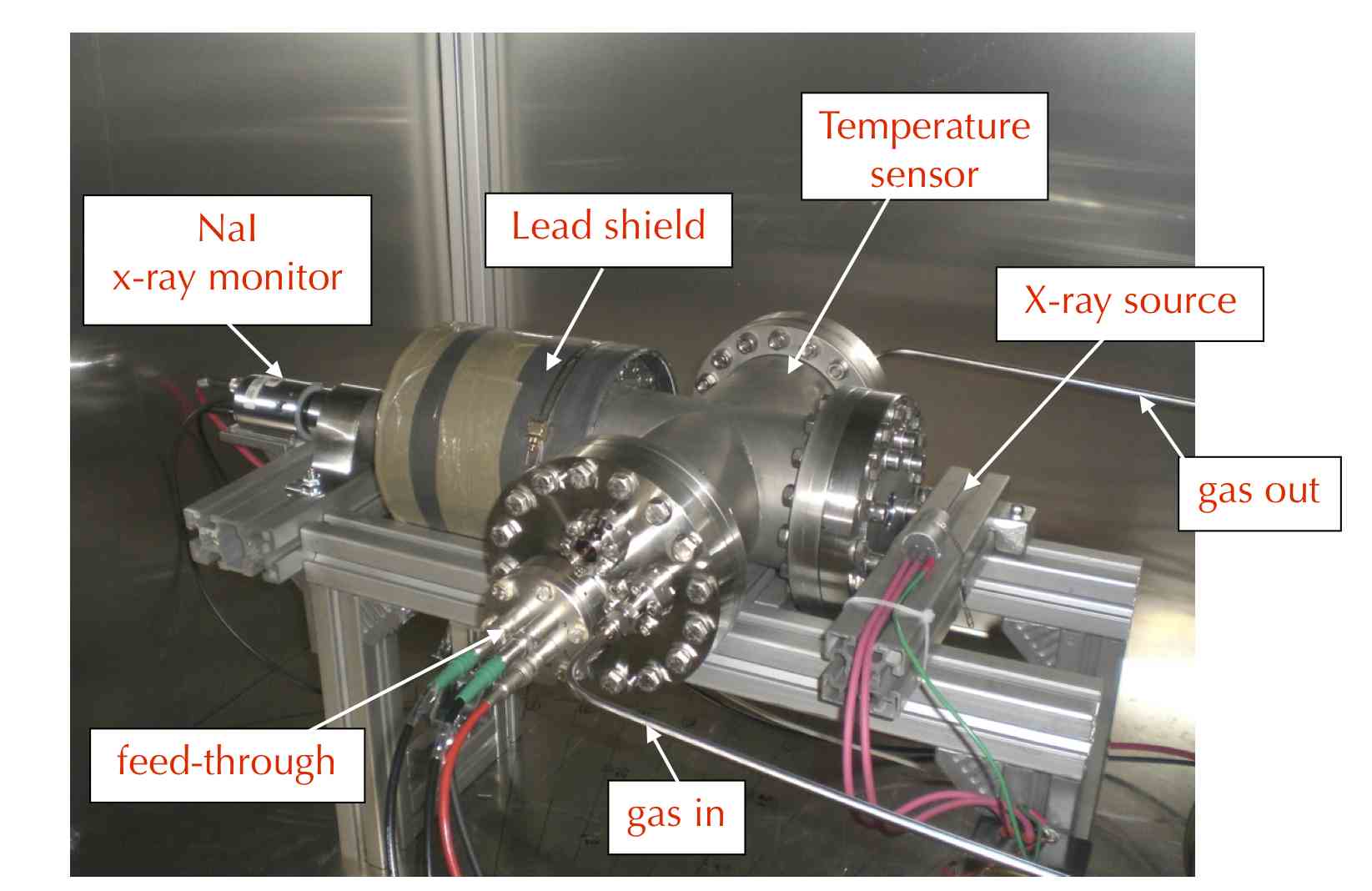}\\
$(a)$ & $(b)$ \\
\end{tabular}
\caption{\label{fig:ageing-setup}
Schematics $(a)$ and picture $(b)$ of the ageing measurement setup.}
\end{center}
\end{figure}

A $2.5$~cm long portion of the central wire is irradiated with a MOXTEK Magnum $40~$keV reflection source
 X-ray gun able to provide $>10^{11}$ X-rays/sec/sterad (see Figure~\ref{fig:ageing-setup} for a sketch of the measurement set-up).
The stability of the source is monitored by means of a NaI X-ray detector preceded by a $0.6$~mm diameter lead collimator. 
Its rate was measured to be linear with the X-ray source current.

Due to the lightness of our gas mixture most of the energetic X-rays go through the sensitive volume undetected. A fraction of X-rays ionizes the gas 
in the cell region and contributes to the DC current. 

\begin{figure}[!hb]
\begin{center}
\begin{tabular}{cc}
\includegraphics[width=0.5\columnwidth]{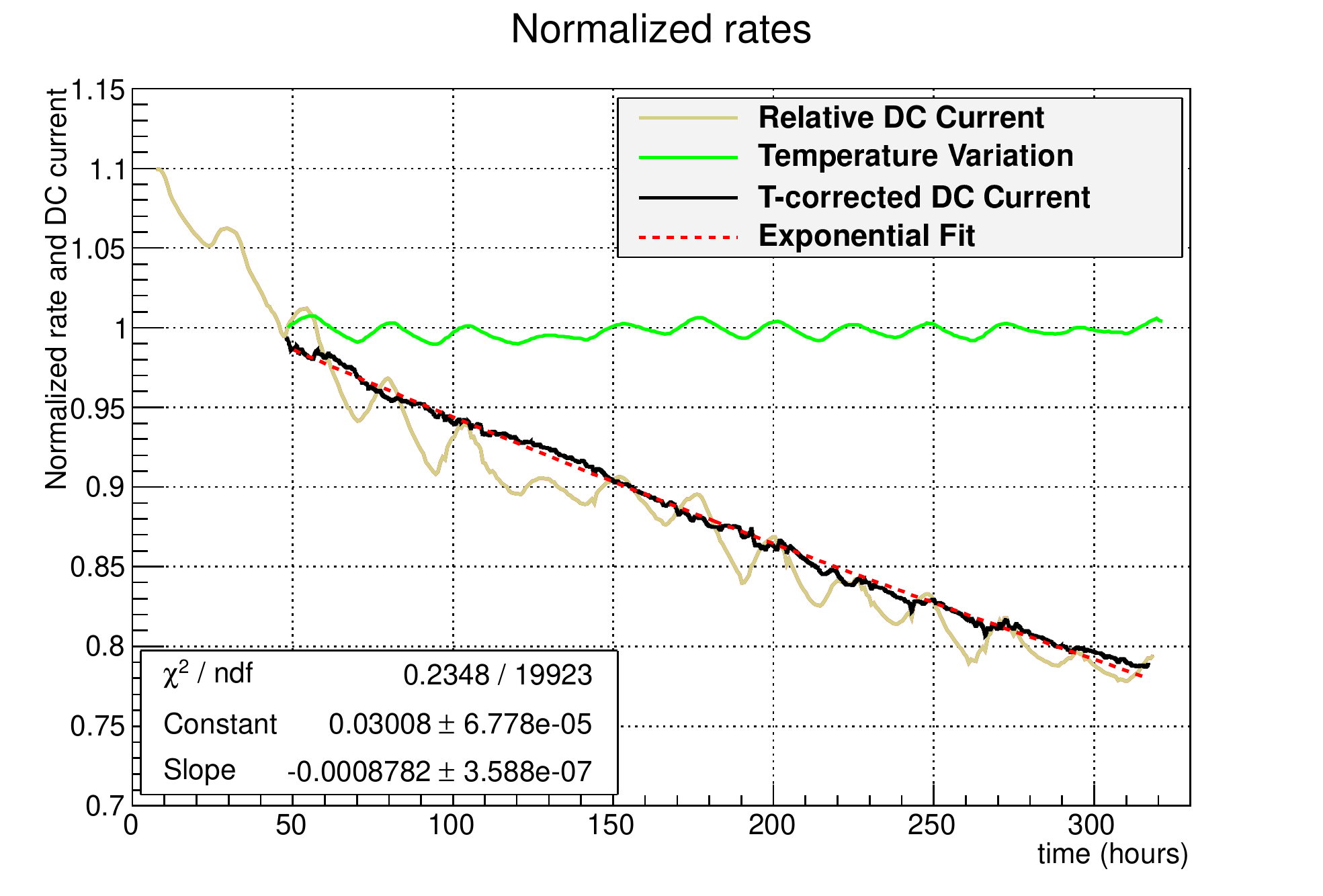} &
\includegraphics[width=0.5\columnwidth]{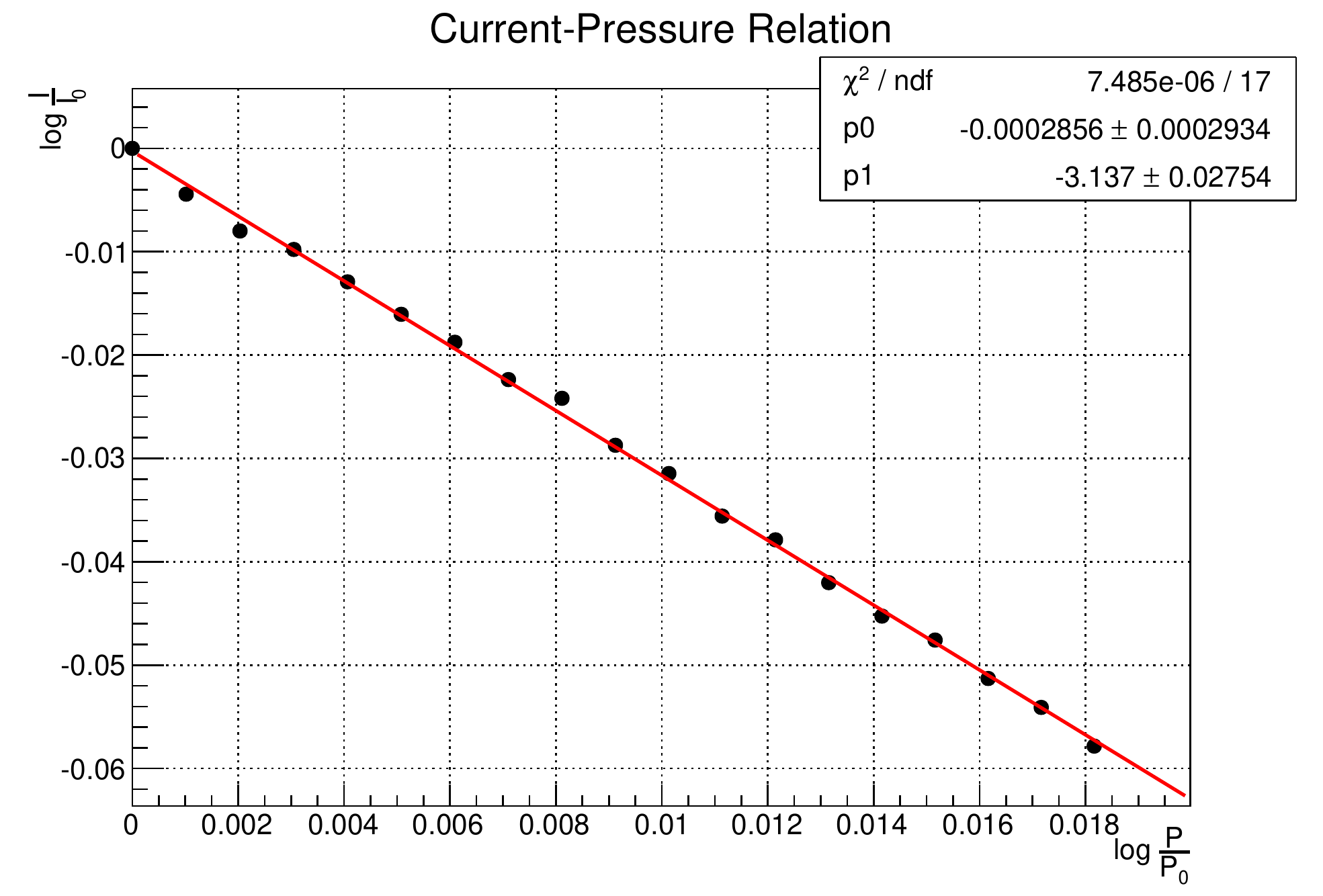} \\
$(a)$ & $(b)$ \\
\end{tabular}
\caption{\label{fig:ageing-result}$(a)$ Variation of the current collected by the anode wire as a function 
of the time. 250 hours of accelerated ageing correspond to 1 PSI DAQ year. $(b)$ The current-pressure relation used to correct
for the temperature oscillations.}
\end{center}
\end{figure}

We conducted the ageing test at an initial DC current of 120~nA/cm, {\em i.e.} 20 times
the maximum current foreseen in normal experimental conditions. This represents also the ageing acceleration factor: in ten days we collect the equivalent charge of one entire running year.
We operated the DC prototype at $-1250$~V corresponding to a gain of ${\sim} 10^4$. 
X-rays and $\alpha-$particle signals are clearly visible on the oscilloscope with no need of preamplification.
The prototype was irradiated for 15 days continuously, and Figure~\ref{fig:ageing-result} shows the gain loss as a function of the time.
Daily temperature oscillations induce a density variation which, in turn, is reflected in the gain oscillation observed.
We measured the gain-density relation by modifying the system pressure (see Figure~\ref{fig:ageing-result}$b$) and applied 
the corresponding factor to correct the temperature-induced oscillations (black line in Figure~\ref{fig:ageing-result}$a$).
Variations of the X-ray source were measured to be well below $1\%$. Every day we checked for possible sparks and discharges.

\begin{figure}[!hb]
\begin{center}
\includegraphics[width=0.9\columnwidth]{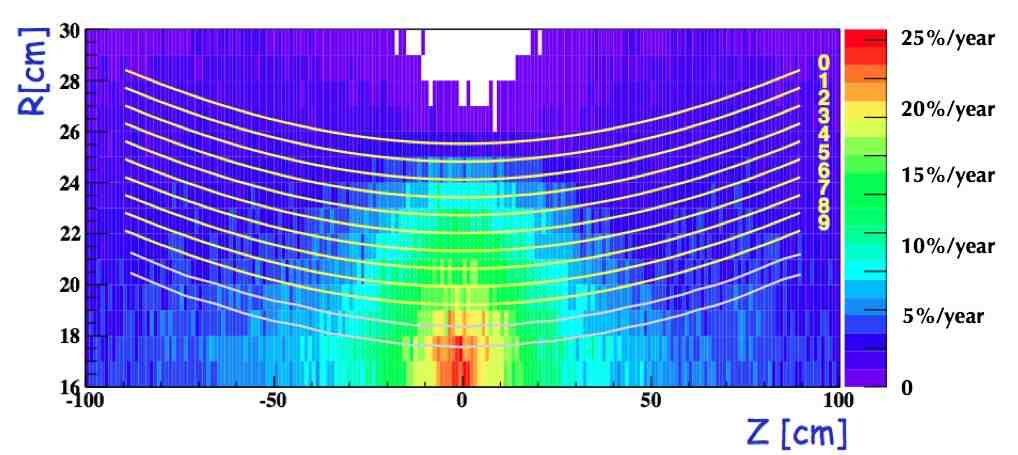}
\caption{\label{fig:2Dage}Gain drop in 1-year od DAQ time at $7 \times 10^7~\mu^+$/sec.}
\end{center}
\end{figure}

These preliminary tests showed that a yearly gain drop of $<25\%$ is expected at the hottest spot of the innermost DC wire; 
the large fraction of the DC is subject to a $<10\%$ gain drop per year (see Figure~\ref{fig:2Dage}).
This represents a good working point, where furhter optimization, such as material selection and different gas flow rate, is possible.

\subsubsection{Monte Carlo simulation}
\begin{figure}[hbct]
\begin{center}
\includegraphics[width=10cm]{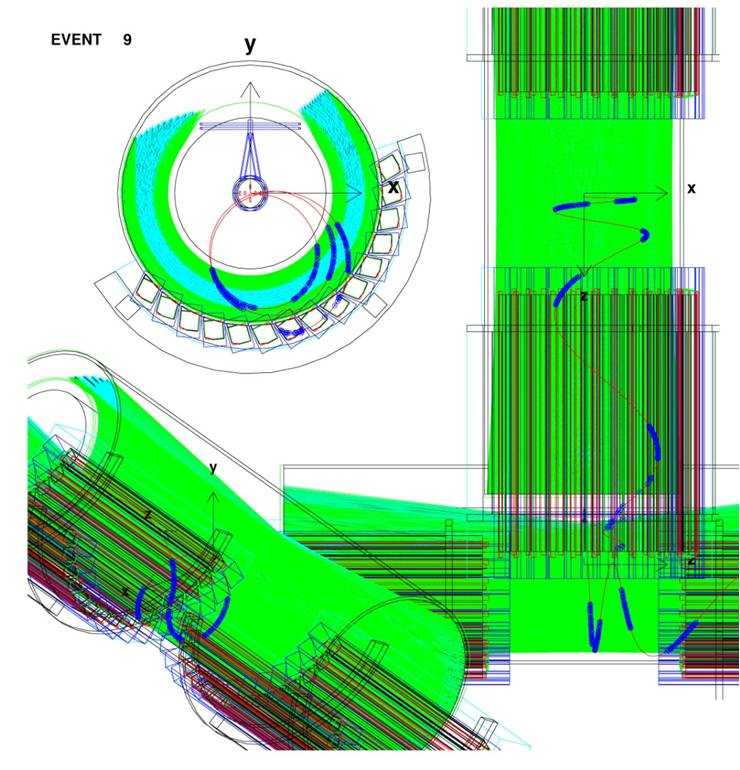}
\caption{\label{fig:DRAGOsimtrack} 
Simulation of a 52.8 MeV/c positron track in the new drift chamber
}
\end{center}
\end{figure}

The response of the new chamber to 52.8 MeV/c positrons from $\meg$ decay was studied by means of a full Monte Carlo simulation program.  We show in Fig. 
\ref{fig:DRAGOsimtrack} the output of a simulated positron track in the spectrometer.

The most probable number of wires hits is $\sim$ 60, a factor of 3 larger than the present MEG DC system. 
This represents a big improvement for the reconstruction efficiency and for the momentum and angular resolutions,
as well as a formidable help for pattern recognition in a high occupancy environment. 

\begin{figure}[hbct]
\begin{center}
\includegraphics[width=10cm]{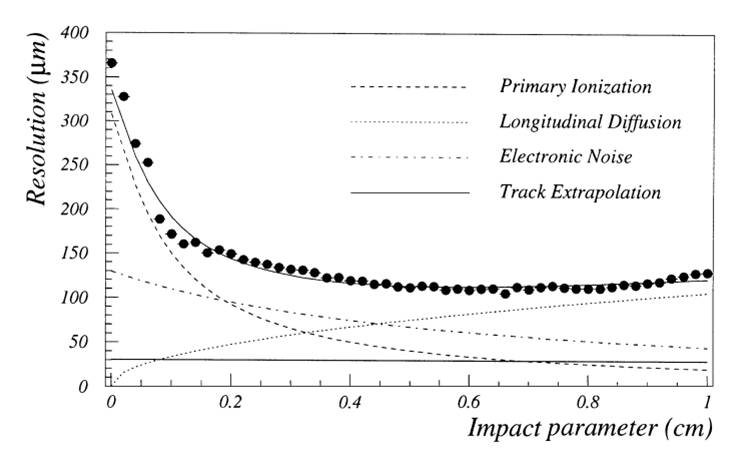}
\caption{\label{fig:KLOEresolution} 
Resolution on the drift measurement as a function of the impact parameter measured by the KLOE experiment \cite{KLOEproto}. The estimated individual contributions
to the final measured values are shown in the picture.
}
\end{center}
\end{figure}

\begin{figure}
\begin{center}
\includegraphics[width=0.9\columnwidth]{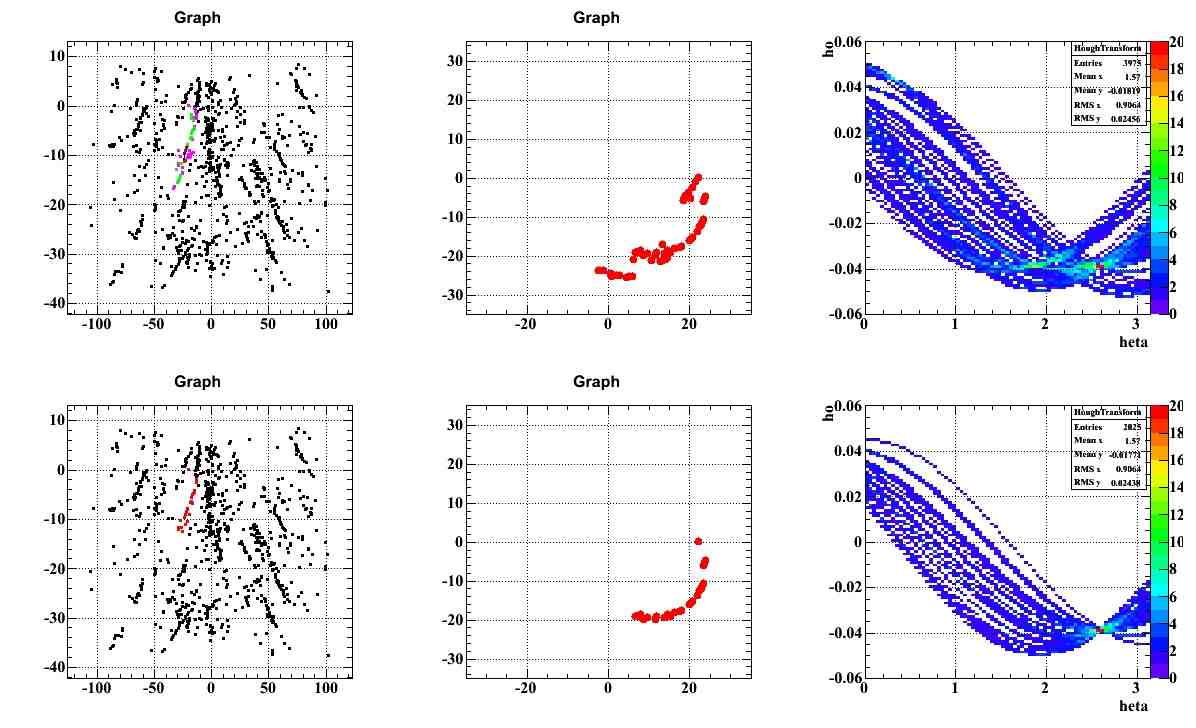}
\caption{\label{fig:hough} A simple hit-finder, track-finder algorithm is capable of finding the signal track among the hits
recorded in one event.}
\end{center}
\end{figure}

The resolution in the measurement of the drift distance to the anode wires is extrapolated from the measurements performed by 
the KLOE experiment \cite{KLOEproto} (which are shown in Fig.\ref{fig:KLOEresolution}) and confirmed by measurements in a
test set-up (see section~\ref{sec:tritubo}).
With reference to KLOE results in Figure~\ref{fig:KLOEresolution}, both the 
electronic (indicated as  "electronic noise"  in the picture) and the
"primary ionization" components can be reduced by using a fast electronic chain and  the "cluster timing technique"
as explained in the next section.
A preliminary hit reconstruction algorithm finds the point of closest approach to the wire, from the arrival time of the first 
avalanche cluster, with a resolution of about 120~$\mu$m independently of noise level, while charge division gives 
the rough $z$-coordinate along the wire in the range $1.3\div 5.5$~cm with noise of $1$~mV$\div 3$~mV respectively. 
With this information, and the wire stereo angle, a simple track finder
looks for groups of hits clustering into straight lines in the $\phi$ vs $z$ plane. 
The Hough Transform of the conformal X/Y mapping of the hit clusters found is used to remove outliers
(see Fig.~\ref{fig:hough}).
We performed the tracks reconstruction by assuming an average resolution of 
$120$~$\mu$m in the measurement of the drift distance to the anode wires with a worsening similar to that in 
Figure~\ref{fig:KLOEresolution} as a funtion of the impact parameter.  
The momentum and angular resolutions obtained from the simulation are:
\begin{equation}
\label{eq:DRAGOresthin}
\Delta\phie \Big|_{\phi=0} = 3.7 \rm{mrad}; 
\hspace{3mm} \Delta\thetae = 5.3 \rm{mrad}; 
\hspace{3mm}\Delta\ppositron = 130~\rm{keV}
\end{equation}
for a 140 $\mu$m target thickness, with 15$^\circ$ slanting angle.

The positron reconstruction efficiency (tracks reconstructed in the DC with a corresponding hit on the TC) 
is larger than $85\%$ due to the chamber's ability to track the positrons up to the DC-TC interface. Furthermore
read-out preamplifiers, cables and the structure supporting the wires are placed in regions which are off the positron paths
(the efficiency is $40\%$ for the present MEG positron spectrometer).

\subsubsection{Fast read-out and the ‘cluster timing’ technique}
\label{sec:DRAGOreadout}
\begin{figure}[hbct]
\begin{center}
\includegraphics[width=7cm]{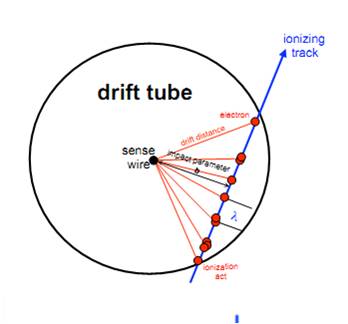}
\caption{\label{fig:DRAGOclustertiming} 
Bias in the minimum distance measurement for low number of clusters per cm
}
\end{center}
\end{figure}

The use of a low Z gas mixture such as He/Isobutane (90/10) is essential for minimizing  the effects of multiple scattering.
The average  number of ionization clusters produced by the passage of a $52.8~$MeV positron in this gas mixture is 
lower than in the present MEG DCs (about 12.5 per cm of track instead of $\sim 30/$cm), leaving a bias in the measurement of the 
distance of closest approach (impact parameter) of a particle from the anode wire (see Fig.~\ref{fig:DRAGOclustertiming}). 
We propose to use the cluster timing technique \cite{Tas2007} to eliminate this bias and aim at reaching possible resolutions even below the 120 $\mu$m previously assumed. This technique, as opposed to the traditional determination of the impact parameter, which  uses only the arrival time of the first cluster, consists in measuring the timing of all the individual clusters and produce a bias free estimator using also the timing 
of the clusters following the first one.

\begin{figure}[tbc]
\includegraphics[width=0.35\textwidth,clip=TRUE]{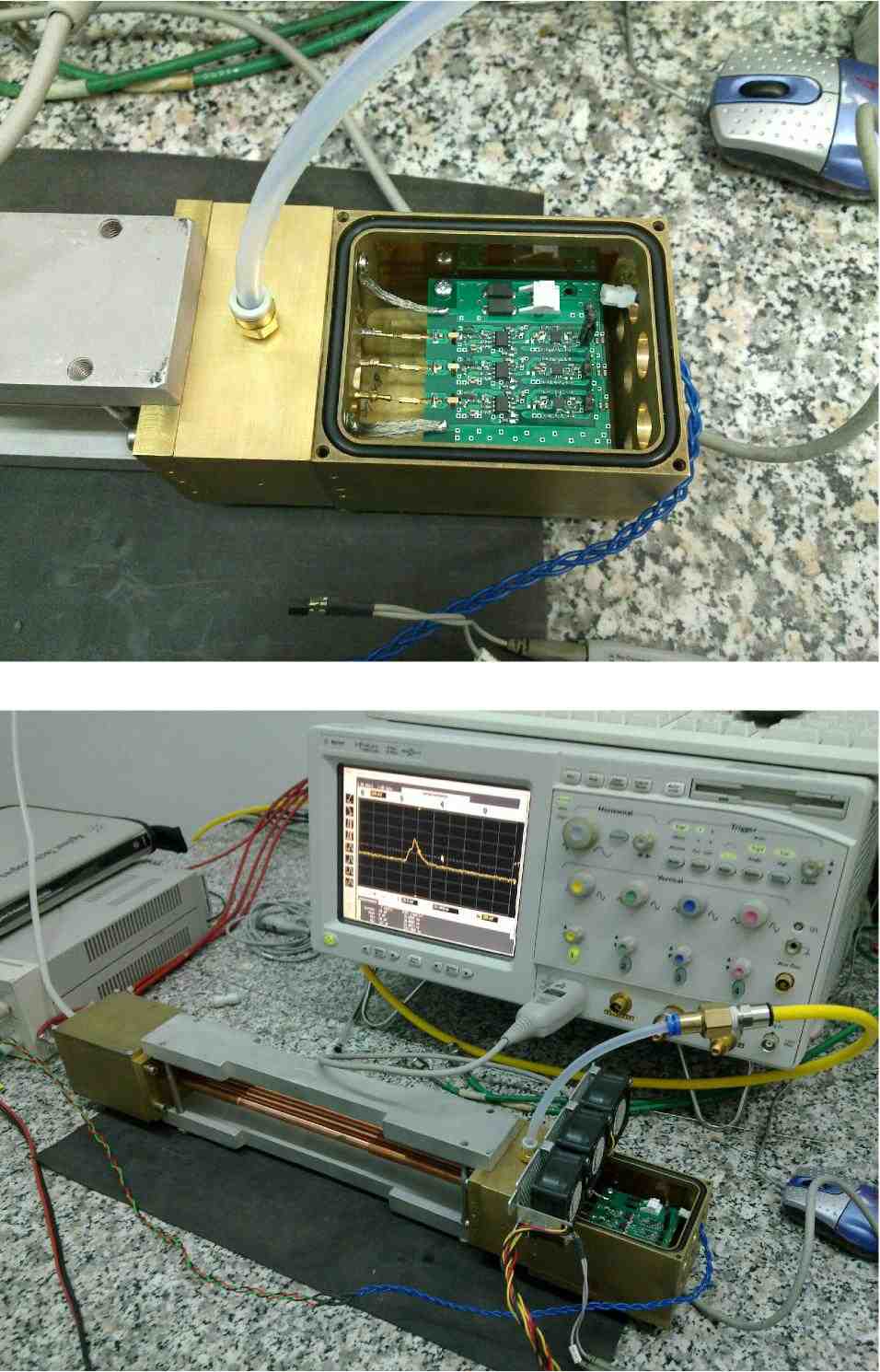}
\includegraphics[width=0.4\textwidth,clip=TRUE]{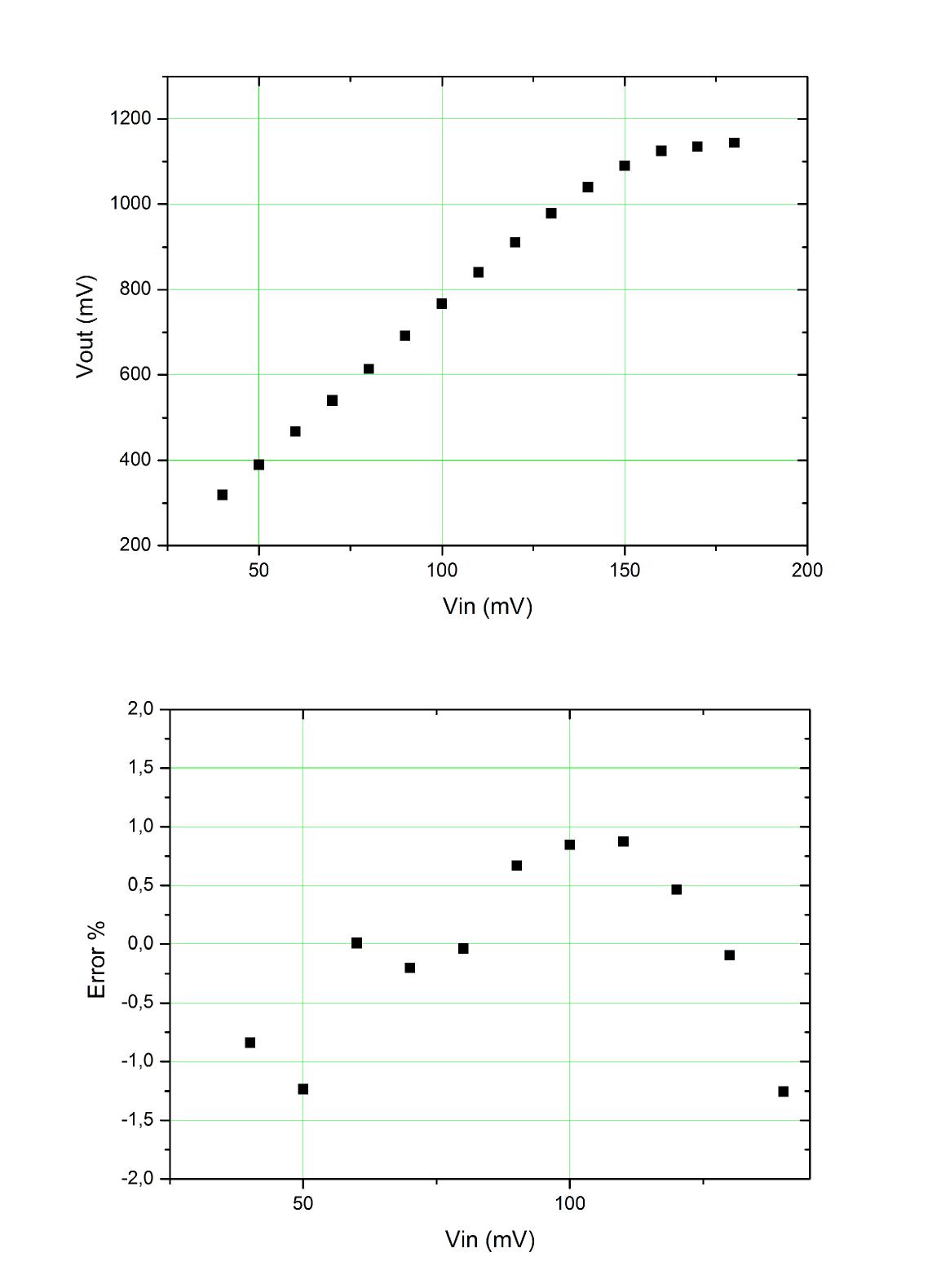}
\caption{\label{fig:FEfig45} 
Test set-up photograph (left) and measured non linearity (right) of the front-end prototype
}
\end{figure}

\begin{figure}[hbct]
\includegraphics[width=0.35\textwidth,clip=TRUE]{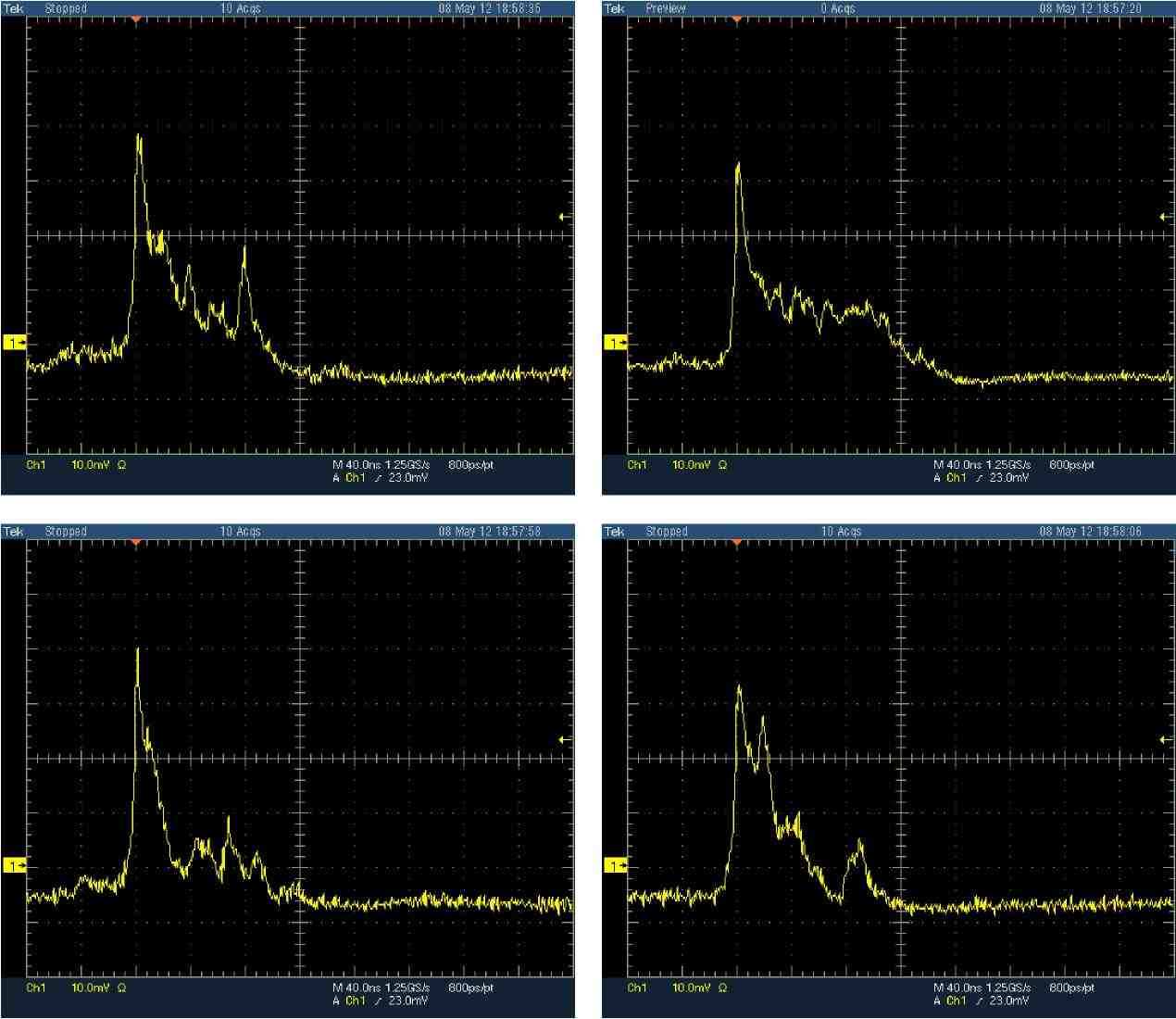}
\includegraphics[width=0.4\textwidth,clip=TRUE]{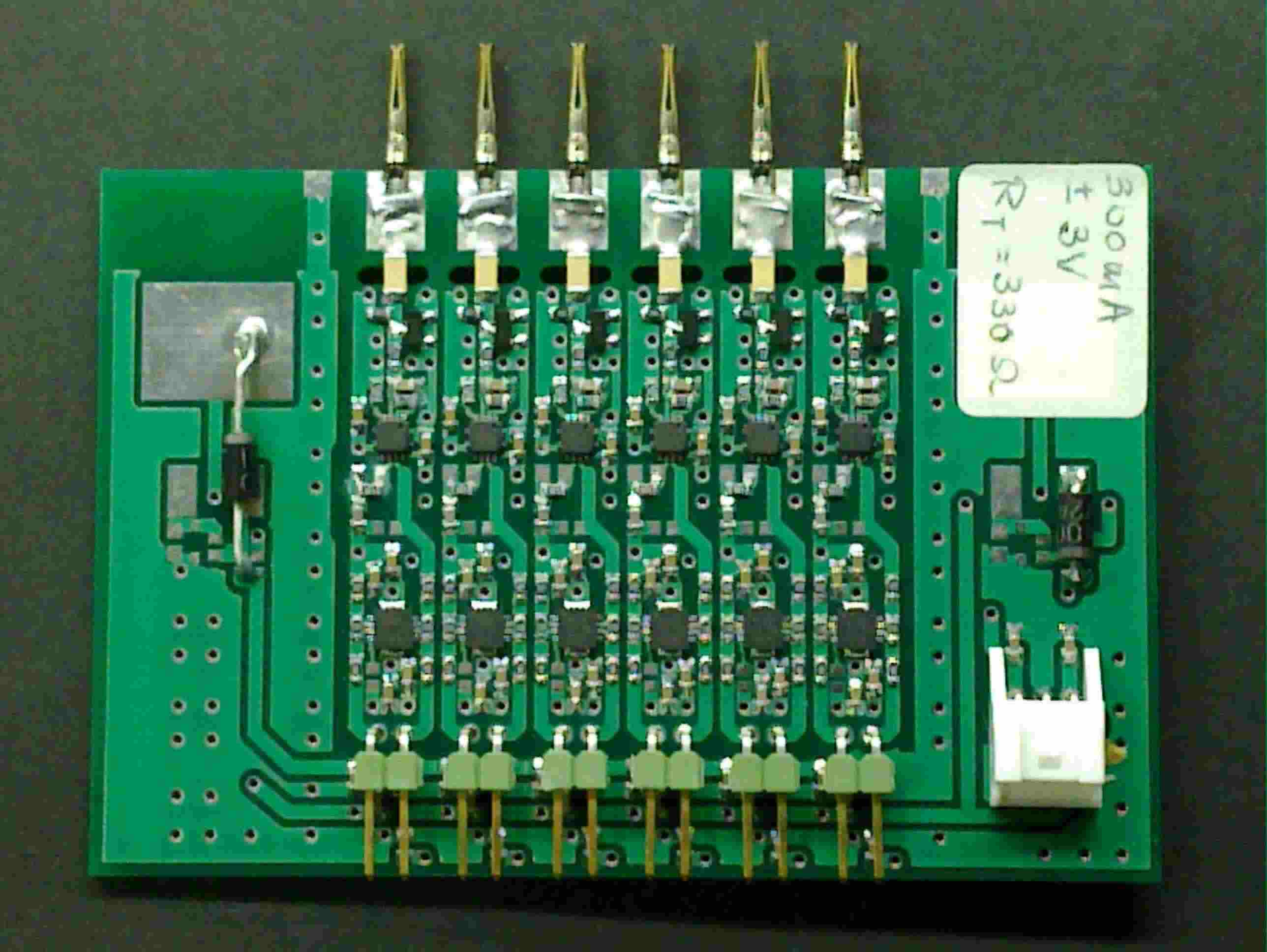}
\caption{\label{fig:FEfig67} 
Time distribution of clusters (left) and PCB (right) of the 6-channel front-end prototype
}
\end{figure}

The cluster counting/timing technique needs very fast frontend electronics for signal acquisition,
the temporal separation between signals produced by the different ionizations clusters being a few nanoseconds; 
therefore, in order that the acquired signal shows temporally separated pulses without overlapping, it is necessary to have frontend 
electronics characterized by a large bandwidth.
\begin{figure}[tbc]
\begin{center}
\includegraphics[width=7cm]{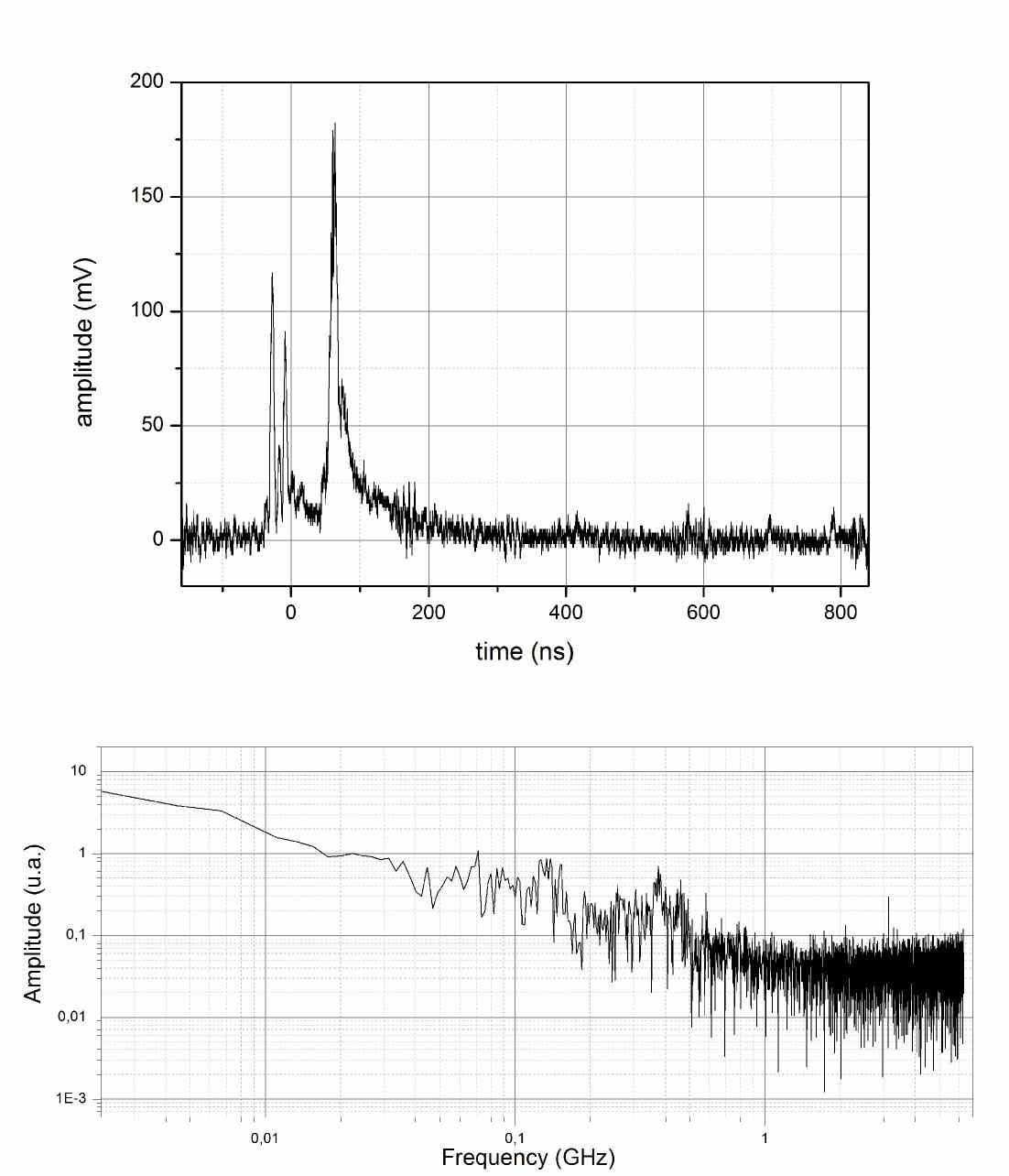}
\caption{\label{fig:FEfig1} 
Typical waveform (above) and Fourier transform(below) for an 8 mm diameter drift tube.
}
\end{center}
\end{figure}
An analysis of the spectral density of signals, done using a single 8 mm diameter drift tube with a $90\%-10\%$ mixture 
shows that signal bandwidth 
is on the order of 1 GHz \cite{baschirotto}. Fig. \ref{fig:FEfig1} shows both a typical waveform acquired using that detector and 
its Fourier transform. 
This result is used to set the overall bandwidth that the frontend electronics must ensure.
\begin{figure}[hbct]
\begin{center}
\includegraphics[width=7cm]{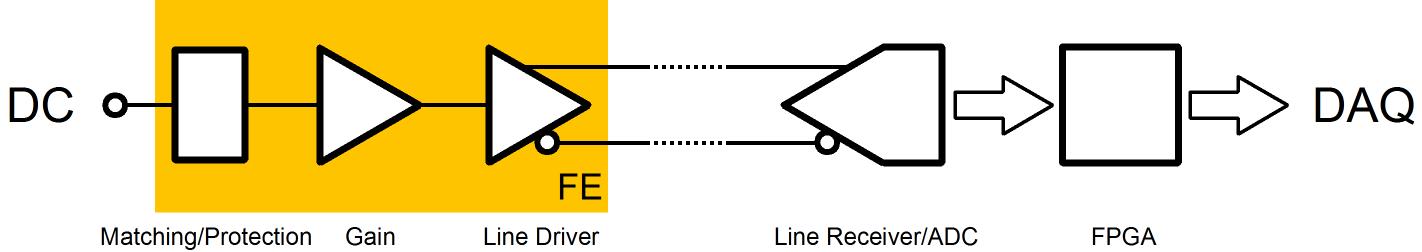}
\caption{\label{fig:FEfig2} 
Low noise and low distortion front-end scheme
}
\end{center}
\end{figure}
We are developing a multistage low noise and low distortion frontend (Fig. \ref{fig:FEfig2}) that provides a total voltage gain on the order of 
10 with a suitable bandwidth. 
The output of the frontend is differential to improve the noise immunity and is connected to the digitization stage through a twisted pairs line. 
To implement the frontend we use commercial components such as fast operational amplifiers that provide a gain-bandwidth product on the order of 1 GHz.
Through a careful survey of available devices on the market, a preliminary frontend was developed using a low noise and low distortion high speed 
operational amplifier (AD8099 from Analog Devices) as the first stage and a wideband low noise and low distortion fully-differential amplifier 
(THS4509 from Texas Instruments) as the second stage/line driver. 
\begin{figure}[hbct]
\begin{center}
\includegraphics[width=10cm]{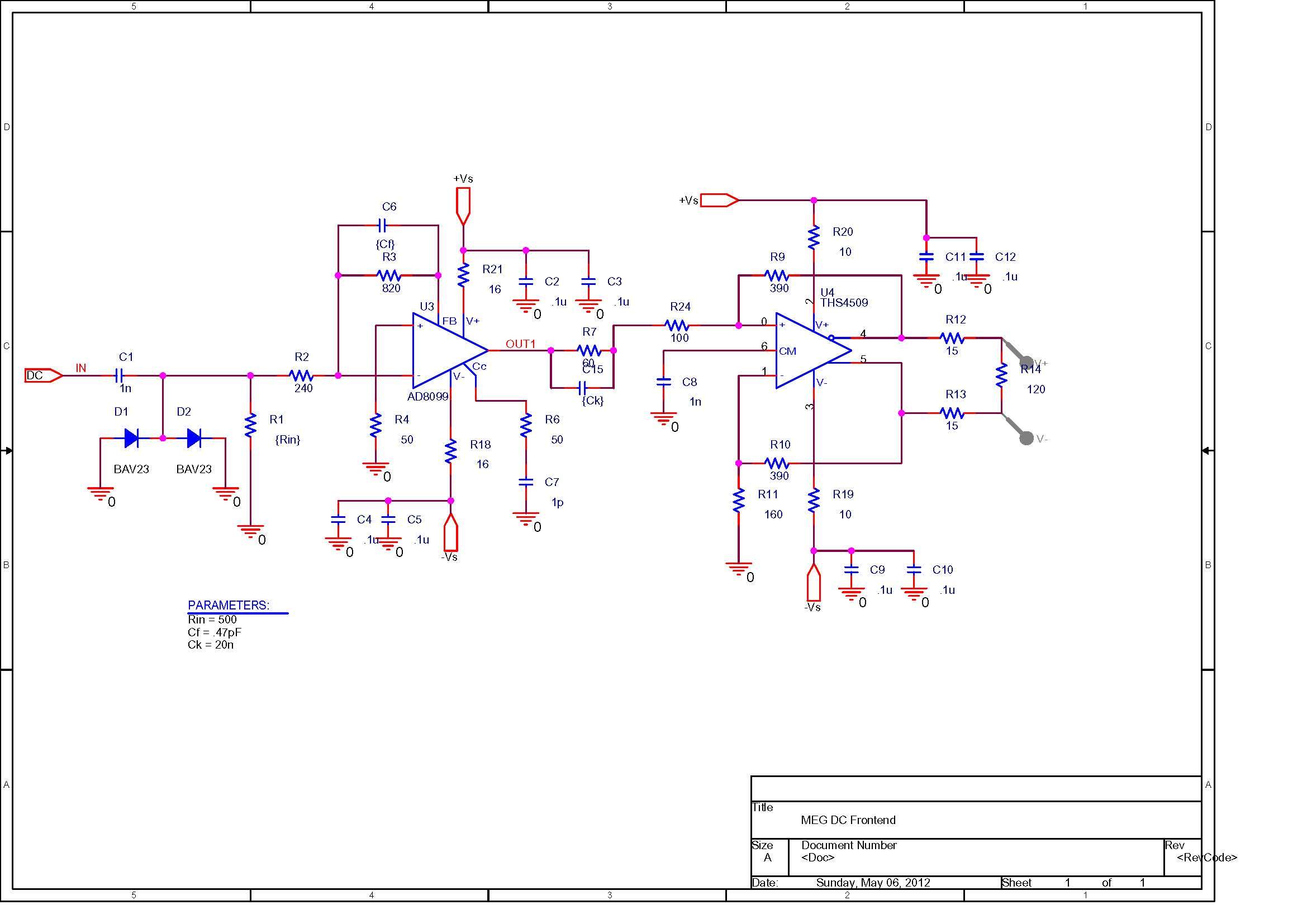}
\caption{\label{fig:FEfig3} 
Schematics of the front-end developed for the new MEG drift chamber
}
\end{center}
\end{figure}
A schematic diagram of this frontend is shown in Fig. \ref{fig:FEfig3}. 
A 3-channels frontend based on that schematic was assembled and tested using both signals from a pulse generator and from a setup of 3 aligned 8 mm
diameter drift tubes 30 cm long (Fig. \ref{fig:FEfig45} left). 
With this setup an integral non-linearity $<$ 1.5\% was measured (Fig. \ref{fig:FEfig45} right).
In the usual operative conditions ($90\% - 10\%$ mixture, 1.5 kV high voltage supply) the acquired signal shows a total drift time/number 
of clusters per pulse compatible with that expected ($\sim 200 \div 250 $ ns / $\sim10$) and an R.M.S. noise amplitude $< 2 $ mV (Fig. \ref{fig:FEfig67} left).
A very compact version of PCB for that frontend with a 7 mm pitch, was designed (Fig. \ref{fig:FEfig67} right) and successfully tested.

The DRS digitizer, developed for MEG,  which can work at digitizing frequencies up to 2.3 GHz, will be used for digitizing the chambers signal. Modifications of the current
boards are necessary, as will be discussed in section \ref{sec:Trigger and DAQ}, in order to increase the input bandwith, as requested by
the cluster timing technique, and to cope with the increased number of channels needed by the upgrade.

\subsubsection{Tests for spatial resolution studies}
\label{sec:tritubo}
The set-up in figure~\ref{fig:FEfig45} was used to give an estimate of the spatial resolution attainable, as a function of 
the gas mixture. Three $8$~mm drift tubes are vertically aligned, the central one being slightly displaced ($\sim 500~\mu$m) on 
one side. In this configuration, for vertical cosmic rays, it is possible to show that the combination of the drift distances in 
the tree tubes 
\begin{equation}
\pm \Delta = (d_1 + d_2)/2 - d_2
\label{eq:tritubo}
\end{equation}
gives the displacement of the central cell up to a sign ambiguity due to the particle 
crossing trajectory with respect to the three anode wires (double-peak distribution). 
The width of the two peaks is related to the single hit resolution by the formula
\begin{equation}
\sigma_\Delta = \sqrt{ \frac{2}{3} } \sigma_{\rm single hit}.
\end{equation}

\begin{figure}
\begin{center}
\includegraphics[width=1.0\columnwidth]{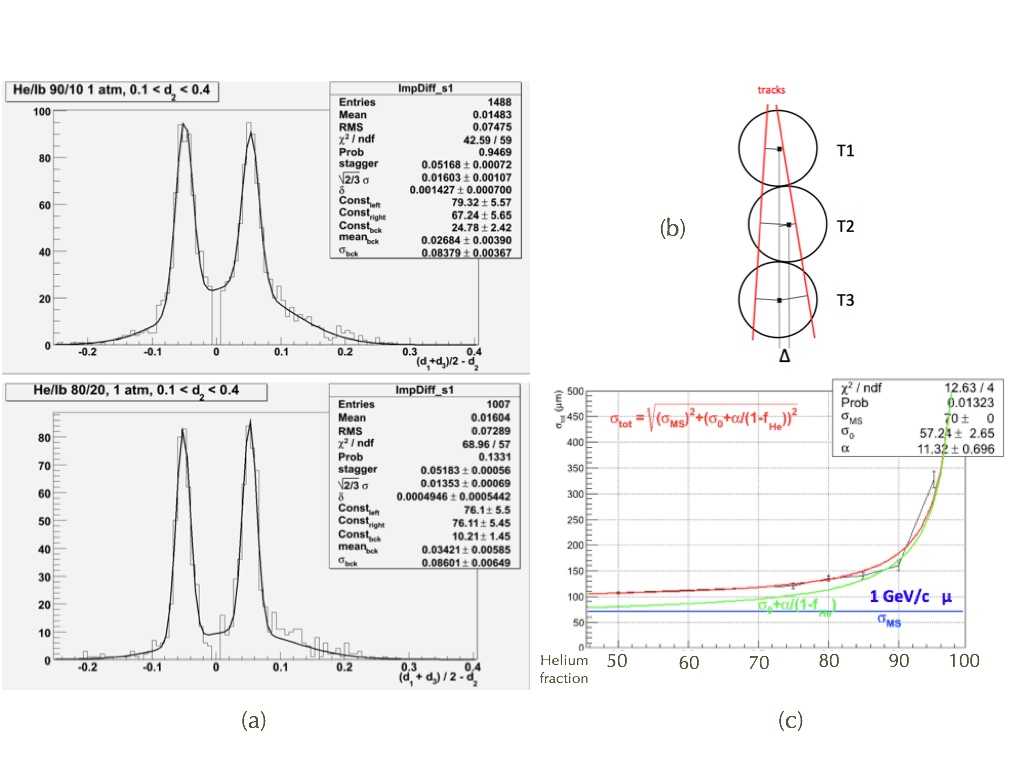}
\caption{\label{fig:risultati-tritubo}$(a)$ Distribution of the double peak in equation~\ref{eq:tritubo} for two representative
gas mixtures. $(b)$ A sketch of the three-tube set-up. $(c)$ Statistical deconvolution of the contribution of multiple scattering
on the three-tube set-up copper walls.}
\end{center}
\end{figure}

Figure~\ref{fig:risultati-tritubo} shows the double-peak distribution for two representative gas mixtures, while the single hit 
resolutions are reported in Table~\ref{tab:risultati-tritubo}, where the contribution of multple scattering 
on the CR muons from the copper tubes is included. The resolution ranges from $150~\mu$m for a $90:10$ mixture to $120~\mu$m for a 
$75:25$ mixture. In these measurements only information from the first cluster is used.
\begin{table}
\begin{center}
\begin{tabular}{cc}
Mixture & Resolution \\
(He:iC$_4$H$_{10}$)  & ($\mu$m)\\ \hline
5:95 & 330 $\pm$ 20\\ 
10:90 & 160 $\pm$ 11\\
15:85 &  140 $\pm$ 7\\
20:80 & 135 $\pm$ 7\\
25:75 & 120 $\pm$ 6\\
50:50 & 110 $\pm$ 7\\
\end{tabular}
\caption{\label{tab:risultati-tritubo} Summary of the resolutions obtained with the three-tube set-up. The single hit resolutions
contain the contribution of CR multiple scattering on the four 200~$\mu$m thick copper walls.}
\end{center}
\end{table}

\begin{figure}[hbct]
\includegraphics[width=0.5\textwidth, clip=true]{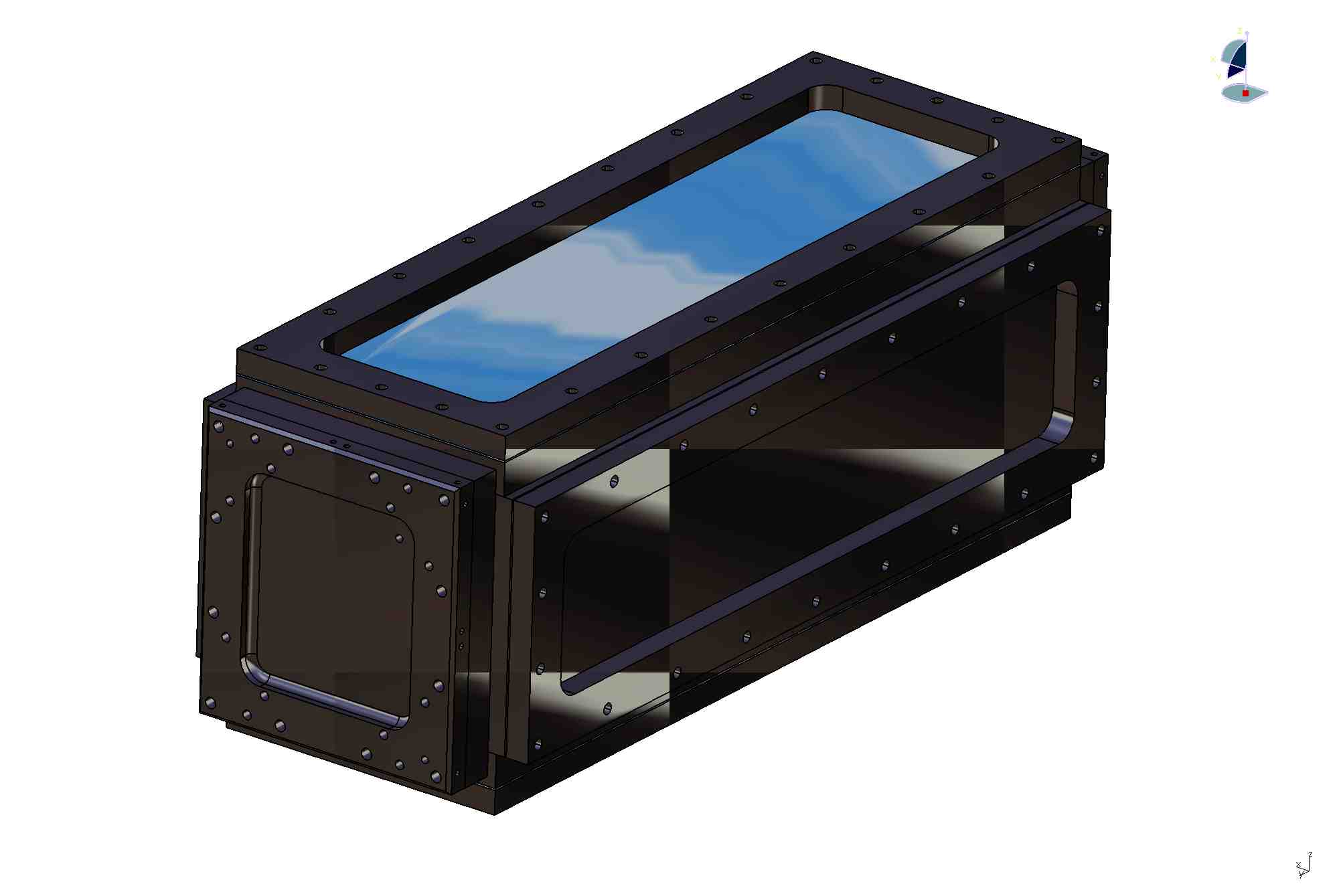}
\includegraphics[width=0.3\textwidth, clip=true]{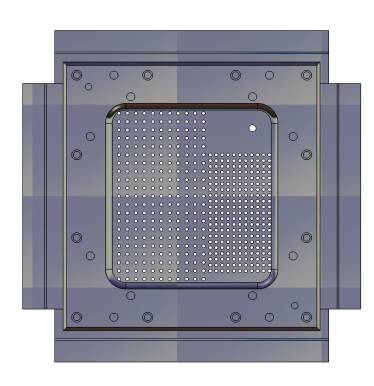}
\caption{\label{fig:PROTOTYPEbody} 
the chamber body (left) and the end cap with the hole pattern (right)}
\end{figure}

A larger prototype was designed and built to perform extended resolution studies:
Its body is  $50 \times 20 \times 20 \; \rm{cm}^3$, made of aluminum and hosts $8 \times 8$ square cells 
(see Fig. \ref{fig:PROTOTYPEbody})
with 7 mm side. 
The top and bottom longitudinal sides of the prototype are made of $50~\mu$m Kapton 
windows to minimize multiple scattering effects. 
For each cell a 25 $\mu$m gold plated tungsten 
sense-wire is surrounded by eight 80 $\mu$m field wires, also made of gold-plated tungsten. 
We use the sense wires feed-throughs made for the BaBar\cite{Babar} drift chamber, whereas for the
field wires, due to the limited space available, we designed new ones with smaller diameter (0.15 mm internal hole,
1.2 mm overall diameter) feed-throughs.
Extensive studies on the  achievable spatial resolutions  with the  proposed chamber geometry were made by using the GARFIELD package \cite{garfield}.
We plan to equip the central region of the detector  with the newly designed high bandwidth readout electronics
described in the previous section whereas the remaining wires will be
read-out by using the KLOE electronics.
Signals will be digitized by the MEG DRS digitizer.

\begin{figure}[hbct]
\begin{center}
	\includegraphics[width = 0.9\linewidth]{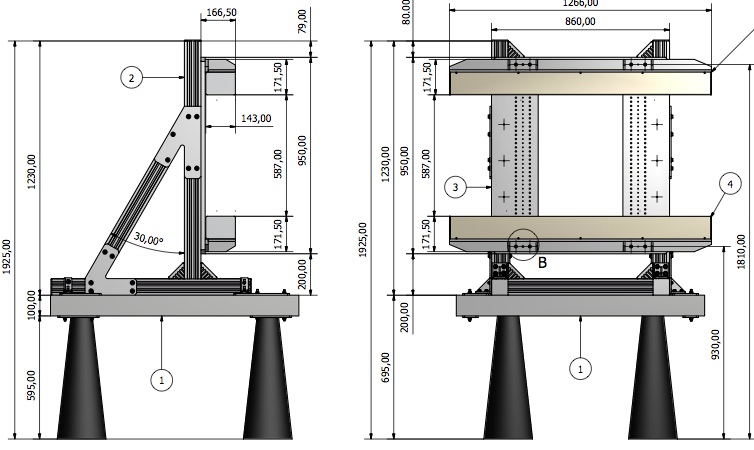}
\end{center}
\caption{Technical design of the telescope assembly for the test of detector prototypes.}
\label{fig:tele}
\end{figure}

In parallel, we are building a telescope for cosmic rays as a test facility of the performances of our prototype.  This telescope consists, as shown in Fig.\ref{fig:tele}, of an assembly of four double-side silicon layers of SVT, the former vertex detector of the BaBar experiment. 
With strip sizes of 50 and 100 $\mu$m respectively for the transverse and the longitudinal view, this
apparatus is suited to provide a reference for position reconstruction for the prototype detector. The single hit resolution is of the order of 20 $\mu$m for the transverse and 40 $\mu$m for the longitudinal view which correspond  in the case of vertical crossing, to an accuracy $\sim 10 \, \mu$m (one order of magnitude better than the expected position resolution of the prototype detector) on the intercept of the cosmic ray tracks with the  detector plane.  
Two Aluminum boxes, designed to provide proper servicing (clean air, temperature control, darkness) to silicon hybrids, host two layers each and are located above and beneath the test detector. To ensure adequate mechanical stability, these modules are stuck to a section-bar frame equipped with machine-controlled holes; this allows us to move the two boxes away, depending on
the size of the detector prototypes, with comparable precision. The whole structure is eventually mounted on the top of a 15-cm thick granite table. 

The trigger of through-going cosmic muons is accomplished by the coincidence of two scintillation counters, which are also used to provide the time reference for the measurement of the electron drift in every detector cell. An old-fashioned DAQ system, once used in BaBar to handshake with silicon hybrid chips, was replaced by custom designed boards (conceived as an interface with the hybrids) coupled with an acquisition front-end PC through commercial FPGA evaluation boards.

\subimport{../06_Positron_Timing_Counter/}{Positron_Timing_Counter.tex}

%% file: 06_Positron_Timing_Counter/Positron_Timing_Counter.tex
\subsection{Pixelated Timing Counter}
%
%

\label{sec:timingcounter}
\subsubsection{Foreword and Status of the Art}
The present timing counter showed an excellent intrinsic time resolution of 40\, ps
in beam tests before the experiment.
The base design was based on multiple, tilted and parallel thick scintillator bars 
to make more uniform the response 
to circular positron tracks and improve the signature with multiple hits on contiguous bars. 
But, the best operating time resolution on the experimental floor has been measured to be worse,
65\,ps, which includes the electronics time jitter. The main issues that cause this worsening are summarized below:
\begin{itemize}
\item the spectrometer magnetic field increases the PMT TTS (5\%) and reduces the gain (up to factor 30): the consequent lower pulse amplitude causes larger time-walk effects: the statistical fluctuation of the residuals of the time walk corrections are calculated to contribute for about 20\,ps;
\item the scintillation pulse width is increased by the large $z$ projection of the tracks that gives an excess time spread of 20\,ps;
\item the electronic time jitter contributes for about 40\,ps. 
\end{itemize}
The sum of these contributions accounts for the above mentioned operating timing resolution. 
Further, since our PMTs work at the far edge of the performance vs. single event rate 
(1\,MHz per PMT),
the expected increase of a factor 3 of the muon decay rate requires 
a segmentation of at least the same factor with respect to the present design, 
i.e., 90 bars and 180 PMTs, in order to preserve the proper PMT working point. 
An improved concept of the upgraded timing counter that overcomes these limits must have: 
\begin{itemize}
\item magnetic field insensitive photo-sensors with lower TTS and higher quantum efficiency; 
\item a design that is insensitive to the Z projection of tracks; 
\item a higher level of segmentation;
\end{itemize} 
Here we present a new pixelated timing counter that meets these requirements.

\subsubsection{Concept of New Detector}
\begin{figure}[htb]                                                                                                                                                                  
\begin{center}
\includegraphics[width=14cm]{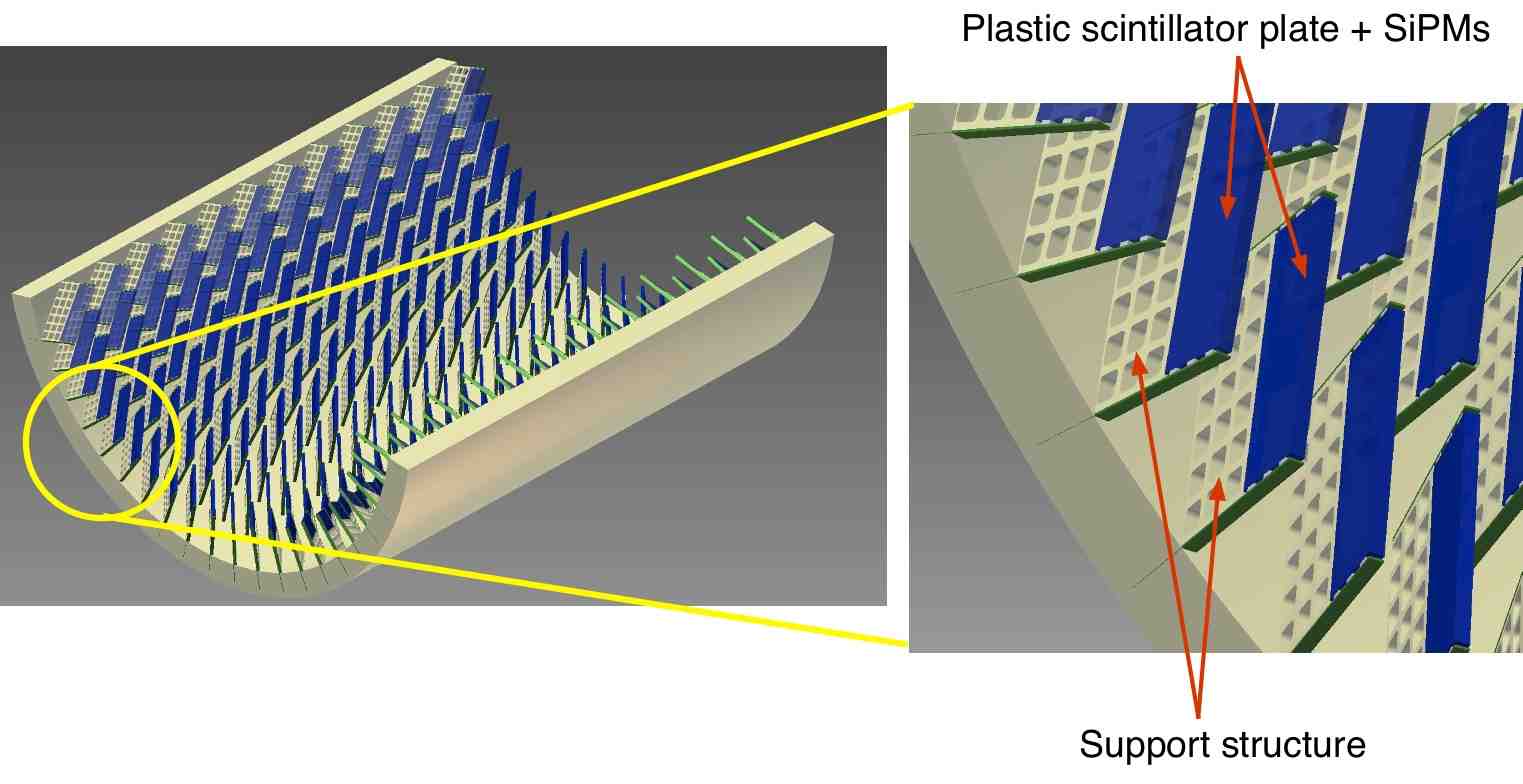}
\caption{\label{fig:Schematics of pixelated timing counter}
   Pixelated timing counter composed of many small scintillator plates. 
}
\end{center}
\end{figure}

The pixelated timing counter is composed of two sets of semi-cylindrical shape scintillation detectors similarly to the current timing counter, 
    but each detector is highly segmented as shown in Fig.\,\ref{fig:Schematics of pixelated timing counter}.
Each segment pixel is a small ultra-fast scintillator plate with silicon photomultiplier (SiPM) readout.
There are a lot of advantages in this detector concept over the current timing counter as follows.

\begin{itemize}

\item The single plate is expected to have a good timing resolution since ambiguity in the positron path length inside the plate 
and also in the scintillation light propagation time to the photo-sensor is small.
It should be noted that a good performance was already proved with a counter with a similar configuration 
in early studies for the PSI $\mu$SR detector\,\cite{StoykovNDIP11_slides}\cite{StoykovNDIP11}.
\item Most of the signal positrons passes through more than one pixels. 
   Proper averaging of the times measured at the hit pixels gives more precise estimation of the positron impact time.

\item The hit rate at each segment pixel is lower than 1\,kHz even at a high beam rate of $10^8\,\mu/\mathrm{sec}$.  
   The pileup probability is quite low. 

\item The multiple pixel hits can provide additional track information.

\item For the current timing counter, a positron sometimes leaves double hits in a single timing counter bar,
   which produces a tail component in the timing response function.
   This problem will not happen in the pixelated timing counter.

\item The proposed photo-sensor (SiPM) is insensitive to magnetic field. Note that the detector is placed in the bore of the spectrometer magnet COBRA.

\item The detector is operational in the COBRA bore filled with helium gas
   in contrast to the current detector with PMTs, which is now housed in a helium-tight plastic bag
   constantly flushed with dry air.

\item Flexible detector layout is possible since the position and angle of each pixel module can be adjusted individually.
 
\end{itemize}

\subsubsection{Pixel Module Design}

\begin{figure}[htb]                                                                                                                                                                  
\begin{center}
\includegraphics[width=10cm]{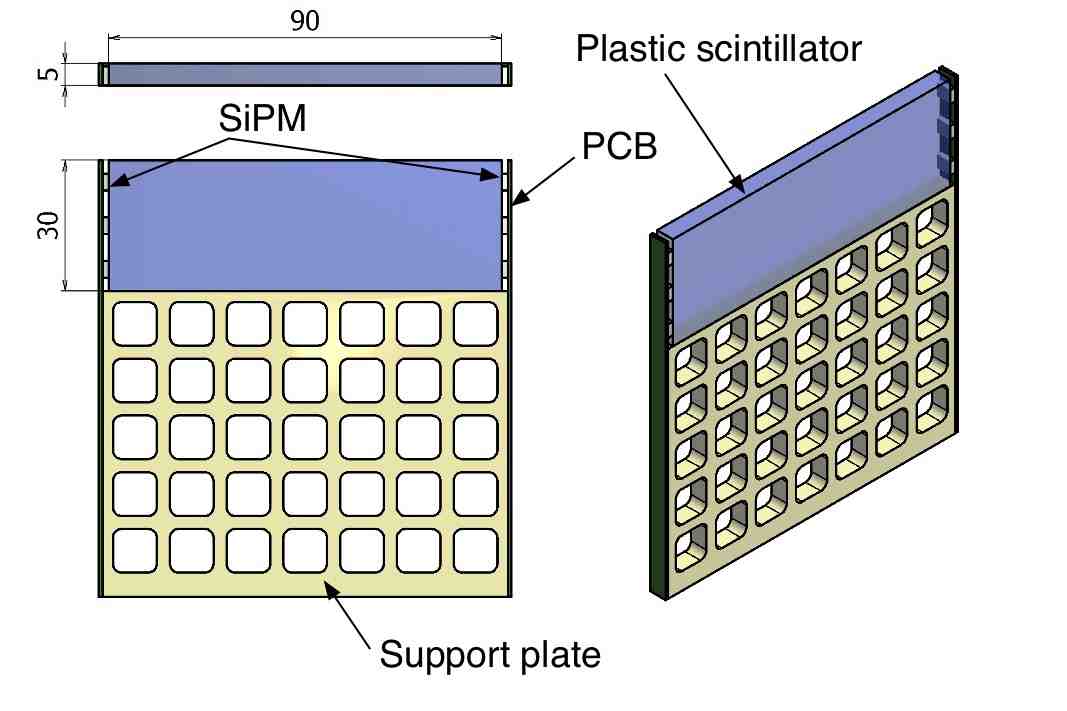}
\caption{\label{fig:pixel module design}
   Possible design of the single pixel module.
}
\end{center}
\end{figure}

Fig.\,\ref{fig:pixel module design} shows a possible design of the single pixel module.
The geometry of the scintillator plate is not fully optimized yet, but the typical dimension can be
$30\,(\mathrm{width}) \times 90\,(\mathrm{length}) \times 5\,(\mathrm{thickness})\,\mathrm{mm}^3$.
The scintillation light is collected by three or four SiPMs at either end of the scintillator plate.
The SiPMs at each end are connected in series and 
the summed signal is directly sent to a waveform digitizer module, which is described in Sec.\,\ref{sec:Trigger and DAQ}.
The positron impact time for the single pixel is obtained by averaging the times measured at both sides.

\subsubsection{Scintillator}
\label{sec:pix TC Scintillator}
\begin{figure}[htb]                                                                                                                                                                  
\begin{center}
\includegraphics[width=10cm]{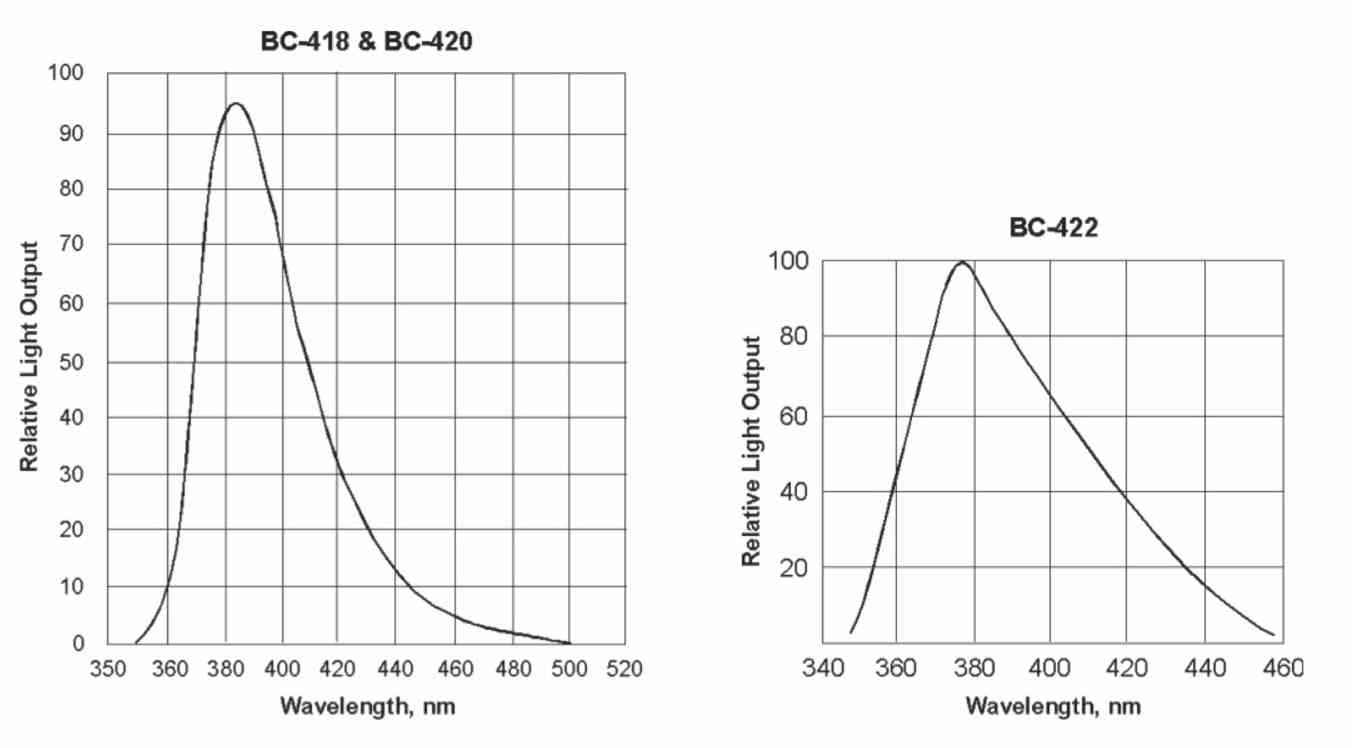}
\caption{\label{fig:BC4XX emission spectra}
   Scintillation light emission spectra of Saint-Gobain BC418, BC420 and BC422\,\cite{Saint-Gobain-BC4XX}.
}
\end{center}
\end{figure}

The choice of the scintillator material is crucial to achieving the best time resolution.
The parameters to be considered are light yield, rise time, decay time and emission spectrum of the scintillation.
The possible candidates of the scintillation materials are the ultra-fast plastic scintillators
from Saint-Gobain listed in Table\,\ref{table:Saint-Gobain PS}\,\cite{Saint-Gobain-BC4XX}.
The scintillation light emission spectra are shown in Fig.\,\ref{fig:BC4XX emission spectra}
BC-422 would be the best for the timing measurement for a small size plate (shorter than several cm) because of its short rise time but with a relatively short attenuation length, 
while BC-418 or BC-420 would be better for a longer plate because of its longer attenuation length and higher light yield.

\begin{table}[tb]
\caption{\label{table:Saint-Gobain PS}
Properties of ultra-fast plastic scintillators from Saint-Gobain (BC-418, 420, 422)\,\cite{Saint-Gobain-BC4XX}. 
The properties of BC-404, which is used in the present timing counter bar, is also shown for comparison.
}
\begin{center}
\begin{tabular}{lccc|c}
   \\{\bf Properties}                 & {\bf BC-418} & {\bf BC-420} & {\bf BC-422} & {\bf BC-404} \\\hline\hline
   Light Output [\% Anthracene]       & 67           & 64           & 55           & 68\\
   Rise Time [ns]                     & 0.5          & 0.5          & 0.35         & 0.7\\
   Decay Time [ns]                    & 1.4          & 1.5          & 1.6          & 1.8\\
   Wavelength of Max. Emission [nm]   & 391          & 391          & 370          & 408\\
   Bulk Light Attenuation Length [cm] & 100          & 110          & 8            & 140\\
\end{tabular}
\end{center}
\end{table}

Single crystal {\it p}-Terphenyl is under study as another candidate for the scintillator material with potentially better timing performance,
which is described in detail in Appendix \ref{sec:p-Terphenyl}.

\subsubsection{Silicon Photomultiplier (SiPM)}
\label{sec:pixTC SiPM}
Silicon photomultiplier (SiPM) is a new type of semiconductor photo-sensor,
which is considered as a possible replacement of the conventional PMT
because of its excellent properties as listed below.

\begin{itemize}
   \item Compact size
   \item Sensitive to single photon
   \item High internal gain ($10^5$--$10^6$)
   \item High photon detection efficiency peaked at $\lambda \sim 450\,\mathrm{nm}$ 
   \item Insensitive to magnetic field
   \item Low bias voltage ($<100\,\mathrm{V}$)
   \item Excellent time resolution ($<100\mathrm{ps}$ for single photo-electron)
   \item low power consumption
   \item No avalanche fluctuation (excess noise factor $\sim$1--1.5)
\end{itemize}

\begin{figure}[htb]                                                                                                                                                                  
\begin{center}
\includegraphics[width=6cm]{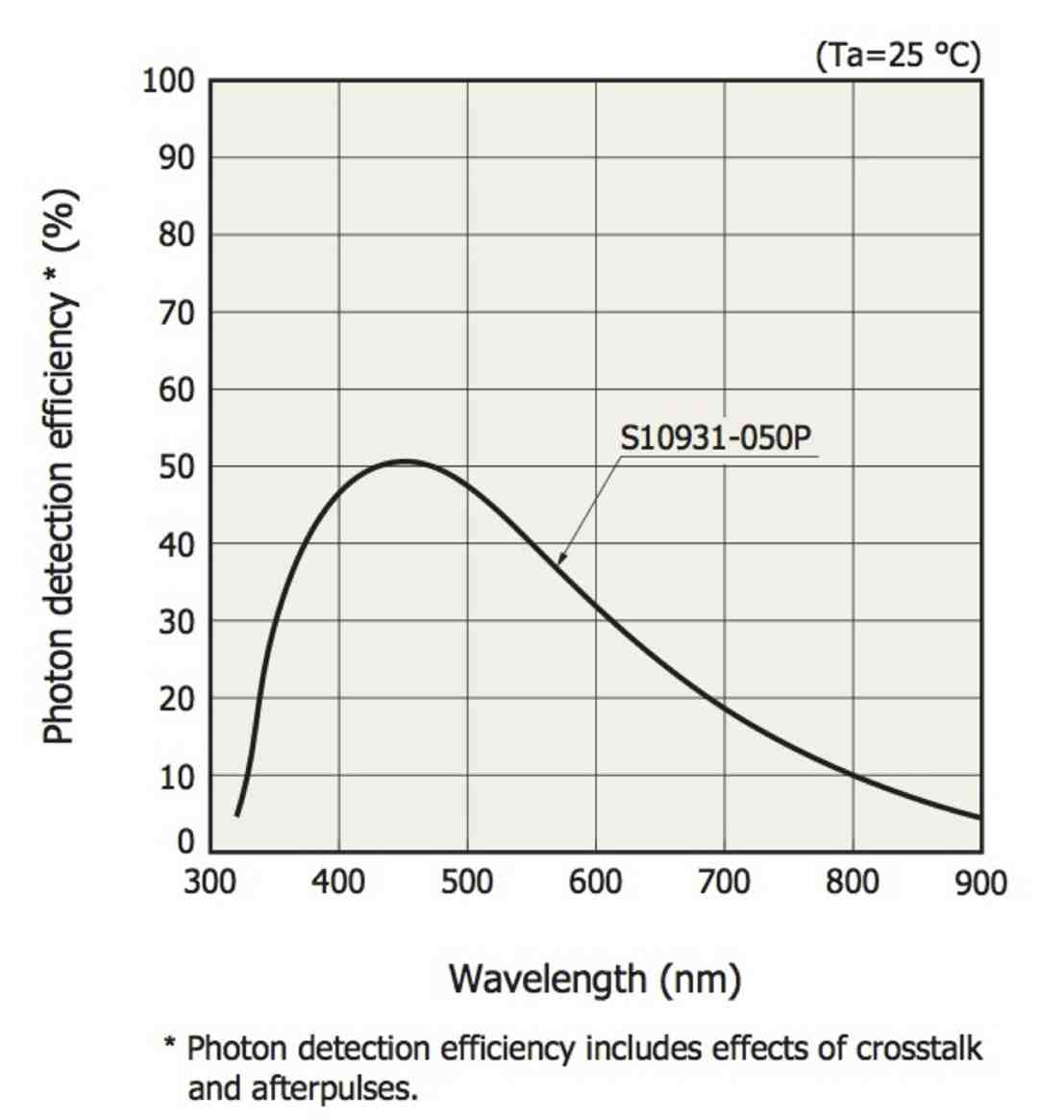}
\caption{\label{fig:MPPC S10931 PDE}
Photon detection efficiency (PDE) as a function of wavelength for Hamamatsu MPPC S10931-050P\,\cite{MPPC:S10931-050P}.
}
   
\end{center}
\end{figure}

It should also be the best solution to the scintillation readout for the pixel module of this detector.
The candidates for the SiPM for the detector are 
Hamamatsu S10931-050P\,\cite{MPPC:S10931-050P}, FBK-AdvanSiD ASD-SiPM3S-P-50\,\cite{FBK-AdvanSiD},
KETEK PM3350\,\cite{KETEK-PM3350}.
The specifications of Hamamatsu MPPC S10931-050P are listed in Table\,\ref{table:MPPC S10931 spec} as an example.
The photon detection efficiency (PDE) as a function of wavelength is shown in Fig.\,\ref{fig:MPPC S10931 PDE},
which is reasonably matched to the scintillation emission spectra of the Saint-Gobain plastic scintillator shown in Fig.\,\ref{fig:BC4XX emission spectra}.

\begin{figure}[htb]                                                                                                                                                                  
\begin{center}
\includegraphics[width=6cm]{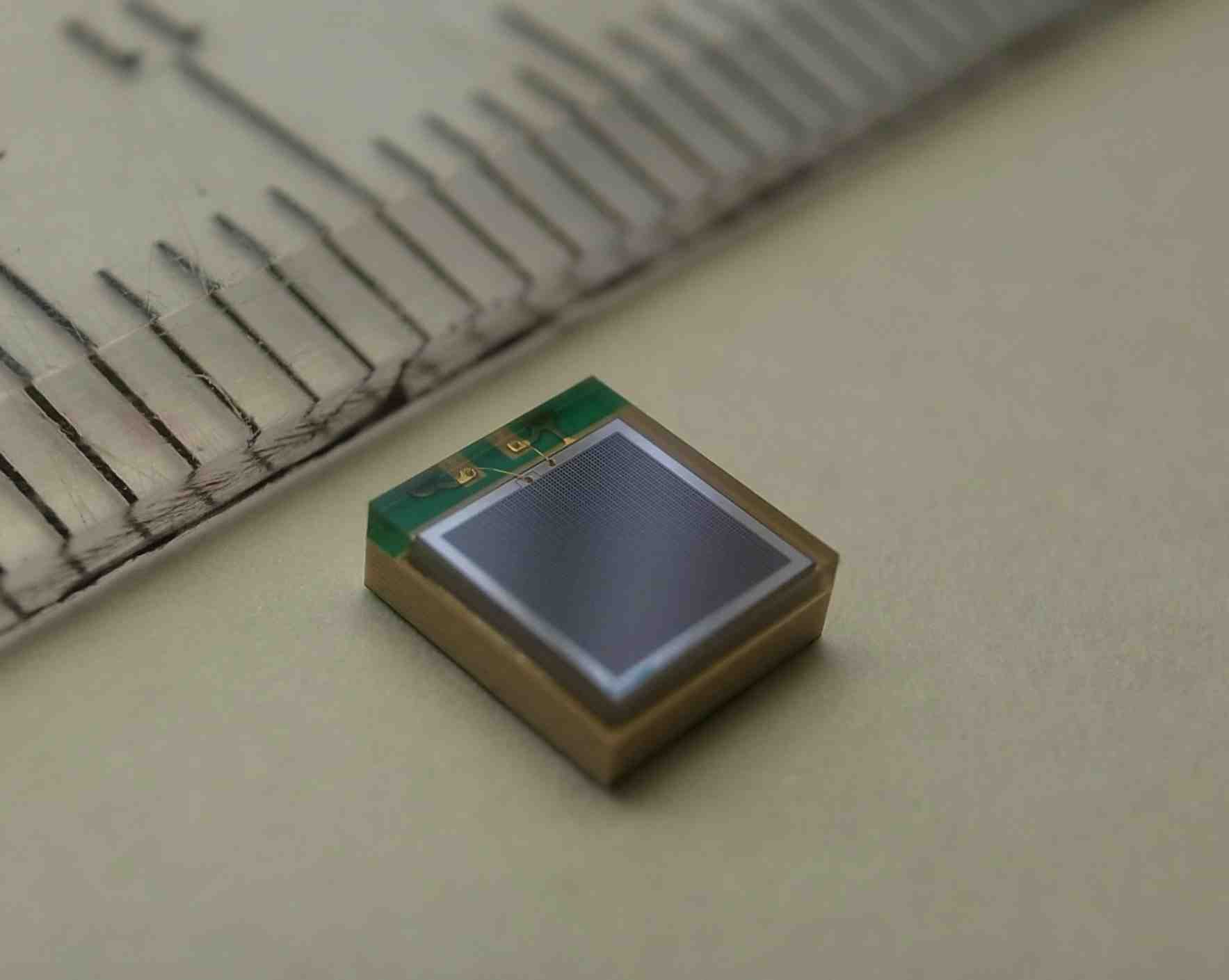}
\caption{\label{fig:MPPC S10931}
Hamamatsu MPPC S10931-050P.
}
   
\end{center}
\end{figure}

\begin{table}[tb]
\caption{\label{table:MPPC S10931 spec}
Specifications of Hamamatsu MPPC S10931-50P\,\cite{MPPC:S10931-050P}.
}
\begin{center}
\begin{tabular}{|l|c|}
   \hline
   Active area & $3\times 3\,\mathrm{mm}^2$\\
   Pixel pitch & $50\times 50\,\mu\mathrm{m}^2$\\
   Number of pixels & 3600 \\
   Geometrical fill factor & 61.5\%\\
   Peak wavelength & 440\,nm \\
   Operating voltage & $70\pm10\,\mathrm{V}$\\
   Temperature coefficient of breakdown voltage & 56\,mV/$^\circ$C\\
   Gain  & $7.5\times 10^5$\\
   Dark count rate & 6\,Mcps\\
   \hline
\end{tabular}
\end{center}
\end{table}

\subsubsection{Support Structure}
The requirement for the precision of the pixel alignment is not so severe.
A precision of 1\,mm, which correspond to the misalignment in time of a few ps, is good enough.
Possible design of the support structure for the pixelated timing counter is seen in Fig.\,\ref{fig:Schematics of pixelated timing counter}.
Each pixel module is supported by a 5\,mm-thick and 40\,mm-high plastic plate with the same length as the pixel module,
which is fixed to a 40\,mm-thick semi-cylindrical base made of plastic.
Several holes are punched on the support plate to minimize undesirable multiple scattering of the positron.
The possible candidate of the support structure material is Delrin plastic (polyoxymethylene).

\subsubsection{Test with Pixel Counter Prototype}
\label{sec:Test with Pixel Counter Prototype}
A good performance of a scintillator counter with a similar configuration
was already demonstrated in the early studies for the PSI $\mu$SR facility\,\cite{StoykovNDIP11_slides}\cite{StoykovNDIP11}.
Nevertheless, a couple of prototype counters are built in order to measure and 
to optimize the performance of the single pixel counter 
with the realistic configuration in our experiment. 

\begin{figure}[htb]                                                                                                                                                                  
\begin{center}
\includegraphics[width=7cm]{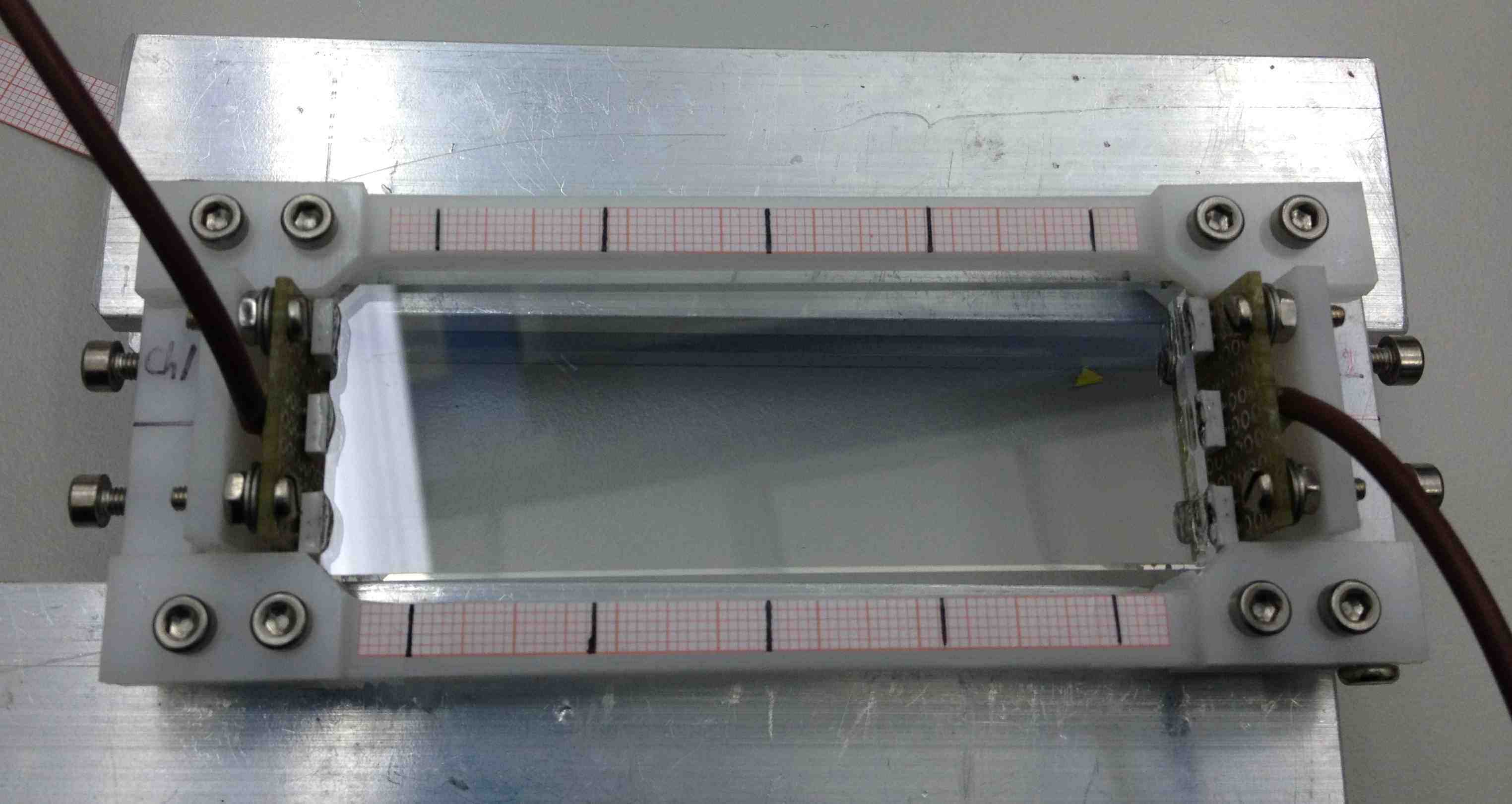}
\includegraphics[width=5cm]{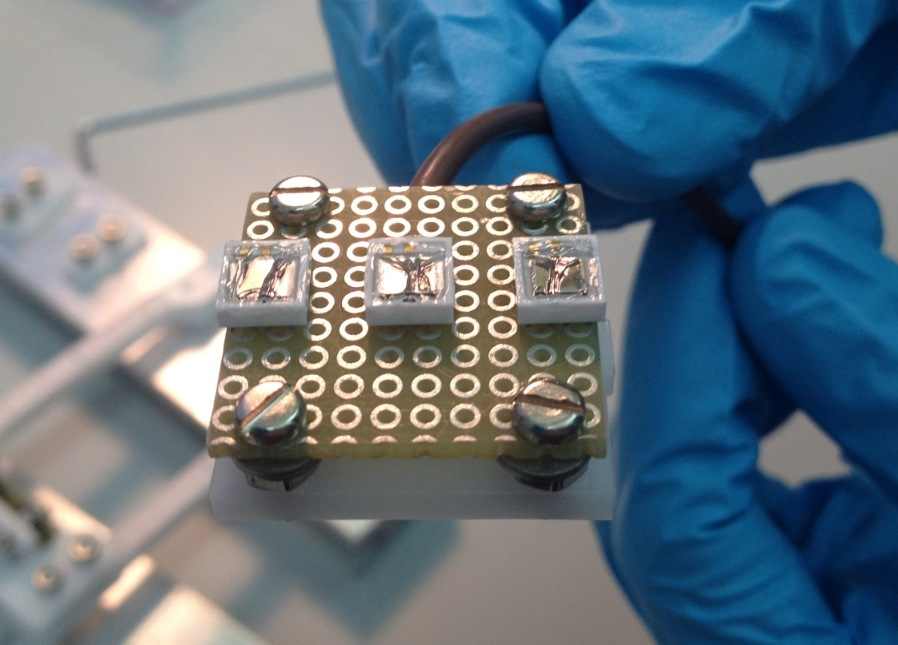}
\caption{\label{fig:prototype counter}
(Left) pixel counter prototype with a plastic scintillator (BC422, $30\times90\times5\,\mathrm{mm}^3$) and six MPPCs (Hamamatsu S10362-33-050C).
(Right) three SiPMs mounted on PCB. Optical grease is put on the sensor surface.
}
   
\end{center}
\end{figure}

Fig.\,\ref{fig:prototype counter} shows one of the prototype counters, 
where a plastic scintillator (BC422) of $30\times90\times5\,\mathrm{mm}^3$ 
is mounted in a supporting frame made of Delrin plastic.
Three SiPMs, which are connected in series, are optically coupled to each end of the scintillator plate
with optical grease (OKEN6262A).

\begin{figure}[htb]                                                                                                                                                                  
\begin{center}
\includegraphics[width=12cm]{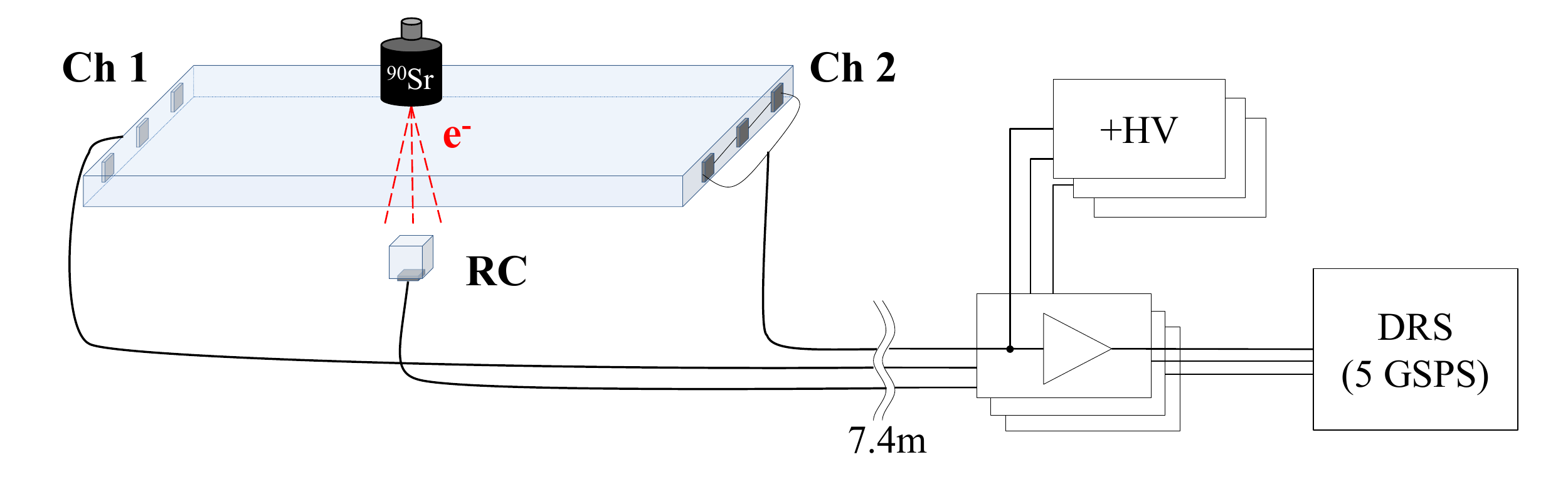}
\caption{\label{fig:prototype test setup}
Typical setup of the measurement with the pixel counter prototype where RC denotes the reference counter.
}
   
\end{center}
\end{figure}

A typical setup of the measurement is shown in Fig.\,\ref{fig:prototype test setup}.
The prototype counter is irradiated by electrons from $^{90}$Sr with an energy up to 2.28\,MeV.
A counter with a small plastic scintillator (BC422, $5\times5\times5\,\mathrm{mm}^3$) readout by a single MPPC, 
whose timing resolution is measured to be 28\,ps in $\sigma$,
is placed behind the prototype counter for the triggering and the collimation of the electrons from $^{90}$Sr.
The output signals from the two SiPM chains in the prototype counter are amplified 
with an optimum signal shaping in a voltage amplifier developed at PSI 
and are then readout by a waveform digitizer (DRS4),
which is also in-house developed at PSI\,\cite{Ritt2010486}.
In the final detector the signals from the pixel counters are transmitted
with a coaxial cable as long as 7\,m
directly to the readout electronics  
without any amplification.
The measurements with the prototype counters are
performed with a RG178-type coaxial cable of 7.4\,m long between the SiPMs and the amplifier
to simulate the realistic condition in the final detector.

\begin{figure}[htb]                                                                                                                                                                  
\begin{center}
\includegraphics[width=8cm]{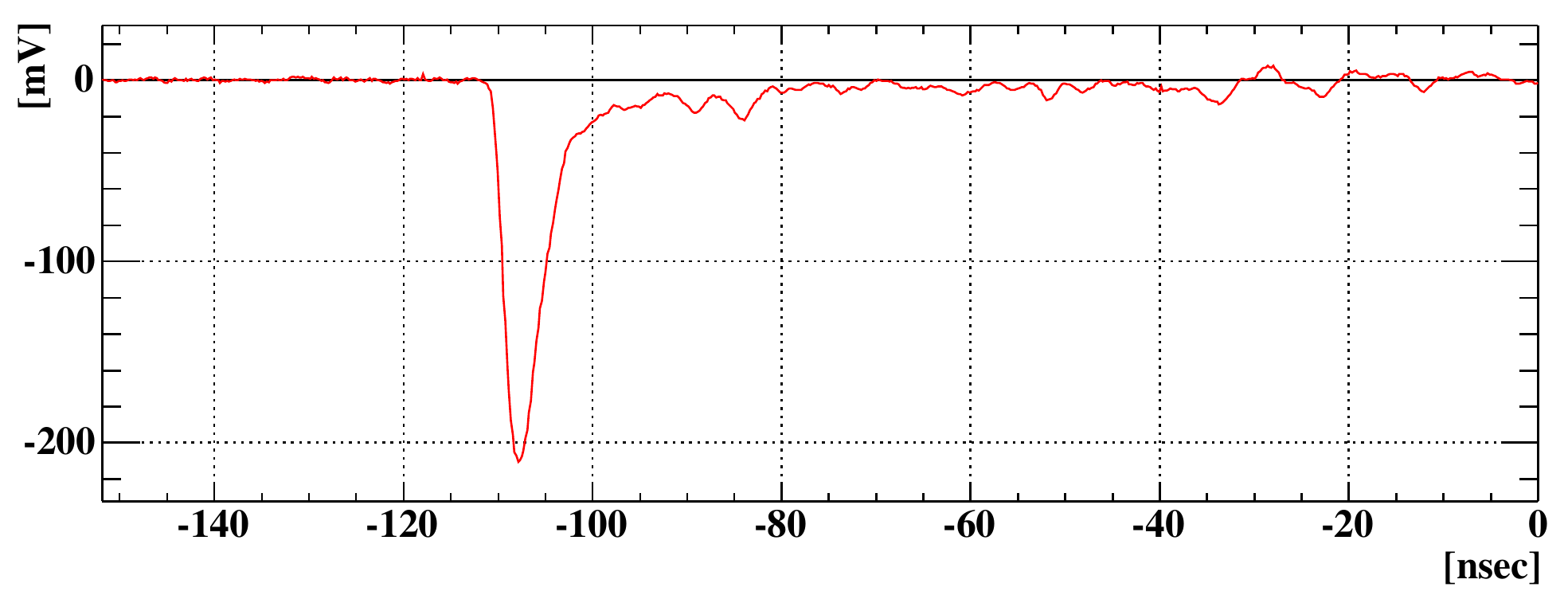}
\caption{\label{fig:prototype waveform}
Typical waveform from the prototype counter (BC422 of $30\times90\times5\,\mathrm{mm}^3$, six MPPCs).
}
\end{center}
\end{figure}

The waveforms as shown in Fig.\,\ref{fig:prototype waveform} are analyzed offline 
to estimate the electron impact time.
The waveform time is extracted by the constant-fraction method. 
The electron impact time is then calculated by taking the difference 
between the average of the waveform times measured at both sides ($t_0$ and $t_1$)
and the time measured at the reference counter ($t_\mathrm{ref}$), 
namely $(t_0 + t_1)/2-t_\mathrm{ref}$, without any correction.

We are trying to find an optimal configuration of the single pixel counter by testing the prototype counters 
with different conditions such as scintillator types and geometries, reflector types and number of SiPMs.
Hamamatsu MPPC S10362-33-050C is used for the following measurements 
although we plan to compare the performance of SiPMs from different manufacturers.

\begin{figure}[htb]                                                                                                                                                                  
\begin{center}
\includegraphics[width=7cm]{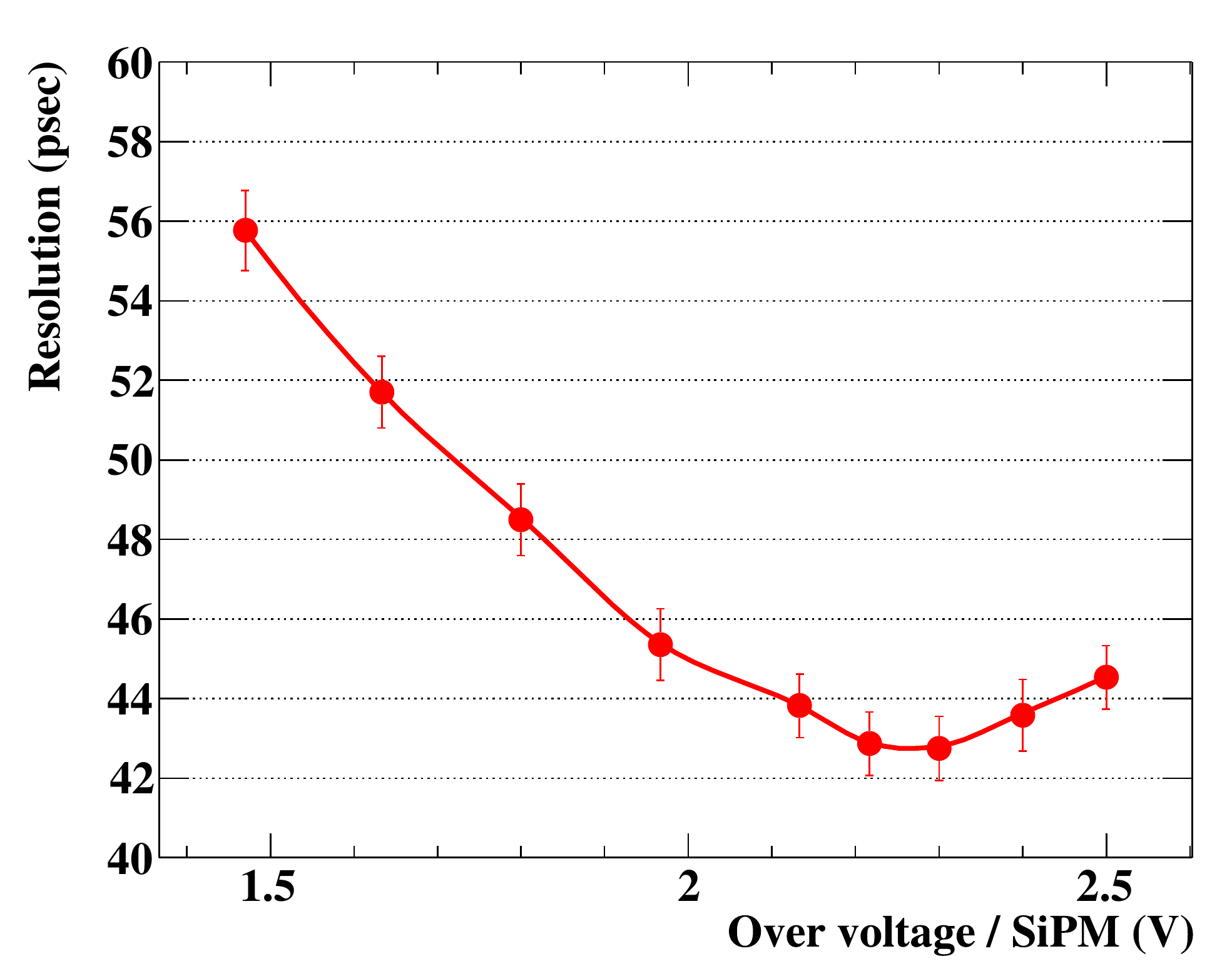}
\caption{\label{fig:resolution bias dependence}
Typical over-voltage dependence of the timing resolution ($\sigma$) of a prototype counter
(scintillator: BC422, $30\times60\times4.5\,\mathrm{mm}^3$, SiPM: 3 MPPCs on each side, reflector: 3M radiant mirror film),
where the contribution from the timing resolution of the reference counter (28\,ps in $\sigma$) is already subtracted.}
\end{center}
\end{figure}

Fig.\,\ref{fig:resolution bias dependence} shows the time resolution in $\sigma$ 
as a function of the over-voltage per single SiPM.
The resolution improves as the over-voltage increases up to about 2.3\,V, 
and then deteriorates at a higher over voltage due to the increased dark noise. 
The following measurements are performed around the optimum over-voltage. 

\begin{figure}[htb]
  \begin{center}
    \subfigure[No reflector]{
     \includegraphics[width=0.48\linewidth]
     {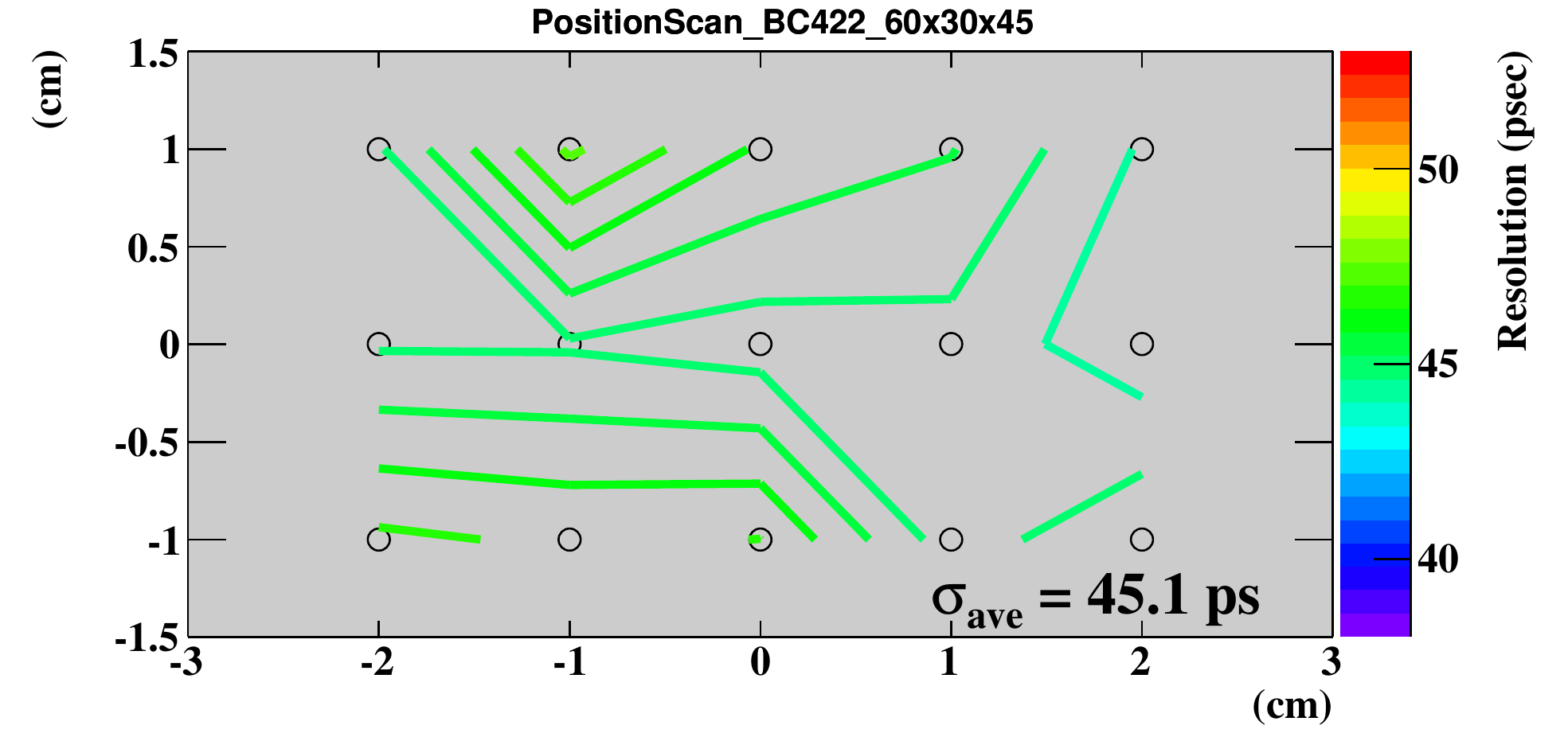}
      }
    \subfigure[Teflon tape]{
     \includegraphics[width=0.48\linewidth]
     {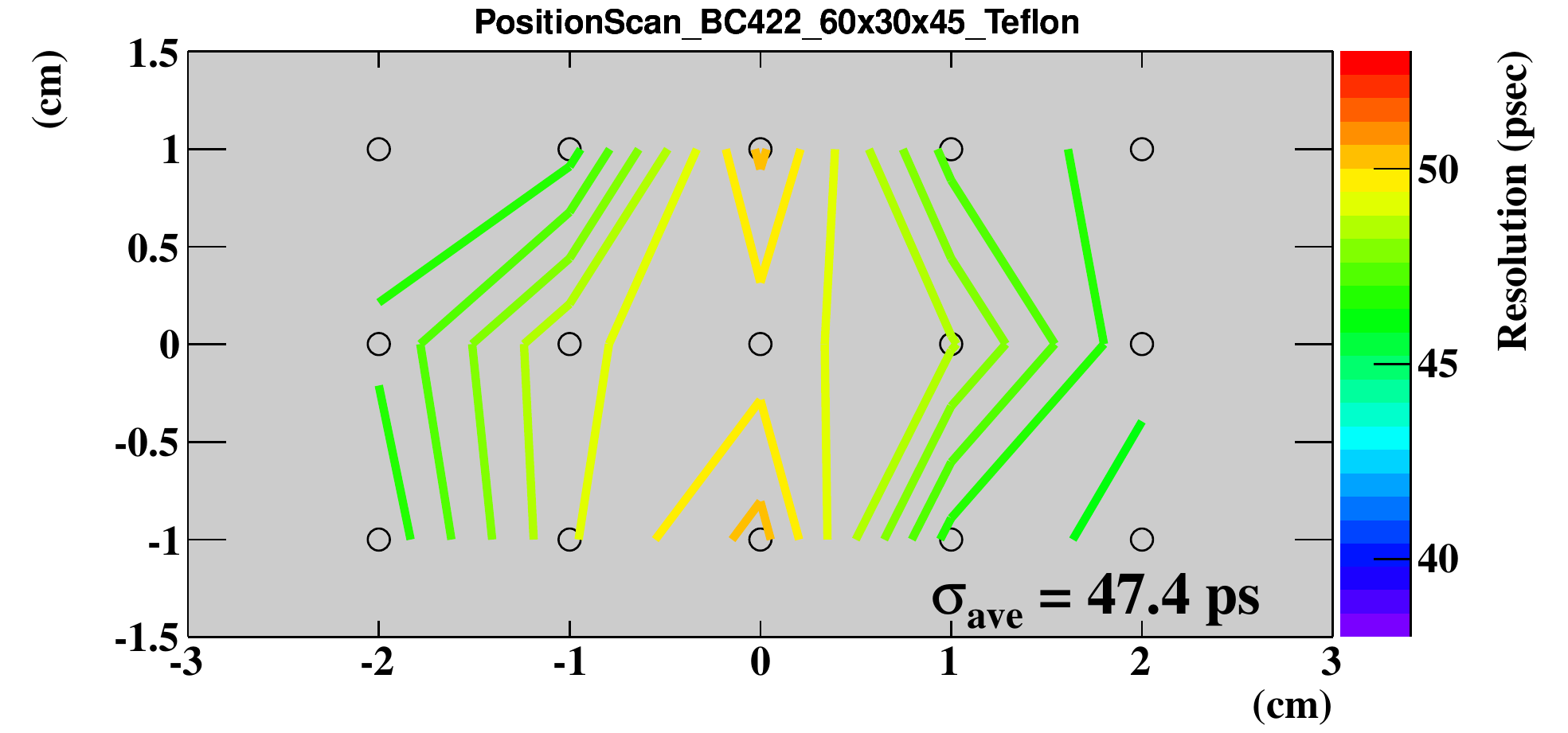}
    }
    \subfigure[Aluminized mylar]{
     \includegraphics[width=0.48\linewidth]
     {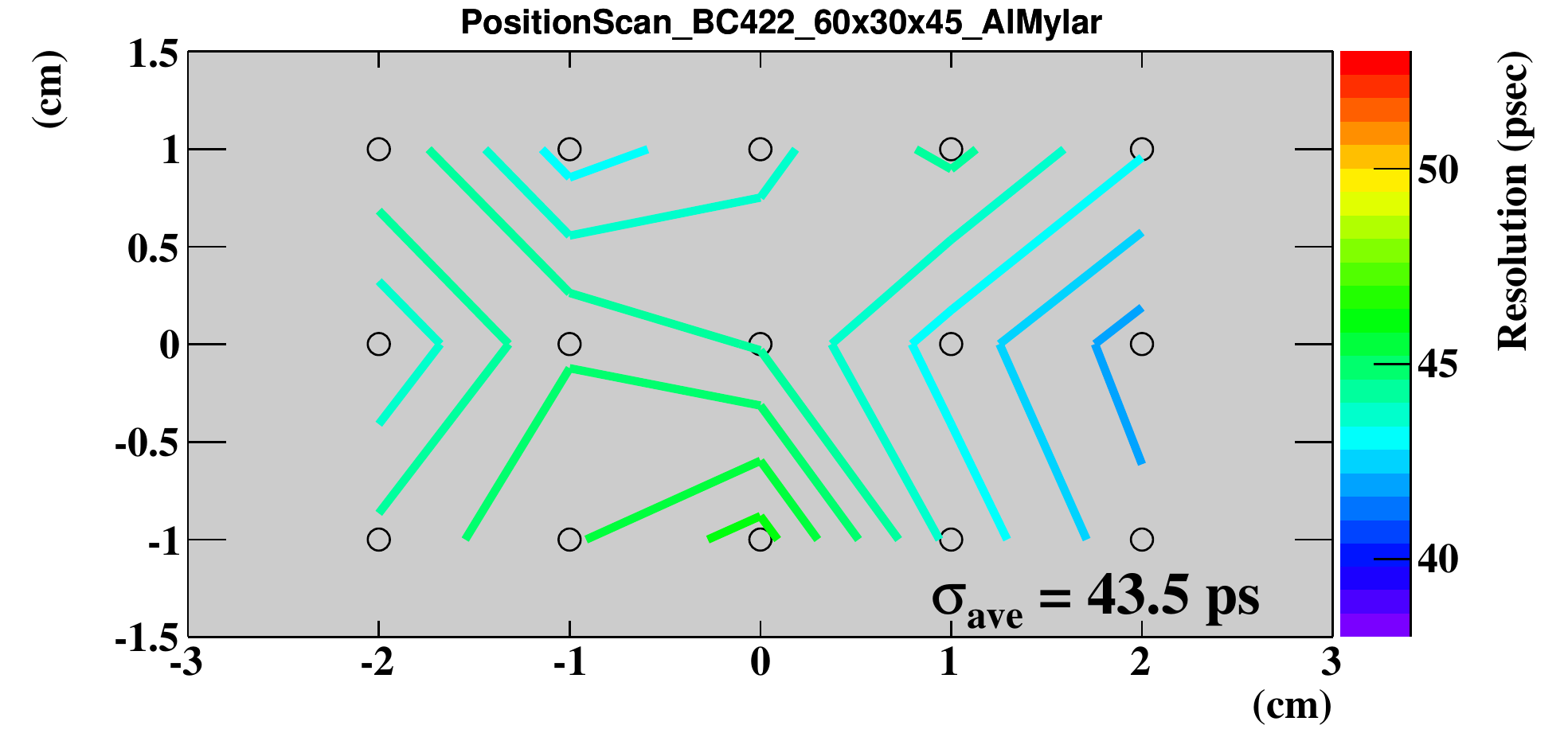}
    }
    \subfigure[3M radiant mirror film]{
     \includegraphics[width=0.48\linewidth]
     {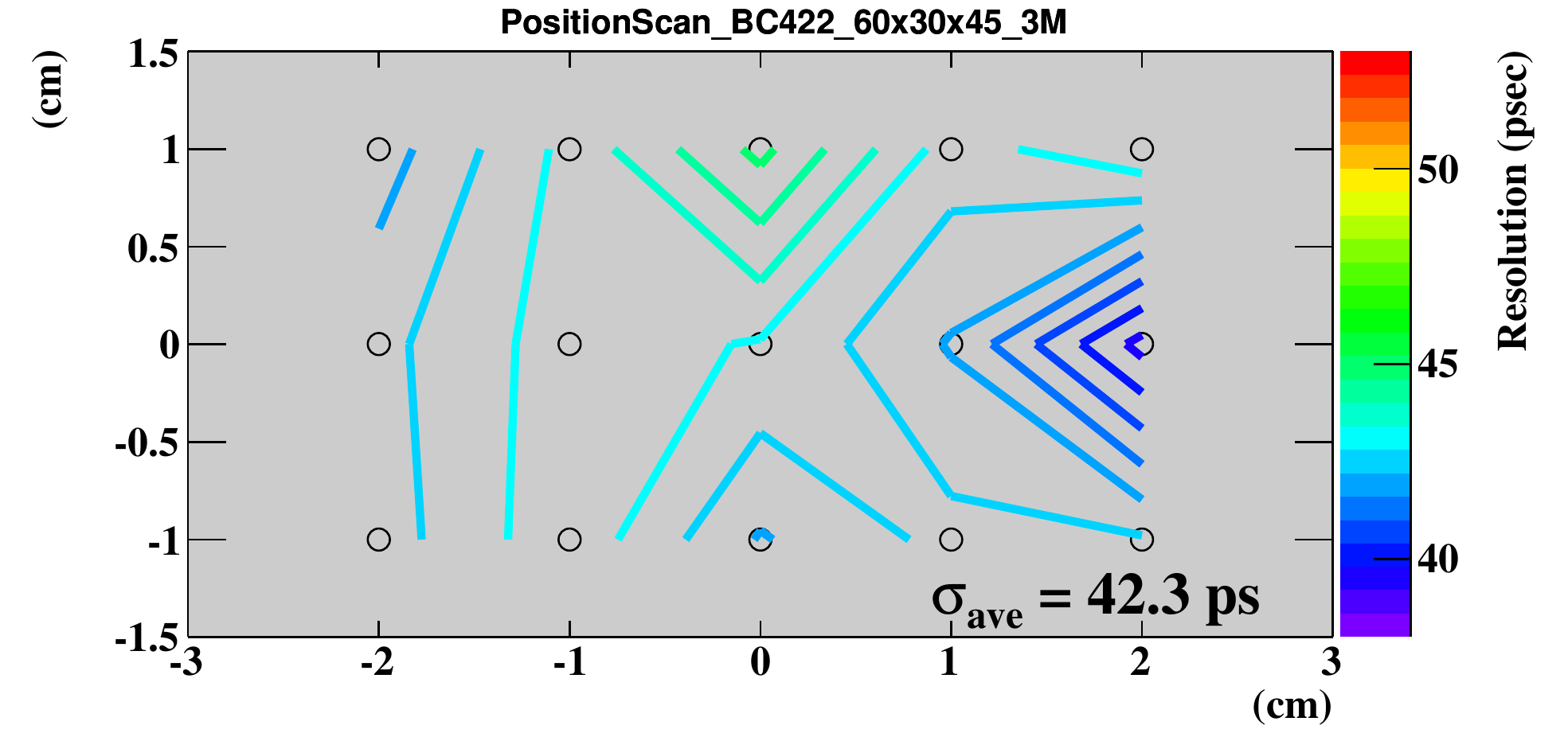}
    }
    \caption{\label{fig:resolution map with different reflectors}
    Single pixel timing resolutions measured with BC422 of $30\times60\times4.5\,\mathrm{mm}^3$ and six SiPMs wrapped with different types of reflectors.
    The resolution is measured at fifteen positions on the scintillator plate as shown as circles in the figure.
    The average timing resolution in $\sigma$ is also shown in each figure.
  }
  \end{center}
\end{figure}

Different types of reflectors such as
no reflector, Teflon tape, 
aluminized Mylar and 3M radiant mirror film
are tested
with a scintillator plate (BC422, $30\times60\times4.5\,\mathrm{mm}^3$).
The timing resolutions are measured at different positions on the scintillator plate.
The measured resolution maps are shown in Fig.\,\ref{fig:resolution map with different reflectors}.
The best timing resolution of 42.3\,ps is obtained with 3M radiant mirror film 
although a little larger position dependence is observed.
This non-uniformity is due probably to the unstable wrapping of the reflector.
A stabler wrapping method is now under study.

\begin{figure}[htb]                                                                                                                                                                  
\begin{center}
\includegraphics[width=8cm]{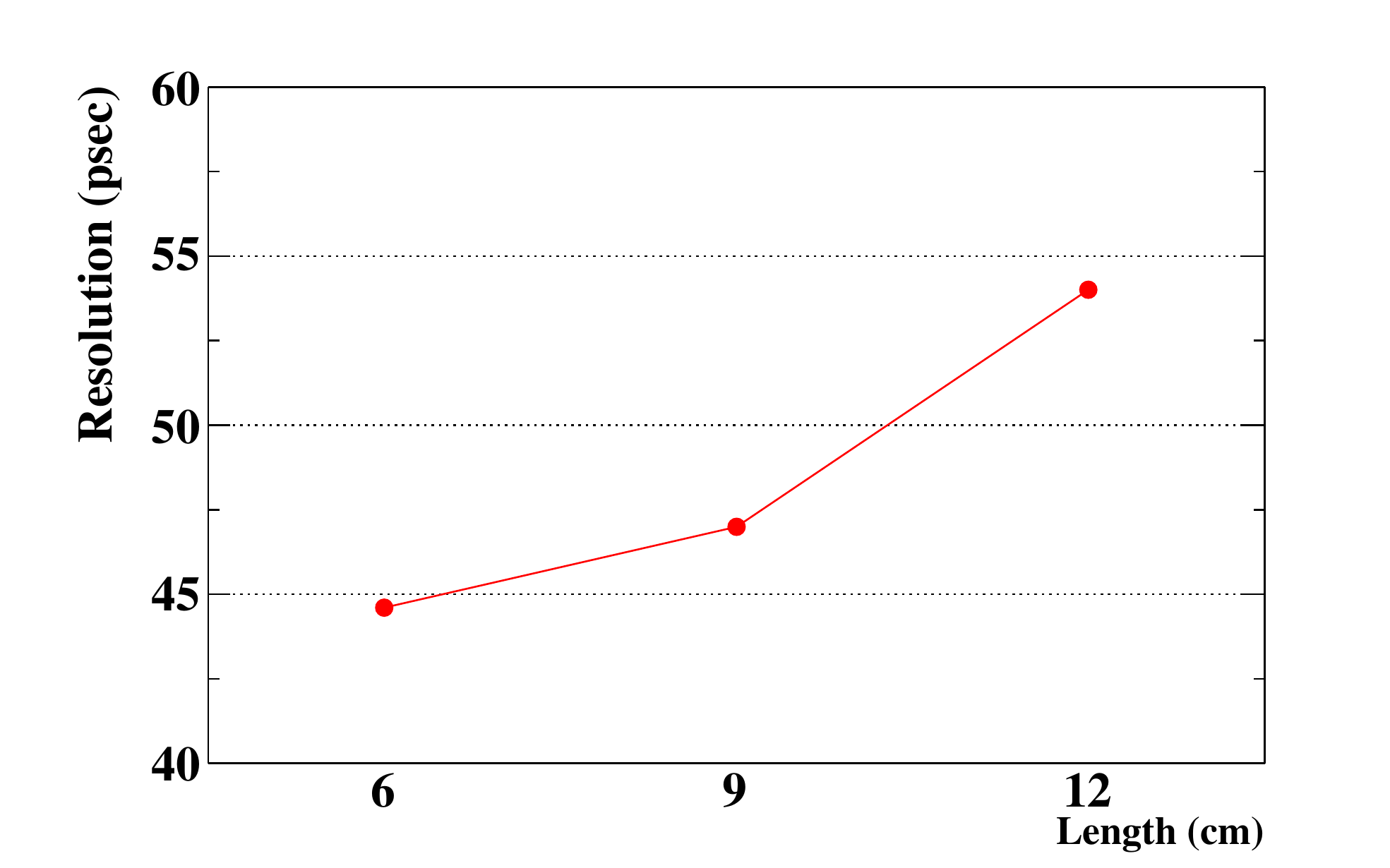}
\caption{\label{fig:prototype length scan}
Single pixel timing resolutions in $\sigma$ measured with scintillator plates (BC422) of three different lengths.
}
\end{center}
\end{figure}

The geometry of the scintillator has to be optimized
by the balance between the cost and the detector performance such as the overall timing resolution and efficiency.
The single pixel timing resolutions are measured with different sizes of scintillator plates (BC422).
Fig.\,\ref{fig:prototype length scan} shows the timing resolutions measured 
for scintillators of three different lengths with six SiPMs and no reflector.  
The deterioration of the timing resolution for longer pixel is not so dramatic at least up to 120\,mm.

We also tested a wider scintillator plate with a width of 40\,mm instead of 30\,mm. 
The average timing resolution is measured to be 54.1\,ps in $\sigma$ 
for a BC422 scintillator plate of $40\times60\times5\,\mathrm{mm}^3$ 
with six SiPMs and no reflector, 
while the resolution of 44.6\,ps is obtained with the 30\,mm-wide pixel at the same condition.
The difference in the resolutions can be mostly explained by the difference 
in the sensor coverage at the cross-sectional area of the scintillator plate.

We also measured the performance of the counter with four SiPMs at each side instead of three.
A scintillator plate of BC422 $30\times60\times5\,\mathrm{mm}^3$ without reflector is used for the measurement.
The timing resolution is found to be improved from 44.6\,ps to 43.3\,ps.
The improvement is slightly worse than expected from the increased sensor coverage.

Another type of scintillator, BC418, which has higher light yield and longer attenuation length 
but a slower rise time as described in Sec.\,\ref{sec:pix TC Scintillator}, is also tested.
The timing resolution for BC418 of $30\times90\times5\,\mathrm{mm}^3$ with six SiPMs and no reflector 
is measured to be 55.9\,ps, which is worse than the timing resolution of 47.0\,ps obtained 
for BC422 at the same condition.

The study on the single pixel configuration is still in progress and 
the above-mentioned results are quite preliminary.
Nevertheless, we already achieved a reasonably good timing resolution for the single pixel counter even with a relatively long scintillator plate.

\subsubsection{Expected Performance}
\label{sec:pixTC expected performance}

\begin{figure}[htb]                                                                                                                                                                  
\begin{center}
\includegraphics[width=4cm]{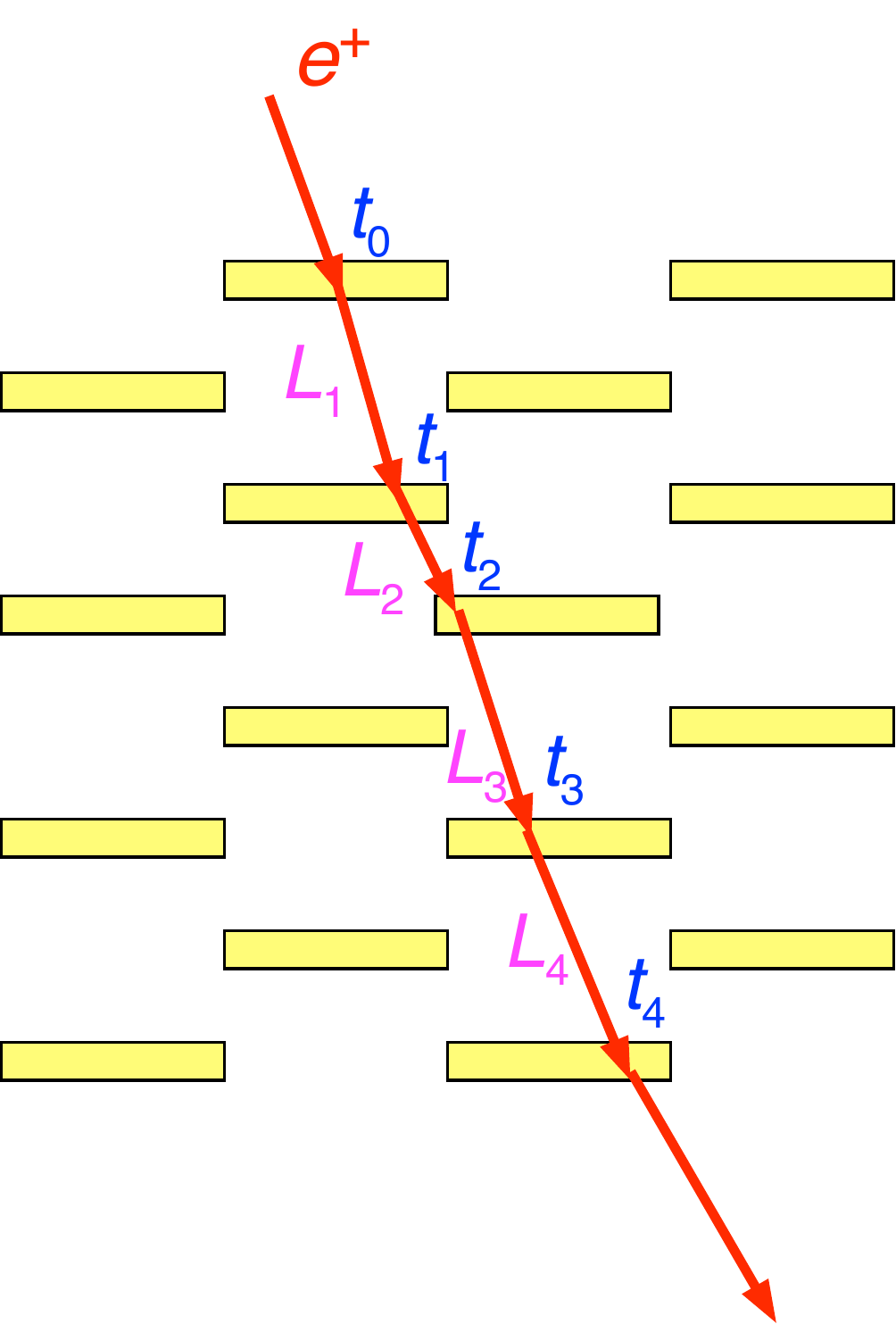}
\caption{\label{fig:MultiHit}
Positron passing through more than one pixels. 
The positron impact time ($t_0$) can be measured by averaging the hit time over the hit pixels after correcting for the traveling time between the hit pixels.
}
\end{center}
\end{figure}

The positron impact time can be precisely measured by averaging the positron hit time over the multiple hit pixels
after correcting for the positron travel time between the pixels as illustrated in Fig.\,\ref{fig:MultiHit}.
The positron path length between the hit pixels is estimated 
using the positron track extrapolated from the track reconstructed by the positron tracker.
Since the positron is tracked by the new tracker up to the entrance of the timing counter,
the uncertainty in the extrapolation of the track to the timing counter should be small.
The accuracy of the estimation of the path length between the pixels is degraded by 
multiple scattering in the hit pixels.

The overall timing resolution for a given number of hit pixels ($N_\mathrm{hit}$) can be expressed as,


\begin{equation}
   \sigma_\mathrm{overall}^2 
   = \frac{\sigma_\mathrm{single}^2}{N_\mathrm{hit}} 
   + \frac{\sigma_\mathrm{inter-pixel}^2}{N_\mathrm{hit}} 
   + \sigma_\mathrm{MS}^2(N_\mathrm{hit}).
   \label{overll timing resolution} 
\end{equation}
$\sigma_\mathrm{single}$ is the timing resolution of the single pixel.
$\sigma_\mathrm{inter-pixel}$ is the contribution from mis-alignment in time 
or time-jitter between pixels,
which is estimated to be 30--40\,ps for the planned electronics
and is expected to be improved in the new system as discussed in Sec.\,\ref{sec:Trigger and DAQ}.
The first two terms are reduced as $N_\mathrm{hit}$ increases.
$\sigma_\mathrm{MS}$ is the contribution from the multiple scattering,
which depends on $N_\mathrm{hit}$.
The positron track deviates from the extrapolation from the track reconstructed 
by the positron tracker due to multiple scattering on the hit pixel.
The angular spread due to multiple scattering on each hit pixel
is estimated to be about 35\,mrad for 5\,mm-thick pixel,
which corresponds to $\sim$5ps uncertainty in the correction 
for the positron flight time between the adjacent hit pixels.
Since the uncertainty is added up every pixel hit, $\sigma_\mathrm{MS}$ increases 
for larger $N_\mathrm{hit}$ 
and the improvement of the overall time resolution is saturated for larger $N_\mathrm{hit}$.

\begin{figure}[htb]                                                                                                                                                                  
\begin{center}
\includegraphics[width=9cm]{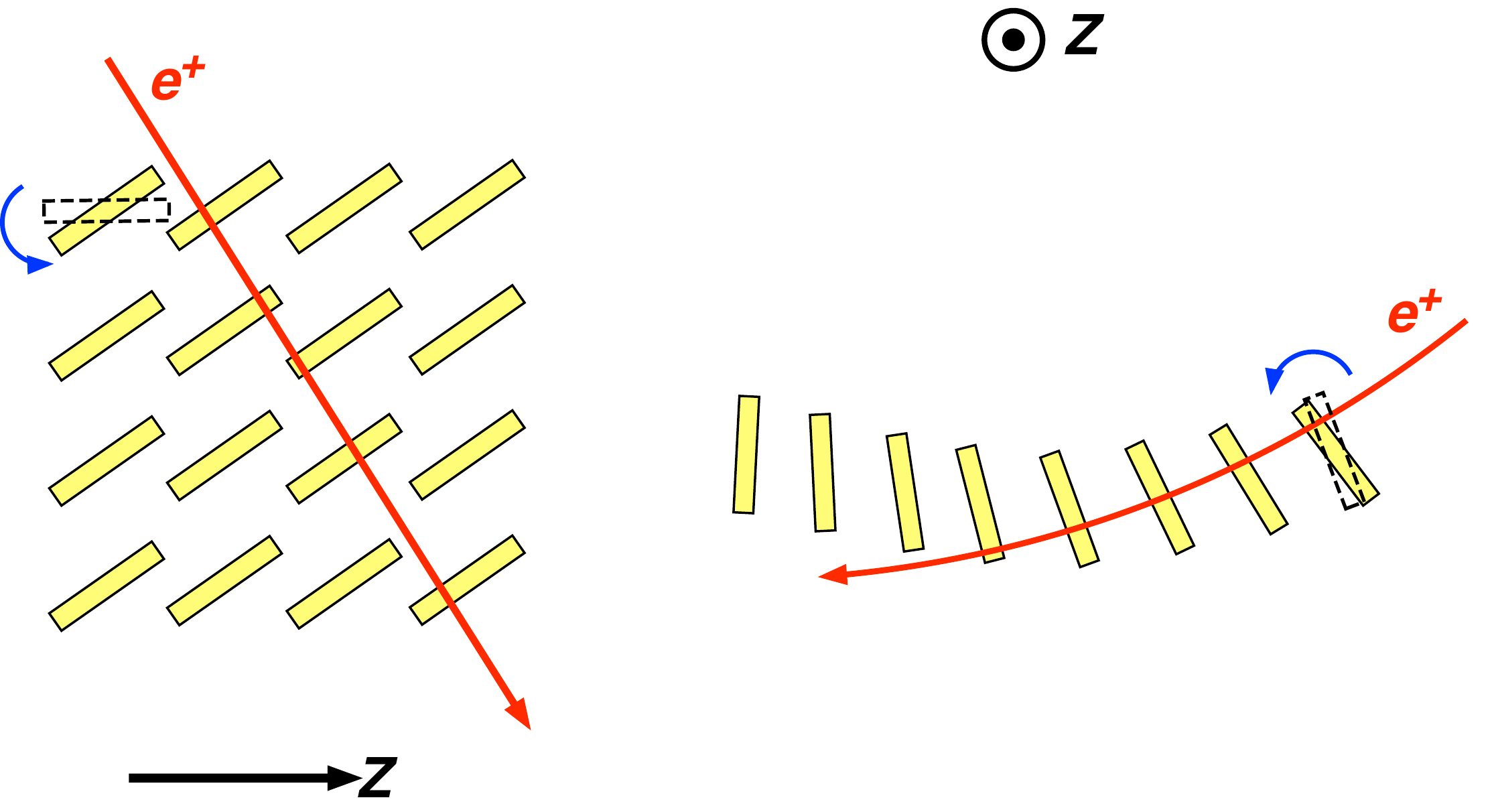}
\caption{\label{fig:gem4 SPX rotation}
Rotation of pixel modules on (left) $z$-$\phi$ plane and (right) $r$-$\phi$ plane.
}
\end{center}
\end{figure}

The overall performance of the pixelated timing counter is evaluated 
by means of a MC simulation based on Geant4, 
where the single pixel timing resolution measured with the prototype counter 
(see Sec.\,\ref{sec:Test with Pixel Counter Prototype}) is taken into account.
In the following simulation studies, no positron tracker is placed in front of the timing counter
in order to establish its performance.
The support structure is not also taken into account, but the effect on the performance is expected to be small.
Each pixel module is rotated on $z$-$\phi$ and $r$-$\phi$ planes 
with $10^\circ$ and $45^\circ$, respectively,
as shown in Fig.\,\ref{fig:gem4 SPX rotation}
such that the angle of the incident signal positron to the scintillator pixel is minimized
and thus the acceptance of the pixel is maximized.
The overall timing resolution is estimated for different geometries 
and layouts of the pixel module to find an optimal configuration.

\begin{figure}[htb]                                                                                                                                                                  
\begin{center}
\includegraphics[width=11cm]{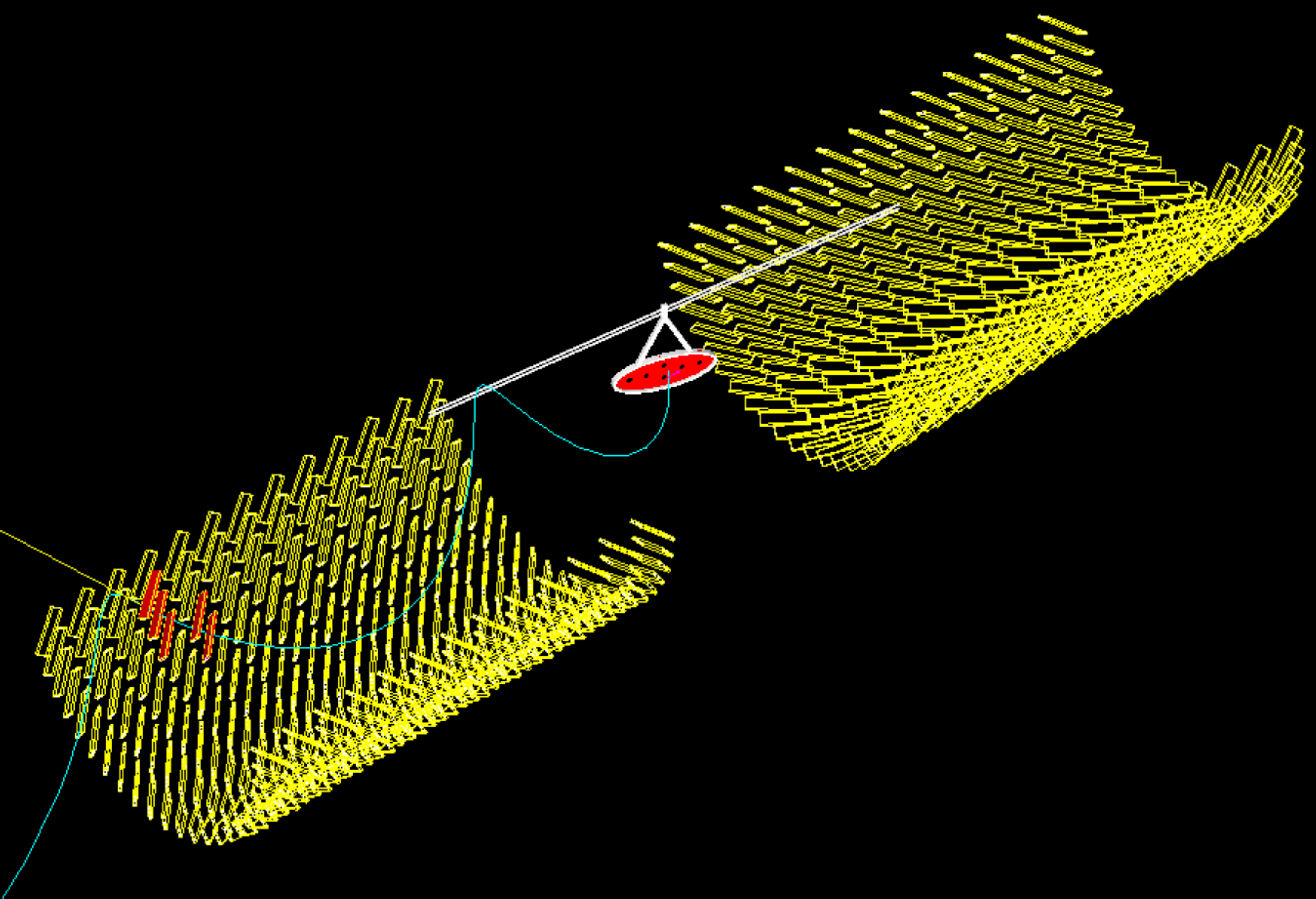}
\caption{\label{fig:MC event display}
Simulated signal positron with the pixelated timing counter.
}
\end{center}
\end{figure}

Fig\,\ref{fig:MC event display} shows a simulated signal positron
with 598 pixel modules with BC422 of $30\times90\times5\,\mathrm{mm}^3$ each
and the pixel spacing of $75\,\mathrm{mm}$ along $z$ and $0.13\,\mathrm{rad}$ along $\phi$.
The signal positron hits five pixel modules in this event.  

\begin{figure}[htb]                                                                                                                                                                  
\begin{center}
\includegraphics[width=8cm]{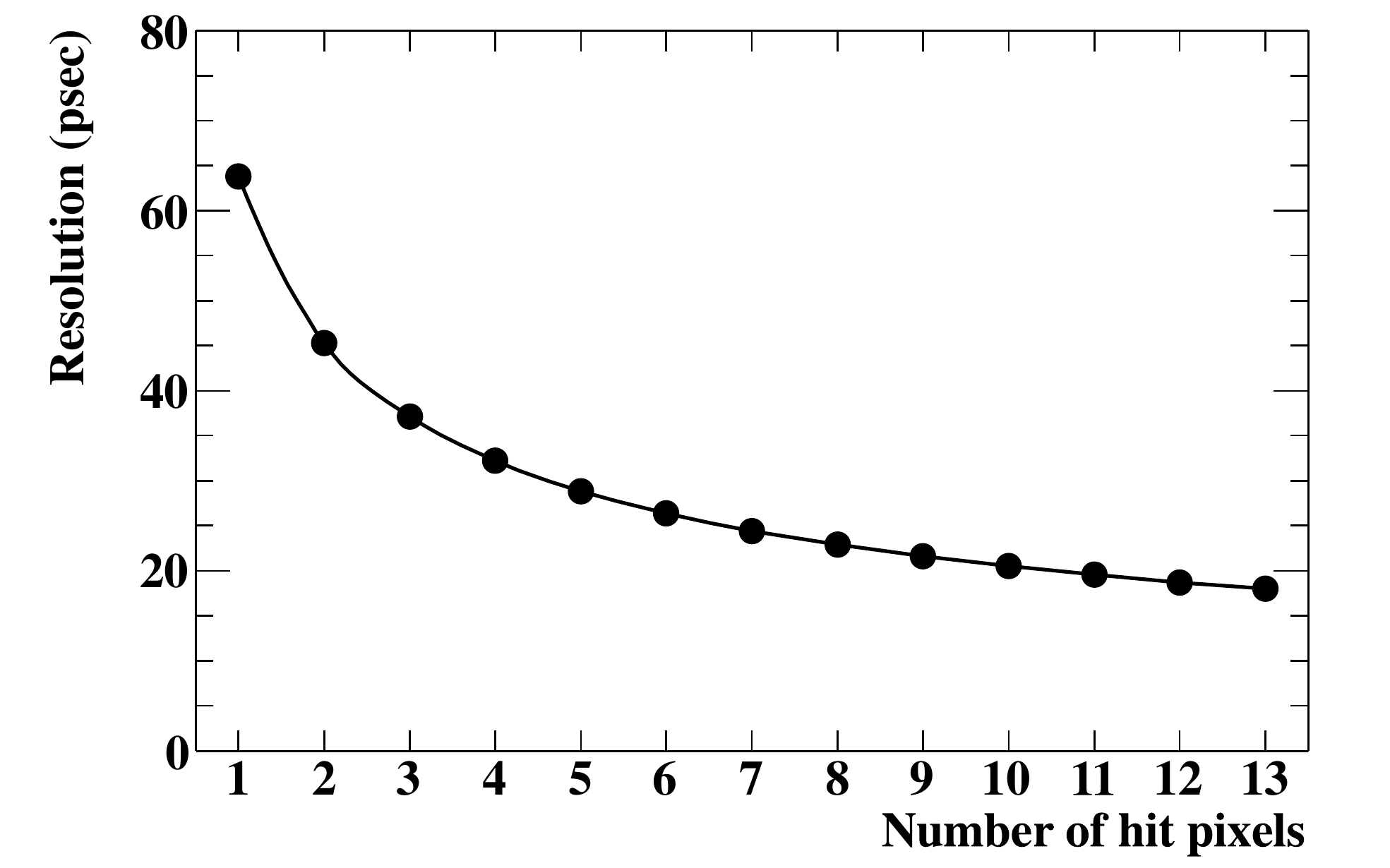}
\includegraphics[width=8cm]{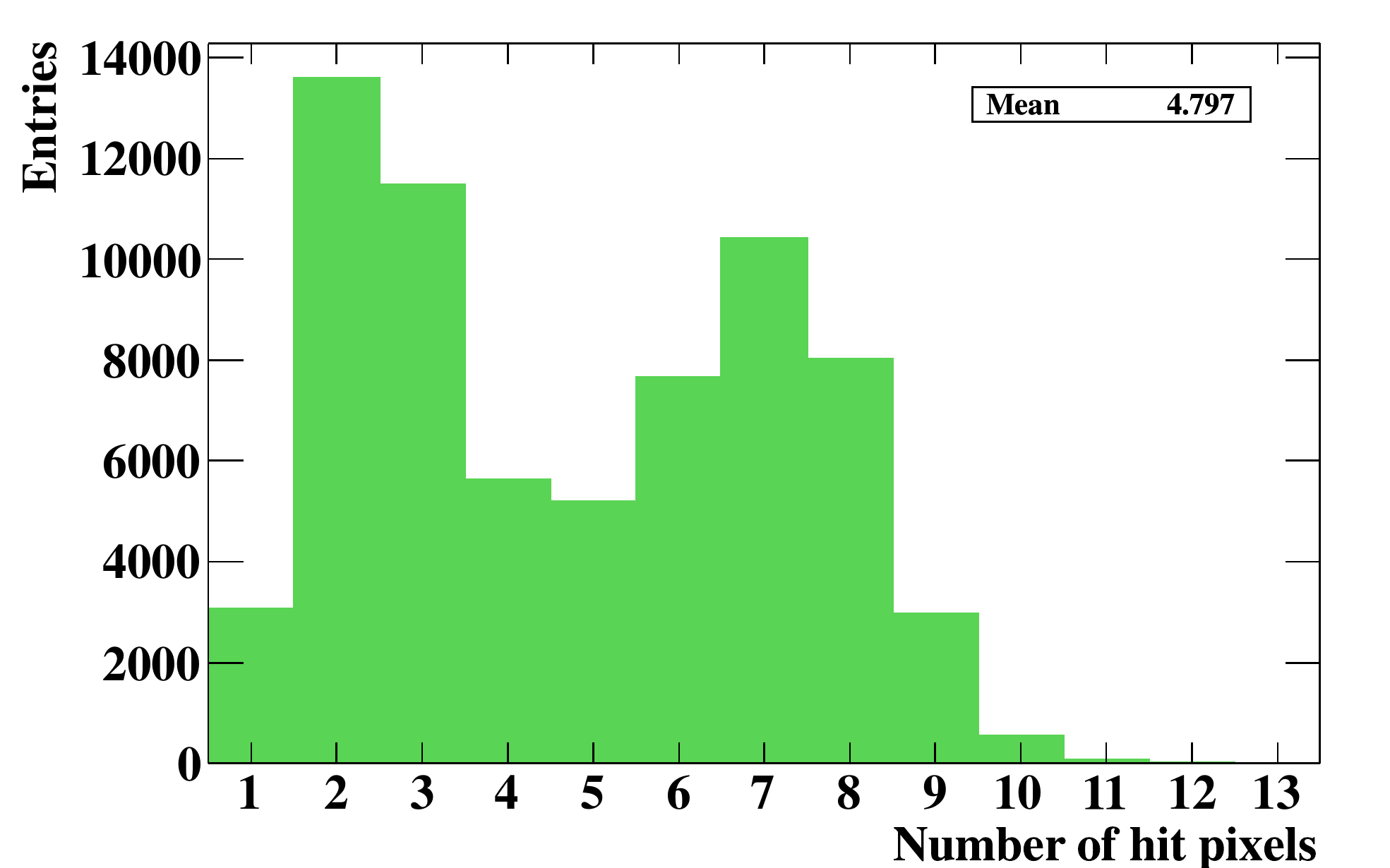}
\caption{\label{fig:Overall time resolution}
(Left) Overall timing resolution in $\sigma$ as a function of the number of hit pixels 
and (right) distribution of the number of hit pixels for signal positron.
}
\end{center}
\end{figure}

The left plot in Fig.\,\ref{fig:Overall time resolution} shows 
how the estimated overall timing resolution improves as the number of hit pixels increases.
The right plot shows the distribution of the number of hit pixels 
for the signal positrons in this setup.
The double-peak structure in the distribution comes from the hit-position 
dependence of the positron incident angle.
The average overall timing resolution is estimated to be 35\,ps in $\sigma$.

\begin{figure}[htb]                                                                                                                                                                  
\begin{center}
\includegraphics[width=8cm]{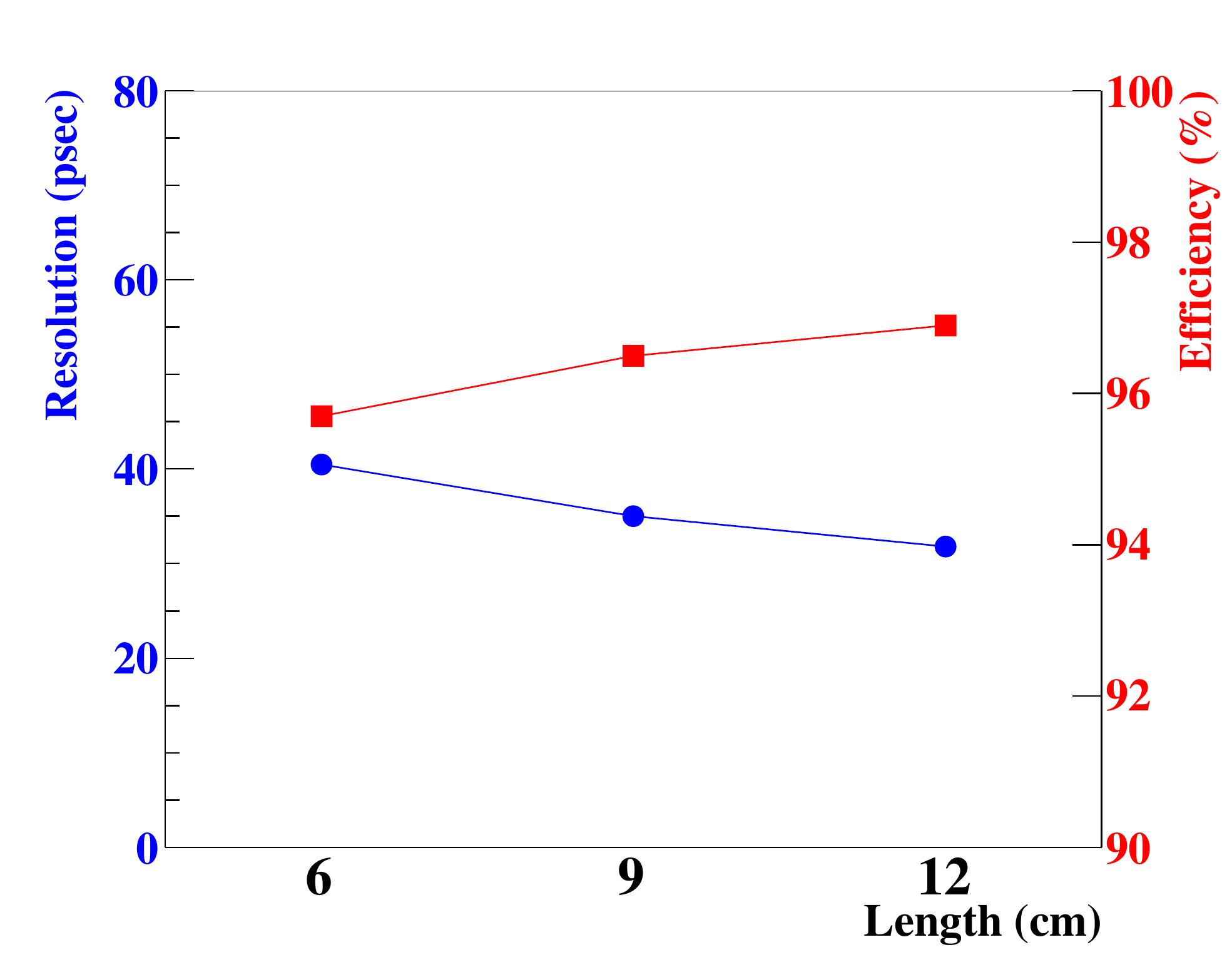}
\includegraphics[width=8cm]{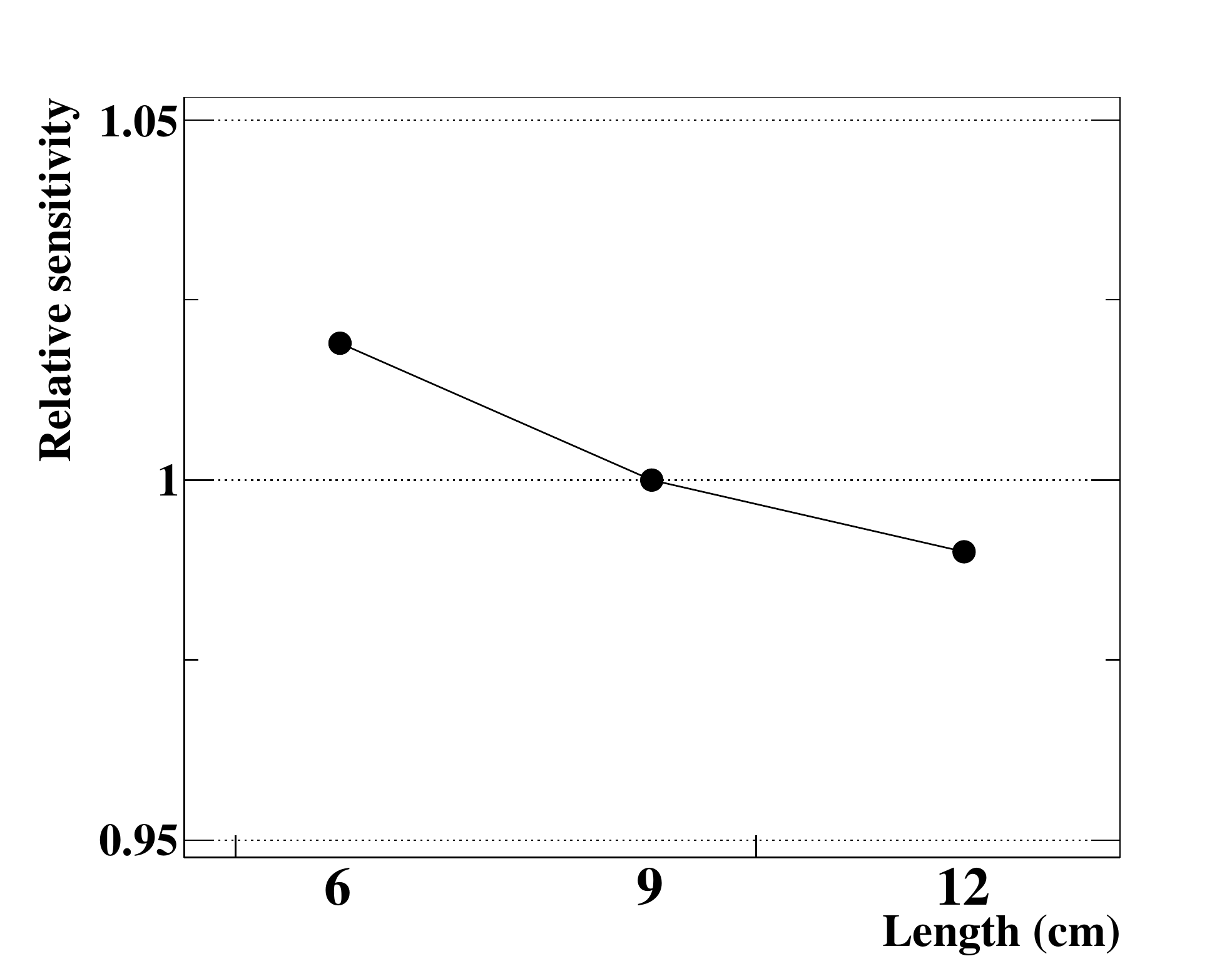}
\caption{\label{fig:sensitivity optimization}
(Left) Overall timing resolution in sigma and efficiency 
and (right) relative sensitivity
as a function of the length of the pixel counter.
}
\end{center}
\end{figure}

The effect of the geometry and the spacing of the pixel module 
on the overall detector performance is studied.
The left plot of Fig.\,\ref{fig:sensitivity optimization} shows the overall timing resolution
and efficiency as a function of the length of the pixel counter 
with the total number of pixel counters fixed to about 600.
The improvement in the overall timing resolution for longer pixel is because of 
the increase of the number of hit pixels. 
The detector performance is compared in terms of a relative sensitivity 
in the left plot of Fig.\,\ref{fig:sensitivity optimization}, 
where the sensitivity is calculated by Punzi's method\,\cite{PunziSensitivity}.
It tends to suggest that longer pixel counter is better from the sensitivity viewpoint.
However we may suffer from the pileup and double-hit in a high rate environment
if the pixel counter is too long. The effect is now under study.

\begin{figure}[htb]                                                                                                                                                                  
\begin{center}
\includegraphics[width=8cm]{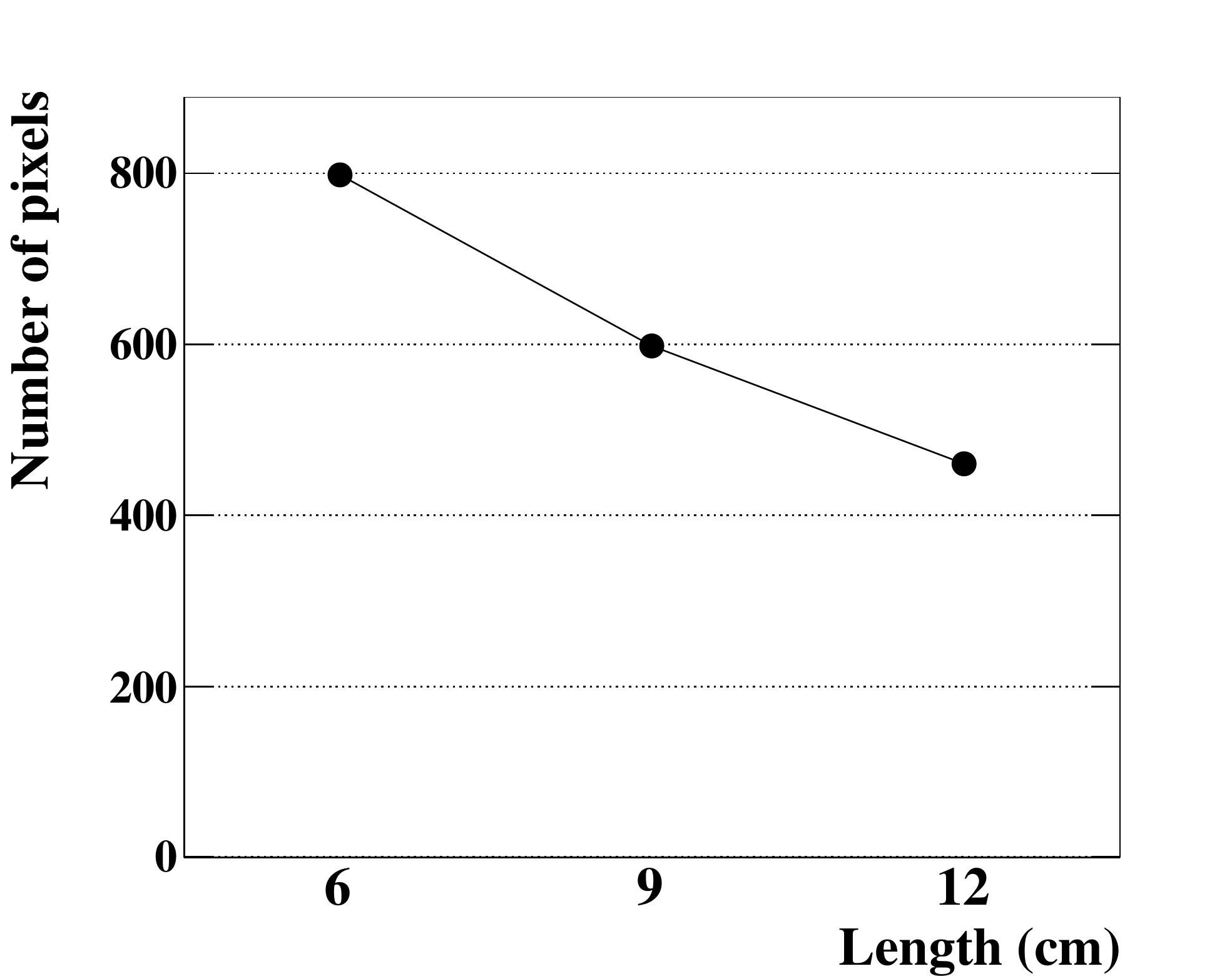}
\caption{\label{fig:number of pixels vs length}
Total number of pixel counters necessary to achieve a given fixed sensitivity 
for three different lengths of the pixel counter.
}
\end{center}
\end{figure}

Another advantage of the longer pixel counter is a possible cost reduction.
Fig.\,\ref{fig:number of pixels vs length} shows the total number of pixels necessary 
to achieve a given fixed sensitivity for the three different lengths of the pixel counter.
The total number of pixels can be significantly reduced with longer pixel counter, 
but again the effect of the pileup and double-hit has to be carefully studied.

\subsubsection{Calibration}
Since SiPM is sensitive to single photoelectron, the gain can be easily calibrated using the single photoelectron peak.
Furthermore the output of the pixel module can be equalized by using the Landau peak from the Michel positrons or cosmic ray muon.

It is also quite important to precisely synchronize all the pixel modules 
although the misalignment in the time of the pixels can be diluted by taking the average over the multiple hit pixels. 
Two schemes are under consideration for the inter-pixel time-alignment.
\begin{itemize}
   \item Michel positrons

     High momentum Michel positrons can also pass through more than one pixel similarly to the signal positrons.
     The multiple hits should allow the time-alignment between adjacent pixels after correcting for the positron travel time between the hits.

   \item Laser

      The time-alignment can also be done by distributing light pulse from a single laser system 
      to all the pixels through optical fibers of the same length.
      The candidate for the light source is Hamamatsu picosecond light pulser (PLP10-040),
      with emission wavelength of 405\,nm, pulse width of 70\,ps (typ.) and peak power of 100\,mW (typ.)\,\cite{HPK:PLP-10}.
      Approximately $10^7$ photons are emitted from the light pulser in a short pulse.
      A fast light pulse with about thousand photons can, therefore, be delivered to each pixel module for the time calibration.

\end{itemize}

\subsubsection{Other Issues}
Other issues for the pixelated timing counter are discussed here.

\begin{figure}[htb]                                                                                                                                                                  
\begin{center}
\includegraphics[width=7cm]{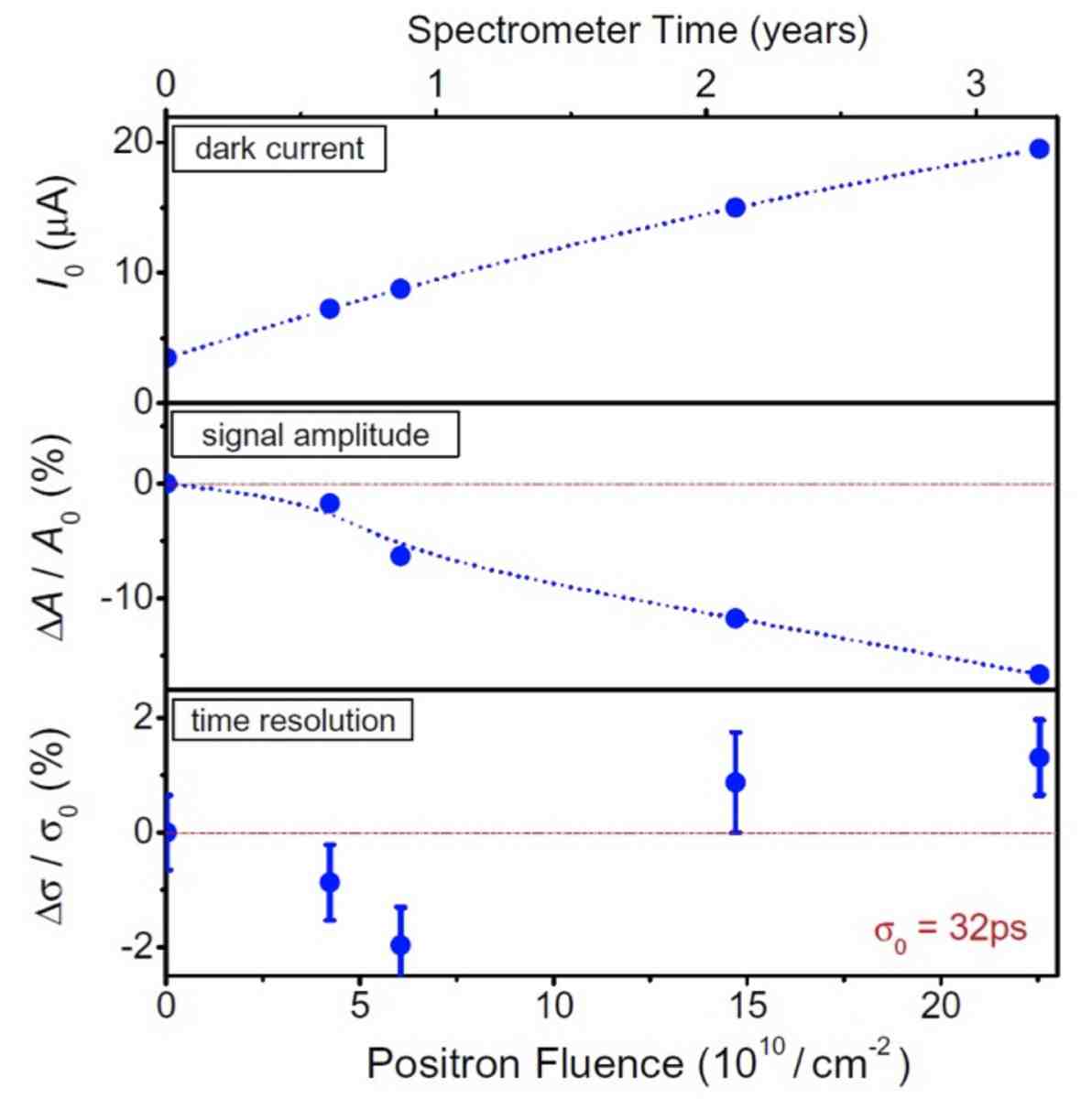}
\caption{\label{fig:muSR MPPC irradiation test}
Results from the irradiation tests of Hamamatsu MPPC (S10362-33-050C) performed by the PSI $\mu$SR group.
Significant increase of dark current (top) and 15\% gain degrease (middle) are observed, while the timing resolution is unchanged (bottom).
Courtesy of Dr. A.~Stoykov of Paul Scherrer Institut.
}
\end{center}
\end{figure}

The modest radiation hardness of SiPM is considered as a weak point of SiPM.
Increase of the dark current and change of the gain of SiPM are typical effects after certain irradiation.
The SiPM in the pixelated timing counter will be irradiated by many Michel positrons during the experiment. 
The integrated fluence of the Michel positrons during three-years running is estimated to be about $5\times 10^9 \,\mathrm{positrons}/\mathrm{cm}^2$.
The PSI $\mu$SR group performed irradiation tests for a scintillator counter with a similar configuration using Michel positrons as shown in 
Fig.\,\ref{fig:muSR MPPC irradiation test}\,\cite{muSR:irradiation-test}.
The SiPMs of the same type as the one which we plan to use coupled to a small plastic scintillator are irradiated by Michel positrons of fluence 
up to $2.5\times 10^{11}/\mathrm{cm}^2$ which is more than one order of magnitude higher than our case.
They observed a significant increase of the dark current by a factor of six and a 15\% gain decrease.   
Interestingly the timing resolution is unchanged even with the highest fluence.
The SiPMs are irradiated also by neutrons ands $\gamma$-rays in our experiment.
The effect is discussed in detail in Sec.\,\ref{sec:LXe MPPC issues}
for the SiPM planned to be used for the LXe detector
and it turns out not to influence the performance of SiPM.
The radiation damage of SiPM should, therefore, not be an issue in our case.

Another issue would be the temperature stability of the SiPM. 
The temperature coefficient of the breakdown voltage for Hamamatsu MPPC S10931-050P is 56\,mV/$^\circ$C.
The MPPC gain at an overvoltage of 1\,V is, therefore, changed by 5.6\% for a temperature change of 1\,$^\circ$C.
The effect of the gain variation on the timing performance should be carefully studied, 
but in the worst case the gain has to be stabilized
either by controlling the SiPM temperature or
by adjusting the bias voltage.
The temperature coefficient of the gain of the KETEK SiPM is smaller than $1\%/^\circ\mathrm{C}$\,\cite{KETEK-PM3350}.
If the timing resolution with the KETEK SiPM is found to be competitive, it can be a good option to resolve the possible issue of the temperature stability.

We plan to connect outputs in parallel from two or three pixels located apart from each other 
in order to reduce the number of channels.
This operation will considerably reduce the number of electronics channels.
The timing performance might be deteriorated by the smearing of the waveform due to the increased output capacitance in the parallel connection.
The longer fall time should be irrelevant because of the low pileup rate, while it would be an issue if the leading edge of the waveform is smeared.
We may also suffer more noise on the waveform with the increased capacitance.

\subsubsection{R\&D Plan}
A good timing performance of the single pixel module up to 120\,mm long 
is already proved experimentally as described in Sec.\,\ref{sec:Test with Pixel Counter Prototype}.
Development of dedicated waveform analysis to obtain the best timing resolution is also planned.
Further optimization of the single pixel performance is still in progress.
Once the geometry of the single pixel module and the readout scheme are optimized, we will build a prototype detector using a small number of pixel modules 
and perform a beam test in order to demonstrate the improvement of the timing resolution with multiple hits.
The schedule of the R\&D and the construction will be described in detail in Sec.\,\ref{sec:Time Schedule}. 

%% file: 07_Photon_Calorimeter/Photon_Calorimeter.tex
%
\section{Gamma detector}
%
\label{sec:photoncalorimeter}
\subimport{./}{requirement.tex}

\subimport{./}{mppc.tex}

\subimport{./}{detector_design.tex}

\subimport{./}{performance.tex}

\subimport{./}{prototype.tex}


\subimport{./}{cost_schedule.tex}

%% file: 07_Photon_Calorimeter/requirement.tex


\subsection{Concept of Upgrade}
\begin{figure}[htb]                                                                                                                                                                  
\begin{center}
\includegraphics[width=10cm]{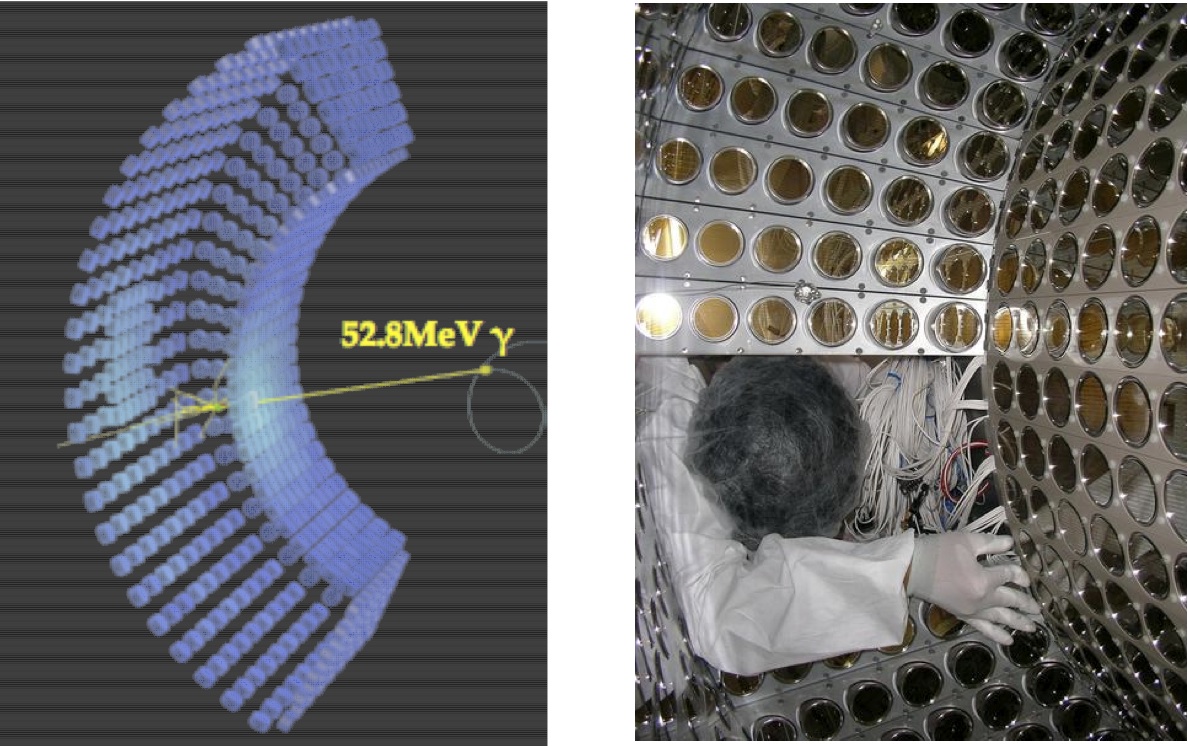}
\caption{\label{fig:current detector}
   Current LXe $\gamma$-ray detector with 900\,$\ell$ LXe surrounded by 846 UV-sensitive PMTs.
}
\end{center}
\end{figure}

The liquid xenon (LXe) $\gamma$-ray detector is a key ingredient of the experiment to suppressing the background in $\mu\rightarrow \mathrm{e}\gamma$ search. 
It is, therefore, crucial to substantially improve the detector performance for the upgrade of the experiment.
The current detector is the world's largest LXe scintillation detector
with 900\,$\ell$ LXe surrounded by 846 photomultiplier tubes (PMTs) submerged in the liquid to detect scintillation light
in the VUV range ($\lambda = 175\pm5\,\mathrm{nm}$) (Fig\,\ref{fig:current detector}).
The 2-inch PMT (Hamamatsu R9869) used in the detector is UV-sensitive 
with a photo-cathode of K-Cs-Sb and a synthetic quartz window.
The quantum efficiency (QE) is about 15\% for the LXe scintillation photon at LXe temperature of 165\,K.

\begin{figure}[htb]                                                                                                                                                                  
\begin{center}
\includegraphics[width=10cm]{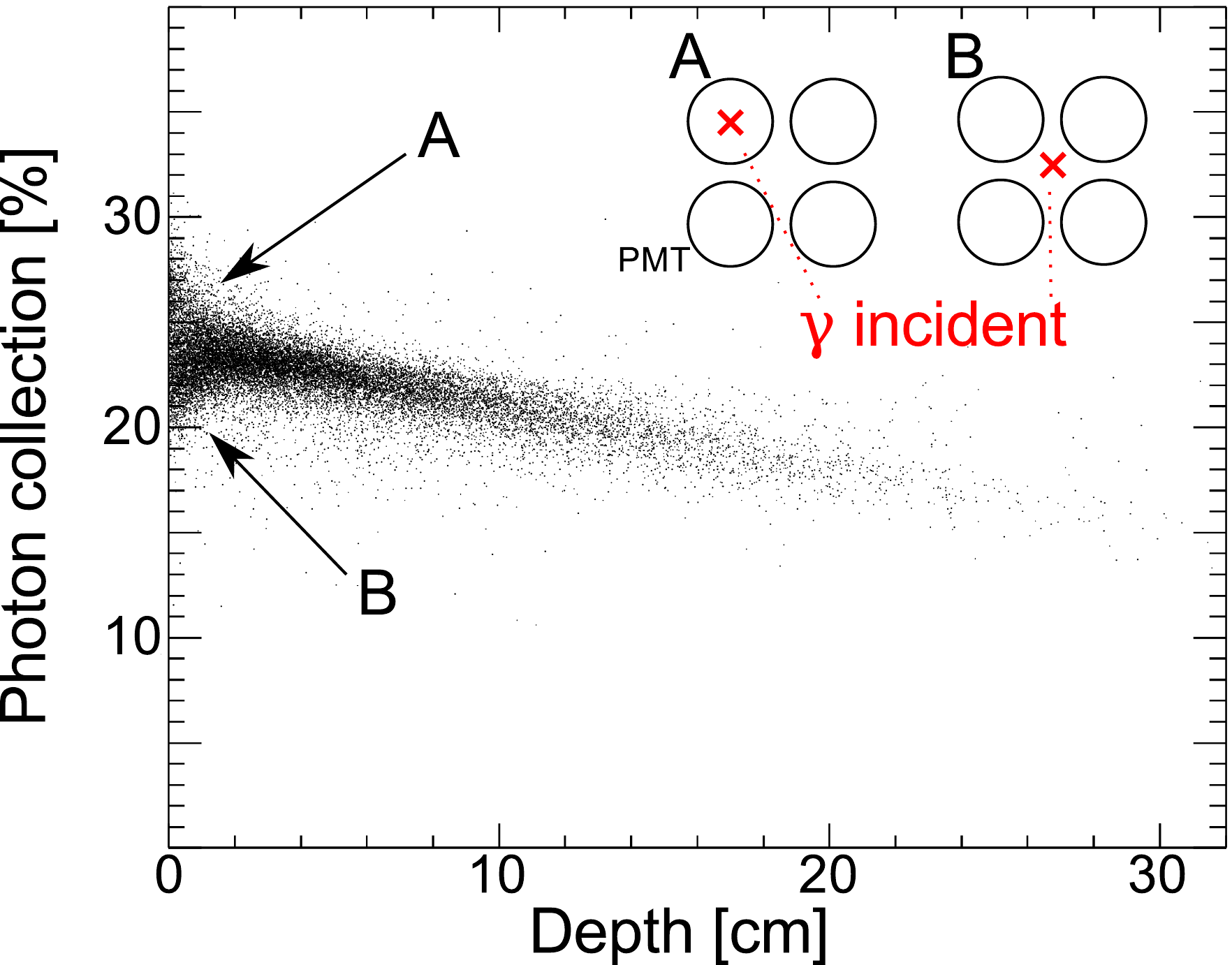}
\caption{\label{fig:current detector non-uniformity}
   Efficiency of the scintillation light collection estimated by MC simulation as a function 
      of the depth of the first interaction for signal $\gamma$-ray of 52.8\,MeV. 
}
\end{center}
\end{figure}

The $\gamma$ entrance face is covered by 216 PMTs with a minimum spacing between adjacent PMTs.
The photo-cathode of the PMT is, however, round-shaped with a diameter of 46\,mm 
which is smaller than the interval of the adjacent PMTs of 62\,mm.
The performance of the current detector is limited due to this non-uniform PMT coverage. 
Fig\,\ref{fig:current detector non-uniformity} shows the efficiency of scintillation light collection 
as a function of the depth of the first interaction for signal $\gamma$-ray of 52.8\,MeV.
The collection efficiency strongly depends on the incident position.
The non-uniform response is partly corrected in the offline analysis,
but it still deteriorates the energy and position resolutions due to event-by-event fluctuation of the shower shape,
    especially for the shallow events.

\begin{figure}[htb]                                                                                                                                                                  
\begin{center}
\begin{minipage}{0.398\linewidth}
\includegraphics[width=\linewidth]{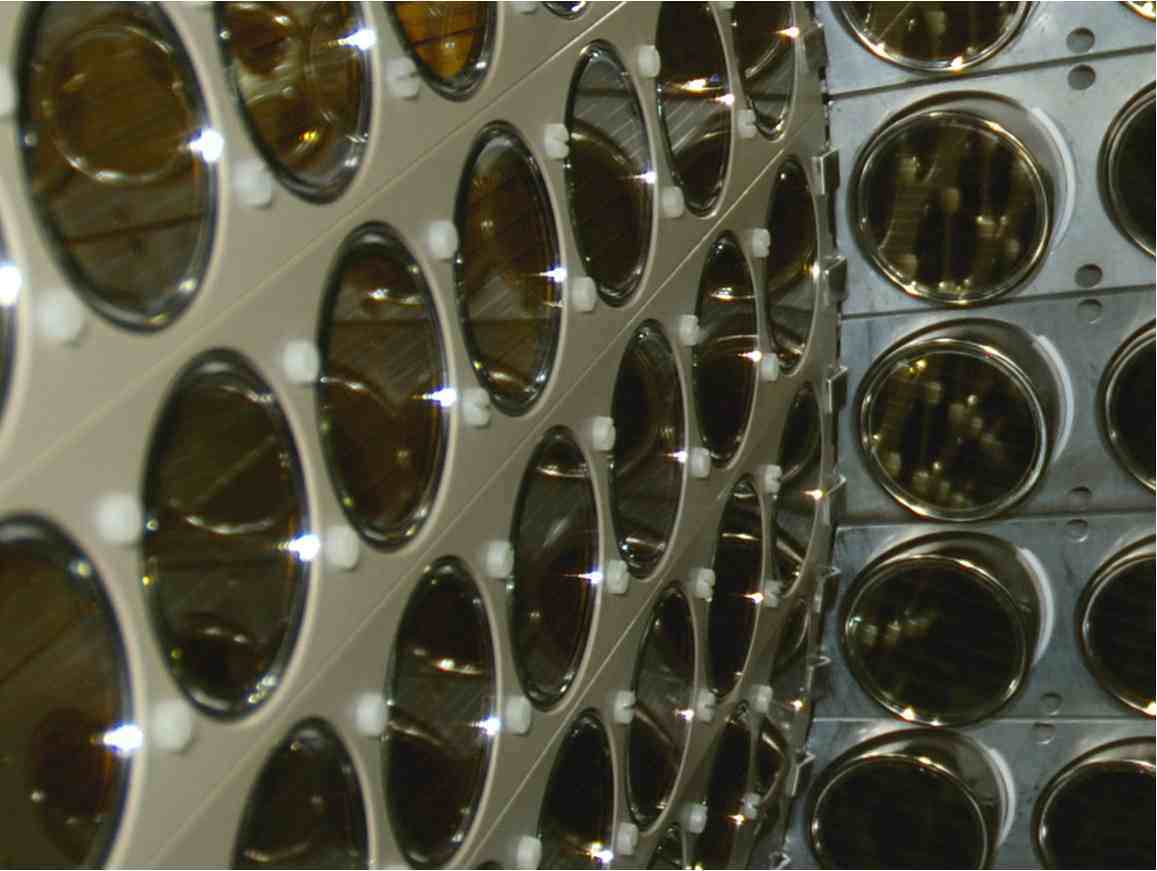}\\(a) Present detector
\end{minipage}
\hspace{1cm}
\begin{minipage}{0.4\linewidth}
\includegraphics[width=\linewidth]{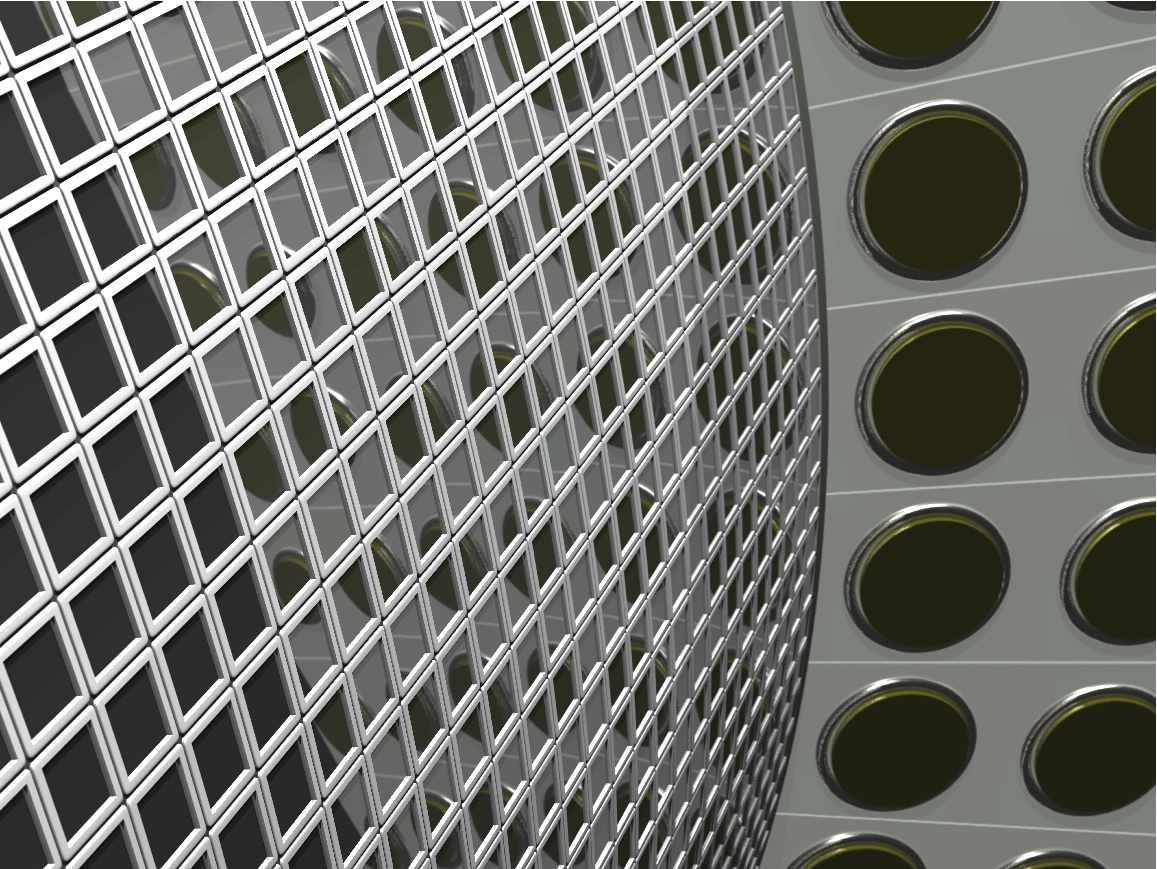}\\(b) Upgraded detector (CG)
\end{minipage}
\caption{\label{fig:MPPCinFrontFace}
Possible replacement of 216 PMTs in the $\gamma$-entrance face with smaller photo-sensors (about 4000 MPPCs with $12\times12\,\mathrm{mm}^2$ area each). 
}
\end{center}
\end{figure}

\begin{figure}[htb]                                                                                                                                                                  
\begin{center}
\includegraphics[width=13cm]{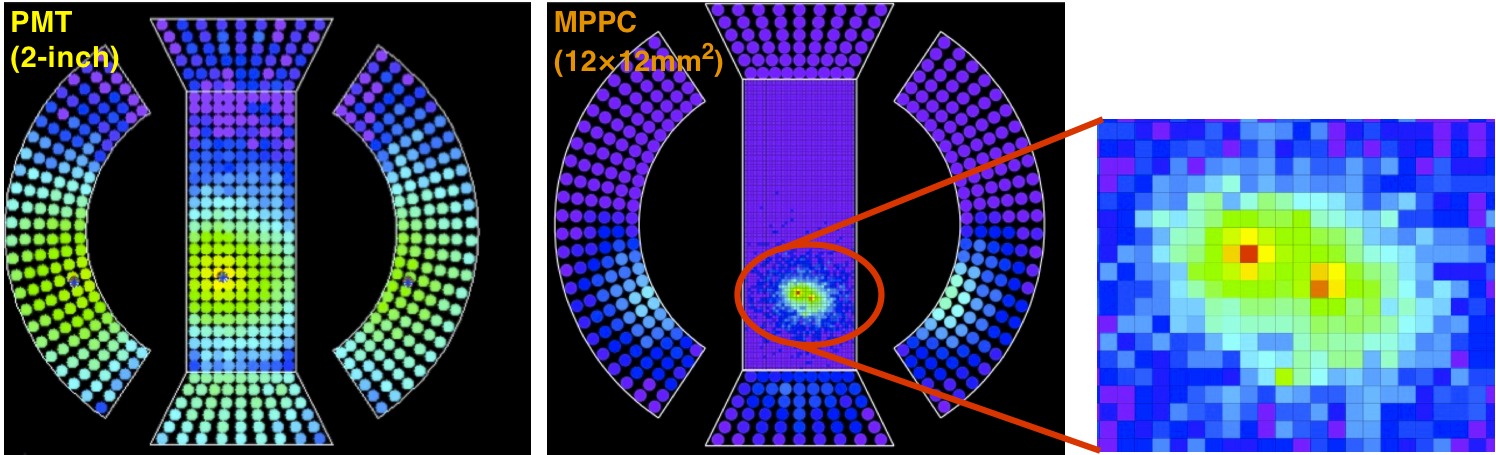}
\caption{\label{fig:comparison of imaging}
   Typical examples of scintillator light distribution seen by photo-sensors 
      in case of (left) PMTs and (right) smaller photo sensors ($12\times 12\,\mathrm{mm}^2$) 
      on the $\gamma$ entrance face.
}
\end{center}
\end{figure}

The main concept of the upgrade of the LXe detector is to reduce this non-uniform response 
by replacing the PMTs of the $\gamma$ entrance face with smaller photo-sensors as shown in Fig.\,\ref{fig:MPPCinFrontFace}.
Fig.\,\ref{fig:comparison of imaging} shows a comparison of how the event would look like
in two cases with the current PMTs and smaller photo sensors ($12\times 12\,\mathrm{mm}^2$)
   on the $\gamma$ entrance face. 
The imaging power is greatly improved with smaller photo sensors.
For example, two local energy deposits in the same shower are clearly separated in the event shown in Fig.\,\ref{fig:comparison of imaging}.
It turns out that both the energy and position resolutions greatly improves 
especially for the shallow events as shown in Sec.\,\ref{sec:expected performance}.

The possible candidates of the smaller photo-sensor as a replacement of the current PMT are
\begin{itemize}
\item SiPM
\item 1-inch square-shape PMT
\item 2-inch flat panel multi-anode PMT, 
\end{itemize}

where the leading candidate is SiPM as discussed in the following sections, 
while the development of the PMT is described in the Appendix section (Sec.\,\ref{sec:LXe PMT}).
The signal $\gamma$-ray traverses the photo-sensors on the entrance face.
The material in front of the active LXe volume can be substantially reduced in case of using SiPM 
which is much thinner than PMT.
The $\gamma$ detection efficiency is estimated to be improved by 9\% as discussed in Sec.\,\ref{sec:expected performance}.

\begin{figure}[htb]                                                                                                                                                                  
\begin{center}
\includegraphics[width=13cm]{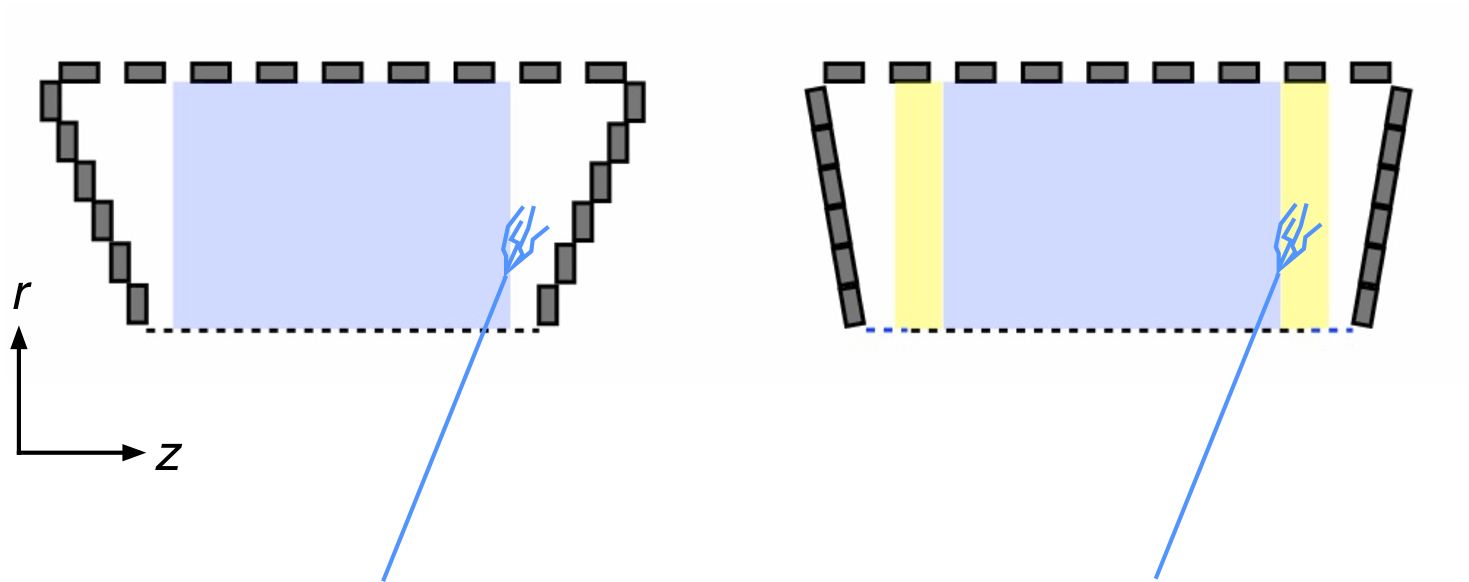}
\caption{\label{fig:PMT layout modification}
Current (left) and modified (right) layouts of the PMTs viewed on a given $r$-$z$ plane.
}

\end{center}
\end{figure}

We plan to use PMTs of the same type as the current one for the other faces than the entrance face.
It turns out by detailed MC studies developed during the current MEG data analysis that 
further improvements are possible by modifying the layout of the PMTs on the lateral faces.
Fig.\,\ref{fig:PMT layout modification} illustrates the modified layout viewed on a given $r$-$z$ plane.

The $\gamma$ entrance face is extended along $z$ to outside of the acceptance 
by 10\% at each side.
The extended volume reduces the energy leakage for the event near the lateral wall. 
The PMTs on the lateral faces are tilted such that all the photo-cathodes lie on the same plane.
This operation minimizes the effect due to shower fluctuation for the events near the lateral wall.
The energy resolution is thus improved especially for the events near the lateral wall.  

%% file: 07_Photon_Calorimeter/mppc.tex

\subsection{Development of VUV-sensitive MPPC}
\subsubsection{MPPC Advantage}

The MPPC (Multi-Pixel Photon Counter) is a new type of photon counting device produced by Hamamatsu Photonics K.K, which is one of the SiPM families.
The MPPC has a lot of excellent features suited for the MEG experiment as discussed in Sec.\,\ref{sec:pixTC SiPM}. 
Moreover, a finer granularity of the scintillation readout with MPPCs
allows us to reconstruct shallow events more precisely.
Less material budget before LXe active region results in a higher detection efficiency.

\subsubsection{Issues}
\label{sec:LXe MPPC issues}
There are a few issues to be addressed to realize a detection of the LXe scintillation light by MPPC.

The first issue is the photon detection efficiency (PDE) for VUV light.
There are two types of the layer structures for the SiPM, p-silicon on a n-substrate (p-on-n) and n-on-p. 
In general, since the ionization coefficient for electrons is higher than that for holes, 
the breakdown initiation probability of electrons is always higher than that of holes. 
Blue light is absorbed close to the SiPM surface, and electrons initiate the avalanche breakdown in the p-on-n case, 
which results in a higher sensitivity in the blue light region.
The MPPC uses the p-on-n structure, which is suitable to detect the blue light.
The PDE of the MPPC for VUV light is, however, nearly zero for commercial products 
since VUV photons can not reach a sensitive layer due to a protection coating layer made of epoxy resin or a silicon rubber
and an insensitive contact layer with no electric field.
An anti-reflection (AR) coating layer is not optimized to 
the refractive index of LXe at the scintillation light wavelength (175\,nm).

The second issue is the MPPC size.
The current largest single MPPC commercially available is $3\times3\,\mathrm{mm}^{2}$, which is still too
small to cover the inner face of the LXe detector because too many readout channels will be necessary for the small sensor.
It is desirable to develop a large area MPPC with $10\times 10\,\mathrm{mm}^{2}$ or larger.
However, the larger size of MPPCs could causes a larger dark count rate, larger gain non-uniformity, 
and larger capacitance (a longer tail in the waveform, a larger noise etc.).  

A UV-sensitive MPPC is under development in collaboration with Hamamatsu Photonics 
to be used for the upgrade LXe $\gamma$-detector.
We will describe the performance of the prototype sample in the following sections.

\subsubsection{Setup}
\begin{figure}[h]
\begin{center}
  \includegraphics[width=.25\textwidth,angle=0]{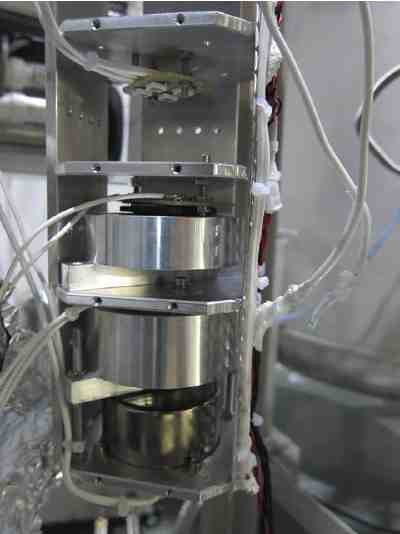}
  \includegraphics[width=.4\textwidth,angle=0]{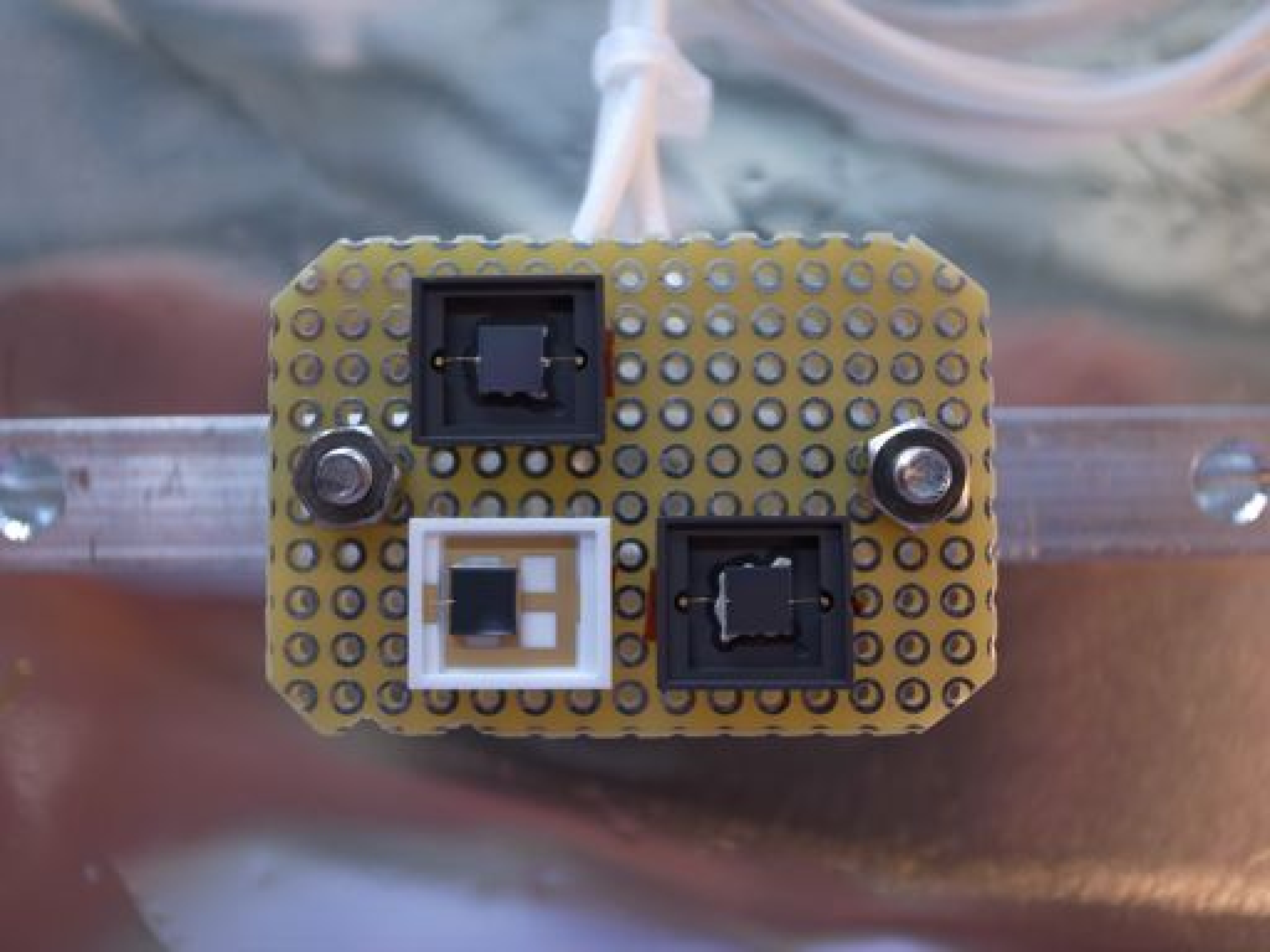}
\caption[]{Test setup for the UV-sensitive MPPC prototype. The left picture shows three stages to mount MPPC samples in the cryostat. 
The right figure shows three MPPC samples mounted on a stage. The size of the sample in this picture is $3\times 3\,\mathrm{mm}^{2}$, 
}
 \label{fig:mppcsetup}
\end{center}
\end{figure}

We built a test facility at PSI
which consists of a $2\,\ell$-LXe cryostat with a pulse tube cryocooler and a xenon gas purifier 
to measure basic characteristics of MPPC samples in LXe.
There are three stages in the cryostat, each of which can mount a few MPPC samples as shown in Fig.~\ref{fig:mppcsetup}. 

We install $\alpha$ sources for the absolute PDE measurement, LEDs for calibration and  
a UV-sensitive PMT for triggering on $\alpha$ events inside the cryostat. 
The MPPC mounting stages are surrounded by a wall 
with a special coating to suppress the reflection of the VUV scintillation light.
About one thousand photons from an $\alpha$ source reach the active area of the MPPC ($3\times3\,\mathrm{mm}^{2}$).

\subsubsection{PDE}
\label{sec:PDE}
Prototypes ($\sim$40 samples up to now) optimized for VUV detection are produced by Hamamatsu,
which have no protection coating, a thinner contact layer, 
or optimized AR coating with different parameters.

\begin{figure}[h]
   \begin{center}
      \includegraphics[width=.6\textwidth,angle=0]{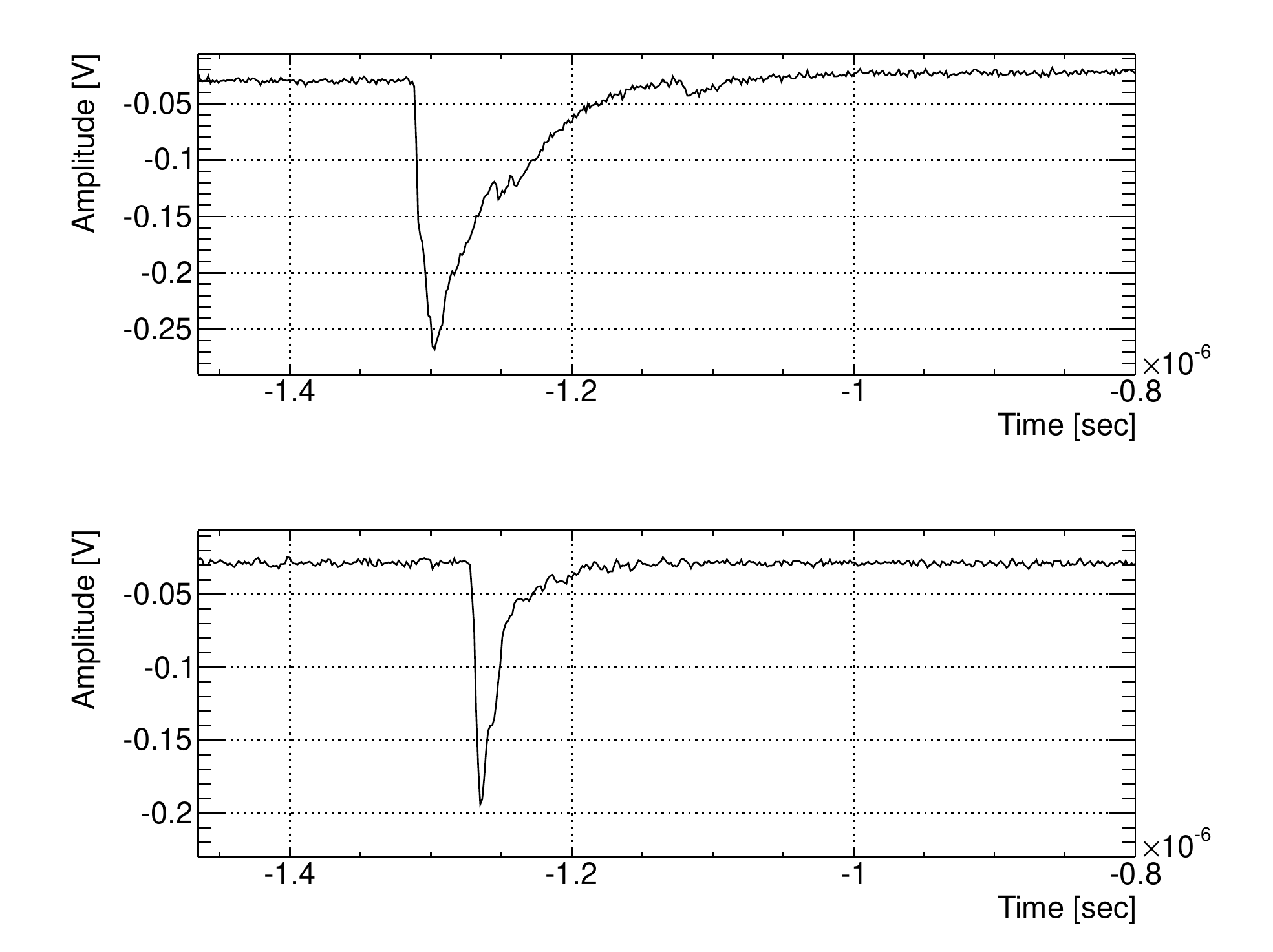}
      \caption[]{A MPPC signal waveform (upper) and a PMT signal waveform (lower) for the same $\alpha$ event taken with a waveform digitizer. 
        A sampling frequency of the digitizer is 700MHz.}
      \label{fig:mppc_pmt_waveform}
   \end{center}
\end{figure}

We succeeded to detect the LXe scintillation photons from $\alpha$ event by using the prototype sample.
Fig.~\ref{fig:mppc_pmt_waveform} shows signal waveforms 
from the MPPC sample (upper figure) and a UV-sensitive PMT (lower one)
for the same $\alpha$ event. 

The number of detected photoelectrons for $\alpha$ event is calculated from the ratio of the observed charge to that obtained 
for a single photoelectron event.
The PDE is then estimated from the ratio of the detected number of photoelectrons to the expected number of incoming scintillation photons from $\alpha$ event.
This PDE still contains contributions from cross-talk, after-pulse, and the infrared component of the LXe scintillation light.
The contribution from the infrared component is estimated to be $\sim1\,\%$ indirectly by using the signal observed with a commercial MPPC (S10362-33-100C) 
, is supposed to be insensitive to VUV component.
%

\begin{figure}[h]
   \begin{center}
      \includegraphics[width=.45\textwidth,angle=0]{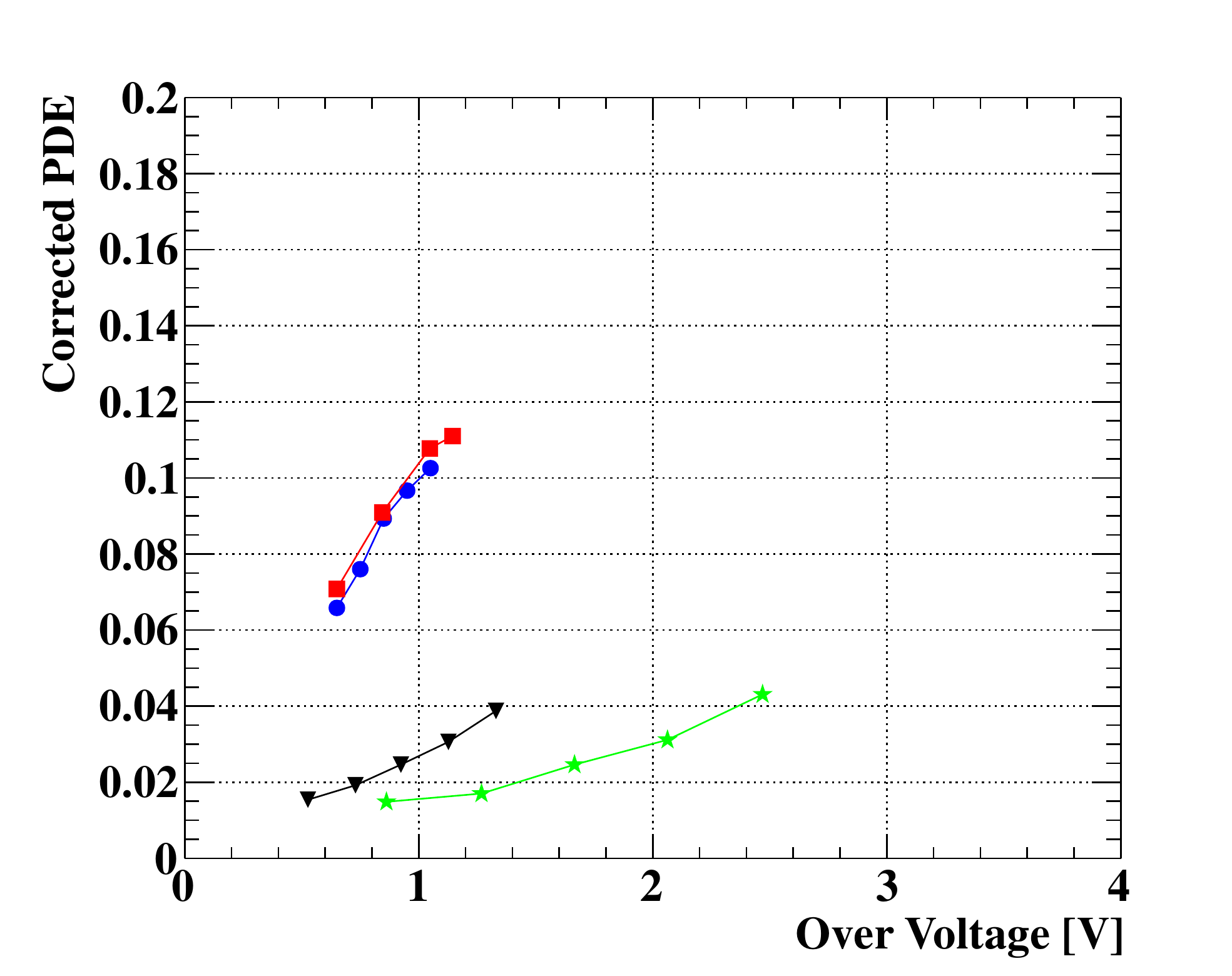}
      \includegraphics[width=.45\textwidth,angle=0]{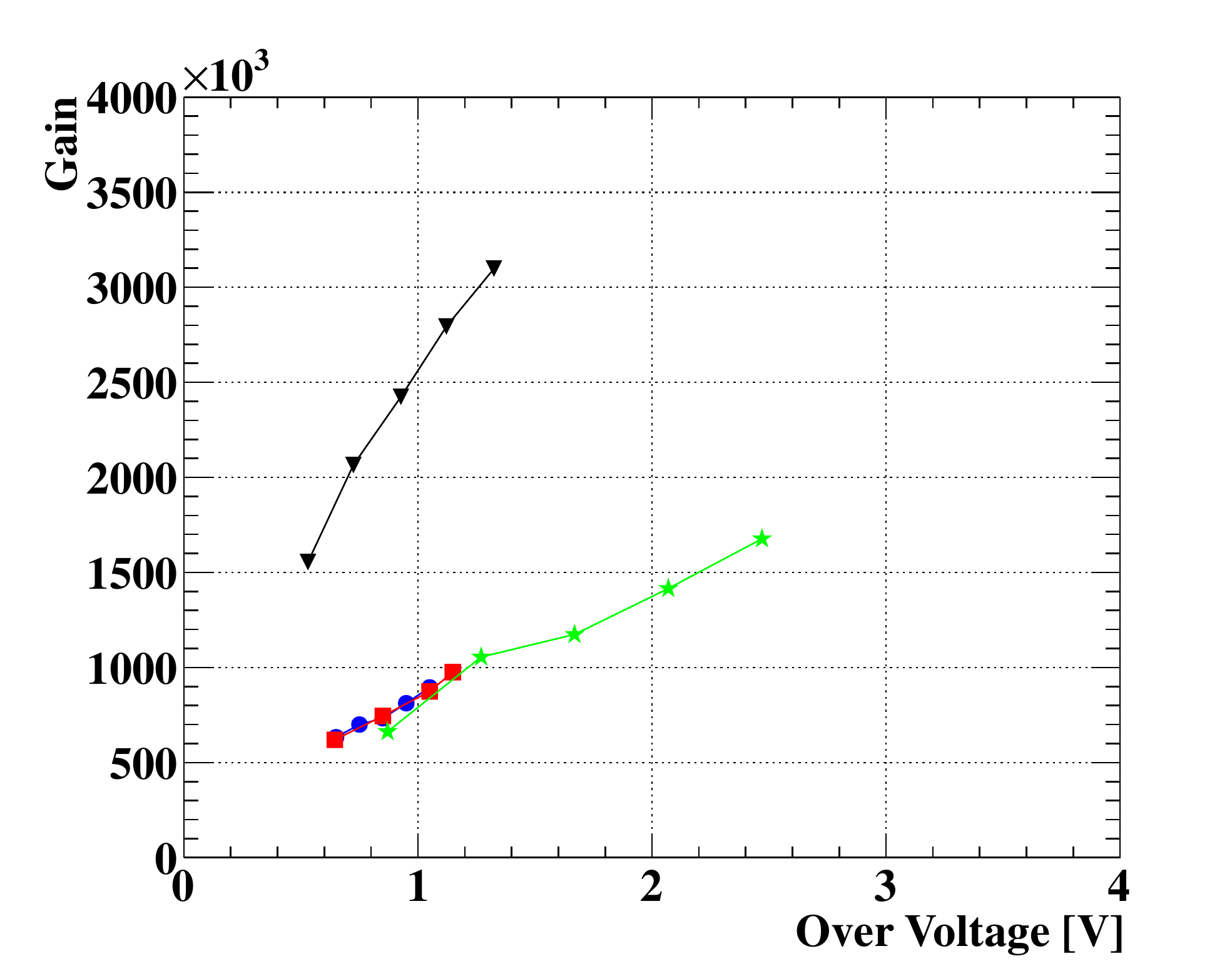}
      \caption[]{(Left) Measured PDEs and (right) gains as a function of the over-voltage.
      In the right plot, the sample with black triangles shows a higher gain and larger slope. It is due to the different pixel pitch 
(100\,$\mu$m for the black triangle and 50\,$\mu$m for the others). 
      }
      \label{fig:MPPC_PDE}
   \end{center}
\end{figure}

Fig.~\ref{fig:MPPC_PDE} shows the measured PDEs for four MPPC samples after correcting for the contributions from cross-talk and after-pulse.
There is roughly 30\% uncertainty in the PDE measurement which comes from the uncertainty 
in the estimation of the cross-talk + after-pulse probability.
The best sample shows a PDE higher than 10\% PDE in LXe. 
It is somewhat smaller than the QE for the UV-sensitive PMT of the current detector ($\sim$16\%).
However since the sensor coverage on the inner face is increased by 50\% using MPPCs, 
the total photoelectron statistics would not be changed.
Note that the energy resolution of the current detector is not limited by the photoelectron statistics.

The measured gains are also shown in Fig.~\ref{fig:MPPC_PDE}.
This can be measured directly by using a single photoelectron charge distribution. 
All the samples show the gains higher than $10^6$ at a reasonable overvoltage, which is high enough for our purpose.

\subsubsection{Transit Time Spread}
The transit time spread (TTS) of commercially available MPPC is about 100\,psec.
It is better than that of PMT, which is typically about 750\,psec.
The single photoelectron time
resolution of MPPC is, therefore, better than that of PMT. 
However TTS of VUV MPPC has never been measured.
The attenuation length VUV light in silicon is very short ($\sim5$\,nm); VUV photons
therefore generate electron-hole pairs near the surface of silicon. The statistical
fluctuation of the travel time of electrons reaching the multiplication layer would be a source of
the time-jitter.

The time resolution is evaluated by measuring the time difference of two MPPCs looking at the same $\alpha$-event.
Since the number of photoelectrons observed in the test-setup is about 100, the width of
the time-difference is limited by the photoelectron statistics. In order to
estimate TTS, the contribution of the photoelectron statistics is deconvolved from
the time-difference distribution. As a result, TTS is estimated to be about 150\,psec.
The contribution of TTS to the time resolution of the MEG LXe detector is, therefore,
negligible because the total number of photoelectrons for the signal $\gamma$-ray is the order
of $10^5$.

\subsubsection{Temperature Dependence}

\begin{figure}[htb]
   \begin{center}
      \includegraphics[width=.45\textwidth,angle=0]{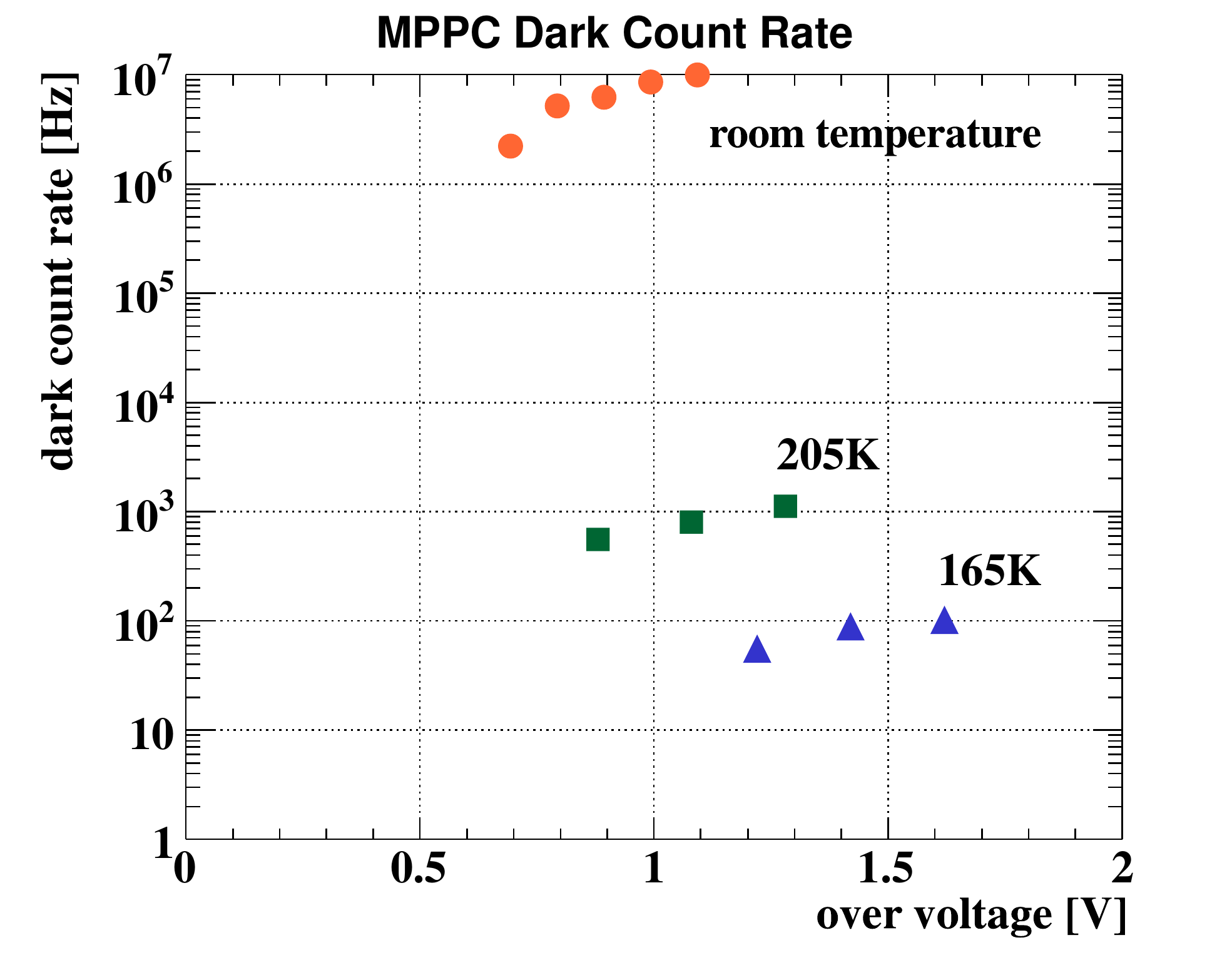}
      \includegraphics[width=.5\textwidth,angle=0]{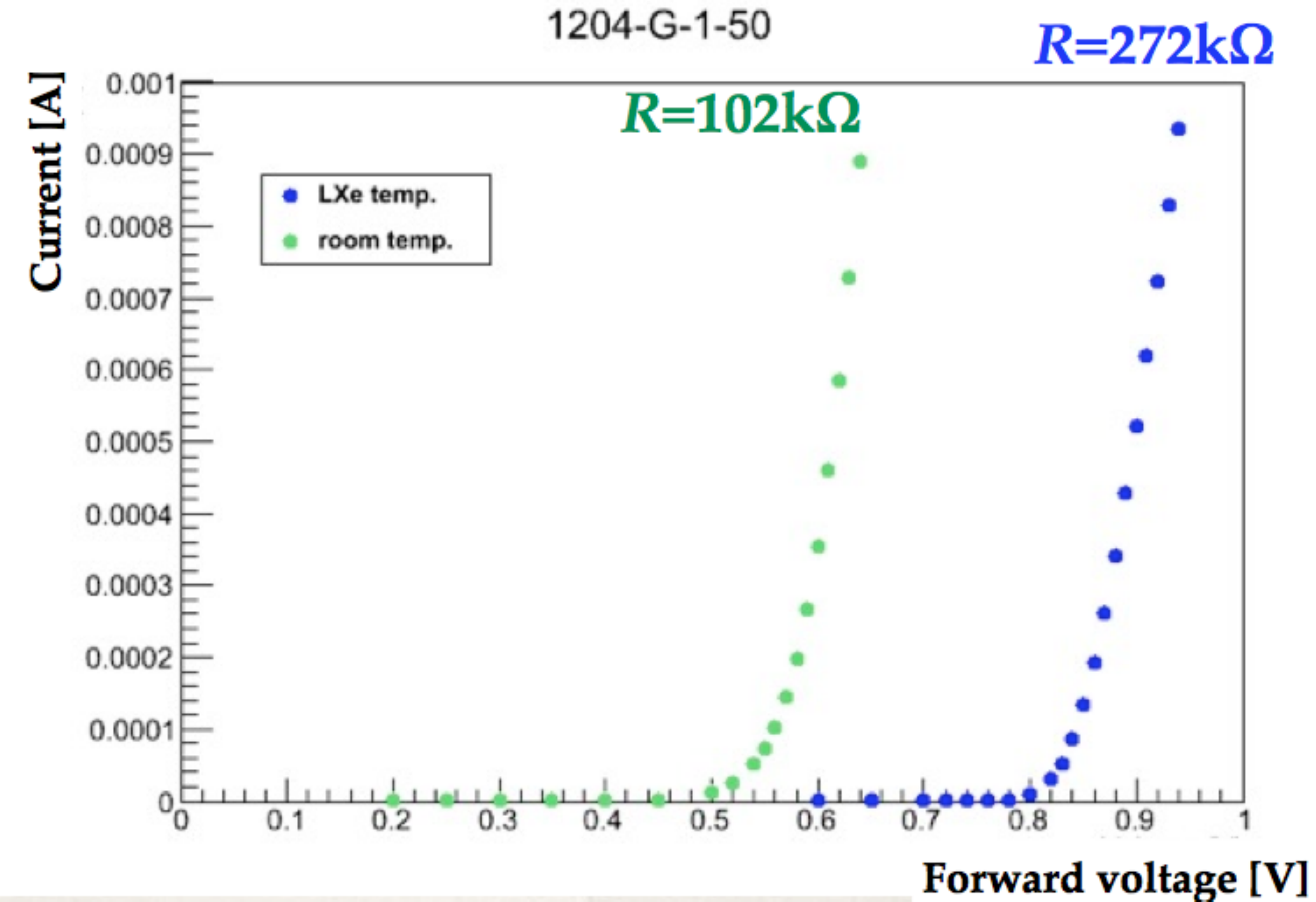}
      \caption[]{(Left) The dark count rate measured at different temperatures (room temperature, 205\,K and 165\,K). 
            (Right) Resistance of poly-silicon quench resistor measured at LXe and at room temperatures by applying forward voltages. }
      \label{fig:MPPC_dark}
   \end{center}
\end{figure}

Thermally generated free carriers in a depleted layer produce dark counts.
The typical dark count rate is
100\,kHz-10\,MHz per mm$^{2}$ at room temperature.
It is known that the dark count rate is suppressed by five orders of magnitude at LXe temperature (165\,K)
\cite{JanicskoCsathy:2010bh}.
Our test measurements confirm that the dark count rate is reduced down to 1\,Hz -100\,Hz for $3\times3\,\mathrm{mm}^2$ samples 
at LXe temperature as shown in the left figure in Fig.\ref{fig:MPPC_dark}.

Polysilicon resistors are commonly used for quenching the breakdown in MPPC. 
The resistivity of the polysilicon changes with temperature.
The right figure in Fig. \ref{fig:MPPC_dark} shows the measurement of the quench resistance at LXe and at room temperatures.
The resistance at LXe temperature is measured to be more than a factor of two higher than that at room temperature.
There is a linear correlation between the resistance of the quench resister and the falling time of MPPC signal; 
a higher resistance leads to a longer waveform tail. 
The quenching resistance has to be reduced for the use at LXe temperature.

\begin{figure}[h]
   \begin{center}
      \includegraphics[width=.5\textwidth,angle=0]{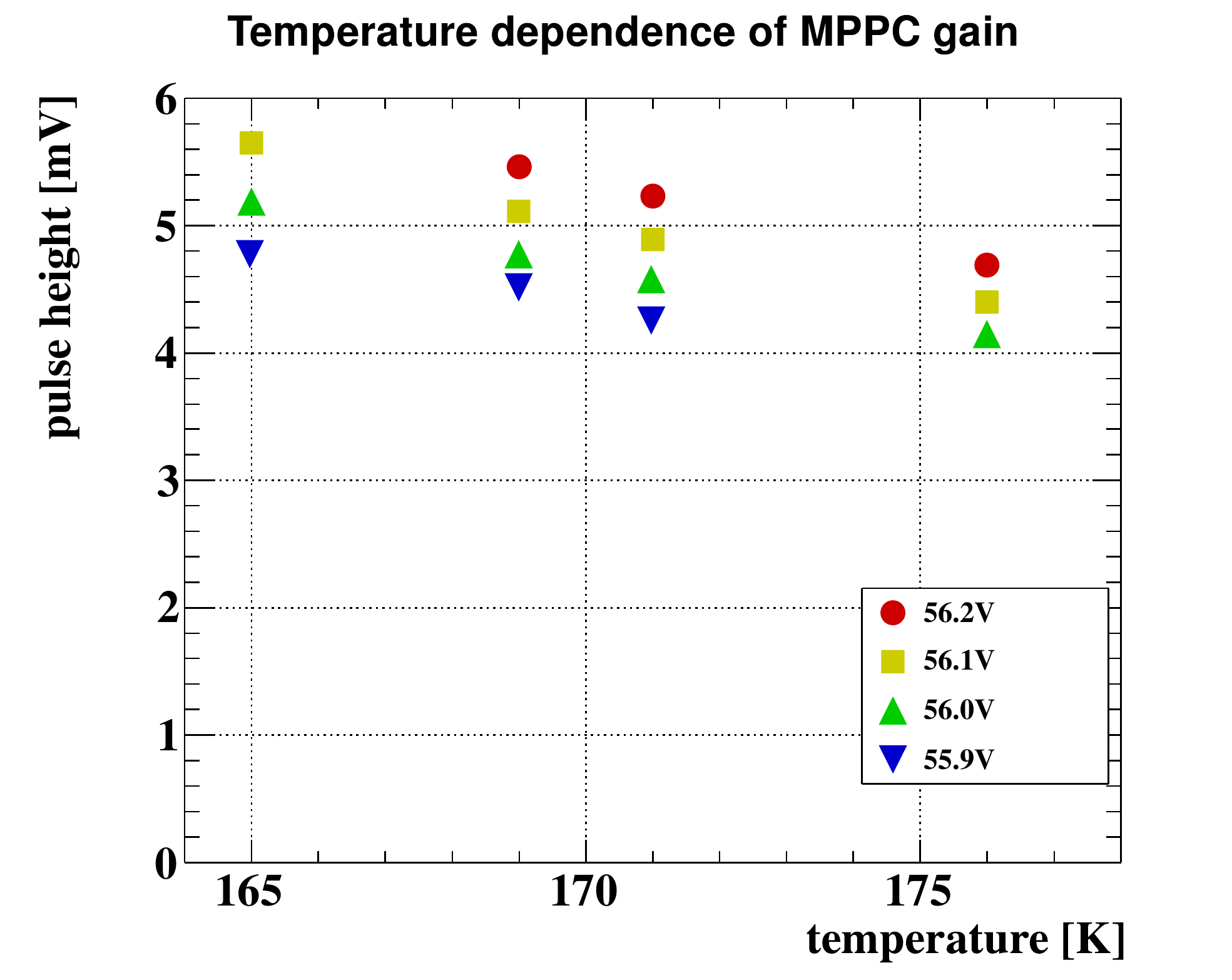}
      \caption[]{Temperature dependence of MPPC gain.}
      \label{fig:MPPC_gaintemp}
   \end{center}
\end{figure}

The breakdown voltage of MPPC is known to have a relatively large temperature coefficient 
and the gain and PDE of MPPC can, therefore, be easily shifted depending of the temperature. 
It could be an issue of the stability of the detector performance.
The temperature coefficient of the gain is measured for the UV-sensitive MPPC.
Fig.\ref{fig:MPPC_gaintemp} shows the temperature dependence of the amplitude of the single photoelectron signal.
The LXe temperature is controlled by changing the pressure in this measurement.
The measurement is performed at the pressures of 0.115, 0.135, 0.15, and 0.19 MPa 
which correspond to temperatures of 165,169,171 and 176K, respectively.
At each temperature condition, the gain measurement is done for different bias voltages.
We measure the temperature coefficient of the gain to be $-2\,\%/\mathrm{K}$ independently of the bias voltage.
On the other hand, the LXe temperature stability of the current detector is measured to be smaller than 0.15K (RMS),
which is most likely to be dominated by the precision of the temperature measuring device.
The fluctuation of the MPPC gain in the final detector is, therefore, expected be smaller than 0.3\% (RMS). 
The temperature dependence of the PDE is not yet measured, 
but is likely to have the same temperature coefficient as the gain. 
This means that the fluctuation of the overall gain of the MPPC in the final detector should be smaller than 0.6\%,
which is smaller than the expected energy resolution of the final detector as described in Sec.\,\ref{sec:expected performance}.

\subsubsection{Radiation Hardness}

Radiation produces defects in the silicon bulk or at the Si/SiO$_{2}$ interface of SiPM. 
As a result, some parameters of SiPM such as the breakdown voltage, leakage current, 
 dark count rate,  gain, and PDE may change after irradiation. 
There were a lot of previous works to study the radiation hardness of the SiPMs 
by irradiating the SiPM with $\gamma$-rays, neutrons, protons, or electrons.

An increased dark count rate was observed at more than 10$^{8}$\,neutrons/cm$^{2}$, 
and a loss of single p.e. detection capability was observed at more than 10$^{10}$\,neutrons/cm$^{2}$\,\cite{Matsumura:2006zz}. 
From the neutron flux measured at the MEG experimental area, the total neutron fluence is estimated to be less than 1.6$\times$10$^{8}$\,neutrons/cm$^{2}$
in the MEG upgrade.

Increased leakage current was observed with a $\gamma$-ray irradiation of 200\,Gy\,\cite{Matsubara:2006zz}, 
while the $\gamma$-ray dose in the MEG upgrade is estimated to be 0.6\,Gy. 

This means that the radiation damage should not be an issue for the MPPC for the MEG upgrade.

\subsubsection{Linearity}

\begin{figure}[htb]
   \begin{center}
      \includegraphics[width=.5\textwidth,angle=0]{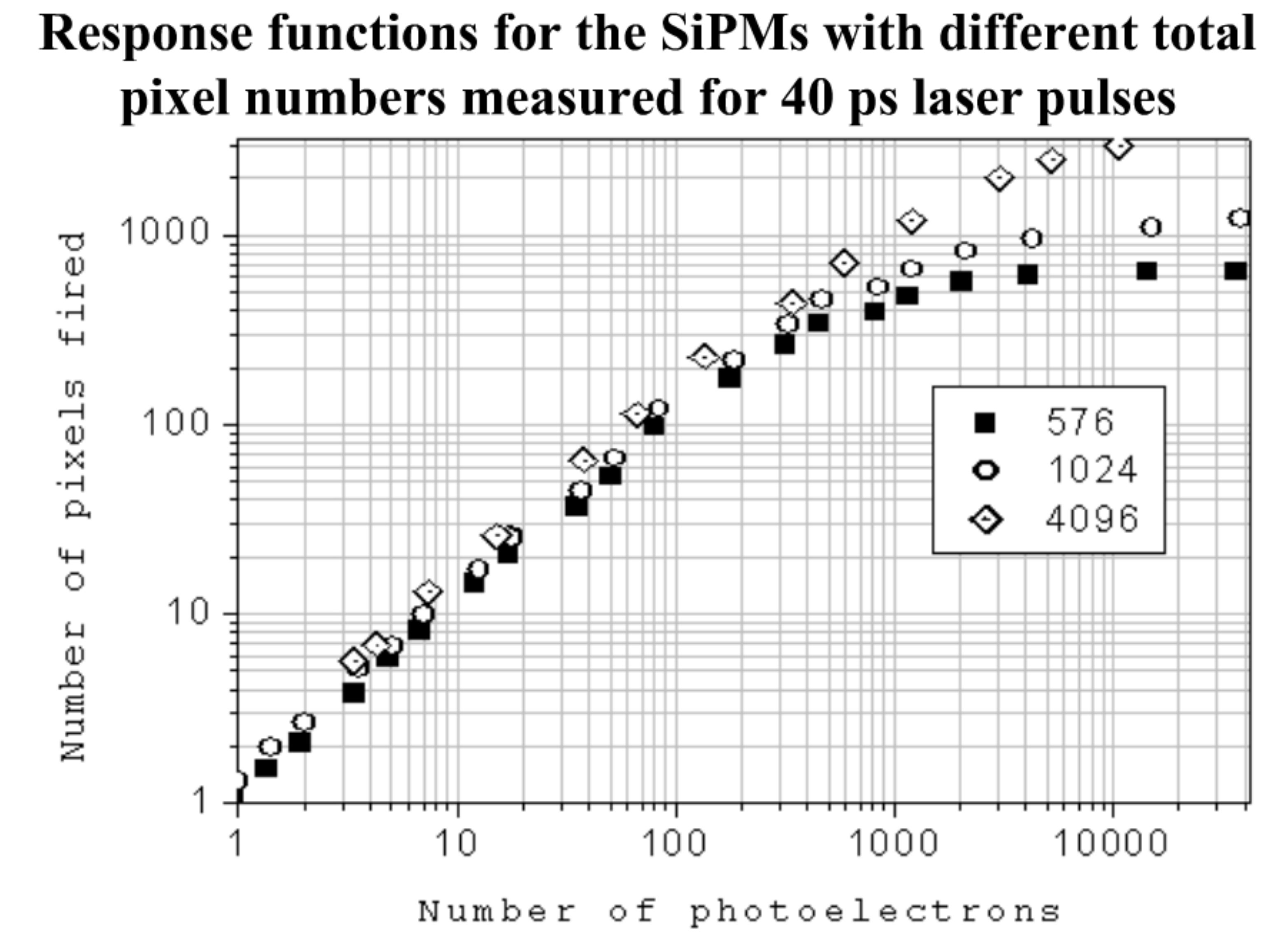}
      \includegraphics[width=.45\textwidth,angle=0]{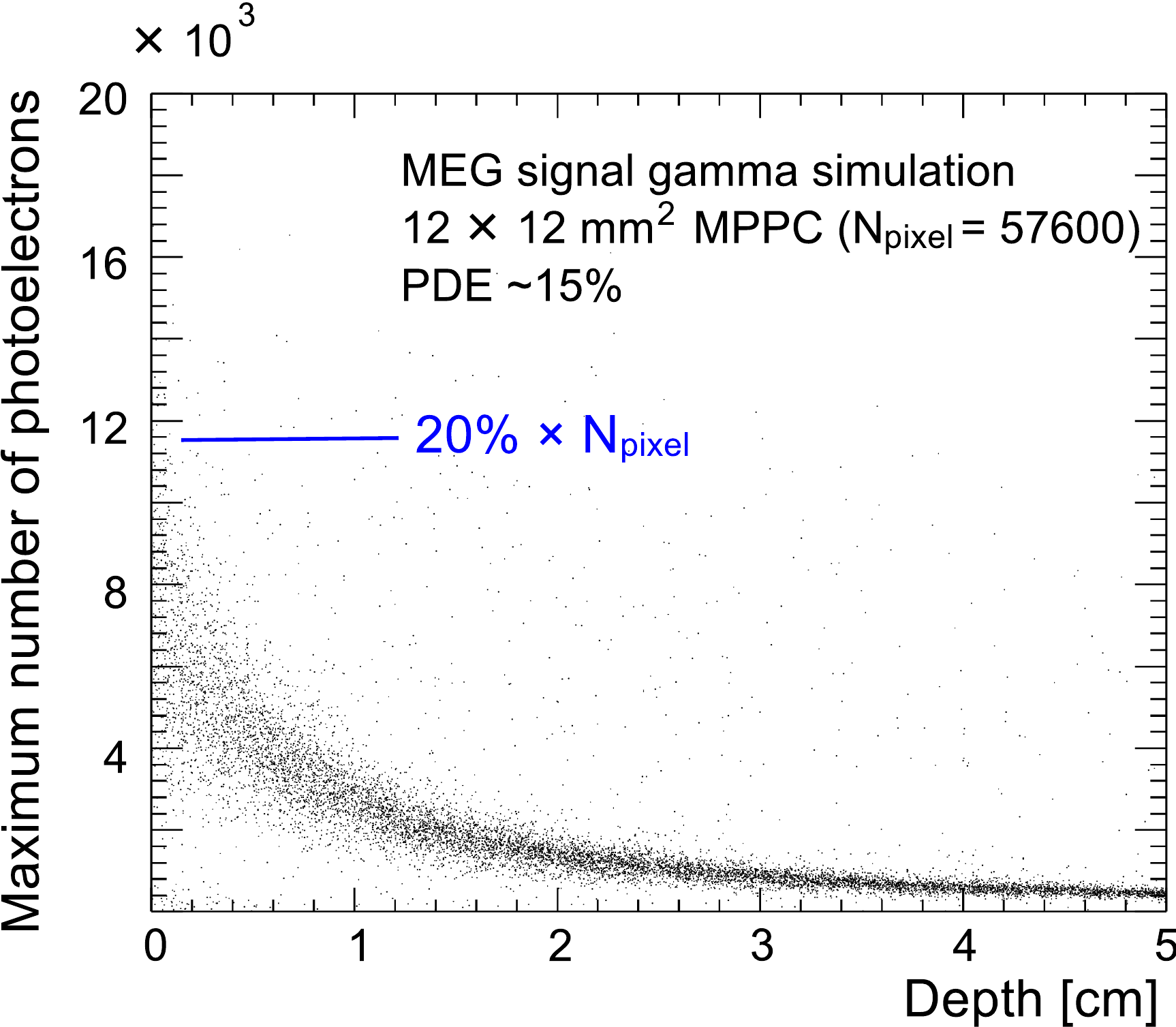}
      \caption[]{(Left) Response functions for the SiPMs with different total pixel numbers measured for 40\,ps laser pulses\,\cite{Andreev:2004uy}.
      	 (Right) The number of photoelectrons expected with 12$\times$12\,mm$^{2}$ MPPC in the MEG Monte Carlo simulation  }
      \label{fig:MPPC_linearity}
   \end{center}
\end{figure}

SiPM shows a non-linear response when the number of incident photons is large compared to the number of pixels of the device.
The optimal condition is that the number of incident photons be much smaller than the number of pixels without any localization.
Fig.\ref{fig:MPPC_linearity} (left) shows the measured response functions for the SiPMs of $1\times1$\,mm$^{2}$ with the different total numbers of pixels 
illuminated by 40\,ps laser pulses\,\cite{Andreev:2004uy}. 
For the LXe detector in the MEG upgrade, the expected number of photoelectrons reaches up to 
12000\,p.e. on 12$\times$12mm$^{2}$ sensor area ($\sim20$\% of the total number of pixels, 57600) 
for very shallow signal events. 
The expected non linearity is, therefore, not a big issue and can be corrected by a careful calibration.

\subsubsection{Large Area Samples}

\begin{figure}[htb]
\begin{center}
  \includegraphics[width=.3\textwidth,angle=0]{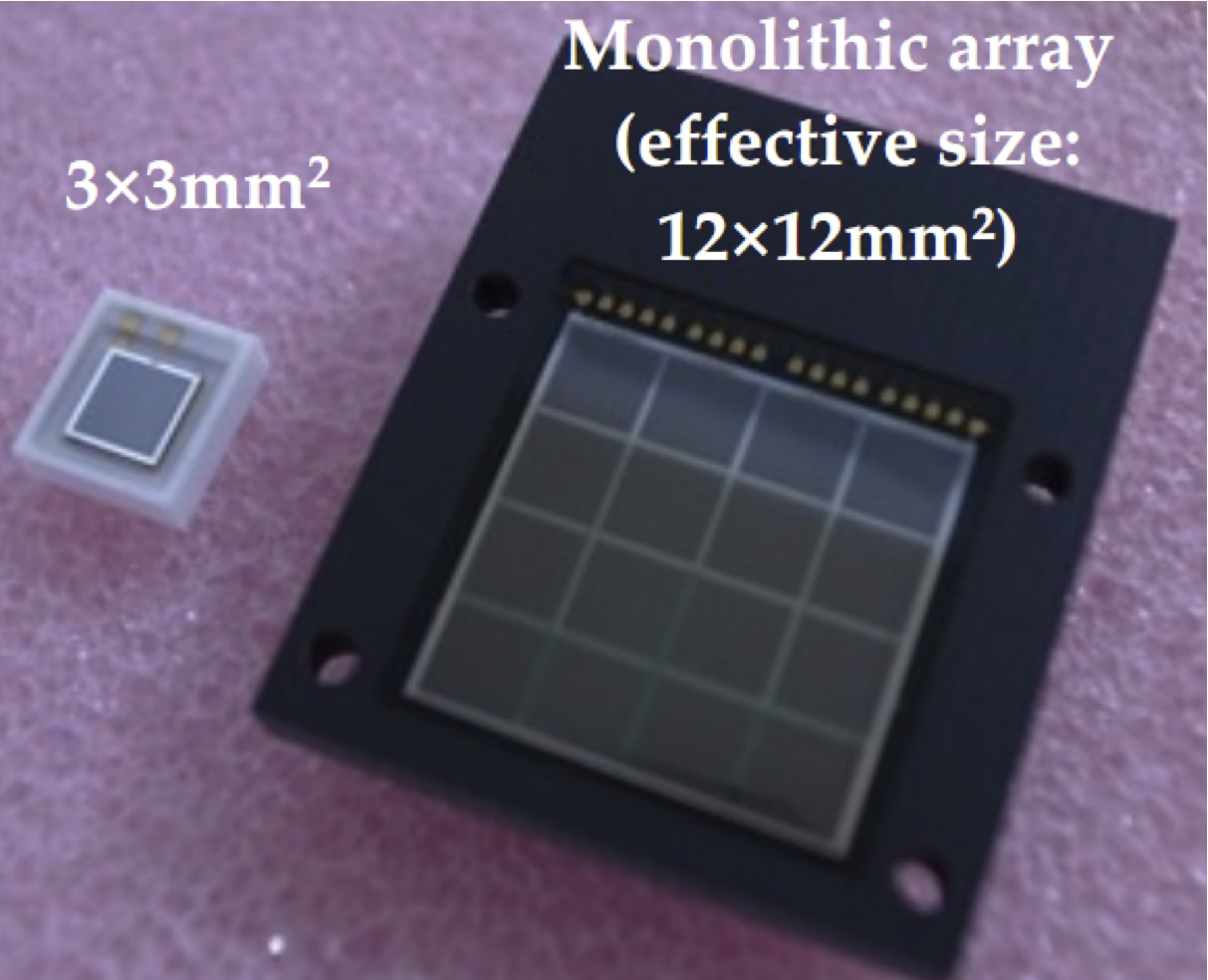}
  \includegraphics[width=.4\textwidth,angle=0]{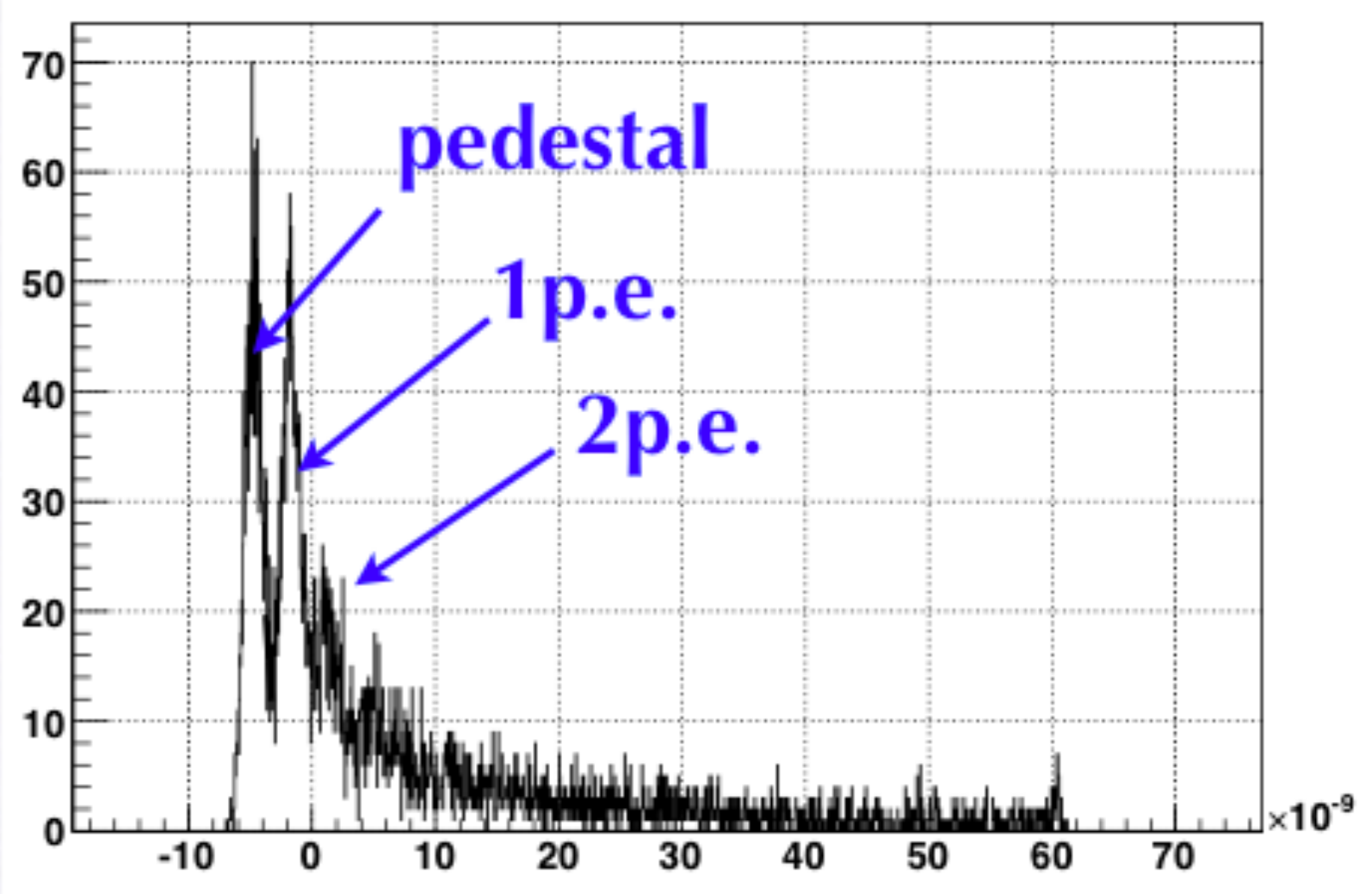}
 \caption[]{(Left) Monolithic array MPPC produced by Hamamatsu (S11827-3344MG). 
 (Right) Charge distribution of dark counts measured by the monolithic MPPC. 1\,p.e. peak is clearly resolved.  }
 \label{fig:monoMPPC}
\end{center}
\end{figure}

The current largest MPPC ($3\times3\,\mathrm{mm}^{2}$) is still too small for the MEG LXe detector,
and we need at least $\sim$ 10$\times$10mm$^{2}$ to replace the PMTs. 
When we construct a larger size of sensors, we have to pay attention to a possible increase
of the dark count, gain uniformity over the sensor area, and an increase of the sensor capacitance. 
We study the possible issues of the large area MPPC by using a commercial monolithic array MPPC (non UV-sensitive), 
which is composed of 4$\times$4 separate sectors with $3\times 3\,\mathrm{mm}^{2}$ active area each
(Hamamatsu S11827-3344MG, the left picture in Fig~\ref{fig:monoMPPC}).
All 16 sectors are connected in parallel so that the module can work as a single large-area MPPC with an active area of 
12$\times$12\,mm$^{2}$. 
Single photoelectron peak is clearly resolved in the charge distribution observed for the dark noise (the right figure in Fig~\ref{fig:monoMPPC}).
The dark count rate of the module is measured to be 720\,Hz at LXe temperature,
which is low enough for the MEG LXe detector. 
We observe a longer fall-time and a larger noise caused by the increase of the sensor capacitance compared to the $3\times 3\,\mathrm{mm}^{2}$ MPPC.

Recently, Hamamatsu has provided us with large area MPPC samples with VUV sensitivity. 
The MPPC parameters are the same as that for the previous small sample with the best PDE (see Sec.\,\ref{sec:PDE}),
and the size is 12$\times$12\,mm$^{2}$.
This sample is being tested in our test setup, and the $\alpha$ source events are successfully observed.
The measurements are still going on, and the careful checks for this sample will be done later,
The characteristics of the large-area samples are still being investigated in detail
but the preliminary measurement shows that the PDE is consistent with the best small samples. 



\subsubsection{Possible Further Improvements of MPPC Performance}
Hamamatsu Photonics has recently developed a new type of MPPC with a significantly improved performance\,\cite{HamamatsuIEEE2012}.
The new features are the followings.

\begin{itemize}
   \item Double metal trace structure
   \item Metal quench resistor
   \item Optimized layer structure 
\end{itemize}

The double layer structure of the metal trace line connecting pixels 
enables more uniform and lower resistance without loosing the geometrical fill factor. 
This leads to a significant improvement of the inter-pixel jitter from 300\,ps to 150\,ps.

A poly-silicon quench resistor is used for the conventional MPPC.
The drawbacks of the poly-silicon quench resistor are that 
it has a large negative temperature coefficient 
and the per-pixel variation of the resistance is relatively large 
due to grain boundaries of the resistor structure.
We, therefore, suffer from high resistance of the quench resistor especially at low temperature, 
which causes slow pixel recovery and a long signal tail.
A metal quench resistor is used in the new MPPC to overcome these problems.
The uniformity of the resistance is found to be improved by more than a factor of two.
The temperature coefficient is measured to be smaller by a factor of five.

The probability of the after-pulsing is greatly reduced by an optimized layer structure in the new MPPC.
Thanks to the dramatic suppression of the after-pulsing, an operation at a much higher overvoltage is now possible.
This will improve the performance of the MPPC such as the gain, PDE, temperature stability and timing resolution.

The performance of the new MPPC is summarized in Table\,\ref{Table:Hamamatsu IEEE2012}.
All the above-mentioned features can be applied to the MEG MPPC too and thus further improvement of the performance is anticipated.

\begin{table}[tb]
   \caption{\label{Table:Hamamatsu IEEE2012}
   Performance improvement for the new type of MPPC\,\cite{HamamatsuIEEE2012}. 
}
\begin{center}
\begin{tabular}{lcc}
                                      & {\bf Standard MPPC} & {\bf New MPPC}  \\\hline\hline
                    Dark Count        & 2.7\,Mcps           & 1.7\,Mcps \\
                    After-pulsing     & $>100\%$            & 3\% \\
                    PDE               & 38\%                & 43\%\\
                    Overvoltage range & 1.5-2.5\,V          & 2-3.5\,V\\
                    Timing resolution & 250\,ps            & 140\,ps\\
\end{tabular}
\end{center}
\end{table}

\subsubsection{MPPC in Other Experiments}
There are examples of particle physics experiments that have already successfully implemented a large
number of MPPCs as photo sensors and achieved high detector performance.
We can learn various aspects of mass production and quality control of MPPCs from these experiments.
The T2K experiment is the first experiment to develop a dedicated MPPC (S10362-13-050C) for their purpose. 
About 64,000 MPPCs of this type are used in the ``near detector complex'' (ND280) 
where a magnetic field of 0.2\,T exists and the space for the sensor is limited. 
They developed a system and technique to characterize a large number of MPPCs and successfully manged a quality control\,\cite{Yokoyama:2010qa}.

Another expample is the First G-APD Cherenkov Telescope (FACT), which is an imaging atmospheric Cherenkov telescope 
using 1440 MPPCs (S10362-33-50-C). They succeeded to improve the sensitivity of their instrument, which is currently using PMTs,
thanks to the lower operation voltage, robust performance, and higher photon detection efficiency of MPPCs\,\cite{Anderhub:2010eh}. 


%% file: 07_Photon_Calorimeter/detector_design.tex

\subsection{Detector Design}
\subsubsection{Design of Sensor Package and Assembly}
\label{sec:Design of Sensor Package and Assembly}
\begin{figure}[htb]                                                                                                                                                                  
\begin{center}
\includegraphics[width=10cm]{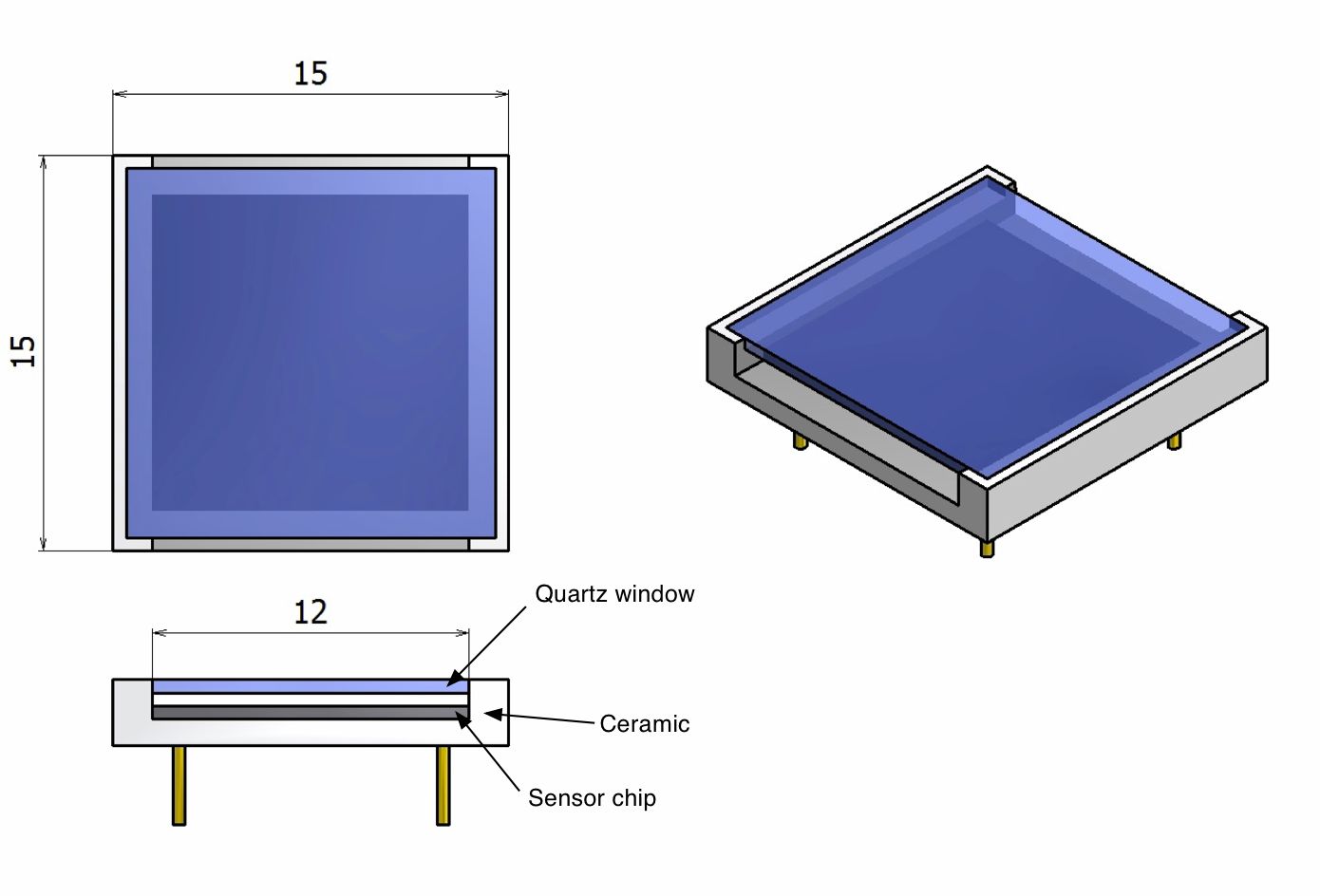}
\caption{\label{fig:MPPC package design}
   Possible design of MPPC package.
}
\end{center}
\end{figure}

Fig.\,\ref{fig:MPPC package design} shows a possible design of the UV-enhanced MPPC package.
The sensor chip with an active area of $12\times 12\,\mathrm{mm}^2$ 
is glued on a ceramic base of $15\times 15\,\mathrm{mm}^2$.
The ceramic is chosen as a base material because the difference in the thermal expansion rate compared to silicon
is relatively small at LXe temperature. 

\begin{figure}[htb]                                                                                                                                                                  
\begin{center}
\includegraphics[width=8cm]{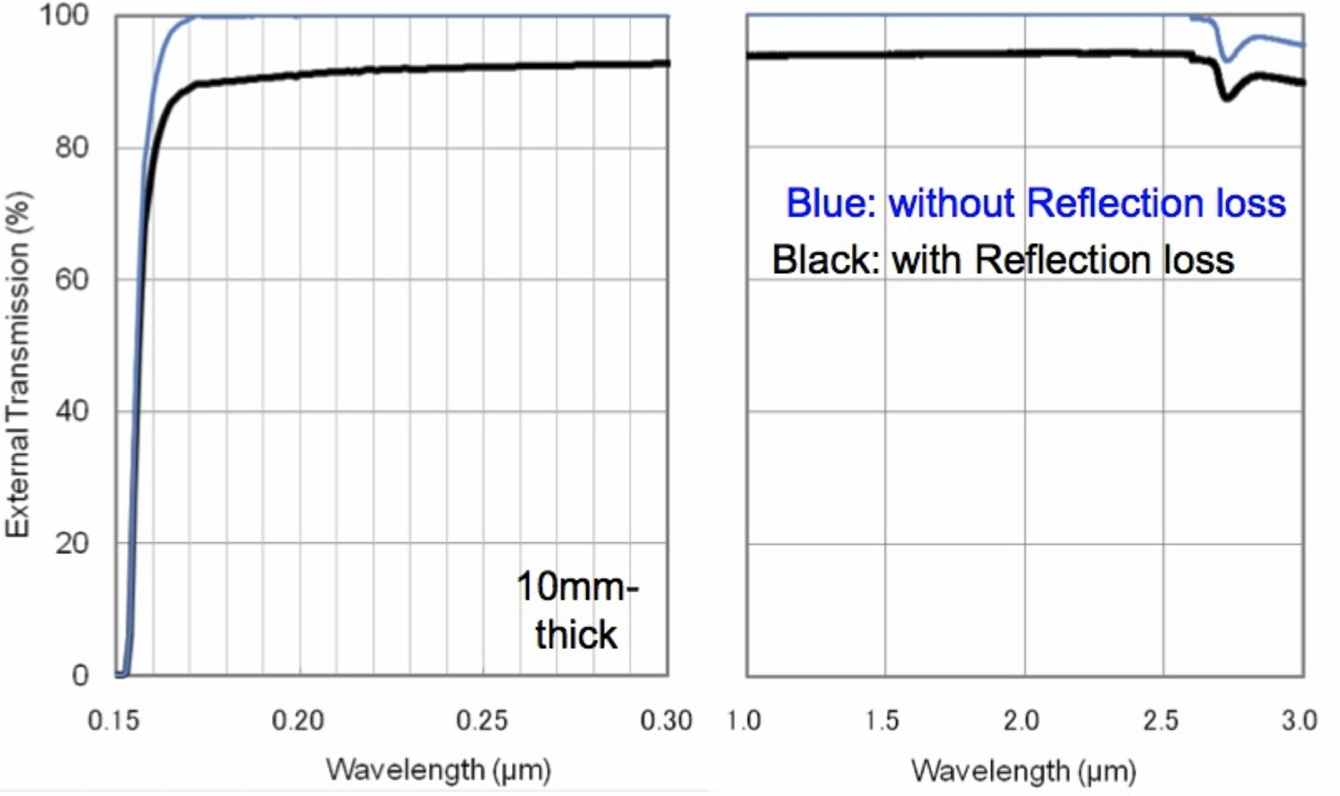}
\caption{\label{fig:AQ2 transmission}
   Transmission efficiency as a function of wavelength for high quality VUV-transparent quartz (Asahi-Glass AQ2).
}
\end{center}
\end{figure}

The sensor active area is covered with a thin quartz window for protection.
The window is not hermetic; there is a gap between the sensor
and the window where LXe penetrates in during the detector operation.
High quality VUV-transparent quartz such as Asahi Glass AQ2 and Shinetsu SUP-P700 is used for the window.
Fig.\,\ref{fig:AQ2 transmission} shows the transmission efficiency as a function of wavelength for Asahi Glass AQ2,
    where we can see that the transmission is quite high for the LXe scintillation light of 175\,nm.
It should be noted that the reflection loss is smaller than shown in the figure
since both sides of the quartz window touch LXe
which has the refractive index close to that of the quartz window
(LXe: 1.64, quartz: 1.60).  

\begin{figure}[htb]                                                                                                                                                                  
\begin{center}
\includegraphics[width=10cm]{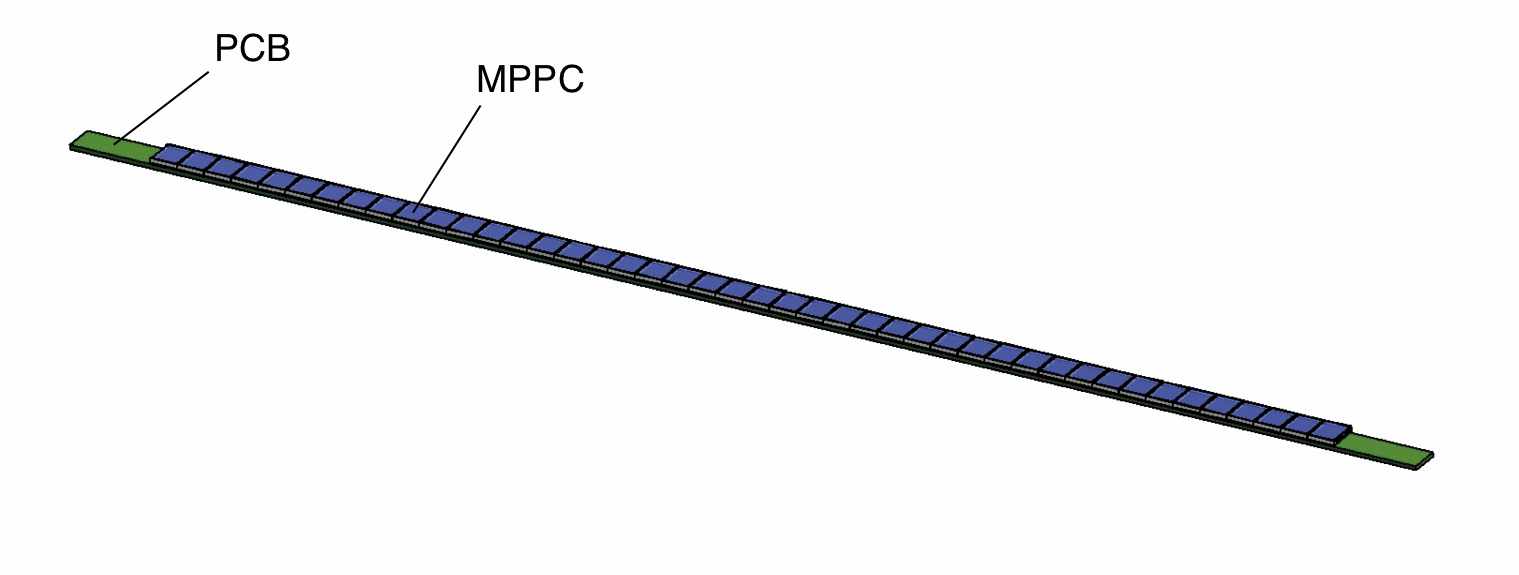}
\caption{\label{fig:MPPC PCB strip}
  MPPCs mounted on PCB strip.
}
\end{center}
\end{figure}

\begin{figure}[htb]                                                                                                                                                                  
\begin{center}
\includegraphics[width=7cm]{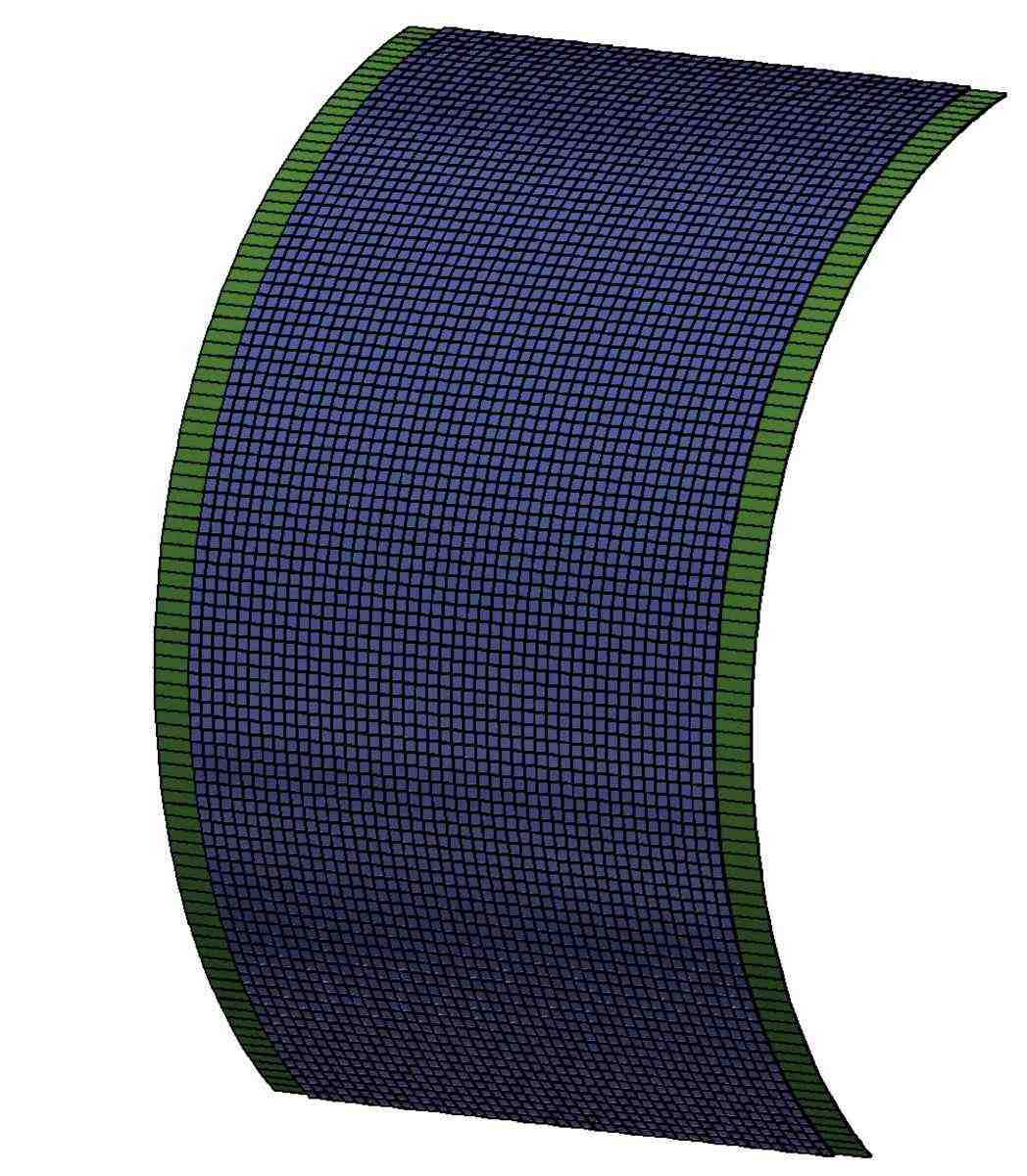}
\caption{\label{fig:MPPC slab assembly}
   Assembly of 93 PCB strips on the inner wall of the $\gamma$ entrance window.
}
\end{center}
\end{figure}

The MPPCs are mounted on a PCB strip as shown in Fig.\,\ref{fig:MPPC PCB strip}.
Each PCB strip has 44 MPPCs in a line along $z$-direction and 93 strips are arrayed along $\phi$-direction 
on the inner wall of the detector cryostat (Fig.\,\ref{fig:MPPC slab assembly}).
The total number of MPPCs is 4092.
The MPPC package has four electrode pins (two pins connected to the sensor chip and two dummy pins just for stable alignment)
 and is plugged in socket pins on the PCB.
 This mounting scheme allows easy replacement of the MPPC module in case it is necessary.

The signals from the MPPCs are transmitted on signal lines of the PCB 
which are designed to be well shielded from both outside and adjacent channels and to have 50\,$\Omega$ impedance.
Similar PCBs are used in the feedthrough of the cryostat as described in Sec.\,\ref{sec:cable and feedthrough}.

\begin{figure}[htb]                                                                                                                                                                  
\begin{center}
\includegraphics[width=12cm]{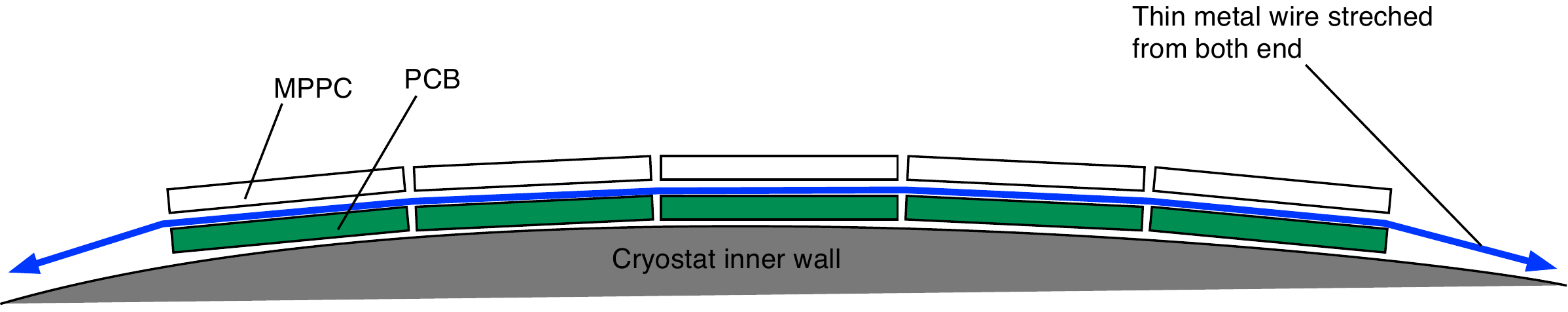}
\caption{\label{fig:PCB strip alignment}
   Possible scheme of alignment of the MPPC PCB strips (cross-sectional view on $x$-$y$ plane at a given $z$). 
   The PCB strips are pressed to the cryostat inner wall by thin metal wires stretched along $\phi$-direction.
}
\end{center}
\end{figure}

It is important to precisely align the PCB strips on the inner wall of the detector cryostat
and to minimize the gap between the strip and the wall 
since LXe in this gap deteriorates the $\gamma$-ray detection efficiency
and causes an undesirable low energy tail in the energy response function of the detector.
We can have mechanical structures on the cryostat to fix the strips, but only at both ends of the strip, outside the acceptance.
Even small distortions of this thin and long PCB strip could, therefore, be an issue. 
Fig.\,\ref{fig:PCB strip alignment} illustrates a possible scheme to firmly fix the PCB strips on the cryostat wall, 
   where the PCB strips are pressed to the wall
by using several thin metal wires stretched along $\phi$-direction. 

\subsubsection{Signal Transmission}
\label{sec:cable and feedthrough}
It is not an easy task to transmit about 4000 MPPC signals to DAQ electronics without introducing noise or distortion. 
We have to pay attention to pickup noise, cross-talk, and limited space in the cryostat and the feed-through etc.
In order to overcome such issues, we are developing a multi-layer PCB with coaxial-like signal line structure.
It is planned to be used for both the PCB for MPPC mounting and the vacuum feed-through of the cryostat. 

\begin{figure}[htb]                                                                                                                                                                  
\begin{center}
\includegraphics[width=8cm]{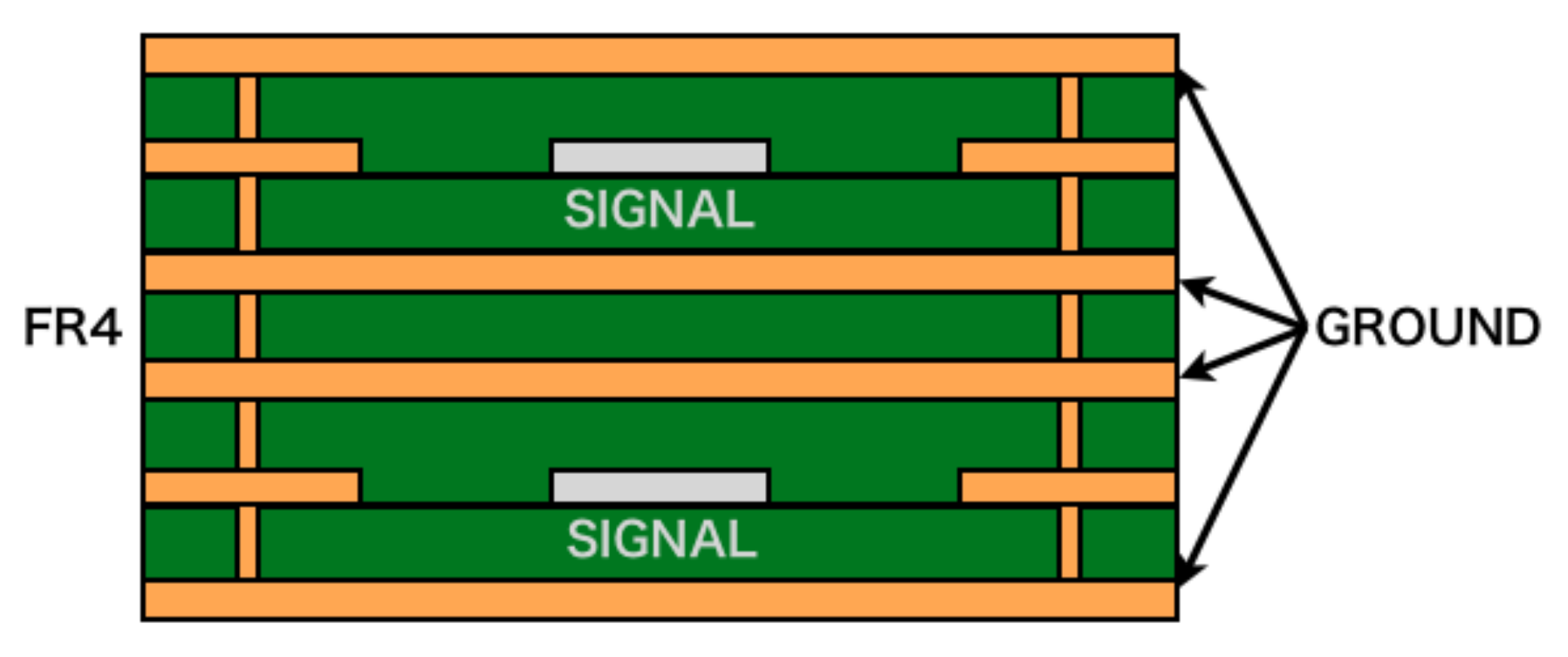}
\caption{\label{fig:PCBlayout}
   A cross-sectional view of the possible PCB design in which a signal line is shielded by surrounding ground lines and ground layers. 
  The total thickness of the PCB board is 1.6\,mm.}
\end{center}
\end{figure}

As described in Sec.\,\ref{sec:Design of Sensor Package and Assembly},
44 MPPCs are mounted on a PCB strip and 22 signal lines embedded in the strip transmit signals to each end.
Fig.\,\ref{fig:PCBlayout} shows the possible layer structure of the PCB.
There are two signal layers, each of which contains 11 signal lines in 15\,mm width.
Each signal line is surrounded by ground patterns to minimize cross-talk effect and to be well shielded from outside. 
In total, six layers (two layers for signal, and four layers for ground) are used.
The dimensions of the layers are adjusted to have 50\,$\Omega$ impedance. 

\begin{figure}[htb]                                                                                                                                                                  
\begin{center}
\includegraphics[width=10cm]{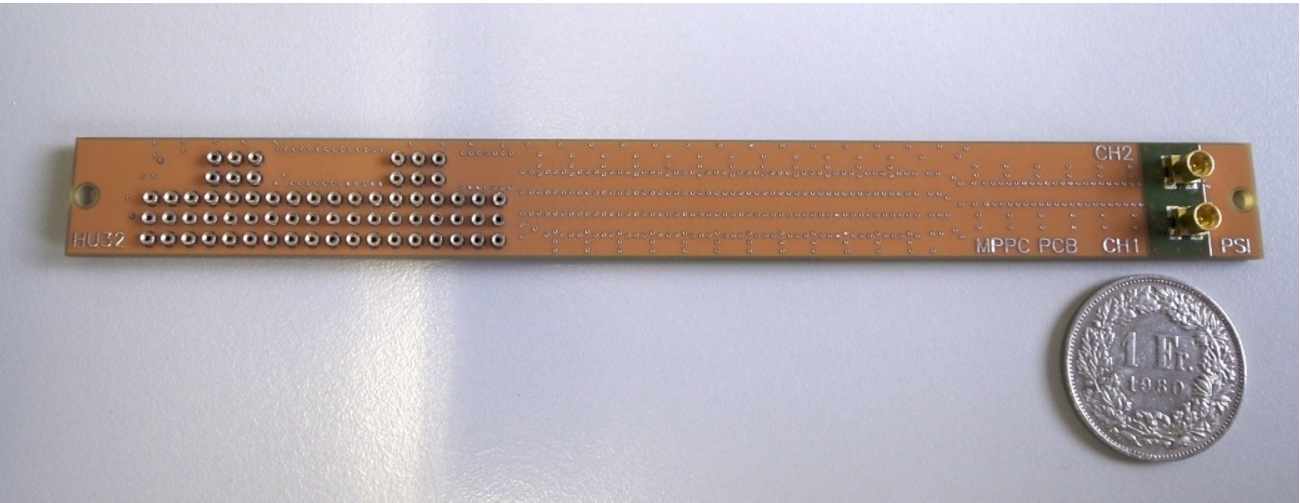}
\caption{\label{fig:prototype of MPPC mounting PCB}
Prototype of the PCB for MPPC mounting.}
\end{center}
\end{figure}

A prototype of the PCB for MPPC mounting is developed as shown in Fig.\,\ref{fig:prototype of MPPC mounting PCB}.
The length of 15\,cm is much shorter than that of the final PCB strip such that it can be tested in the small LXe test facility.
However, the signal line in the PCB has the same coaxial-like layer structure as the final one.
The total length of the signal line is about 35\,cm, which is also the same as that of the final one.
Similarly to the final PCB, MPPCs are plugged in socket pins on the PCB
and MMCX (micro-miniature coaxial) connectors are used at the end of the signal line.
The prototype PCB is being tested in the LXe test facility using real MPPC signals.
The test results will be described in Sec.\ref{sec:pcbtest}.

The signal lines on the PCB strip are connected to (real) thin coaxial cables
either by means of connectors or by soldering them at the edge of PCBs. 
Then the signals are transmitted to the feed-throughs using 
thin coaxial cables with a length of at least 3\,m.
One of the possible candidates for the coaxial cables is Radiall MIL-C-17/93-RG178\cite{Radiall},
which is currently used for the drift chamber in the MEG experiment.

\begin{figure}[htb]                                                                                                                                                                  
\begin{center}
\includegraphics[width=10cm]{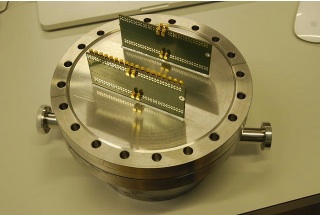}
\caption{\label{fig:feedthru} Prototype of the PCB-type vacuum feedthrough for the MEG LXe detector. 
   }
\end{center}
\end{figure}

The current LXe detector has in total 10 DN160CF flanges for signal and HV cables of 846 PMTs. 
Since only the 216 inner PMTs will be replaced with 4092 MPPCs and other 630 PMTs will remain in the detector, 
more feedthrough ports will be necessary. 
A PCB-type feedthrough similar to the PCB for MPPC mounting is considered as a possible candidate,
which can realize high density signal transmission through vacuum walls and low noise environment.
A prototype of the feedthrough is made as shown in Fig.\,\ref{fig:feedthru}.
We will soon conduct basic tests with the prototype
(vacuum tightness, signal transmission quality) both at room temperature and in LXe.

In the final detector, the signal from the feedthrough is transmitted to the electronics using 10\,m coaxial cables.

\subsubsection{Test with Cables and Prototype PCBs} 
\label{sec:pcbtest}
A test measurement is performed in order to check the effects of the prototype of the PCB for MPPC mounting, 
and the Radiall coaxial cable (3\,m long) which is a candidate for the MPPC signal cable inside the cryostat
as described in Sec.\ref{sec:cable and feedthrough}).
The pulse height, the pulse width and the rise time of the MPPC signal
are measured using a commercial MPPC (S10362-33-050C) with the following four different conditions.

\begin{enumerate}
  \item short cable, without the prototype PCB
  \item Radiall cable (3\,m), without the prototype PCB
  \item short cable, with the prototype PCB
  \item Radiall cable (3\,m), with the prototype PCB
\end{enumerate}

Fig.\,\ref{fig:ov-a} shows the measured pulse heights of single photoelectron events at four different conditions.

\begin{figure}[htb]                                                                                                                                                                  
\begin{center}
\includegraphics[width=8cm]{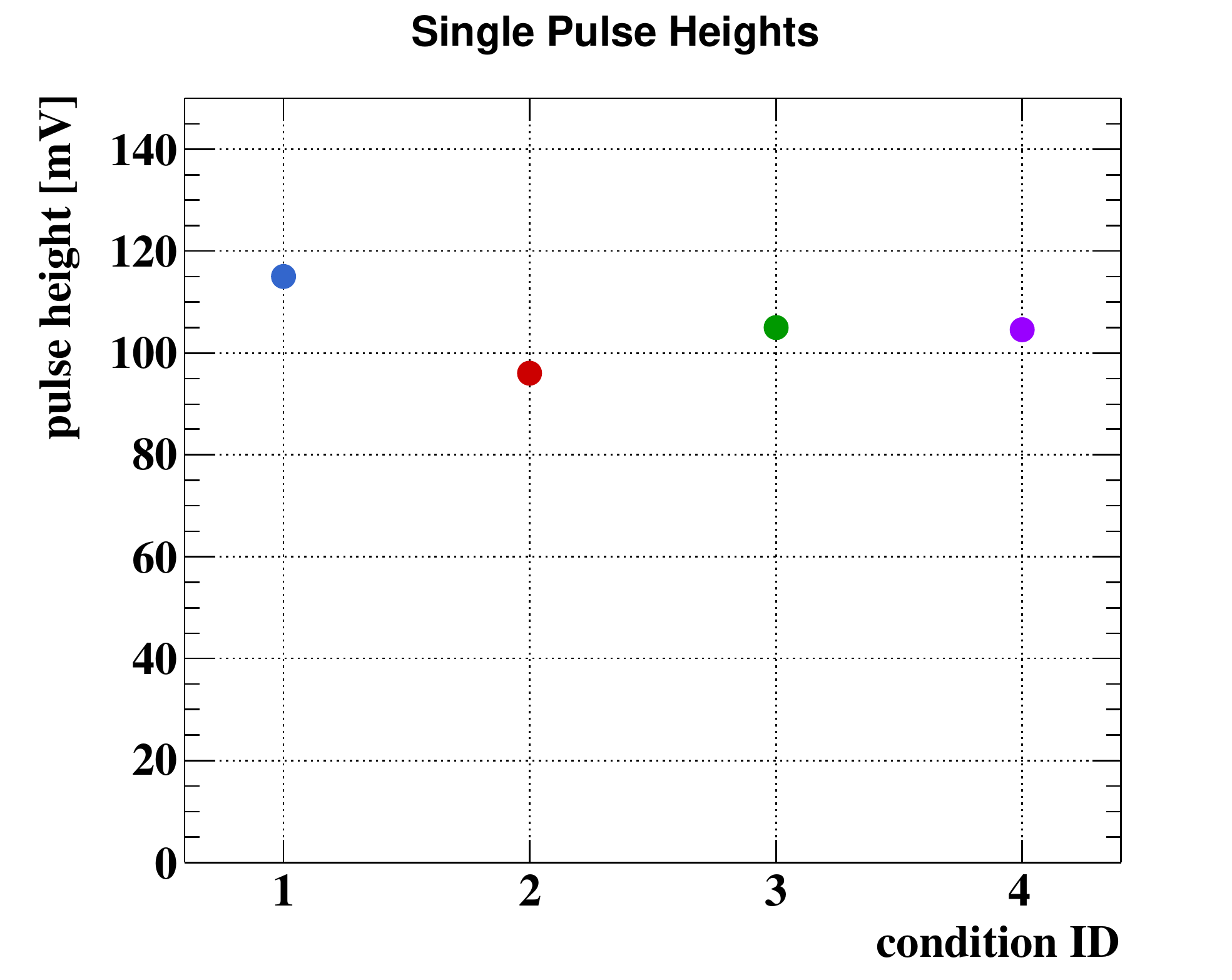}
\caption{\label{fig:ov-a}
Pulse heights of single photoelectron events at four different conditions. }
\end{center}
\end{figure} 

No significant deterioration is observed with the prototype PCB nor with the Radiall cable.
One notice is that condition 4 was expected to give the worst in these configurations, but it was not the case in this measurement.
It indicates that there is some systematic uncertainty in this measurement 
at the level of 10\%.

We plan to send the MPPC signal from the feedthrough to the readout electronics with $\sim$10\,m-long coaxial cable without any amplification.
In order to check the effect of the long cable, we measure the MPPC pulse height, pulse width and the rise time 
with the Radiall coaxial cables, which is also a candidate for the cable outside the LXe detector cryostat, with different lengths 
between the feedthrough and the preamplifier.

\begin{figure}[htb]                                                                                                                                                                  
\begin{center}
\includegraphics[width=5.2cm]{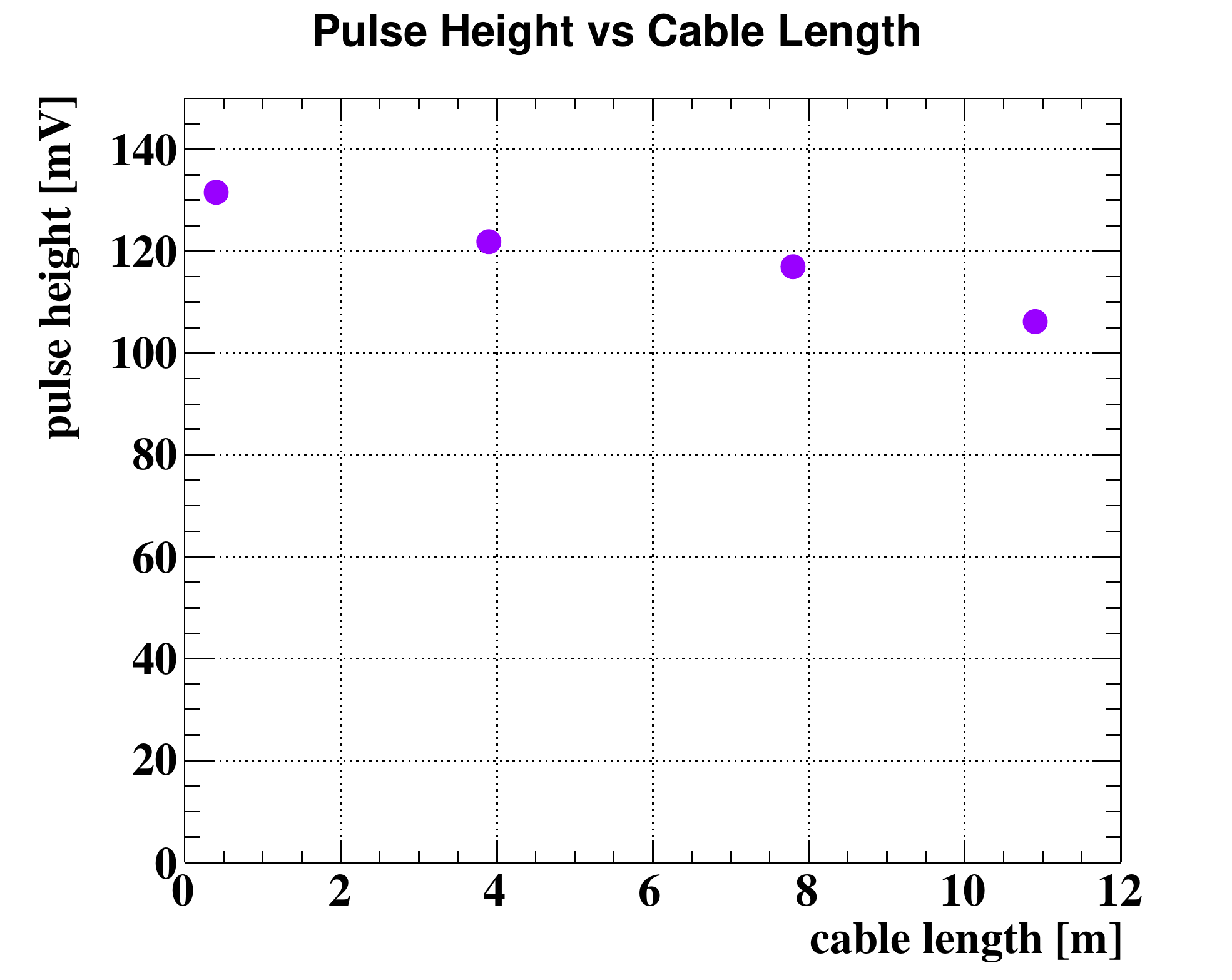}
\includegraphics[width=5.2cm]{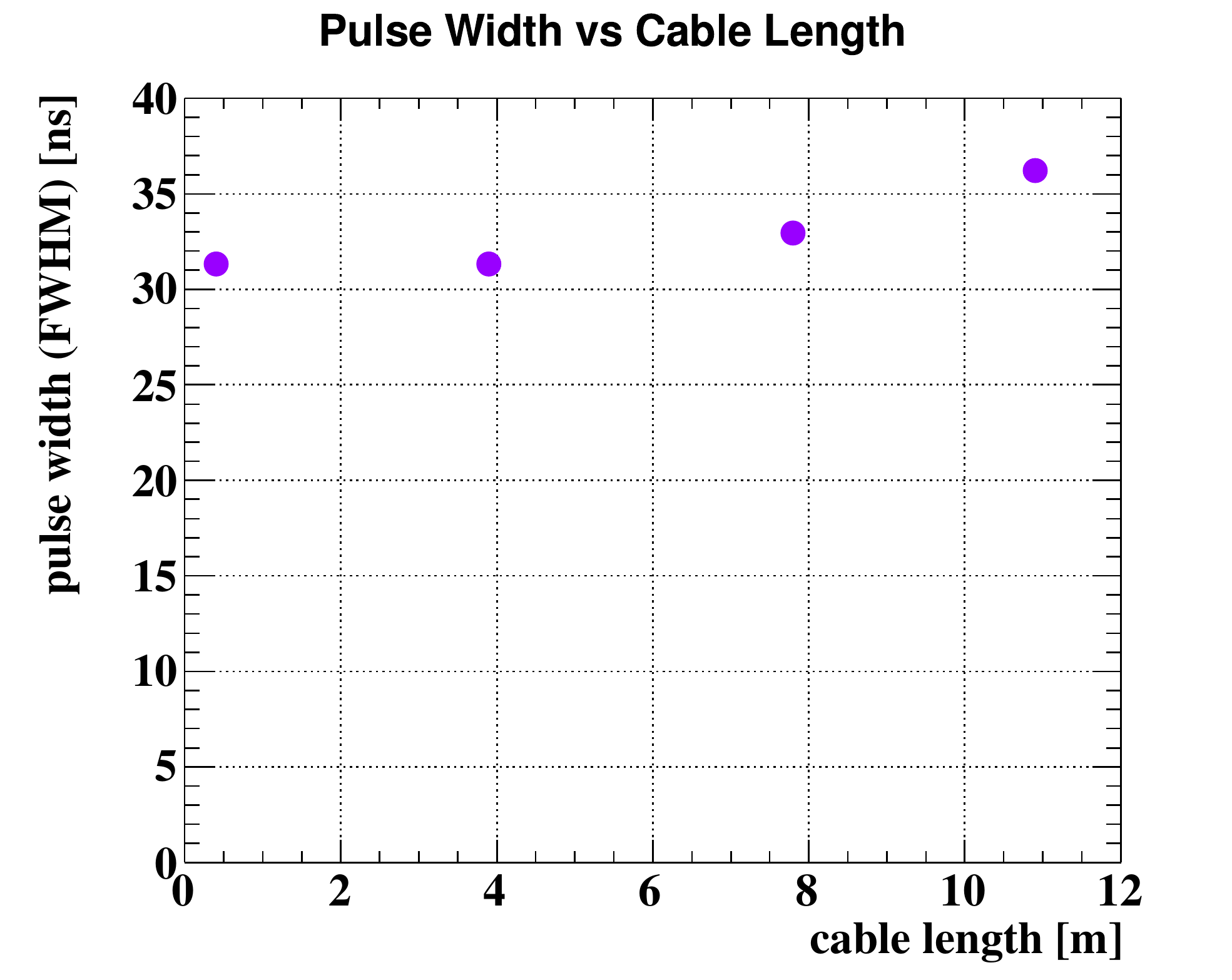}
\includegraphics[width=5.2cm]{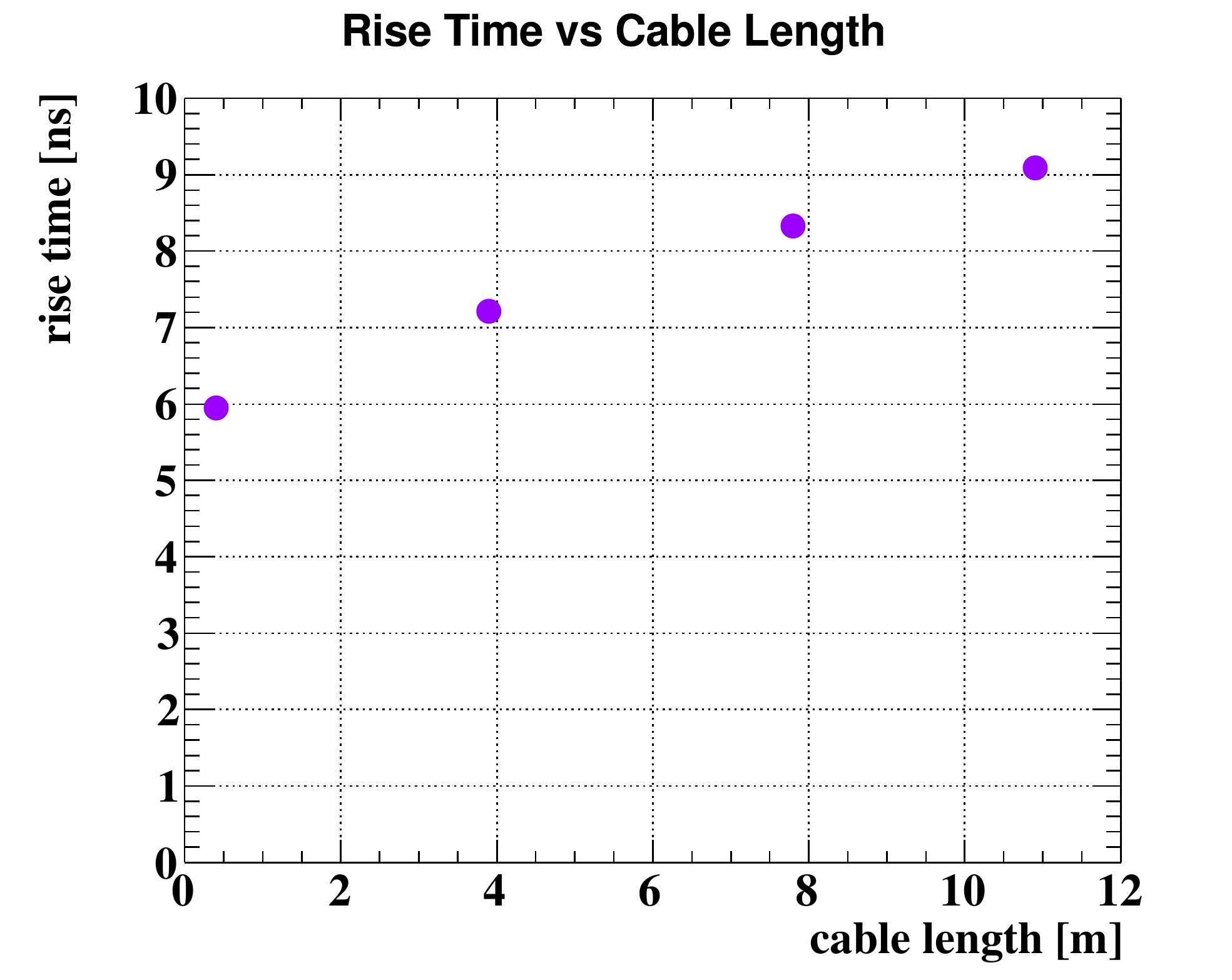}
\caption{\label{fig:ov-a}
(Left) Pulse height of single photoelectron events, (middle) the pulse width, and (right) the rise time
as a function of the cable length.}
\end{center}
\end{figure} 

The pulse height deterioration is not very large ($\sim$25\%) with 10\,m cable, and we believe this can be recovered using a higher gain at 
pre-amplifier etc. 
The transmission of the MPPC signal using 10\,m long cable without any amplification seems possible.

\subsubsection{Readout Electronics}
Both the PMTs and MPPCs signals will be readout by DRS4 boards.
The size of each readout board is 10$\times$16\,cm$^2$, which has 16 input channels with differential DRS driver.
Amplifiers are mounted on the board with three switchable gain settings.
The difference between adjacent gain stages is about 5 to 10.
The different gain stages can then be switched at any time.
The higher readout-gain mode is used to detect single photo-electrons for the calibration of MPPCs,
while the low readout-gain mode is used to take physics data 
where a large dynamic range is needed.

No amplifier is installed between the MPPC and the DRS4 bouard.
The bias voltage for MPPC, which is typically 70\,V, is supplied from the readout boards through the signal cable.

The signals readout by the DRS boards are used also for the trigger; 
splitters or additional readout boards are not anymore needed for triggering.
All the DRS boards to readout the channels from the LXe detector will be inserted into two 19-inch racks.
The place to put the racks would be either near the feedthroughs or at a distance
connected with 10\,m cables.
The use of 10\,m cable doesn't seem an issue as discussed in Sec.\,\ref{sec:pcbtest}, 
but the final decision will be made after measuring the noise-level and the
signal-attenuation using the prototype detector.

\subsubsection{Cryogenics}
The current cryostat will be re-used for the LXe detector upgrade. 
In order to cover the increase of the external heat inflow due to $\sim 4000$ signal cables for the MPPCs, 
the cooling power of the refrigerator has to be increased 
either by installing more powerful pulse tube cryocooler 
or by adding another cryocooler of the same type as the current one.
We will evaluate the realistic heat-load through the prototype test. 
R\&D for cooling power upgrade of the refrigerator will be continued in parallel.

%% file: 07_Photon_Calorimeter/performance.tex
\subsection{Expected performance\label{sec:expected performance}}
The expected performance of the upgrade of the LXe detector is evaluated using MC simulation.

\subsubsection{Simulation}
A full MC simulation code based on Geant4 \cite{Geant4} was developed to compare the performance of the present detector design and the upgrade design.
In the simulation, scintillation photon propagation is simulated by Geant4.
The reflection of scintillation photons on the MPPC surface was simulated using the complex refractive index of pure silicon crystal. 
The reflectance is about 60\% typically. 
In the simulation, the index-number of hit pixel and the arrival time of each scintillation photon are recorded. 
They are used to make distributions of avalanches in each MPPC. 
The dark-noise, optical-crosstalk, after-pulsing, saturation and recovering are modeled based on a real measurement and incorporated in the simulation.  
The waveform of MPPC is simulated by convolving the single photo-electron pulse and the time distribution of avalanches. 
A simulated random electronics noise is added assuming the same noise level as the present readout-electronics.

The event reconstruction analyses are basically the same as those for the present detector, 
while the parameters such as waveform integration window and corrections for light collection efficiency 
depending on the conversion position are optimized for the new design.
The non-linear response of the MPPC due to the pixel saturation (see
Fig.\,\ref{fig:MPPC_linearity}) is taken into account, resulting in a non-linear energy
response of the detector.
However it is found that the effect on the energy reconstruction is negligible
because the fraction of the number of photoelectrons observed by each MPPC is small.

\subsubsection{Results}

\begin{figure}[tbc]
\begin{center}
\begin{minipage}{0.4\linewidth}
\includegraphics[width=\linewidth]{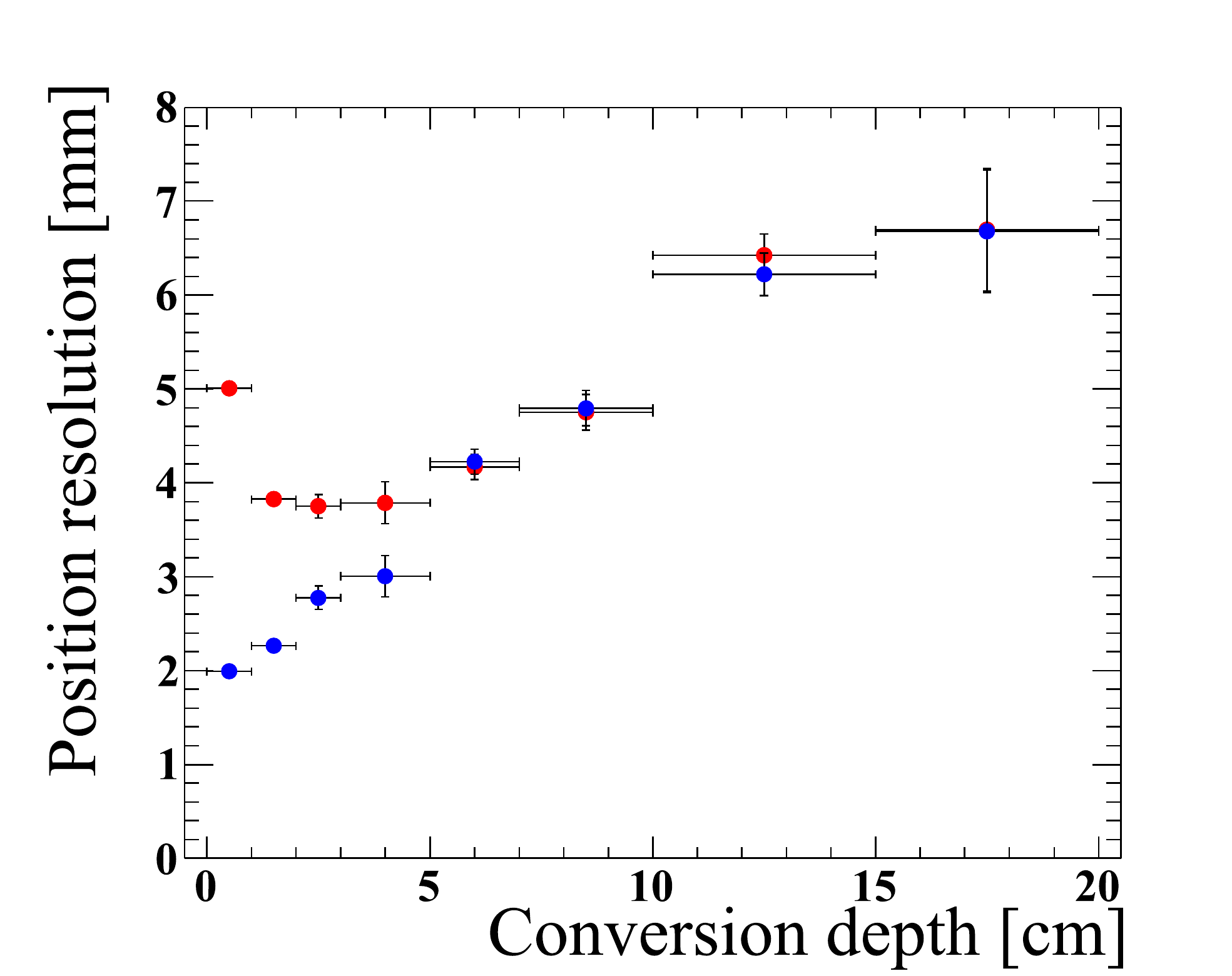}\\(a)
\end{minipage}
\begin{minipage}{0.4\linewidth}
\includegraphics[width=\linewidth]{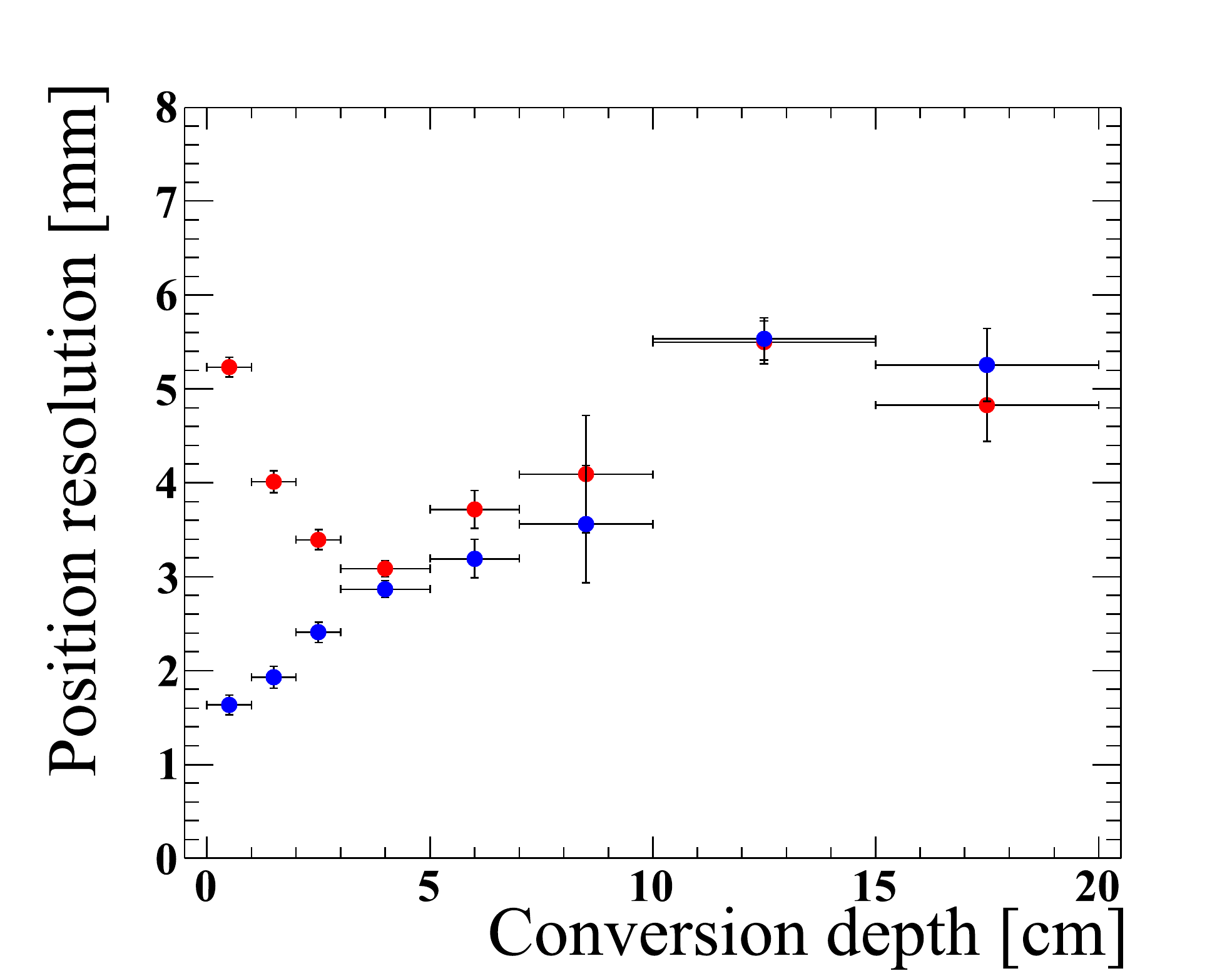}\\(b)
\end{minipage}
\caption{\label{fig:xec_posreso}
Position resolutions in the horizontal (a) and vertical (b) directions as
a function of the first conversion depth. The resolutions in the present detector
configuration are shown in red markers, and those in the upgraded detector configuration
are shown in blue markers.}
\end{center}
\end{figure}

\begin{figure}[tbc]
\begin{center}
\begin{minipage}{0.4\linewidth}
\includegraphics[width=\linewidth]{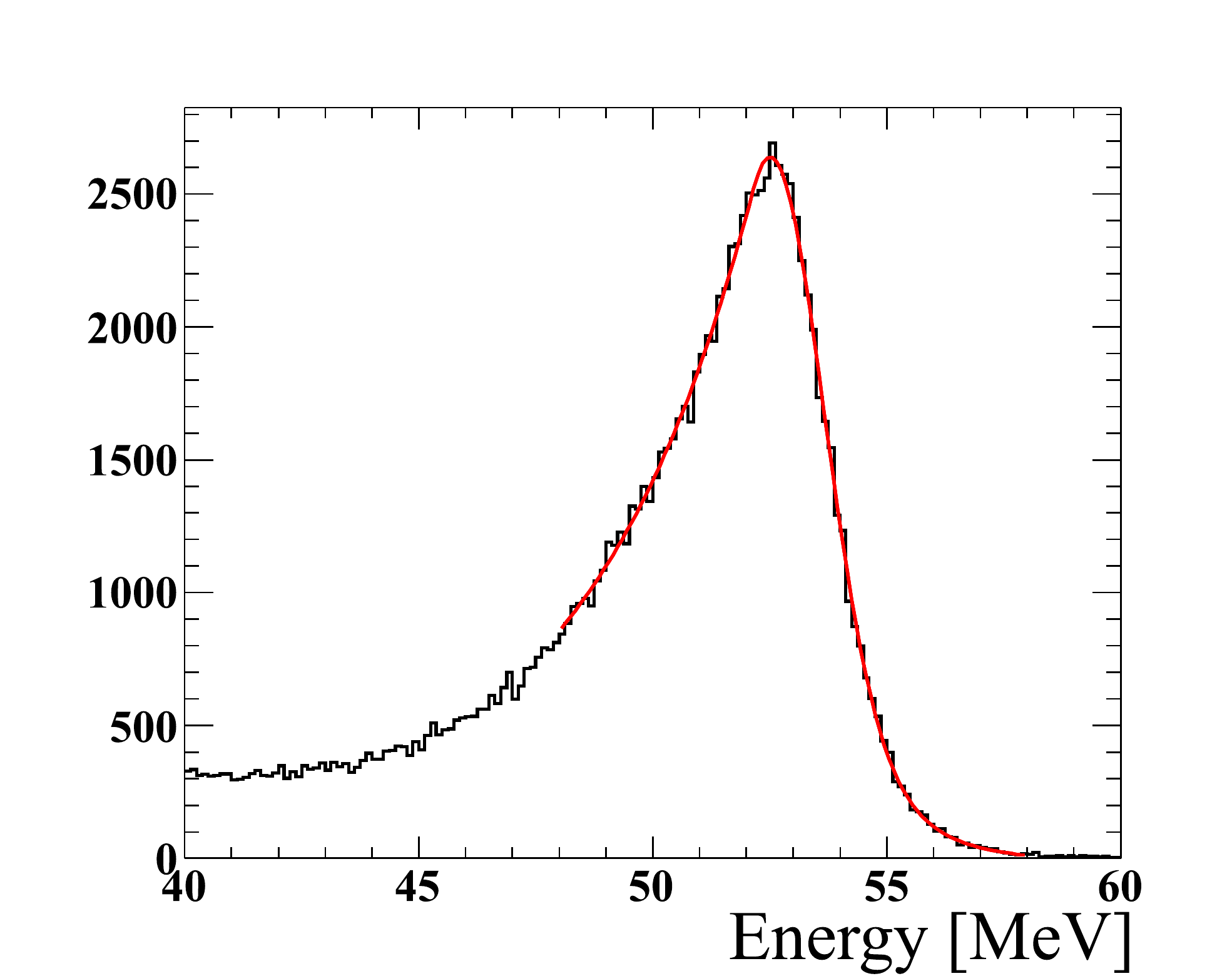}\\(a) Present configuration ($w < 2$~cm)\\
\includegraphics[width=1.07\linewidth]{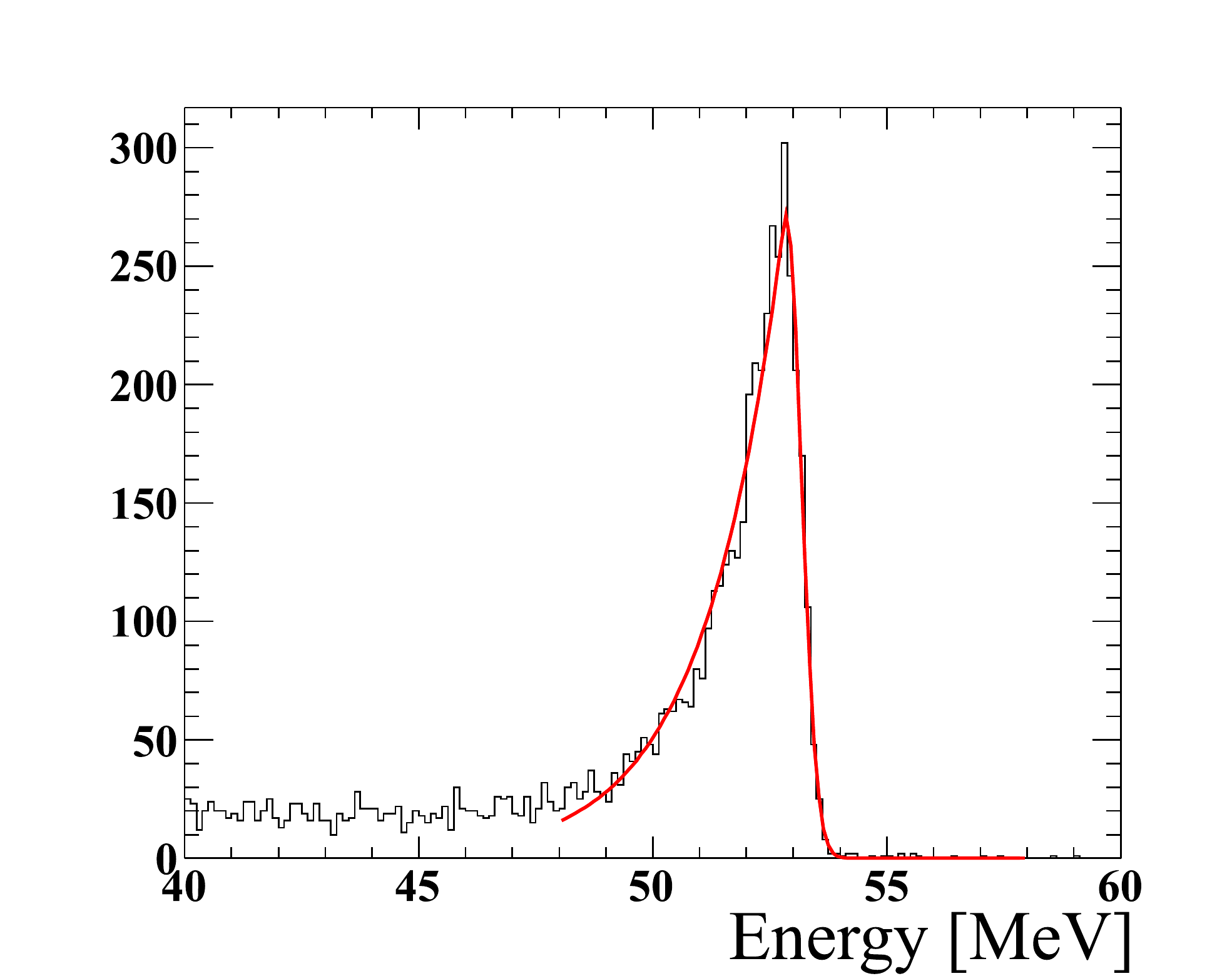}\\(c) Upgraded configuration ($w < 2$~cm)
\end{minipage}
\begin{minipage}{0.4\linewidth}
\includegraphics[width=\linewidth]{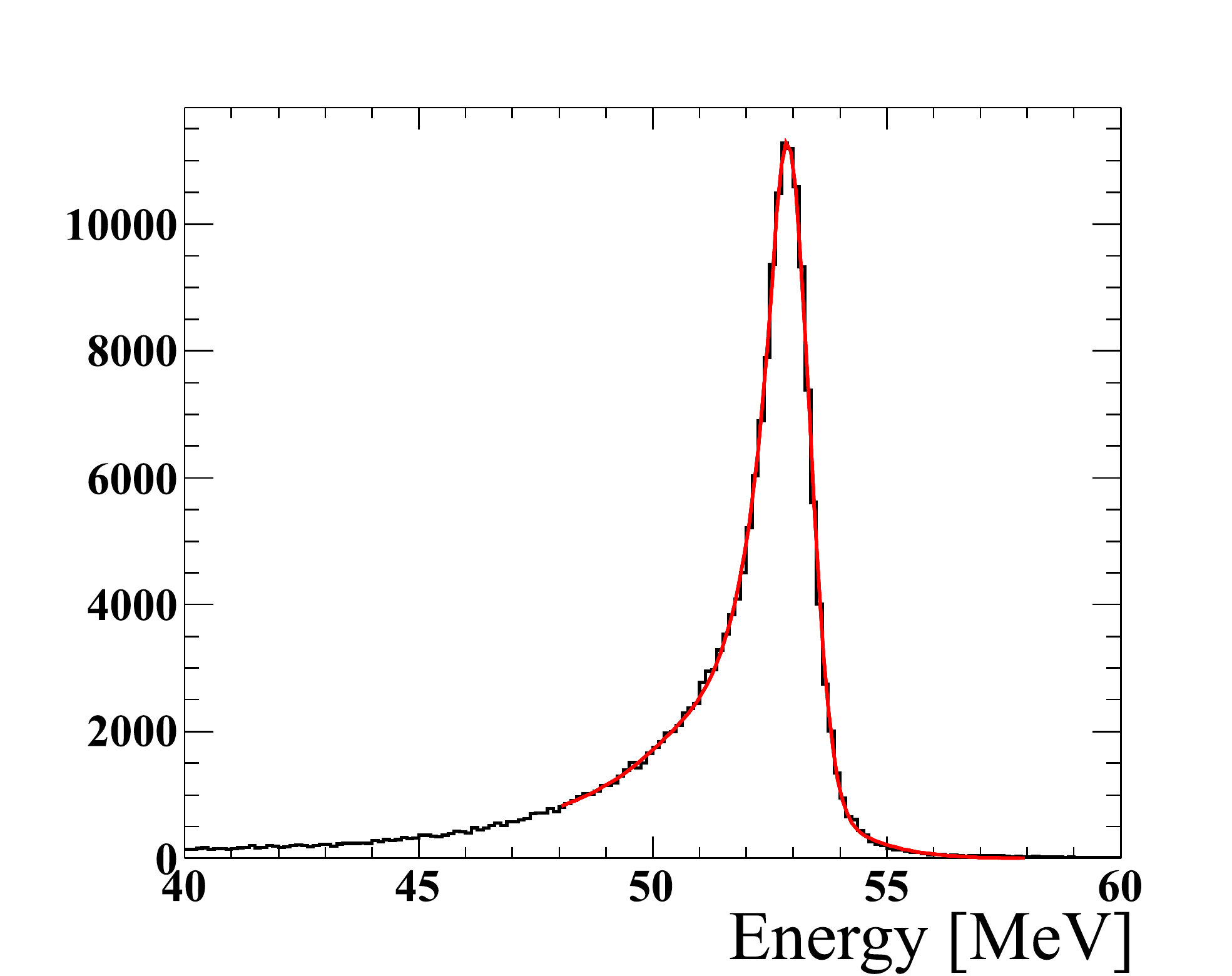}\\(b) Present configuration ($w \geq 2$~cm)\\
\includegraphics[width=1.07\linewidth]{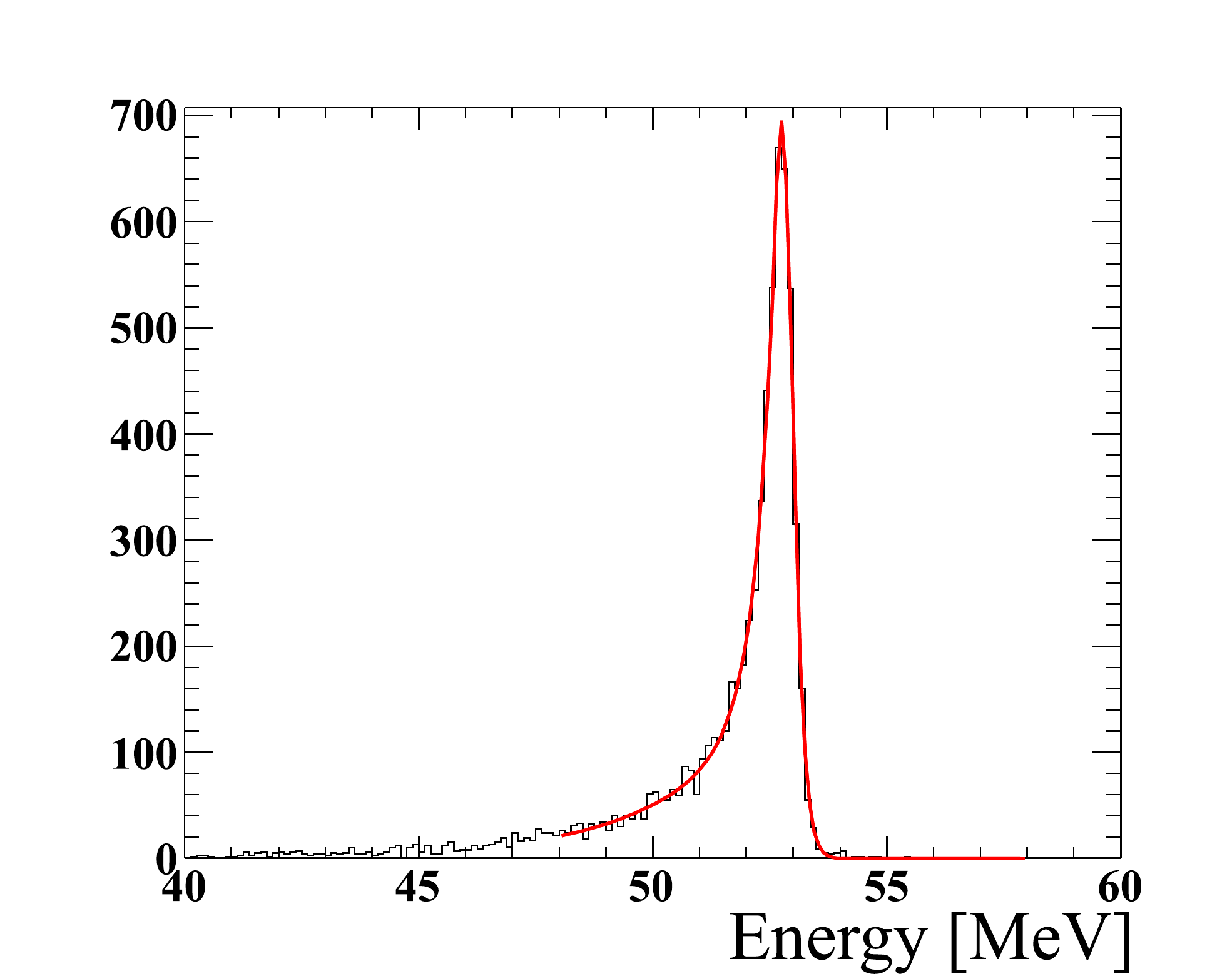}\\(d) Upgraded configuration ($w \geq 2$~cm)
\end{minipage}
\caption{\label{fig:xec_enereso} Energy responses of the LXe detector in the simulation
   with present (a,b) and the upgraded (c,d) detector configurations. Responses for the shallow (a,c) and
   deep (b,d) conversion events are shown separately.}
\end{center}
\end{figure}

\begin{figure}[tbc]
\begin{center}
\includegraphics[width=.4\linewidth]{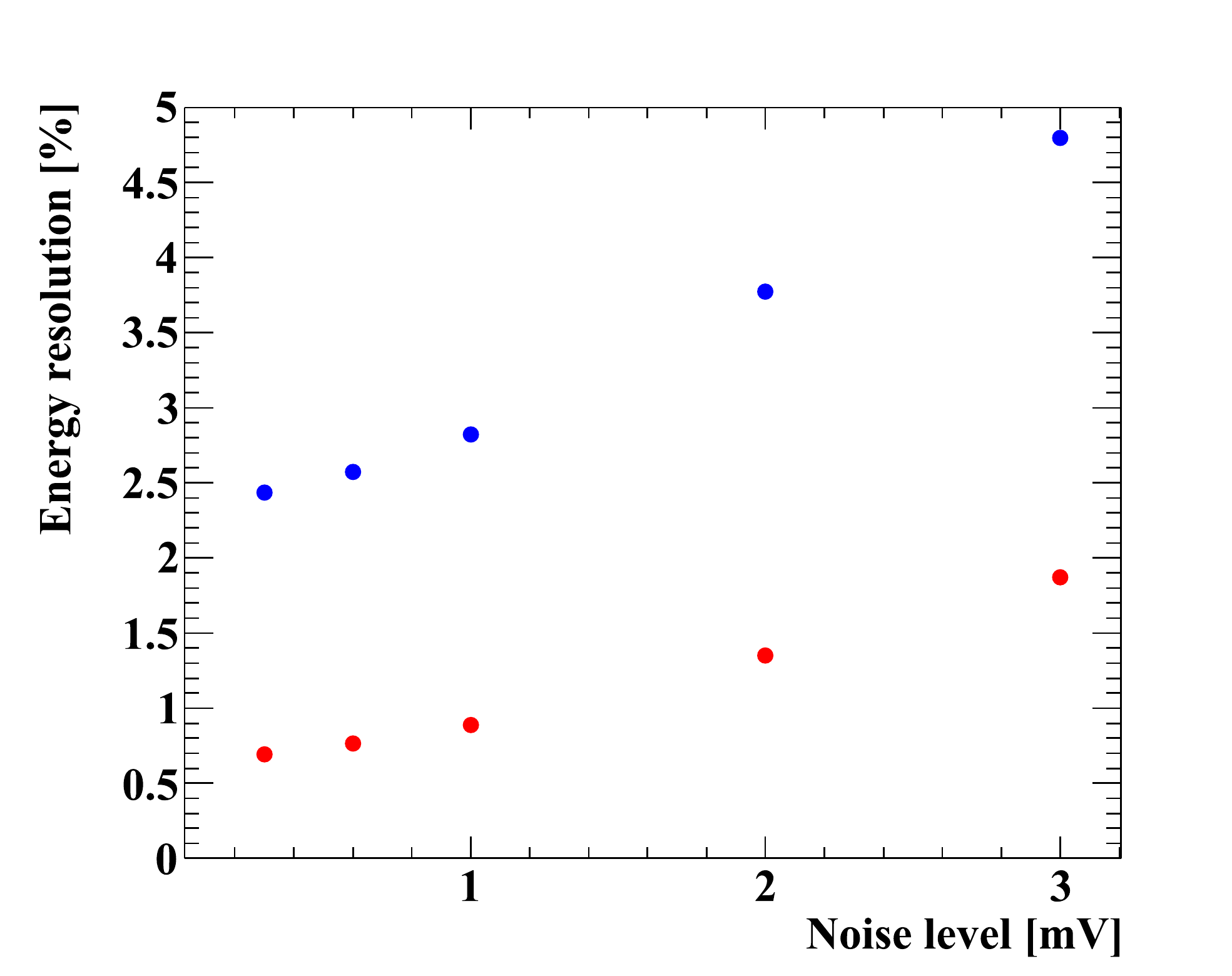}
\caption{\label{fig:xec_enereso_noise}
Energy resolutions as a function of added noise level in the
simulation.
Effective sigmas of the upper edge (red markers) are obtained by scaling the HWHMs of the upper edge.
FWHMs using both edges are shown in blue markers.
}
\end{center}
\end{figure}

\begin{figure}[tbc]
\begin{center}
\begin{minipage}{0.4\linewidth}
\includegraphics[width=\linewidth]{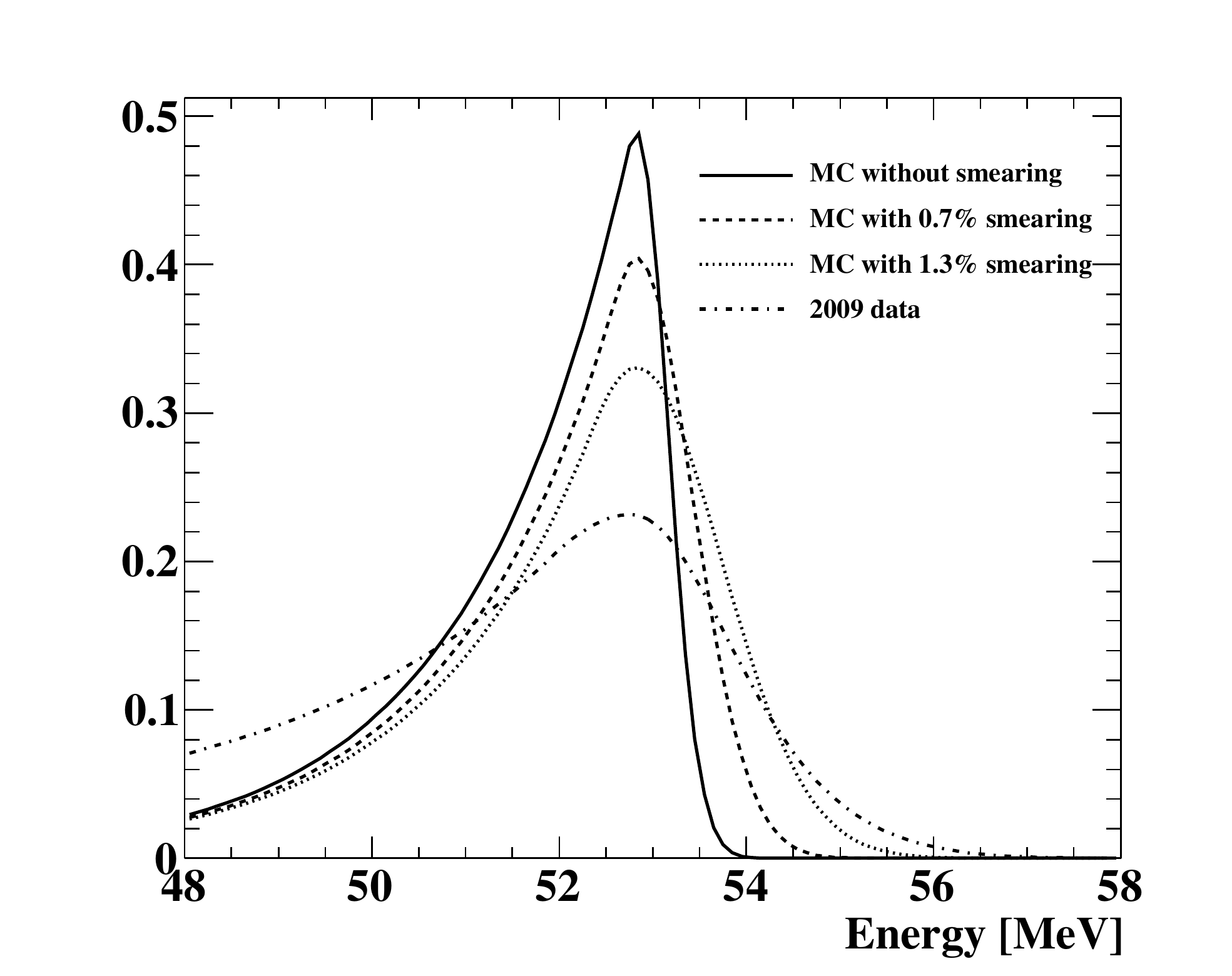}\\ (a)
\end{minipage}
\begin{minipage}{0.4\linewidth}
\includegraphics[width=\linewidth]{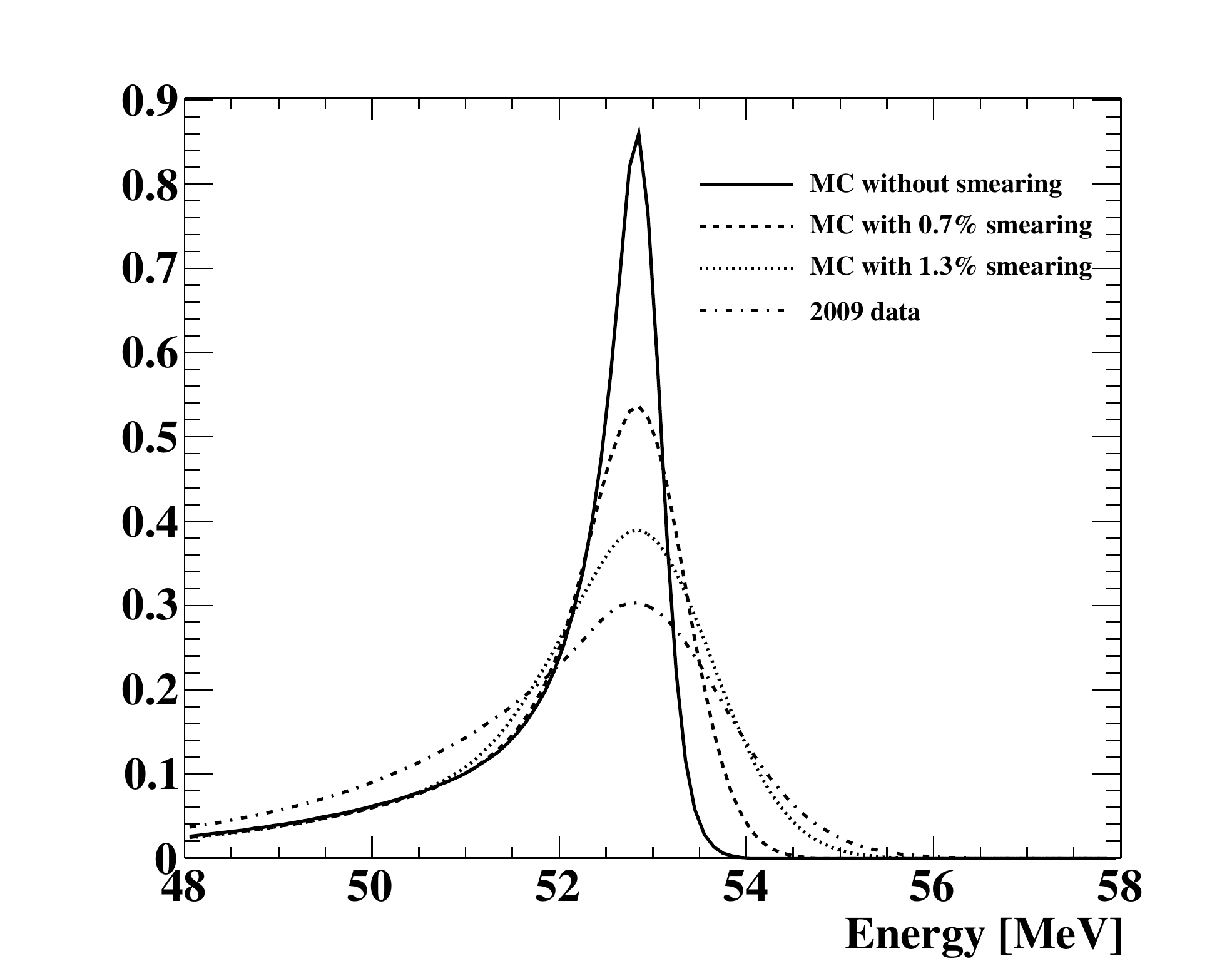}\\ (b)
\end{minipage}
\caption{\label{fig:xec_energy_smear}
Energy response functions with various assumptions of additional fluctuation (0, 0.7 and
1.3\%) and that of 2009 data.
}
\end{center}
\end{figure}

Fig.\,\ref{fig:xec_posreso} shows the position resolutions for signal $\gamma$-rays as a
function of the reconstructed conversion depth ($w$). In the present detector configuration, the
position resolution is worse in the shallow part because of the size of PMTs. The
resolution is much better in the upgraded detector configuration.
The energy resolution is also much better in the shallow part with the upgraded detector
configuration than that with the present detector configuration as shown in
Fig.\,\ref{fig:xec_enereso} due mainly to the more uniform photon-collection efficiency.
The low energy tail is smaller in the upgraded detector configuration because of the lower
energy leakage. The resolution is also better in the deepest part because of the modification
of the angle of the lateral PMTs.
The effect of the electronics noise is studied by changing the random noise level in the
simulation. It is confirmed that the effect of the noise is small 
(Fig.\,\ref{fig:xec_enereso_noise}) if the noise level stays at the same level as the present detector (0.3-0.4\,mV).
The measured energy resolution of the present detector (1.7\% for $w>$2~cm) is worse than
that in the simulation (1.0\% for $w>$2~cm). 
The reason is not fully understood while the source of the difference can be related
to the behavior of PMTs (e.g.~gain stability, angular dependence and so on) or optical
properties of liquid xenon (e.g.~effect of convection). If the former is the case, the
difference can become smaller in the upgraded configuration.
On the other hand, if the latter is the case, the difference could remain.
Figure\,\ref{fig:xec_energy_smear} shows the energy responses with different assumptions; 1:
when the additional fluctuation compleltely vanish in the upgraded detector, 2: when a part of fluctuation remain
which corresponds to 1.2\% resolution in the present detector configuration. The
resolution of 1.2\% was acheived with the MEG LXe large prototype detector., 3: when the
fluctuation remain which makes the resolution of the present detector to be 1.7\%.

The time resolution of the calorimeter can be limited by six components;
the transit time spread (TTS) of photo-sensors, the statistical fluctuation of scintillation photons, 
the timing jitter of the readout-electronics, the electoronics noise,
the resolution of the photon conversion point and the finite size and fluctuation of the energy deposits in the LXe. 
The most of them are common for both in the present and upgraded detectors, however the
effect from TTS and electronics noise are different because of the different
photo-sensors.
The effect of TTS is neglibible because it scales as a function of the number of
photoelectron, and the light output of liquid xenon is large.
The effect of the electronics is larger in the upgraded detector than in the present
detector because the leading time of a MPPC pulse for liquid xenon scintillation signal is slower than that of a PMT pulse.
In order to estimate the effect, the time resolution of the upgraded detector for signal
$\gamma$ rays is measured in the simulation. The evaluated time resolution with a preliminary
waveform and reconstruction algorithms is 84~psec, where the gain of $2 \times10^7$
(including the preamplifier gain)
and 0.3~mV noise RMS are assumed.
In the preliminary analysis, sum-waveforms of 16 adjacent MPPCs are used.
The analysis will be improved to use waveform of single MPPC near the $\gamma$ interaction
point.
The time resolution could be improved because the small size of MPPCs compared to PMTs
makes the fluctuation of the travel-time of scintillation photons from the $\gamma$ conversion point to a
photo-sensor smaller.

\subsubsection{High intensity}

\begin{figure}[tbc]
\begin{center}

\begin{minipage}{0.48\linewidth}
\includegraphics[width=\linewidth]{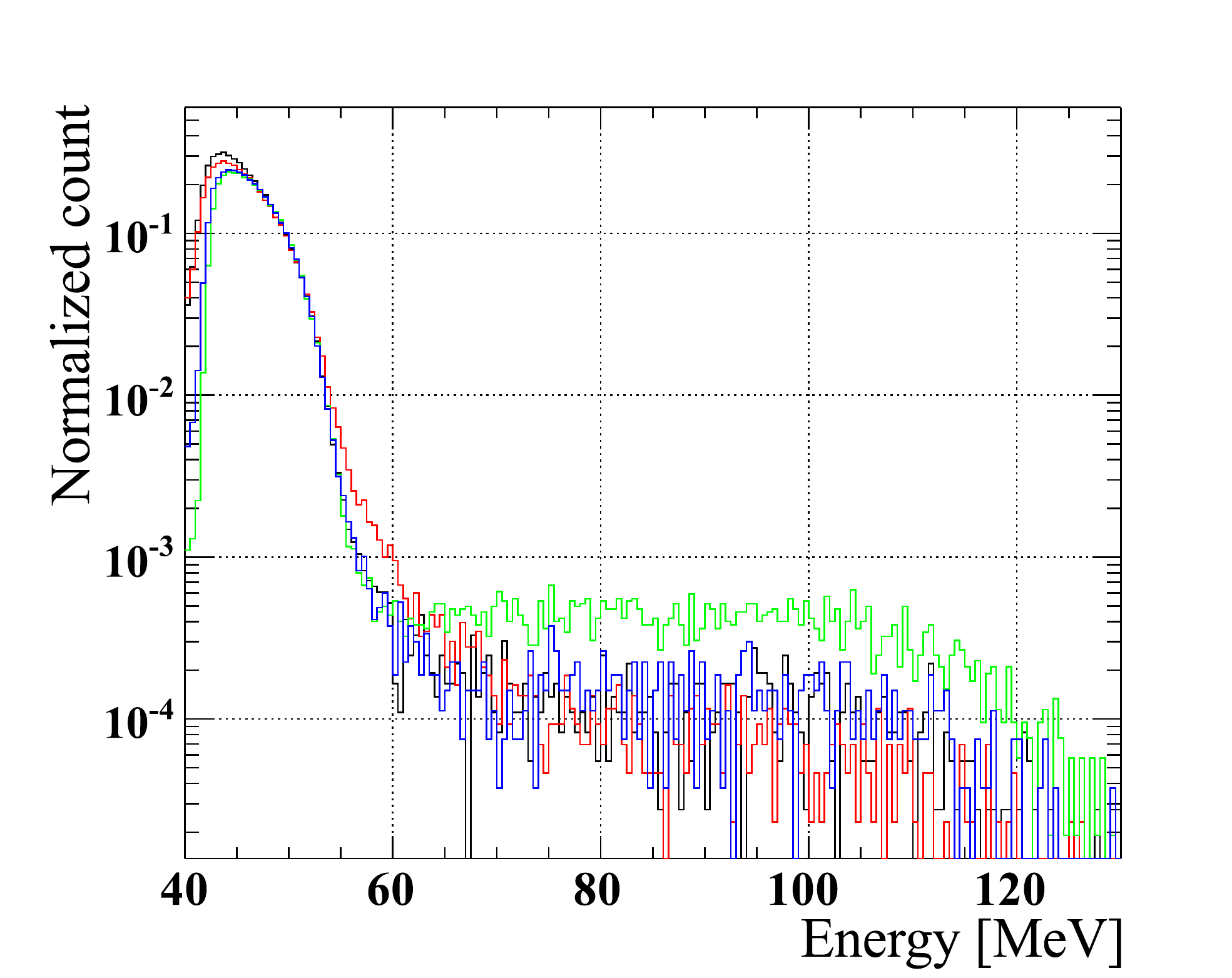}\\(a)
\end{minipage}
\begin{minipage}{0.48\linewidth}
\includegraphics[width=\linewidth]{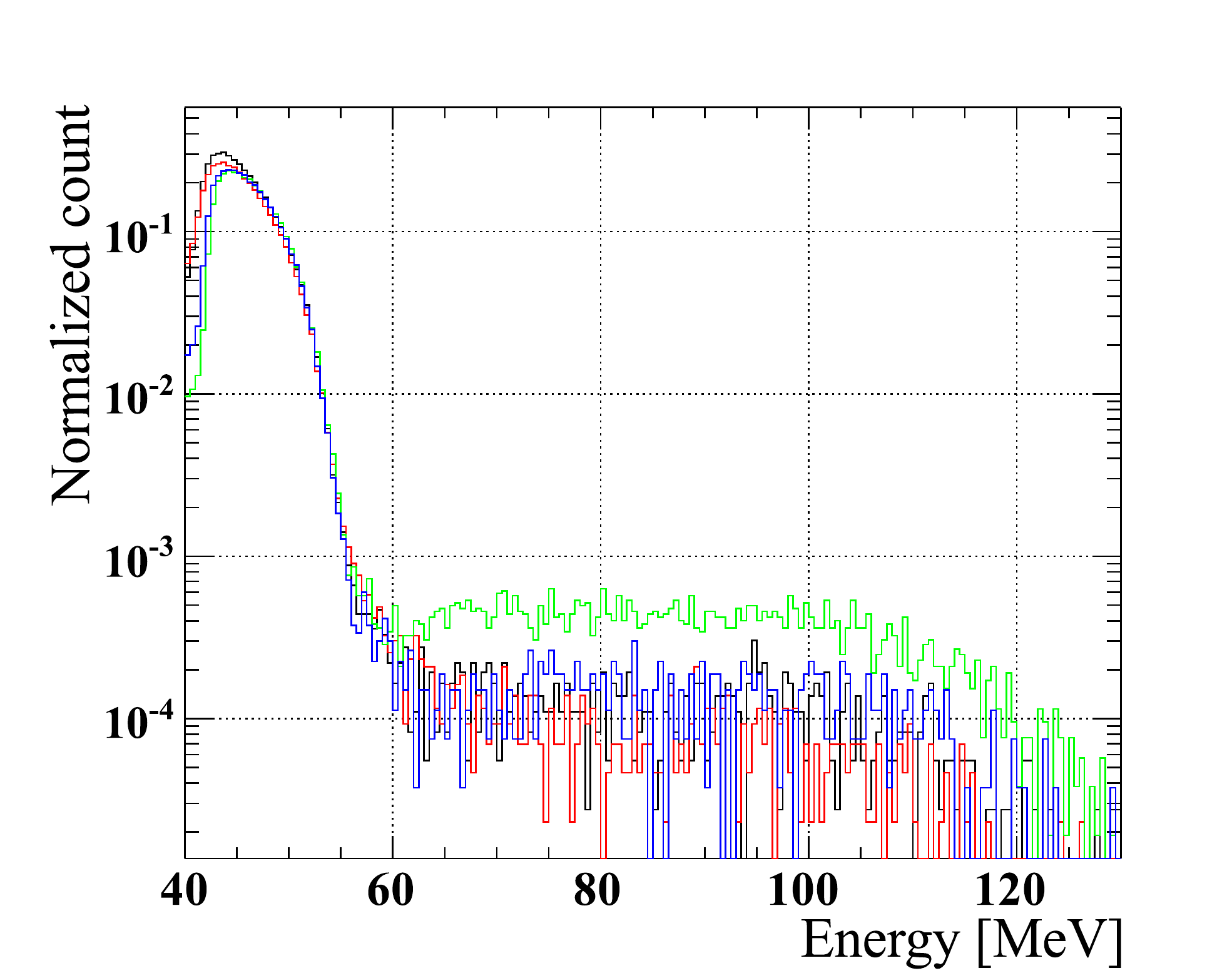}\\(b)
\end{minipage}

\caption{\label{fig:xec_gammabg_intensity} Reconstructed energy spectrum obtained for different beam intensities.
The horizontal axis shows energies without unfolding pileup gamma rays (a), the same after
unfolding and subtracting the energy of pileup gamma rays (b), respectively, in GeV.
Green, black, blue and red lines show spectrum at muon stopping rate of 1.0, 3.0, 3.3 and
8 $\times 10^7$\,$\mu$/s, respectively.
The spectra are normalized by the number of events from 48 to 58\,MeV; the scaling factors are consistent with the muon
stopping rate on the target.
A difference in the low energy part below 0.045\,GeV is due to different effective trigger
thresholds; a difference in the high energy part is due to the different
ratio between $\gamma$-rays backgrounds and the cosmic-ray backgrounds.
}
\end{center}
\end{figure}

The higher background $\gamma$ rate with the higher muon intensity in the upgraded experiment should not be a
problem for photo-sensor operation. 
On the other hand, the background rate in the analysis energy region
would be increased due to pileup. In the analysis for the present detector, the energies of pileup gamma rays are unfolded using the waveform and light
distribution on the inner face. 
In 2011 data acquisition we took data with different beam
intensities, 1.0, 3.0, 3.3 and 8 $\times 10^7$\,$\mu$/s.
Figure\,\ref{fig:xec_gammabg_intensity} shows the $\gamma$-ray spectrum normalized by the
number of events from 48 to 58\,MeV; the scaling factors are consistent with the muon
stopping rate on the target.
The shapes of spectra are almost identical in the analysis region between 48 and 58\,MeV after
subtracting the energies of pileup $\gamma$ rays.
Since the same analysis can be used also for the upgraded detector, a higher beam rate
will not cause additional background rate due to pileup.

%% file: 07_Photon_Calorimeter/prototype.tex

\subsection{Prototype Detector}

\begin{figure}[htb]                                                                                                                                                                  
\begin{center}
\includegraphics[width=14cm]{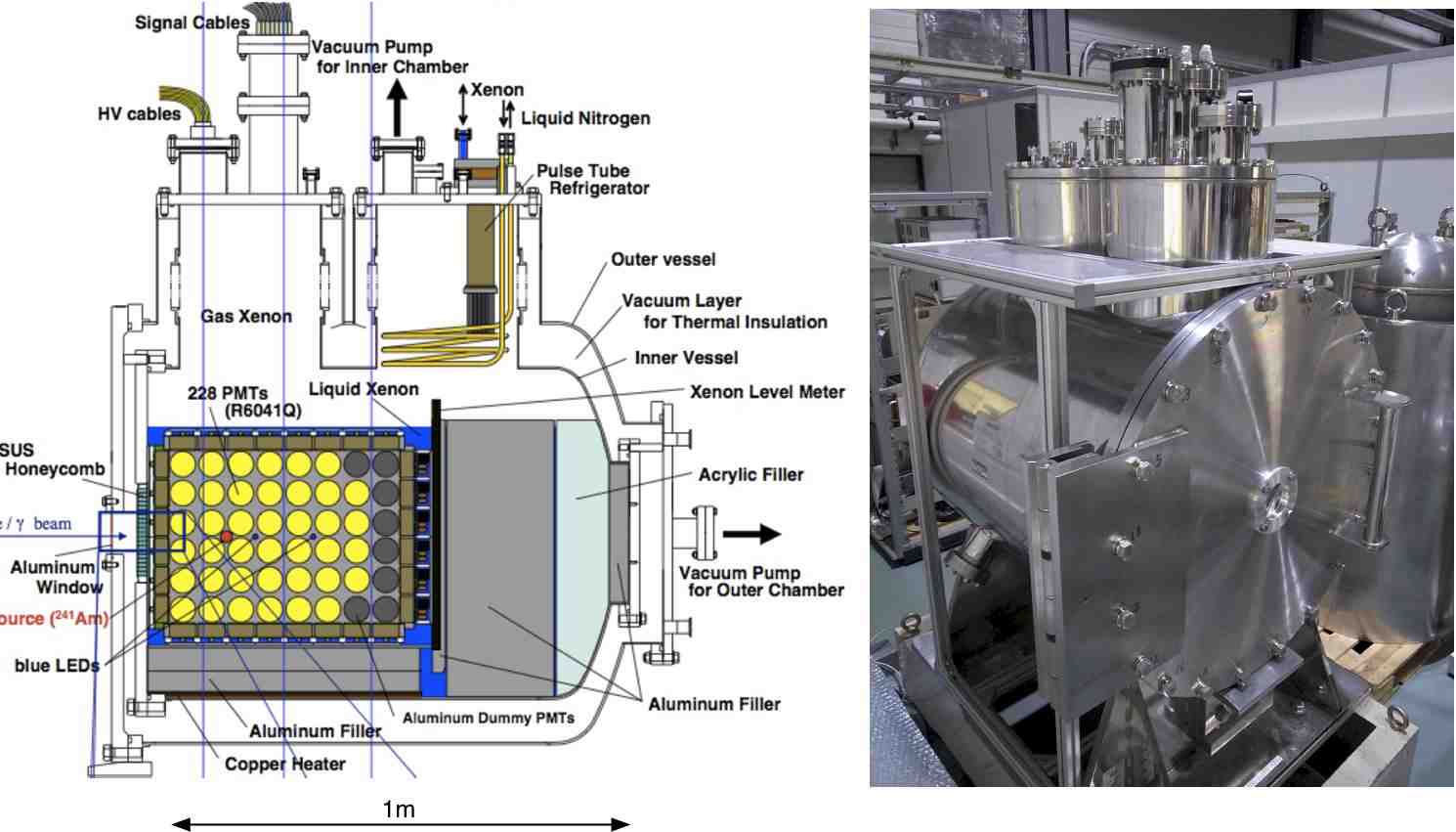}
\caption{\label{fig:LP}
  Prototype of the current LXe detector. The PMTs on the $\gamma$ entrance face will be replaced with the UV-MPPCs 
  and the depth will be somewhat shortened for the prototype of the upgrade detector. 
}
\end{center}
\end{figure}

A prototype detector is planned to be built to demonstrate the performance of the LXe detector 
with MPPC readout on the $\gamma$ entrance face.
The cryostat and some other resources, which were used for the prototype test of the current LXe detector, will be recycled (Fig.\,\ref{fig:LP}). 
A box-shaped LXe with an active volume of approximately 70\,$\ell$ is surrounded by 576 UV-MPPCs 
with $12\times 12 \mathrm{mm}^2$ active area each on the $\gamma$ entrance face
and 180 PMTs on the other faces ($6\times 6$ PMTs for each face).  
Since the depth of the active volume is the same as the full-scale detector,
this prototype detector is considered as a fraction of the acceptance of the full-scale detector.
Either spare PMTs of old version or the PMTs used in the current detector are used in the prototype.
For the former case, the construction of the prototype detector can be started before the current experiment is finished. 

We plan to study the performance of the prototype detector
with the same techniques as for the current detector, namely using $\gamma$-rays with various energies
up to 129\,MeV from $(p,\gamma)$-reaction with a Cockcroft-Walton accelerator
, $\pi^-p$ charge exchange and radiative capture reactions.
All components to be developed for the upgrade detector such as the cables, feedthroughs and readout electronics will be tested.

%% file: 07_Photon_Calorimeter/cost_schedule.tex

%% file: 08_Trigger_and_DAQ/Trigger_and_DAQ.tex
\section{Trigger and DAQ}
\label{sec:Trigger and DAQ}
%
%

\subsection{Requirements}

The proposed upgrade of the MEG detector requires two main improvements of the
existing DAQ system. On one side more channels are needed both for the DAQ and
the trigger systems, and on the other side a higher bandwidth of the waveform
digitizing system is required in order to use the algorithms based on cluster 
timiing. While the increase of DAQ channels is moderate and could be fulfilled with the
existing system just by using a few more crates, the bandwidth of the old system
is limited by the analog front-end of the DRS4 waveform digitizing boards. A
cheap and simple scheme has originally been chosen, which uses only a passive
transformer to convert the single-ended signal from the drift chambers into the
required differential signal needed by the DRS4 chip. This scheme however limits
the bandwidth to about 200 MHz.

\subsection{Proposed DAQ boards---WaveDREAM}

We propose a new DAQ board, which combines both the waveform digitizing
technology using the DRS4 chip as well as the trigger and splitter functionality
of our current system. This way a much more compact system can be constructed,
which is necessary given the limited space around our detector.

\begin{figure}[htb]
 \centerline{\hbox{
  \includegraphics[width=.8\textwidth,angle=0] {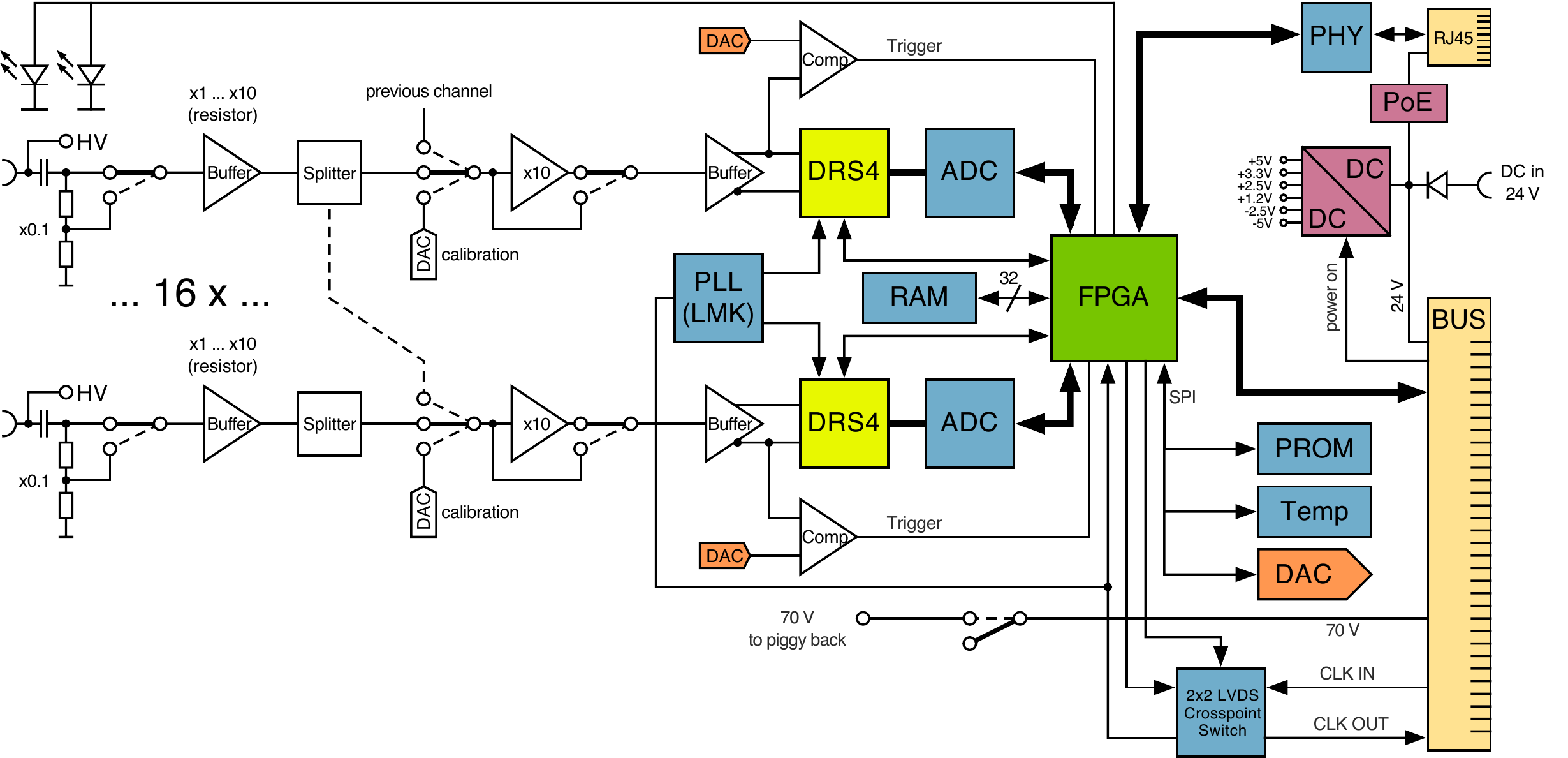}
 }}
 \caption[]{Simplified schematics of the proposed DAQ board}
 \label{DAQ:board}
\end{figure}

To overcome the bandwidth limitation, a new active analog front-end has already
been designed and tested. This front-end has two switchable gain stages, which
can be combined to obtain a post-amplification by a factor one to about 70. The
post-amplification can be used to increase the signal amplitude coming from the
drift chamber and the SiPM pre-amplifiers which are typical in the order of a
few ten millivolts. By increasing the amplitude to a few hundred millivolts, the
signal-to-noise ratio inside the DRS4 chip is improved, which allows more
accurate charge measurements. Furthermore, the front-end section has been
carefully optimized to give an overall bandwidth above 700 MHz. Figure
\ref{DAQ:board} shows the simplified schematic of the new DAQ board.

Several additional features as compared with the current DAQ electronics will be
implemented on the new board. Each channel will have a dedicated comparator
which can be used to implement a rate counter or a trigger (see subsection
below). The board can be operated in a standalone mode, in which a Gigabit
Ethernet interface can be used to transfer acquired data directly to a PC. This
might be useful for first tests where only a couple of channels will be needed.
For the full system, the DAQ boards will be housed in a custom crate especially
developed for this purpose. This crate can contain up to 16 boards with a total
of 256 channels and a so-called DAQ concentrator board, which collects the data
from the 16 DAQ boards, combines them and sends it via Gigabit Ethernet to a PC.
The usage of a custom crate as compared with a standard VME system brings a cost
saving by a factor of three.

Furthermore, new DRS4 timing calibration methods recently reported on
conferences indicate that a significant improvement in the DRS4 timing
calibration of up to a factor of three can be achieved. We plan to adapt this
new timing calibration to the new boards, which should allow to improve the
current timing resoluiton of 40 ps significantly. This is an important step if
the particle detector resolution improves with the upgrade, in which case the
electronics contribution would become more significant.

The proposed drift chamber requires the recognition of individual clusters
originating from the primary ionization in the chamber gas. In a first phase,
this cluster recognition will be done off-line on a PC, which makes it easy to
develop and optimize the required waveform analysis algorithms. In a second
phase, it is planned to implement the algorithms directly inside the FPGA of the
DAQ board. FPGAs with sufficient resources and DSP capabilities have already
been selected to support this second phase.

The proposed electronics will be a significant improvement over the current one
in terms of channel density and analog bandwidth. It has enough flexibility to
adapt itself in terms of selectable gain and FPGA programming to the needs of
the new drift chamber system, even if the requirements change during the
operation.

\subsection{Trigger}

The functionality of the current MEG trigger system will be integrated into the
new DAQ boards. The DRS4 chips operate in the so-called "transparent mode" where
their input appears directly at the output. There the connected ADCs can
continuously digitize the input to provide data to the trigger. Each ADC
digitizes the input signal with 80 MSPS and 12 bits.  A potentially poorer
determination of the pulse amplitude arising from the slower sampling speed (80
MSPS vs. 100 MSPS), which is relevant for the $\Pgg$-energy reconstruction, will
be eventually compensated by the increased ADC resolution (12 bits vs. 10 bits).
A wider dynamic range is in order if we want to fully exploit the improved
single photoelectron response of new LXe photosensors\footnote{ Let us recall
that the C-shape of our LXe detector and the depth of the $\Pgg$-interaction
vertex are such that the bulk of the inner face is shadowed; therefore most of
the inner PMTs (namely those which are not just in front of the e.m. shower) use
to work in a very few photoelectron regime.}. As a result, we don't expect any
deterioration of the energy resolution at a trigger level.

Concerning the relative $\Pep-\Pgg$ timing, we will benefit from using the DAQ
comparator coupled to each input signal (on both LXe and TC). The estimate due
to comparator latch time can be further refined by implementing look-up tables
on the FPGA to correct for time-walk effects. Also in this case, the overall
resolution is not expected to be spoiled, rather it might be improved up to 1 ns.

Each DAQ board is conceived to operate the same digital algorithm as the current
``Type 1'' boards: sum of all 16 inputs, maximum pulse amplitude and related
index (to provide a coarse determination of $\Pgg$ direction), pulse time.
These values are arranged in a 48-bit bus and passed through LVDS serializers to
trigger concentrator boards, which are housed at the side of each crate next to
the DAQ concentrator boards. These boards are designed to receive and combine
all the pieces of information from DAQ boards, thus operating as upper-level trigger
boards (a role currently being played by ``Type 2'' trigger boards, which do not fit in 
the future design).

\subsection{Data reduction}
The use of a higher beam intensity as well as the higher number of  detector channels
than in the current configuration will have the effect of a sizable increase of the  data throughput,
with severe concerns on both DAQ speed and off-line storage capability.
We envisage using two handles to tackle this problem by reducing:
\begin{itemize}
\item the event size, by operating a substantial on-line zero suppression;
\item the event rate, by exploiting a higher-level trigger based on an on-line track reconstruction.
\end{itemize}

The first point can be addressed thanks to the sparse occupancy of detector
channels. In the case of LXe, for instance, photosensors which are far away from
the $\Pgg$-ray impact point collect very few photons (to be compared to a few
thousands being detected by the ones in front of the shower). As their
contribution to the timing is almost irrelevant, there is no loss of information
by recording the sum of groups (4:1 or 9:1) of them.

The basic idea for the reduction of the trigger rate relies on the possibility
of using individual signals for each wire end, while in the current trigger
configuration we are using overall 32 fanned-in signals (inner and outer wires
for each DCs). With so many pieces of information available at DAQ boards, we
plan to combine the features of FPGAs and associative memory chips to accomplish
both pattern recognition and track filter tasks. This will allow us to strongly
suppress the bulk of the events we are currently recording onto the disk, where
no condition on track selection is asked for.

Of course any reduction tool should be applied upon condition of no
deterioration of signal efficiency and/or resolution.

%% file: 09_Final_Sensitivity/Final_Sensitivity.tex
\section{Final sensitivity}
%

The sensitivity of the upgraded MEG experiment is evaluated by using a maximum likelihood analysis 
technique developed to extract the upper limit (UL) at 90\% C.L. on $\BR(\mu \rightarrow {\rm e} 
\gamma$) in the MEG data analysis \cite{meg2010}. 
This technique is more efficient and reliable than a simple box analysis, 
since all types of backgrounds are correctly folded in the global likelihood function and taken into 
account with their own statistical weights. 

An ensemble of simulated experiments ({\it toy MC}) is created from the probability density functions (PDFs) 
describing the signal shapes and the background distributions for the photon energy ($E_{\gamma}$), 
positron energy ($E_{{\rm e}^+}$), relative timing and relative angles. The enhanced precision of all 
upgraded detectors allows a much better separation of the signal from the background and 
reduces significantly the spill of the gamma and positron background distributions into the signal region, 
which is mainly due to experimental resolution effects. With a much lower accidental background in 
the new detector, the muon stopping rate can be higher than the present one: optimization studies are under 
way, but a muon stopping rate of at least $7 \times 10^{7}\,\mu/\mathrm{sec}$ is envisaged. The increased muon stopping 
rate and the enhanced resolutions are taken into account in estimating the number and the 
distributions of background events expected in the upgraded experiment. 

A representative scenario for the detector resolutions and efficiencies is summarized in Tab. \ref{tab:scenario} 
and compared with the present MEG performance. The efficiency of the positron reconstruction is highly improved 
with respect to the current one, thanks to the high efficiency of the new tracking system (close to 1) 
and to the optimized relative position of the tracker and the timing counter.
  
\begin{table}[htb]
\caption{ \label{tab:scenario}Resolution (Gaussian $\sigma$) and efficiencies for MEG upgrade}
\newcommand{\m}{\hphantom{$-$}}
\newcommand{\cc}[1]{\multicolumn{1}{c}{#1}}
\begin{tabular}{@{}lll}
\hline
  {\bf PDF parameters }                               & \m Present MEG    & \m  Upgrade scenario  \\
  \hline\noalign{\smallskip}
  e$^+$  energy  (keV)                                & \m  306 (core)    & \m  130   \\
  e$^+$ $\theta$ (mrad)                               & \m 9.4            & \m  5.3     \\
  e$^+$ $\phi$   (mrad)                               & \m 8.7            & \m  3.7      \\
  e$^+$ vertex (mm) $Z/Y ({\rm core}) $                & \m 2.4 / 1.2      & \m  1.6 / 0.7  \\
  $\gamma$ energy (\%)  ($w<$2\,cm)/($w>$2\,cm)        & \m 2.4 / 1.7      & \m  1.1 / 1.0 \\
  $\gamma$ position (mm) $u/v/w$                       & \m 5 / 5 / 6      & \m  2.6 / 2.2 / 5 \\
  $\gamma$-e$^+$ timing (ps)                          & \m 122            & \m  84 \\ 
  \hline
  {\bf  Efficiency (\%)}                              &                   & \\
  \hline
  trigger                                             & \m $\approx$ 99   & \m $\approx$ 99 \\
  $\gamma$                                            & \m 63             & \m 69 \\
  e$^+$                                               & \m 40             & \m  88  \\ 
\hline
\end{tabular}\\[2pt]
\end{table}

\begin{figure}[htb]
\centering
\centering
\includegraphics[width= 7 cm]{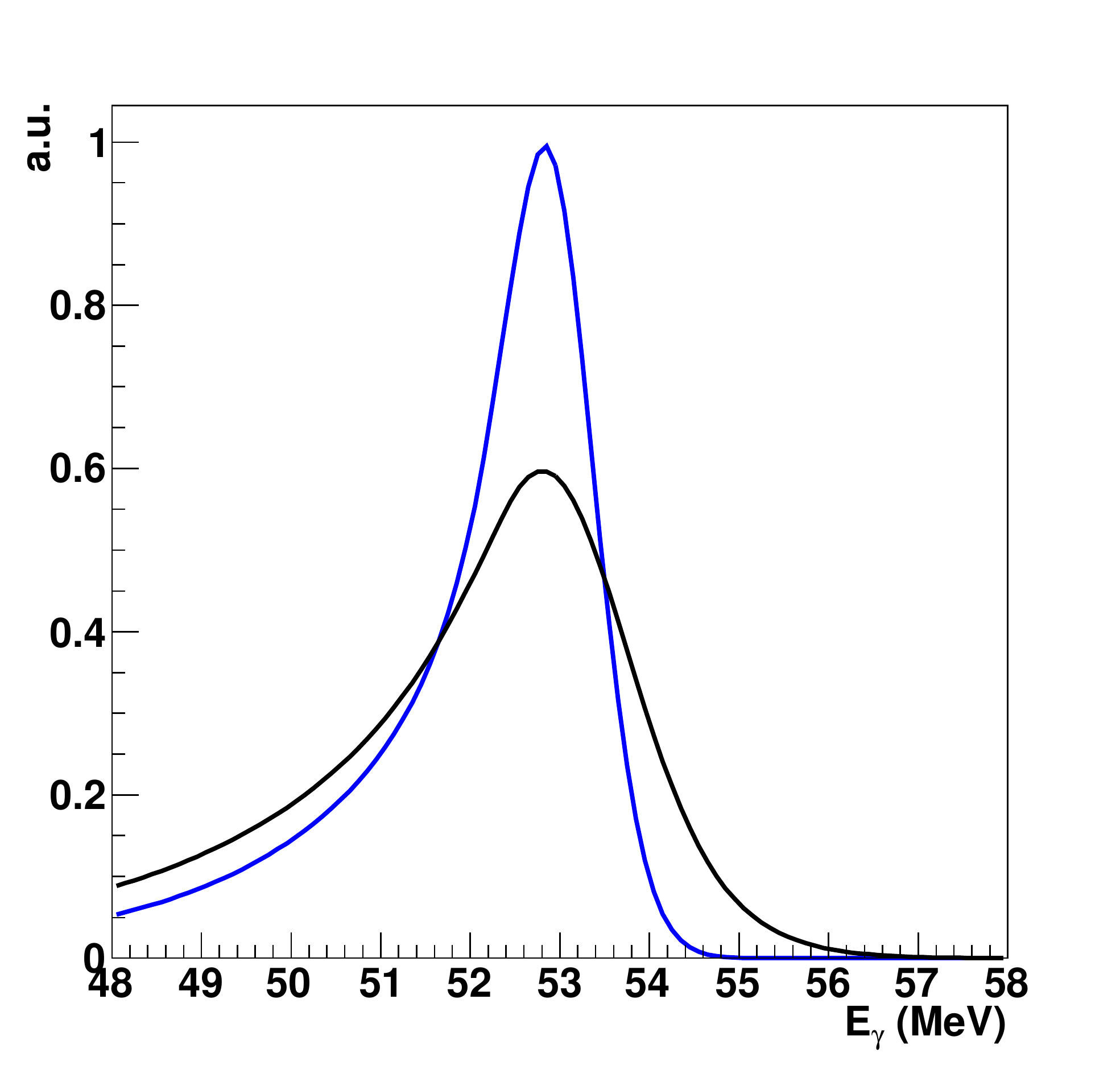}
\includegraphics[width = 7 cm]{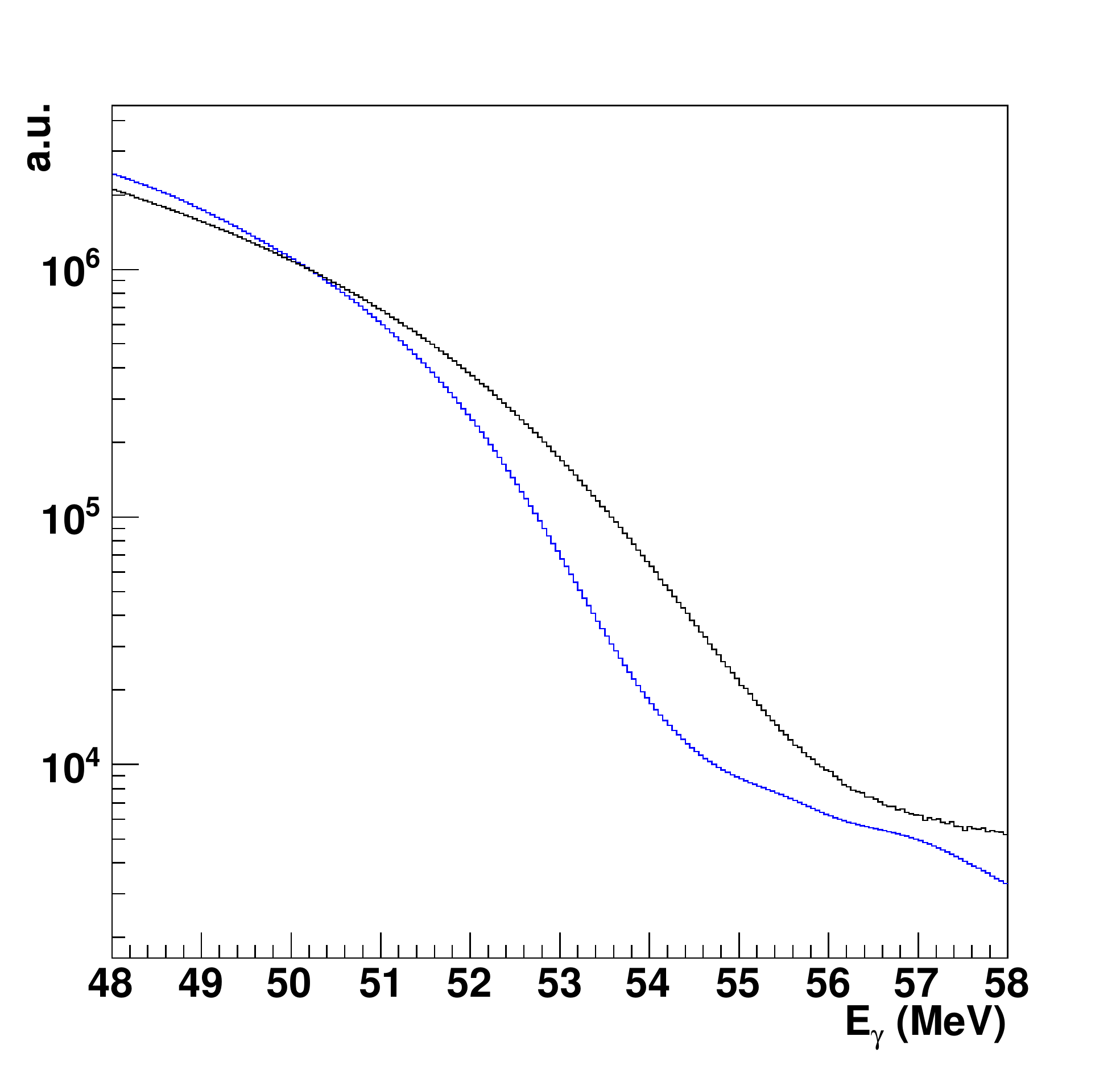}
\caption{\label{fig:senspdf1} Comparison of photon energy PDF for signal events (left) and accidental background 
events (right) based on the resolutions obtained in 2010 data (black) and on the projected value for the upgrade 
(blue). Differences in relative background contributions between RMD, AIF and pileup are also taken into account.} 
\end{figure}

As an example we show in Fig. \ref{fig:senspdf1} the $E_{\gamma}$ PDF for signal and accidental 
background events, as simulated in {\it toy MC}. The expected improvement for the upgrade scenario is visible 
in comparing these PDF (in blue) with the 2010 MEG data PDF (in black). In the $E_{\gamma}$ background 
PDF various contributions are taken into account: radiative muon decay (RMD), photons from positron annihilation in flight (AIF), or from bremsstrahlung on materials in the detector, pile-up events and resolution effects. 
The new configuration of the cylindrical drift chamber, with a smaller amount of material close to the 
electromagnetic calorimeter, reduces the AIF contribution, which is dominant for photon energies 
$\egamma>52$\,MeV, of about $20\%$ with respect to the present MEG detector. The combined effect of the increased 
resolution and of the lower high energy background is clearly visible in the right side of 
Fig. \ref{fig:senspdf1}.

The {\it toy MCs} are generated assuming zero signal events and an average number of radiative and accidental 
events obtained by extrapolating the previous years results and taking into account the new detector 
performance. The number of radiative and accidental events is then left free to fluctuate, according to 
Poisson statistics. 
A maximum likelihood fit is performed  
and an UL on the number of signal events is determined for each {\it toy MC};
; this value is then converted to an UL on 
$\BR( \mu \rightarrow {\rm e} \gamma$) by using the appropriate normalization factor. 
We define as {\it sensitivity} the median of the distribution of the UL obtained on the {\it toy MCs}.

\begin{figure}[htb]
\centering
\includegraphics[width= 9 cm]{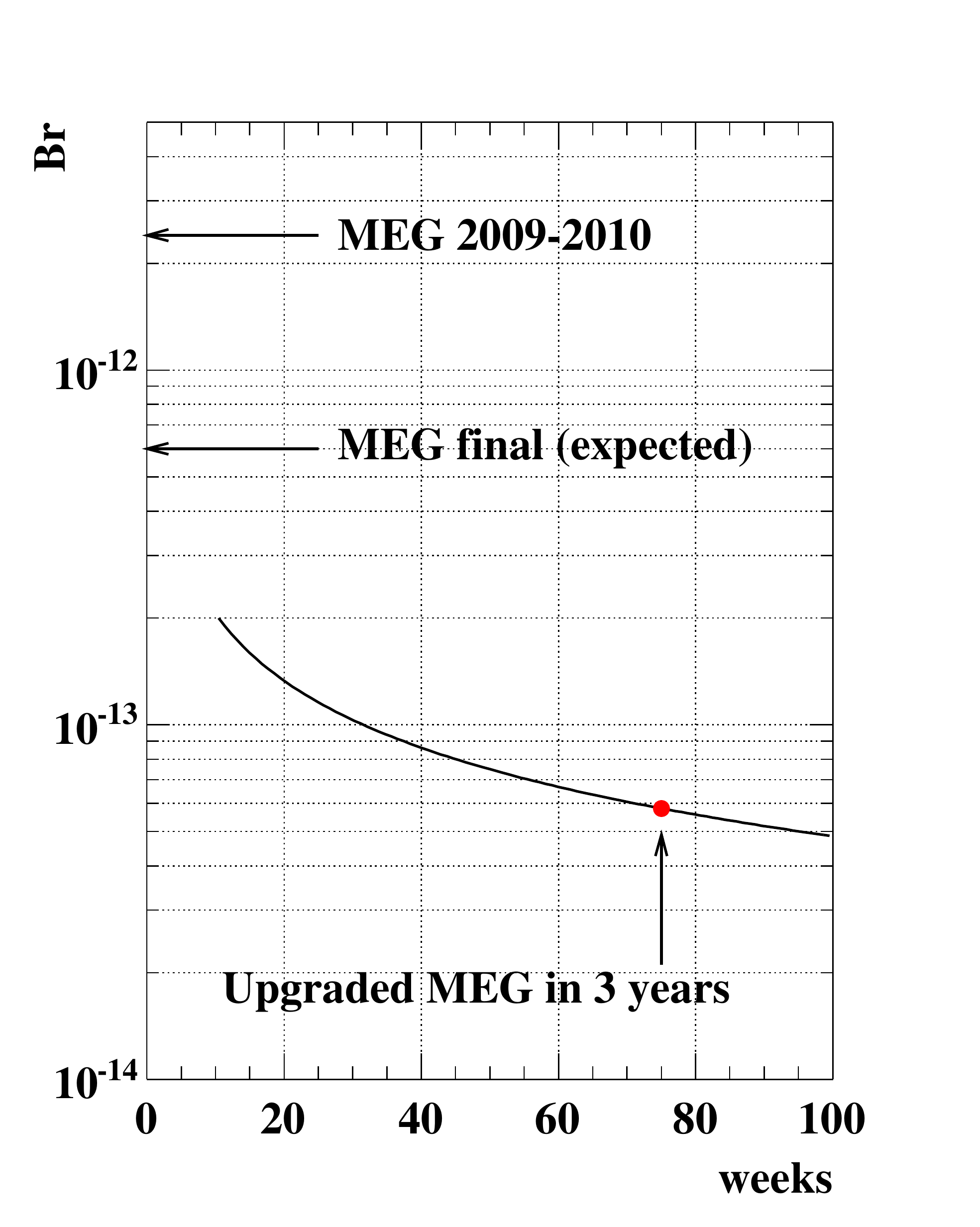}
\caption{\label{fig:sensEv} Expected sensitivity of upgraded MEG as a function of 
DAQ time in weeks. Assuming 175 DAQ days per year, we expect to reach an 
UL on $\BR(\mu \rightarrow {\rm e} \gamma)$ of $\approx 6 \times 10^{-14}$ 
in 3 years of running.}
\end{figure}

In Fig. \ref{fig:sensEv} we show the evolution of the sensitivity as a function of the 
DAQ time (in weeks). 
With a muon stopping rate on target of $7 \times 10^{7}\,\mu/\mathrm{sec}$ and a 
target thickness of 140\,$\mu\mathrm{m}$ and assuming $175$ DAQ days per year,
we expect to reach a final sensitivity of about $6 \times 10^{-14}$ in 3 years of running. 
We note that the sensitivity is calculated based on the improved detector performace shown in Table\,\ref{tab:scenario}, 
but it has an approximately 30\% ambiguity 
according to possible different scenarios in the performance improvement.

%% file: 10_Budget_and_Responsibilities/Budget_and_Responsibilities.tex
\section{Costs and responsibilities}
%

We present here detailed cost estimates for the construction of the different parts proposed for this upgrade,
namely the upgrade of the xenon calorimeter (Tab. \ref{tab:LXe costs}), the new timing counter (Tab. \ref{tab:TC costs}),  the DAQ upgrade (Tab. \ref{tab:DAQ costs})  and  the new drift chamber (Tab. \ref{tab:DRAGO costs}).
The proposed subdivision of responsibilities for the construction of the individual items is shown in Tab. \ref{tab:respo}.

\begin{table}[hbct]

\begin{minipage}[b]{.4\textwidth}
\centering
 \caption{\label{tab:LXe costs} Cost estimate for the liquid xenon calorimeter upgrade }
 \begin{tabular}{|c|c|}
 \hline
 {\bf Item} & {\bf Cost (k\officialeuro)}\\
 \hline
 LXe (180 liters) & 340\\
 \hline
 UV-SiPM & 720\\
 \hline
 Cables & 369 \\
 \hline
 Connectors & 149 \\
 \hline
 Feedthrough & 1 \\
 \hline
 PCB (inner slab) & 1 \\
 \hline
 Refrigerator & 40 \\
 \hline
 PMT holder (lateral faces) & 10 \\
 \hline
 Radiative decay veto & 12\\
 \hline
 {\bf Total} & {\bf 1642} \\
 \hline
\end{tabular}
\end{minipage}\qquad
\begin{minipage}[b]{.4\textwidth}
\centering
 \caption{\label{tab:TC costs} Cost estimate for the new pixelated timing counter}
 \begin{tabular}{|c|c|}
 \hline
 {\bf Item} & {\bf Cost (k\officialeuro)}\\
 \hline
 SiPM & 68\\
 \hline
 Scintillator & 11 \\
 \hline
 Support structure & 20 \\
 \hline
 Cables & 47 \\
 \hline
 PCB & 2 \\
 \hline
 Connectors & 17 \\
 \hline
 Laser for calibration  & 20 \\
 \hline
 Opgtical fibers & 21 \\
 \hline
 {\bf Total} & {\bf 206} \\
 \hline
\end{tabular}
\end{minipage}

\end{table}

\begin{table}[hbct]


\begin{minipage}[t]{.4\textwidth}
\centering
 \caption{\label{tab:DAQ costs} Cost estimate for the DAQ upgrade}
 \begin{tabular}{|c|c|}
 \hline
 {\bf Item} & {\bf Cost (k\officialeuro)}\\
 \hline
 Trigger concentrator boards & 90\\
 \hline
 Clock distribution cables and ancillary boards & 10 \\
 \hline
Total Digitizer Cost (7400 channels) & 815\\
\hline
{\bf Total} & {\bf 915} \\
\hline
\end{tabular}
\end{minipage}
\end{table}

\begin{table}[hbct]

\caption{\label{tab:DRAGO costs} Detailed cost estimate for the new Drift Chamber}
  \begin{tabular}{|l|l|r|}
  \hline
  \multicolumn{2}{|c|}{ {\bf Prototype }(full length wedge – 128 ch.)} & \bf{98 k\officialeuro}\\
  \hline
  \multirow{3}{4cm}{ Mechanics} & {\it endplates (fabrication, machining, drilling)} \,\, & 8\\  \cline{2-3}
  &  {\it supports} & 4\\ \cline{2-3}
  & {\it Testing jig} & 6\\
  \hline
   Materials & {\it wires and supports} & 24\\  
  \hline
  Tools & {\it  soldering, cleaning} & 4 \\ 
  \hline
  Quality Control & {\it mech. tension, alignment} & 8 \\ 
  \hline
   Front End & {\it preampl. + driver, LV power supply  cables + connectors} & 16\\
  \hline
  HV & {\it power supply + distribution} &  28\\ 
  \hline
  \hline
  \multicolumn{2}{|c|}{ \bf Chamber end plates + support structure} & \bf{174 k\officialeuro}\\
  \hline
  Project design & {\it Software licenses + engineering} & 36\\ 
  \hline
  \multicolumn{2}{|l|} {Model construction for end plates} & 12\\
  \hline
   \multicolumn{2}{|l|} {E.P. Fabrication} & 30\\
  \hline
    \multicolumn{2}{|l|} {E.P.Drilling} & 16\\
  \hline
  Measuring & {\it dimensional, stress and deformation checks} & 10 \\
  \hline
   \multicolumn{2}{|l|} {Inner and Outer Cylinders Fabrication and Machining} & 40\\
  \hline
   \multicolumn{2}{|l|} {Support Structure Skeleton} & 18\\
  \hline
  Assembling Jig & {\it incl. glueing and gas sealing flanges} & 12 \\
  \hline
  \hline
  \multicolumn{2}{|c|}{ \bf Chamber Wiring and Construction} & \bf{160 k\officialeuro}\\
  \hline
  \multicolumn{2}{|l|} {Clean Room Utilization} & 20\\
  \hline
  \multicolumn{2}{|l|} {Mechanical Structure} & 10\\
  \hline
  \multicolumn{2}{|l|} {Positioning and Alignment monitoring} & 10\\
  \hline
  \multicolumn{2}{|l|} {Full Length Wiring Machine} & 40\\
  \hline
  \multicolumn{2}{|l|} {Tools} & 10\\
  \hline
  Materials & {\it wires, supports} & 70 \\
  \hline
  \hline
  \multicolumn{2}{|c|}{ \bf Front-end, Electronics, HV system} & \bf{260 k\officialeuro}\\
  \hline
  Front-end (2400 ch.) & {\it preamp., driver, cables, connectors, LV supply} & 200\\
  \hline
  HV System & {\it supply + distribution} & 60\\
  \hline
  \hline
  \multicolumn{2}{|c|}{ \bf Gas system} & \bf{94 k\officialeuro}\\
  \hline
  \multicolumn{2}{|l|} {Flow controller + Mass flow meters + Pressure sensors} & 32\\
  \hline
  \multicolumn{2}{|l|} {Infrared analyzer for hydrocarbons} & 18\\
  \hline
  \multicolumn{2}{|l|} {O$_2$/H$_2$O monitor} & 18\\
  \hline
  \multicolumn{2}{|l|} {Gas distribution} & 16\\
  \hline
  \multicolumn{2}{|l|} {Monitor chamber} & 10\\
  \hline
  \multicolumn{3}{|c|}{ \,}\\
  \hline
  \multicolumn{2}{|c|}{ \bf Total} & \bf{786 k\officialeuro}\\
  \hline
\end{tabular}
\end{table}

\begin{table}[hbct]
\renewcommand\arraystretch{2.0} 
\addtolength{\tabcolsep}{0.2cm}

\centering
 \caption{\label{tab:respo} Subdivision of the construction responsibilities}
 \begin{tabular}{|c|c|c|}
 \hline
 {\bf Item} & \bf{Nation} & {\bf Responsibility}\\
 \hline
 \multirow{1}{*}{ Liquid xenon } & Japan & SiPMs, LXe, Cryogenics \\
 \hline
\multirow{2}{*}{Drift Chamber}  & Italy & End-caps, Wiring, FE electronics, HV \\ \cline{2-2} & Switzerland & 
Mechanical integration, Gas-System \\
 \hline
 \multirow{3}{*}{Timing counter} & Italy & SiPMs, Support structure \\ 
\cline{2-2} 
& Japan & \multirow{2} {*} {Scintillator, Laser system,  Cables/Connectors}  \\
\cline{2-2} &  UCI &  \\
 \hline
\multirow{2}{*} {DAQ}  & Italy & Trigger boards \\ \cline{2-3} & Switzerland & WaveDREAM boards \\
 \hline
\end{tabular}
\end{table}

%% file: 11_Time_Schedule/Time_Schedule.tex
\section{Time Schedule and Man Power}
\label{sec:Time Schedule}
%
\begin{fgantt}
\newganttlinktype{mio}{%
 \ganttsetstartanchor{on bottom=1}%
 \ganttsetendanchor{on top=0.5}%
 \draw [/pgfgantt/link]
 (\xLeft, \yUpper) --
 (\xRight, \yLower)
 node [pos=.5, /pgfgantt/link label anchor] {\ganttlinklabel};
 }
\caption{Overall MEG Upgrade Schedule\label{gantt:MEGUP}}
\scalebox{0.77}{
\begin{ganttchart}[hgrid,vgrid,
    x unit=1.3cm, bar/.style={fill=green, rounded corners=3pt},incomplete/.style={fill=red},group left shift=0., group right shift=0.,milestone/.style={fill=orange}]{13}
     \ganttset{progress label anchor/.style={right=0.1cm},link type=f-s}
     \gantttitle{Year}{13}\\
     \gantttitle{2012}{1}
     \gantttitle{2013}{2}
     \gantttitle{2014}{2}
     \gantttitle{2015}{2}
     \gantttitle{2016}{2}
     \gantttitle{2017}{2}
     \gantttitle{2018}{2}\\
     \ganttbar[bar/.style={fill=blue}]{Design}{1}{4}\\
     \ganttbar[bar/.style={fill=green}]{Construction}{3}{6}\\
     \ganttbar[bar/.style={fill=yellow}]{Engineering Run}{7}{7}\\
     \ganttbar[bar/.style={fill=red}]{Run}{8}{13}
   \end{ganttchart}
 
 }
\end{fgantt}

The overall planned schedule for the upgrade and its implementation is shown in figure~\ref{gantt:MEGUP}. The initial period of design and construction
will be followed by an engineering run in 2015. After that three years of data taking are foreseen.

The time schedule for the final R\&D tests and construction are presented for the new MEG drift chamber (Gantt chart \ref{gantt:DRAGO}), the new Timing Counter(Gantt chart \ref{gantt:TC}), the modifications to the liquid xenon calorimeter (Gantt chart \ref{gantt:LXe}) and the DAQ system (Gantt chart \ref{gantt:DAQ}). 
The starting time of these schedules is the time of preparation of this document, namely end of July 2012.
We may note that some R\&D have already started since some time.

In Table \ref{tab:FTEschedule} we  further show the number of full time equivalent (FTE) researchers for the different construction items as a function of time. 

\vspace{0.7cm}


\begin{fgantt}
\newganttlinktype{mio}{%
 \ganttsetstartanchor{on bottom=1}%
 \ganttsetendanchor{on top=0.5}%
 \draw [/pgfgantt/link]
 (\xLeft, \yUpper) --
 (\xRight, \yLower)
 node [pos=.5, /pgfgantt/link label anchor] {\ganttlinklabel};
 }

\caption{ New Drift Chamber schedule \label{gantt:DRAGO}}

 \scalebox{0.9}{
  \begin{ganttchart}[hgrid,vgrid,
    x unit=1.3cm, bar/.style={fill=green, rounded corners=3pt},incomplete/.style={fill=red},group left shift=0., group right shift=0.,milestone/.style={fill=orange}]{7}
     \setganttlinklabel{f-s}{}
     \ganttset{progress label anchor/.style={right=0.1cm},link type=f-s}
    \gantttitle{Year}{7}\\
    \gantttitle{2012}{1}
    \gantttitle{2013}{2}
    \gantttitle{2014}{2}
    \gantttitle{2015}{2}\\

    \ganttgroup[group label anchor/.style={left=0.2cm}]{Prototype}{1}{2.5}\\
    \ganttbar[progress=100]{Electronics}{1}{1}\\
    \ganttbar[progress=100]{Silicon strip telescope}{1}{1}\\
    \ganttbar[progress=50]{Chambers}{1}{2}\\
    \ganttbar[progress=5]{Chambers tests}{1.5}{2.5}\\

    \ganttmilestone[milestone width=0.3] { Resolution and ageing test results } { 1 } \\

    \ganttgroup[]{Design}{1}{3}\\
    \ganttbar[progress=20]{End plates design}{1}{2.5}\\
    \ganttbar[progress=35]{Wiring instrumentation}{1}{3}\\

    \ganttgroup{Construction}{3}{4}\\
    \ganttbar[progress=0]{Front end electronics}{3}{4.}\\
    \ganttbar[progress=0]{Mechanics}{3.5}{4}\\

    \ganttgroup{Mounting}{5}{7.0}\\
    \ganttbar[progress=0]{Wiring the chamber}{5}{5}\\
    \ganttbar[progress=0]{Assembling at PSI}{6}{5.5}\\
    \ganttbar[progress=0]{DAQ test}{6.5}{6}\\
    \ganttbar[progress=0]{Engineering run}{7.0}{7.0}\\

    \ganttlink[link type=mio]{elem1}{elem5}
    \ganttlink[link type=mio]{elem2}{elem5} 
   \ganttlink{elem4}{elem11}
    \ganttlink{elem7}{elem11}
    \ganttlink[link type=auto]{elem8}{elem13}
    \ganttlink{elem9}{elem13}
    \ganttlink{elem10}{elem13}
    \ganttlink{elem11}{elem13}
    \ganttlink{elem13}{elem14}
    \ganttlink{elem14}{elem15}
    \ganttlink{elem15}{elem16}
  
  \end{ganttchart}
  }

\end{fgantt}

%
%
\begin{fgantt}
\newganttlinktype{mio}{%
 \ganttsetstartanchor{on bottom=1}%
 \ganttsetendanchor{on top=0.5}%
 \draw [/pgfgantt/link]
 (\xLeft, \yUpper) --
 (\xRight, \yLower)
 node [pos=.5, /pgfgantt/link label anchor] {\ganttlinklabel};
 }
\newganttlinktype{pio}{%
 \ganttsetstartanchor{on bottom=0.5}%
 \ganttsetendanchor{on top=0}%
 \draw [/pgfgantt/link]
 (\xLeft, \yUpper) --
 (\xRight, \yLower)
 node [pos=.5, /pgfgantt/link label anchor] {\ganttlinklabel};
 }

\caption{ Timing counter schedule \label{gantt:TC}}


 \scalebox{1.0}{
  \begin{ganttchart}[hgrid,vgrid,x unit=1.3cm, bar/.style={fill=green, rounded corners=3pt},incomplete/.style={fill=red},group left shift=0.,milestone/.style={fill=orange}]{7}

     \setganttlinklabel{f-s}{}
     \ganttset{progress label anchor/.style={right=0.1cm},link type=f-s}
    \gantttitle{Year}{7}\\
    \gantttitle{2012}{1}
    \gantttitle{2013}{2}
    \gantttitle{2014}{2}
    \gantttitle{2015}{2}\\


    \ganttgroup{Single pixel R\&D}{1}{1}\\
    \ganttbar[progress=60]               { Pixel layout \& geometry           } { 1 } { 1 } \\
    \ganttbar[progress=50]               { Readout scheme                     } { 1 } { 1 } \\
    \ganttmilestone[milestone width=0.3] { Proved performance of single pixel } { 1 } \\

    \ganttgroup{Prototype}{1}{3}\\
    \ganttbar[progress=0]      { MPPC } { 2 } {1.5} \\
    \ganttbar[progress=0]      { Scintillator } { 2 } {1.5} \\
    \ganttbar[progress=20]      { Calibration R\&D} { 1  } { 1.5 } \\
    \ganttbar[progress=0]      {Assembly} { 2.5  } { 1.5 + 2/6 } \\
    \ganttbar[progress=0]      {Test (laser/CR/checking source)} { 2.5+ 2/6} { 1.5 + 4/6 } \\
    \ganttbar[progress=20]      {Electronics} { 1  } { 2 + 1/6 } \\
    \ganttbar[progress=0]      {Beam test} { 2.5 + 4/6  } {2.5} \\
    \ganttbar[progress=30]      {Full detector design} { 1  } { 3} \\

    \ganttgroup{Construction}{3}{5}\\
    \ganttbar[progress=0]      { MPPC production             } {4} { 4} \\
    \ganttbar[progress=0]      { Scintillator } { 4 } { 4} \\
    \ganttbar[progress=0]      { Support structure } { 4 } { 4}\\
    \ganttbar[progress=0]      { Electronics                    } { 3 } { 3 } \\
    \ganttbar[progress=0]      { Assembly                    } { 5  } { 5 } \\

    \ganttgroup{Commissioning}{6}{7}\\
    \ganttbar[progress=0] { Installation } { 6} { 5 + 1/2} \\
    \ganttbar[progress=0] { Engineering run         } { 6 + 1/2 } { 7 } 

    \ganttlink[link type=mio]{elem1}{elem3}
    \ganttlink[link type=mio]{elem2}{elem3}
    \ganttlink[link type=pio]{elem3}{elem5}
    \ganttlink[link type=pio]{elem3}{elem6}
    \ganttlink{elem5}{elem8}
    \ganttlink{elem7}{elem8}
    \ganttlink{elem8}{elem9}
    \ganttlink{elem9}{elem11}
    \ganttlink{elem10}{elem11}
    \ganttlink{elem11}{elem14}
    \ganttlink{elem12}{elem14}
    \ganttlink{elem12}{elem15}
    \ganttlink{elem12}{elem16}
    \ganttlink{elem14}{elem18}
    \ganttlink{elem15}{elem18}
    \ganttlink{elem16}{elem18}
    \ganttlink{elem17}{elem18}
    \ganttlink{elem18}{elem19}
    \ganttlink{elem20}{elem21}
  
  \end{ganttchart}
  }

\end{fgantt}
%

\begin{fgantt}
\caption{ LXe detector schedule \label{gantt:LXe}}
\newganttlinktype{mio}{%
 \ganttsetstartanchor{on bottom=1}%
 \ganttsetendanchor{on top=0.5}%
 \draw [/pgfgantt/link]
 (\xLeft, \yUpper) --
 (\xRight, \yLower)
 node [pos=.5, /pgfgantt/link label anchor] {\ganttlinklabel};
 }
\newganttlinktype{pio}{%
 \ganttsetstartanchor{on bottom=0.5}%
 \ganttsetendanchor{on top=0}%
 \draw [/pgfgantt/link]
 (\xLeft, \yUpper) --
 (\xRight, \yLower)
 node [pos=.5, /pgfgantt/link label anchor] {\ganttlinklabel};
 }


 \scalebox{1.0}{
  \begin{ganttchart}[hgrid,vgrid,x unit=1.3cm, bar/.style={fill=green, rounded corners=3pt},incomplete/.style={fill=red},group left shift=0.,milestone/.style={fill=orange}]{7}

\setganttlinklabel{f-s}{}
\ganttset{progress label anchor/.style={right=0.1cm},link type=f-s}
    \gantttitle{Year}{7}\\
    \gantttitle{2012}{1}
    \gantttitle{2013}{2}
    \gantttitle{2014}{2}
    \gantttitle{2015}{2}\\
%

    \ganttgroup{R\&D}{1}{2}\\
    \ganttbar[progress=50]               { UV-sensitive MPPC                     } { 1 } { 1+1 } \\
    \ganttbar[progress=40]               { PCB feedthrough                       } { 1 } { 1+4/6 } \\
    \ganttmilestone[milestone width=0.3] { Established MPPC readout technologies } { 1+1 } \\

    \ganttgroup{Prototype}{1}{4}\\
    \ganttbar[progress=0]                { MPPC production              } { 1+1+1     } { 1+4/6+1 } \\
    \ganttbar[progress=0]                { Feedthrough/cable/connector  } { 1+1     } { 1+4/6 } \\
    \ganttbar[progress=0]                { Cryostat refurbishment       } { 2       } { 2+4/6 } \\
    \ganttbar[progress=0]                { Assembly                     } { 1+1+4/6+1 } { 2+1/6+1 } \\
    \ganttbar[progress=20]                { Electronics                  } { 1       } { 2+3/6 } \\
    \ganttbar[progress=0]                { Beam test                    } { 3+3/6+1   } { 3+1     } \\


    \ganttgroup{Construction}{3}{5.5}\\
    \ganttbar[progress=0]      { MPPC production             } { 4+1 } { 4+1   } \\
    \ganttbar[progress=0]      { Feedthrough/cable/connector } { 4+1 } { 4+1   } \\

\ganttset{progress label anchor/.style={left=1.4cm}}
    \ganttbar[progress=0]      { Refrigerator                } { 3 } { 3   } \\
\ganttset{progress label anchor/.style={right=1.4cm}}
    \ganttbar[progress=0]      { Electronics                 } { 3.5 } { 4   } \\
\ganttset{progress label anchor/.style={left=1.4cm}}
    \ganttbar[progress=0]      { Lateral PMT holder          } { 3 } { 3   } \\
\ganttset{progress label anchor/.style={right=0.1cm}}

    \ganttbar[progress=0]      { Assembly                    } { 5+1 } { 4.5+1 } \\

    \ganttgroup{Commissioning}{6.5}{7}\\
    \ganttbar[progress=0] { Installation } { 5.5+1 } { 5+1 } \\
    \ganttbar[progress=0] { Engineering run } { 6+1   } { 6+1 }

    \ganttlink[link type=mio]{elem1}{elem3}
    \ganttlink[link type=mio]{elem2}{elem3}
    \ganttlink[link type=pio]{elem3}{elem5}
    \ganttlink{elem5}{elem8}
    \ganttlink{elem6}{elem8}
    \ganttlink{elem7}{elem8}
    \ganttlink{elem8}{elem10}
    \ganttlink{elem9}{elem10}
    \ganttlink{elem10}{elem12}
    \ganttlink{elem10}{elem13}
    \ganttlink{elem12}{elem17}
    \ganttlink{elem13}{elem17}
    \ganttlink{elem14}{elem17}
    \ganttlink{elem15}{elem17}
    \ganttlink{elem16}{elem17}
    \ganttlink{elem17}{elem18}
    \ganttlink{elem19}{elem20}

  \end{ganttchart}
  }

\end{fgantt}

\begin{fgantt}

\caption{ DAQ schedule \label{gantt:DAQ}}


\setganttlinklabel{f-s}{}
\ganttset{progress label anchor/.style={right=0.1cm},link type=f-s}

 \scalebox{1.0}{
  \begin{ganttchart}[hgrid,vgrid,x unit=1.3cm, bar/.style={fill=green, rounded corners=3pt},incomplete/.style={fill=red},group left shift=0.]{6}

    \gantttitle{Year}{6}\\
    \gantttitle{2012}{1}
    \gantttitle{2013}{2}
    \gantttitle{2014}{2}
    \gantttitle{2015}{1}\\

\ganttset{progress label text={}}

    \ganttgroup[group label anchor/.style={left=0.3cm}] {WaveDREAM boards}{1}{4}\\
    \ganttbar[progress=0]{First prototype board}{1}{1}\\
    \ganttbar[progress=0]{$2^{nd}$ iteration in case of problems}{2}{1.5}\\
    \ganttbar[progress=0]{Mass production}{2+0.5}{4}\\

    \ganttgroup[]{Concentrator board}{1 + 6/6}{ 1 + 5/6}\\
    \ganttbar[progress= 0]{Prototype}{1 +6/6}{1+3/6}\\
    \ganttbar[progress=0]{Ethernet firmware}{1+6/6}{1+5/6}\\

    \ganttgroup[]{Trigger board}{2}{3 + 4/6}\\
    \ganttbar[progress= 0]{Prototype board: design and contruction}{2}{1+4/6}\\
    \ganttbar[progress= 0]{Prototype board: test}{2+4/6}{2}\\
    \ganttbar[progress=0]{$2^{nd}$ iteration in case of problems}{3}{2+2/6}\\
    \ganttbar[progress=0]{Board mass production}{3+2/6}{3 + 4/6}\\
    \ganttbar[progress=0]{FPGA firmware \& DAQ bus interface}{2}{1+4/6}\\
    \ganttbar[progress=0]{Control signal protocol, clock synchronization}{3}{2+3/6}\\
  
   \ganttlink{elem1}{elem2}
   \ganttlink{elem2}{elem3}
   \ganttlink{elem8}{elem9}
   \ganttlink{elem9}{elem10}
   \ganttlink{elem10}{elem11}

  \end{ganttchart}
  }

\end{fgantt}


\begin{table}[p]
\renewcommand\arraystretch{2.0} 
\addtolength{\tabcolsep}{0.2cm}

\caption{\label{tab:FTEschedule} FTE researchers as a function of time for the different tasks}
  \begin{tabular}{|l|r|r|r|r|r|r|r|r|r|r|}
  \hline
  \multirow {2} {*}  {\bf Item} & \multicolumn{10}{|c|} {\bf Months }\\
  \cline{2-11}
                         & 6    & 12   & 18  & 24  & 30  & 36  & 42   & 48   & 54   & 60\\ \hline \hline
Beamline and Target      & 3    & 3    & 3   & 3   & 3   & 3   & 2    & 2    & 2    & 2 \\ \hline
Drift Chamber            & 11   & 11   & 11  & 12  & 12  & 12  & 7.5  & 3.5  & 2.5  & 2.5 \\  \hline
Timing Counter           & 9.6    & 9.6    & 10.1 & 10.1 & 10.1 & 10.1 & 5.6   & 5.6   & 3    & 3 \\ \hline
Liquid Xenon Calorimeter & 7    & 7    & 9   & 9   & 9   & 9   & 4    & 4    & 4    & 4\\ \hline
Trigger \& DAQ           & 2.5  & 2.5  & 2.5 & 3.5 & 3.5 & 3.5 & 2.5  & 2.5  & 1.5  & 1.5 \\ \hline
Analysis                 & 3    & 3    & 3   & 3   & 3   & 3   & 15.5 & 19.5 & 21.1 & 21.1 \\ \hline
{\bf Total}              & 36.1 & 36.1 & 38.6  & 40.6  & 40.6  & 40.6  & 37.1 & 37.1 & 34.1 & 34.1 \\
  \hline
\end{tabular}
\end{table}

%% file: 12_Summary/Summary.tex
\section{Summary}
%

The MEG experiment upgrade we propose with this document aims at reaching  a sensitivity
down to $\BR \simeq 6 \times 10^{-14}$  in the  search  of the $\meg$ decay. 
This is a very ambitious goal of extreme scientific interest, since a positive observation of the $\meg$ decay would represent the opening
of a new world of phenomena in the field of particle physics. The proposed 
upgrade is  competitive even with future projects like COMET or Mu2e, whose 
budgets are orders of magnitudes higher than what is needed for the MEG 
improvements.

It is however  a really difficult task
since in MEG we have to struggle with the rising wall of the accidental background which quadratically increases 
as a function of the muon stop rate. In order to accomplish it we need to deeply modify the liquid xenon calorimeter
read-out  and build new devices for measuring the positrons kinematical variables. We presented in full detail the selected solutions, their costs, the manpower involved in their realization and described the  R\&D tests that  will be
completed before the end of the year to make us sure that we will be able to reach the planned
sensitivity goal.

The chosen solutions are somewhat conventional since we believe it is important not to exceed a total time of the order of 5 years for the construction and running of the apparatus: a time schedule which also fits well within the international context of cLFV searches.
Less conventional solutions  were however studied and described in the Appendix section.


Finally we would  like to remark that the estimates and extrapolations we 
present in this document are firmly based on the results of a running 
experiment. The MEG international team of physicists has been successfully 
collaborating for a period of about 10 years; the total budget we are 
requiring roughly corresponds to 30\% of the original MEG construction budget.

%% file: 13_Appendix/Appendix.tex
\section{Appendix}
%
 In this section an overview of related non-baseline upgrade studies is reviewed. These consist of complementary and alternative technologies that were considered as areas of potential for an upgrade programme. The scope of such on-going studies is rather broad and spans four main areas of interest:
 
 \begin{itemize}
 \item Tracking - (Active Target ATAR, TPC-based Tacker and a Si-based Vertex Tracker SVT)
 \item Diagnostics - (Active Target ATAR)
 \item Background Suppression - (Radiative Muon Decay Veto Counter RDC)
 \item Generic detector R$\&$D - (LXe PMT Development, Timing Counter generic Scintillation Material Studies)
 \end{itemize}

  The devices described are also at different stages of R$\&$D and development and some could, after development and successful testing and providing no serious background issues are discovered, be integrated into a running upgrade programme. However currently, these devices are not considered as part of the baseline solution presented in this proposal.

\input{ATAR}

\input{RDC}

\input{Positron_Tracker_TPC}

\input{Positron_Tracker_SVT}
\input{ptp}

\input{LXePMT}

%% file: 13_Appendix/ATAR.tex
\subsection{The Active target option}
\subsubsection{Active target concept}
We consider in this section the possibility to complement the new proposed MEG spectrometer with  an active target \cite{ATAR:Elba2012}. The active target has two aims: 1) to improve the determination of the muon decay vertex and consequently to achieve better positron momentum and angular resolutions; 2) to continuously measure the muon stopping rate and therefore provide a direct evaluation of the detector acceptance (or an absolute normalization of our data). The former will be achieved by detecting the emerging positron using a single layer of 250 $\rm{\mu m}$ fast scintillating fibres.  This measurement is challenging since only a 30 keV mean energy deposit is expected from minimum ionizing particles in such a thin scintillating material. Stringent requirements must be satisfied to minimize positron multiple scattering and $\gamma$ background from positron annihilation in the target materials. Therefore the target must have a minimal thickness and consist of low-Z materials. It must also have a  fast response time to sustain the high rates and be able to operate in a high magnetic field  environment (1.3 T). The muons ranging out into the target will provide the measurement of the stopping rate. The high target segmentation is a benefit also for this purpose. Muon and positron signals can be distinguished by their large difference in energy deposit in the detector, as described in section~\ref{sec:experimental}.

\paragraph{The muon decay vertex measurement} 
The position of the muon decay vertex can be determined by detecting the stopping muon or the emerging positron.

\begin{itemize}
\item Muon detection.
In this case a spatial correlation must be used to connect the muon signal detected by the target, with the positron track measured by the spectrometer. The reconstructed positron track is extrapolated back to the target plane selecting only a small area. Only one fibre is expected to be fired in this area, which provides the muon position. This process is however rate limited to $\approx 10^7$ particles/s, above which, muon multiplicities become larger than one and the efficiency of the method rapidly diminishes.
\item Positron detection.
In this case the positron detected by the target is timing correlated with the reconstructed track in the spectrometer, the positron time being provided by fast scintillators. This timing correlation can be used to have an external trigger on the target signal, which is fundamental to extract the expected weak positron signal as explained below.
\end{itemize}

With the proposed target a position resolution of $\sigma_y< 100~\rm{\mu m}$ is achievable and a timing resolution $\sigma_t<500$ psec is expected with a photo-electron statistics of about 10. A detection efficiency larger than $80\%$ is desired. 
 
\subsubsection{Monte-Carlo simulations}
In the current  MEG experimental set-up there is no direct measurement of the muon decay vertex. Muons are stopped in a passive target (target thickness = 205 $\rm{\mu m}$, slant angle 20.5 deg relative to the beam direction) and the positrons from muon decay are detected, after traversing $\sim 40$ cm  He gas (at 1 atm pressure and room temperature). The muon decay vertex is determined by back-projecting the reconstructed positron track to the target plane. The measured performances of the present spectrometer coupled with a passive target are given in Tab.~\ref{tab:res}.

\begin{figure}[hbct]
\begin{center}
\begin{tabular}{cc}
\includegraphics[scale=0.3]{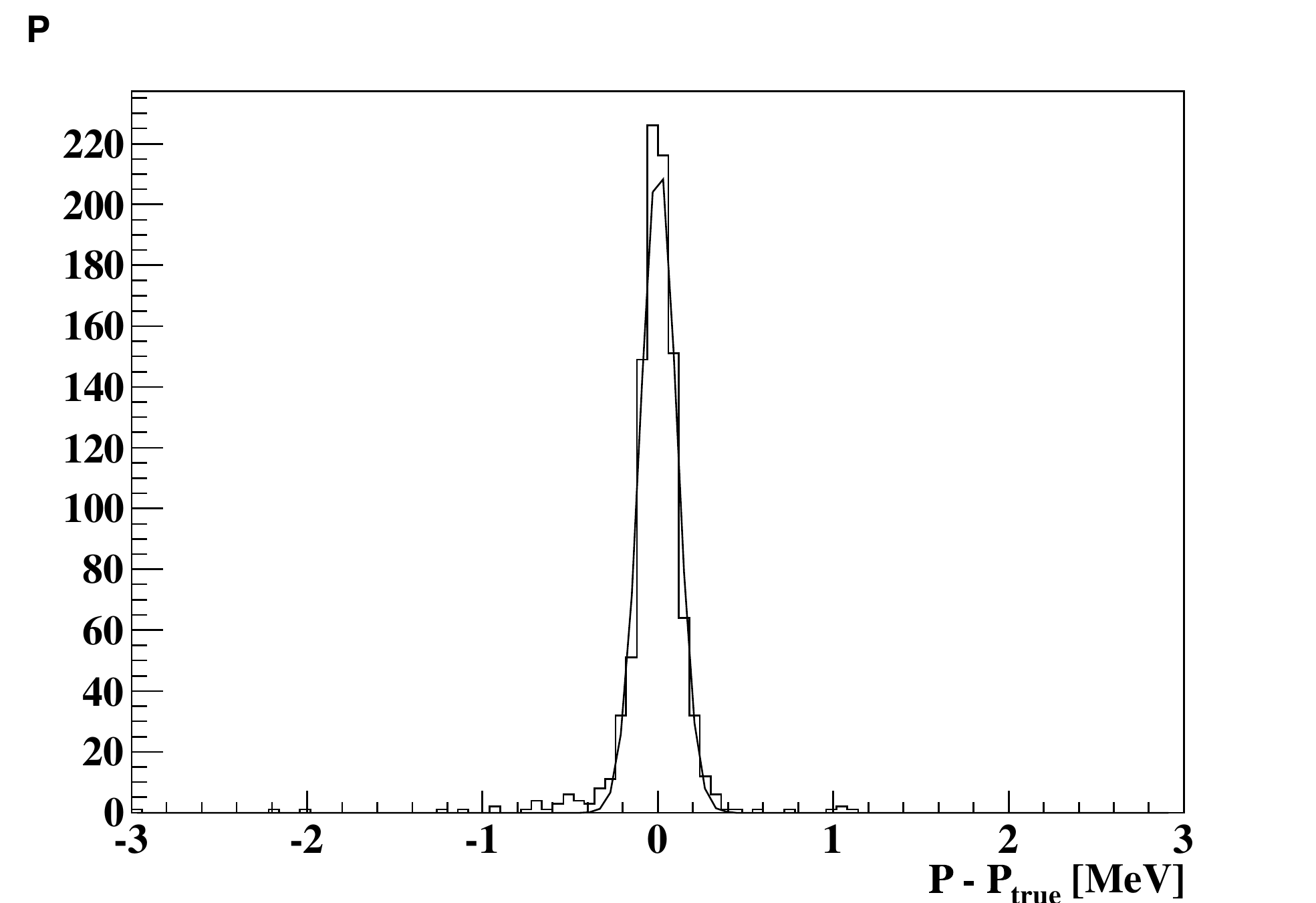} &
\includegraphics[scale=0.3]{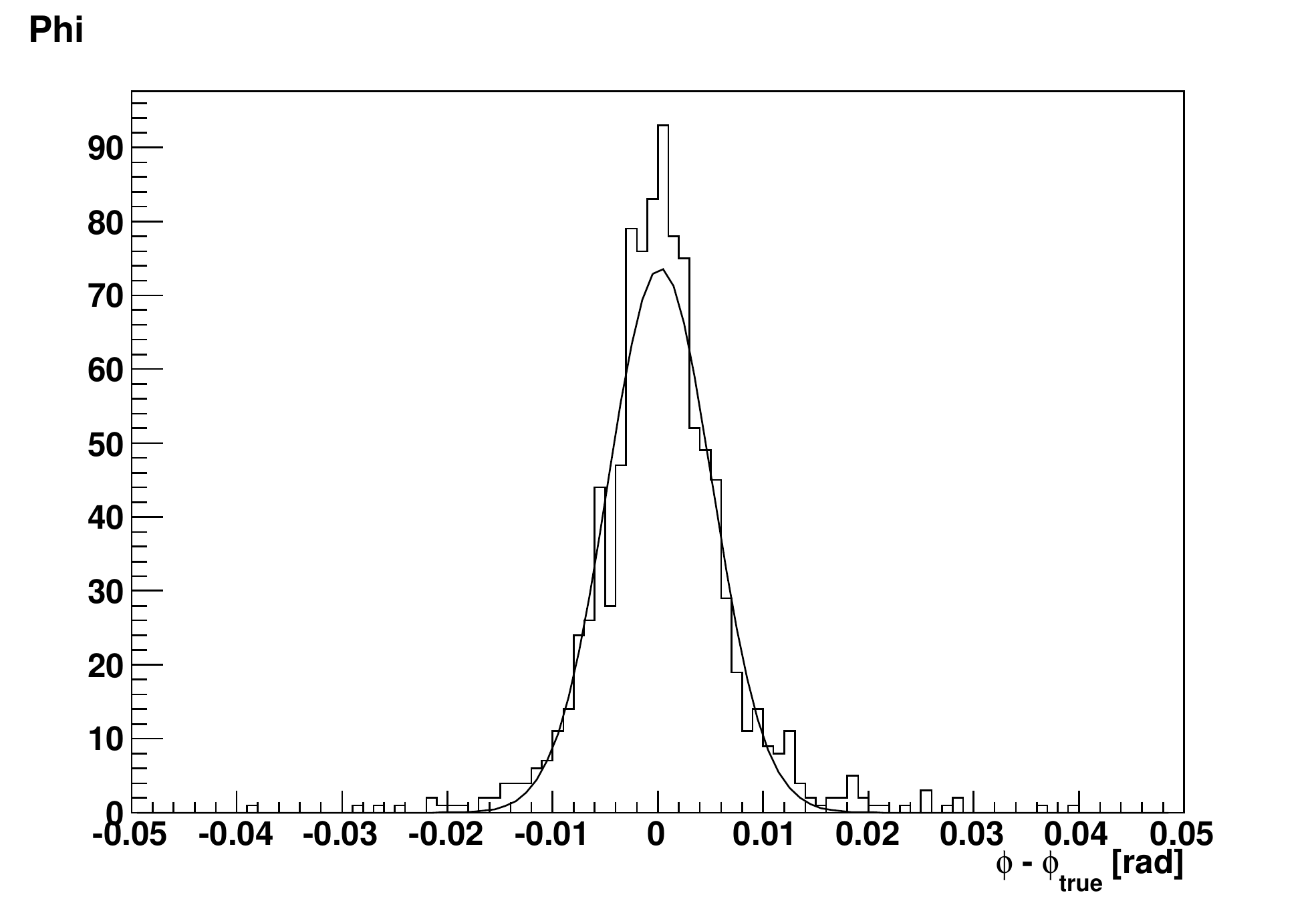} \\
\end{tabular}
\caption{Monte Carlo simulation of the positron momentum $\sigma_p = 90$ keV/c (left) and the phi angle $\sigma_{\phi} = 4.7$ mrad (right) resolutions of  $ \mu^{+} \to e^{+} \gamma$  events using an active target coupled with the new spectrometer.}
\label{fig:atarres}
\end{center}
\end{figure}

If an active target is used in a minimal configuration with only one single layer of horizontally mounted fibres, the measurement of the Y-coordinate can be performed which strongly constrains the positron momentum reconstruction. Fig.~\ref{fig:atarres} shows  typical positron momentum (left) and $\phi$ angle (right) resolution distributions. The improved resolutions are summarized in Tab.~\ref{tab:res}. 
\begin{table}
\begin{center}
\begin{tabular}{cccccccc}
\hline
\hline
Target/  & thickness ($\mu m$)/ & $\sigma_{p}$ &  $\sigma_{\phi}$  & $\sigma_{\theta}$& comment \\
 Spectrometer             &  angle (deg)                &(keV)                          & (mrad)                              & (mrad) &  \\
\hline
Passive/old & 205/ 20.5 & 320  & 11.7 & 9.8 & measured \\
\hline
Passive/new & 205/  20.5 &110  & 6.3 & 5.3  & simulated  \\
Passive/new & 140/ 15 & 110  & 5.3 & 4.8 & simulated \\
Active/new    & 250/ 20.5 & 90   & 4.7 & 5.1 & simulated \\
\hline
\hline
\end{tabular}
\end{center}
\caption{\label{tab:res} Measured positron momentum and angular resolutions with the present spectrometer (first row) . Monte Carlo simulation results  for different  target options coupled to the new spectrometer  in $ \mu^{+} \to e^{+} \gamma$  events are then reported.}
\end{table}

\subsubsection{Experimental set-up for R \& D studies}
\label{sec:experimental}

The active target will be made from an array of 240 ($0.25\times0.25~{\rm mm}^{2}$) multi-clad BCF12 (from Saint-Gobain) square fibres, with a peak emission at 435 nm, a light yield of about $\mbox{8000}~{\rm ph/MeV}$, a trapping efficiency of $\approx7\%$, an attenuation length of 2.7 m and a decay time of 3.2 ns. Each fibre will be coupled to a single SiPM (Hamamatsu S10362-11-100C). The efficiency of this detector to be used was already optimized by using the SiPM with a detection efficiency (PDE)  of 65\% and a gain of $2.4\times10^{6}$ (dark rate = 600 kHz at 0.5 phe).

\paragraph{The positron detection from a muon decay vertex measurement}
The main challenge is to detect minimum ionizing particles (m.i.p.) with high efficiency using $0.25\times0.25~{\rm mm}^{2}$ fibres. A set of measurements of a single fibre (BCF12) coupled to a SiPM (Hamamatsu S10362-11-50C) were undertaken. A Sr$^{90}$ source provides electrons with an end-point energy of 2.28 MeV. A plastic collimator mounted in front of the source ensured that a fraction of electrons goes through the fibre first before being stopped in a thick plastic scintillator (BC400 - diameter 20 mm x length 20 mm), coupled to a photomultiplier (Hamamatsu R5900U). The plastic scintillator signal provides the trigger and suppresses the SiPM  dark current background. An energy threshold of 1.5 MeV was set to select only minimum ionizing particles.

Several fibre thicknesses ($1\times1$, $0.5\times0.5$ and $0.25\times0.25~{\rm mm}^{2}$) were tested. The measured number of photoelectrons scales, as expected, with the deposited energy, in all cases except for the 0.25 mm fibre, where slightly less light was collected due to a non-optimized mechanical and optical coupling. The signal from the SiPM was amplified by a factor 10 using a low-noise ($<10$~mV peak-to-peak) front-end preamplifier board (PSI made). The same board delivers the 70 V bias voltage to the detector. The data was acquired using the DRS4 evaluation board.

\begin{figure}
\centering
\includegraphics[width=0.7\linewidth]{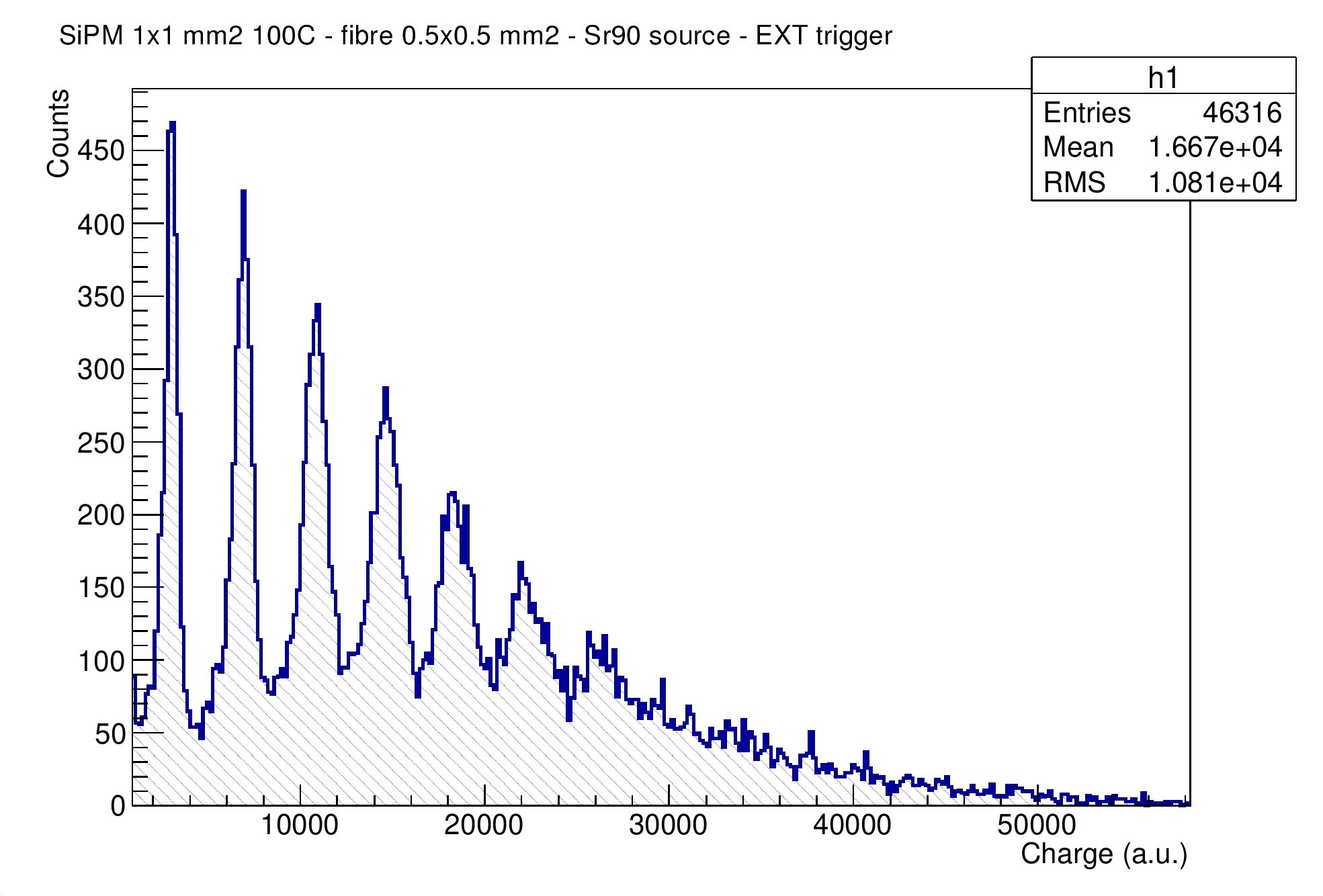}
\caption{Charge spectrum induced by minimum ionizing electrons in a $0.25\times0.25\, {\rm mm}^2$ scintillating fibre coupled to a SiPM Hamamatsu S10362-11-100C.}
\label{fig:025fibre}
\end{figure}

Fig.~\ref{fig:025fibre} shows the charge spectrum obtained as a result of a waveform integration over a fixed time window (15 ns). A mean of about 3 photoelectrons is measured to be compared with the expected 5. A m.i.p. detection efficiency was also measured to be $\epsilon\sim63\%$. \\

\begin{figure}[hbct]
\begin{center}
\begin{tabular}{cc}
\includegraphics[scale=0.37]{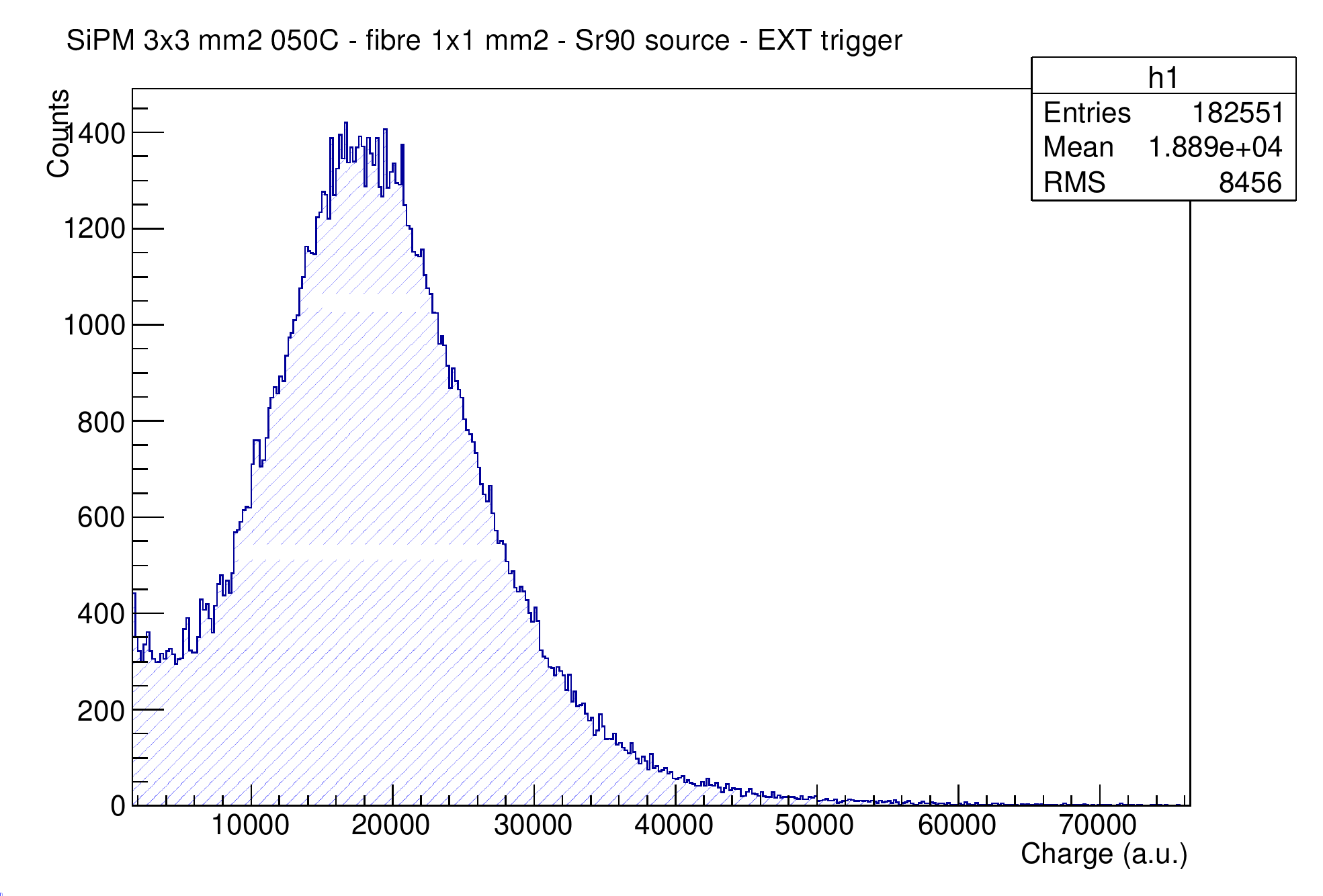} &
\includegraphics[scale=0.37]{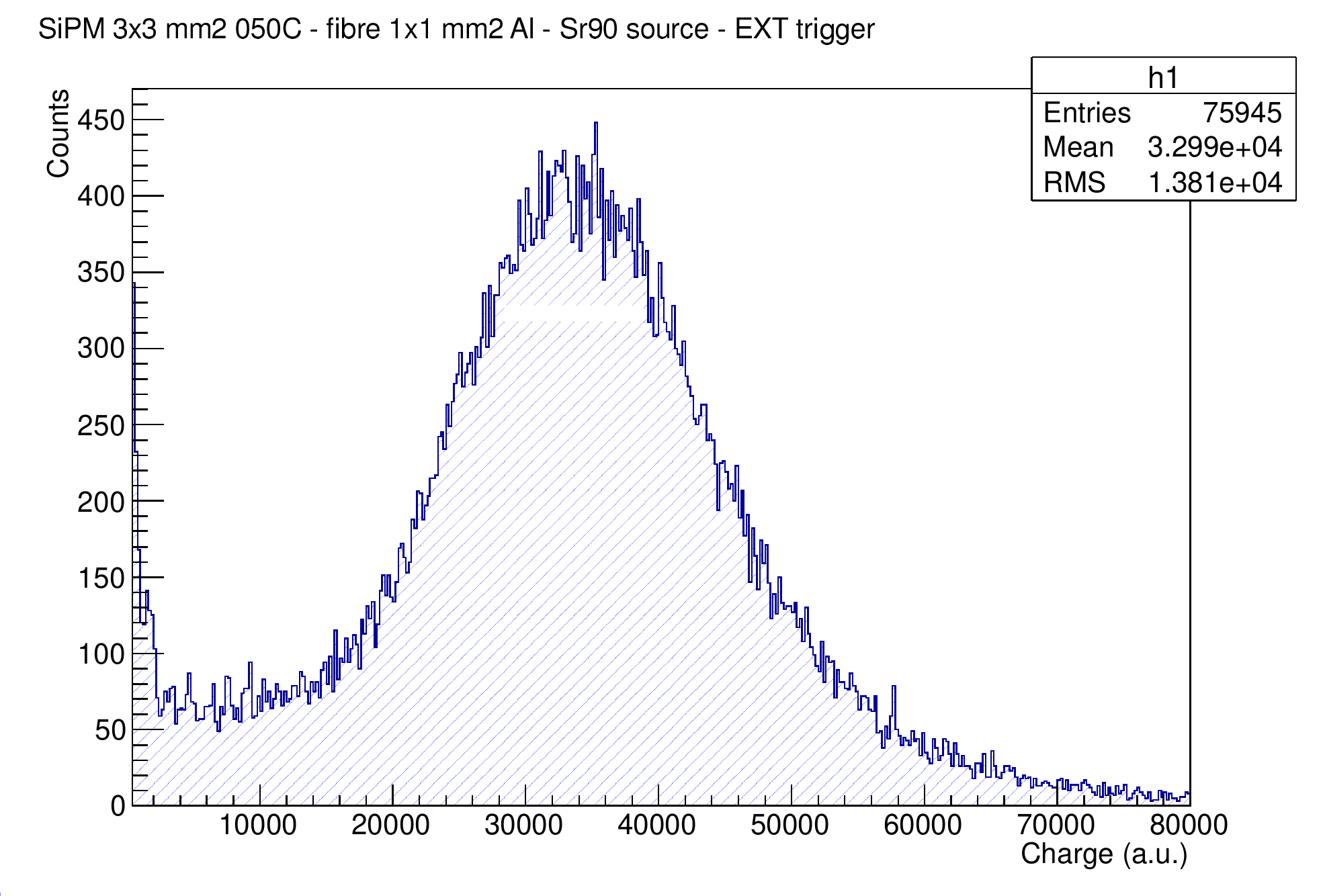} \\
\end{tabular}
\caption{Charge spectrum induced by minimum ionizing electrons in a $1\times1\, {\rm mm}^2$ scintillating fibre coupled to a SiPM Hamamatsu S10362-33-50C to one fibre end, without (left) and with (right) aluminum deposit on the other  end.}
\label{fig:1sputteredAlfibre}
\end{center}
\end{figure}

\begin{figure}
\centering
\includegraphics[width=0.7\linewidth]{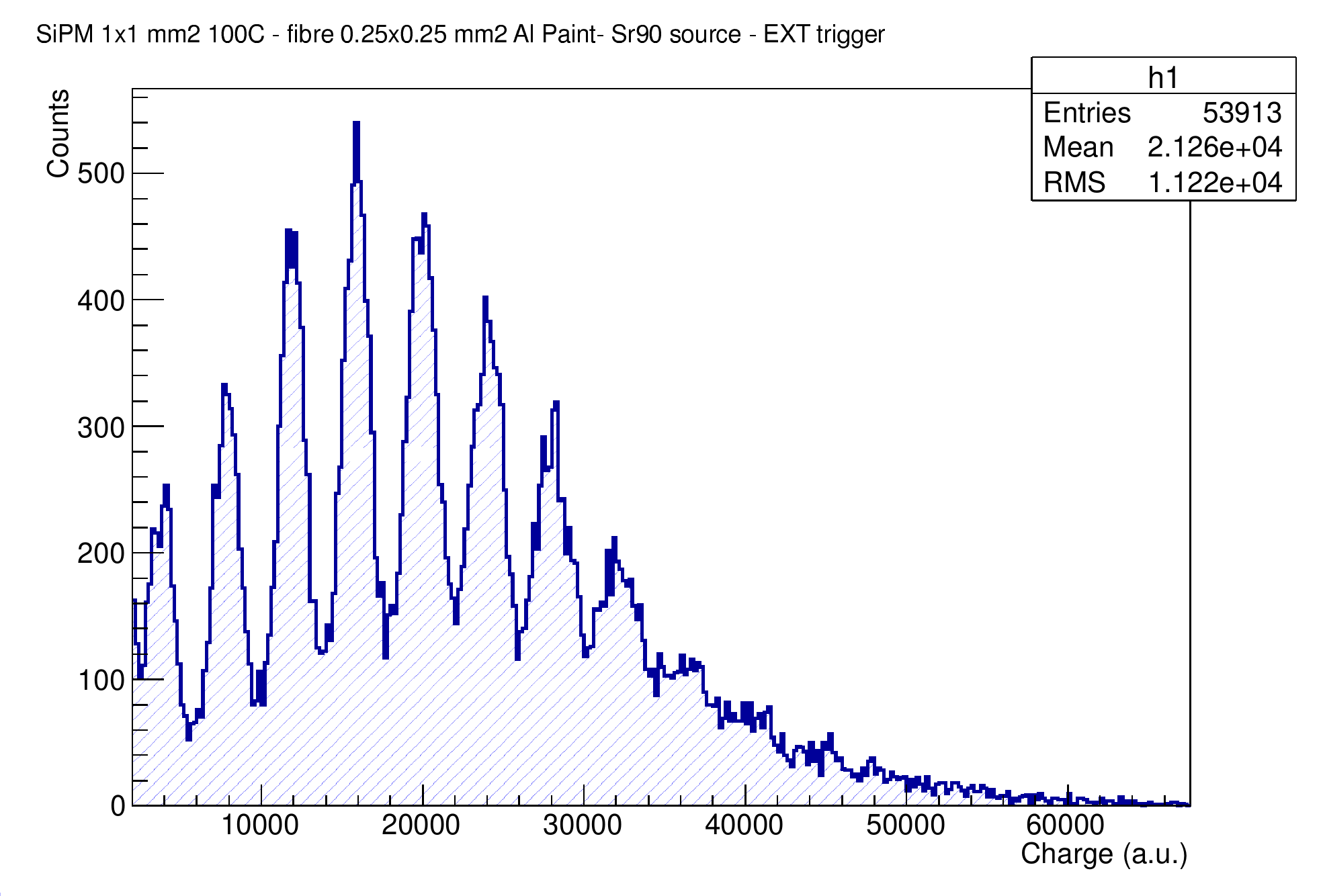}
\caption{Charge spectrum induced by minimum ionizing electrons in a painted Al $0.25\times0.25\, {\rm mm}^2$ scintillating fibre coupled to a SiPM Hamamatsu S10362-11-100C.}
\label{fig:025paintedAlfibre}
\end{figure}

{\bf The fibre optimization}.
\begin{figure}
\centering
\includegraphics[width=1.0\linewidth]{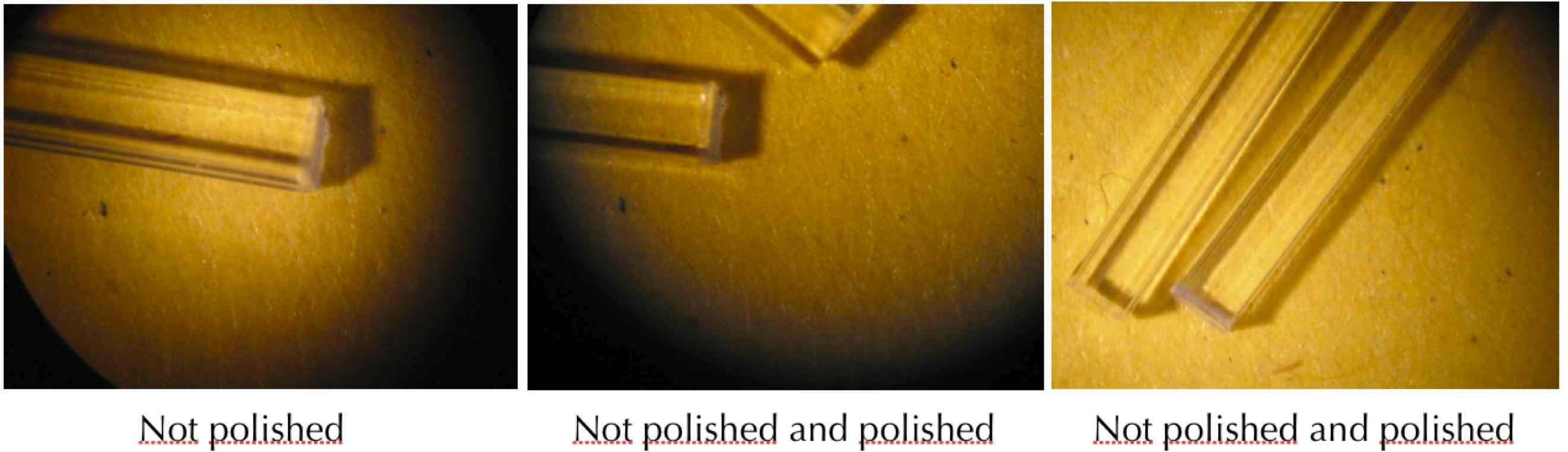}
\caption{Polishing of $1\times1$, $0.5\times0.5$ and $0.25\times0.25~{\rm mm}^{2}$ fibres with diamond head.}
\label{fig:polishing}
\end{figure}
The light collection, for instance,  can be enhanced by depositing an aluminum reflector on the opposite end of the fibre to increase the detection efficiency. Samples of all fibre thicknesses were polished using a diamond milling head. Fig.~\ref{fig:polishing} shows how the fibre surface appears under a microscope. An aluminum deposit was performed by painting or sputtering. The advantages of the sputtering method are that a known and uniform layer thickness can be deposited, in the range 10-1000 nm. Metallic and ceramic material can also be sputtered with the purity of the material kept under control. Unfortunately sputtering is time consuming. For $1\times1~{\rm mm}^{2}$ fibres an increase in the light collection of about $70\%$ was obtained. Fig.~\ref{fig:1sputteredAlfibre} shows the obtained results using a $1\times1~{\rm mm}^{2}$ fibre without (left) and with (right) an aluminum deposit. An increasing in the light collection of about 1.75 was measured for both painted and sputtered Al. A similar result was obtained using painted Al on $0.5\times0.5~{\rm mm}^{2}$ and $0.25\times0.25~{\rm mm}^{2}$ fibres. Furthermore it was observed that the benefit of the Al deposit is strongly reduced or vanished (this is the case e.g. for the smallest fibre) if the fibre surface is not polished. Fig.~\ref{fig:025paintedAlfibre} shows the present best result for a painted Al $0.25\times0.25~{\rm mm}^{2}$ fibre. A mean of 4.5 photoelectrons is measured and a m.i.p. detection efficiency of $\epsilon\sim80\%$ is measured. Some improvement in both light collection and detection efficiency can be expected glueing the fibre directly to the SiPM.

A sputtered $\rm{TiO_2}$ deposit is in preparation. The reflectivity of $\rm{TiO_2}$ is expected to be higher than that of Al. This could also increase slightly the light collection and the efficiency.

The sputtering method is also considered for a thin deposit ($<1 \rm{\mu m}$) on the sides of the fibre for light-tightness. Some samples are in preparation and will be tested it soon.  

\begin{figure}
\centering
\includegraphics[width=0.6\linewidth]{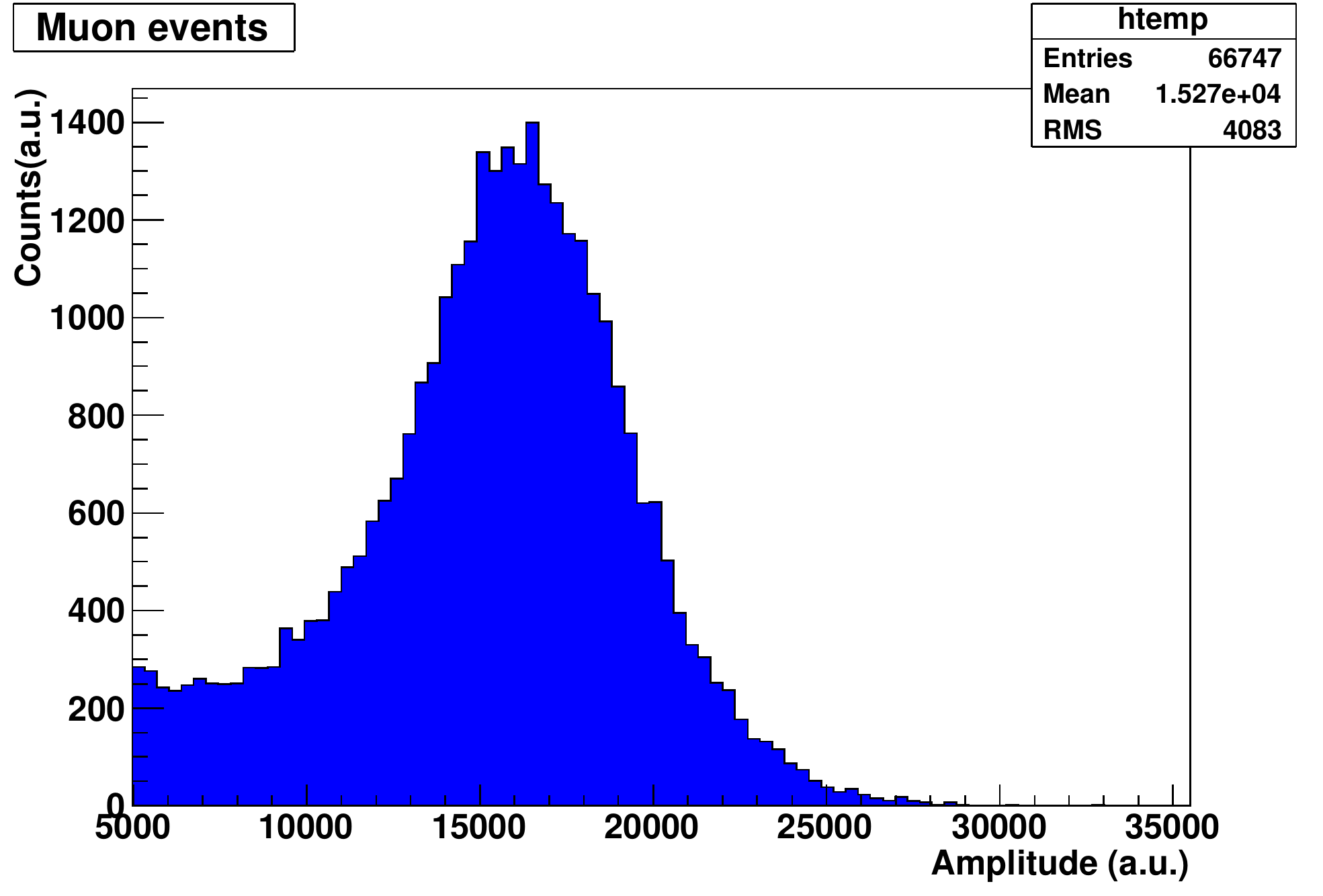}
\caption{Waveform amplitude spectrum induced by muons of 28 MeV/c in a 0.5 mm (diameter) scintillating fibre coupled to a SiPM Hamamatsu S10362-33-50C.}
\label{fig:muampli}
\end{figure}

\paragraph{The muon detection for muon stopping rate measurement}
The muon detection study was performed using a layer of 4 round fibres (BCF-12, diameter 0.5mm) coupled to SiPM (Hamamatsu S10362-33-50C) at the PSI $\pi$E1 beam line, tuned to positive muons of 28 MeV/c (the same as used in MEG). The muon intensity was about $2\times10^6~\mu/{\rm s}$. A typical waveform amplitude distribution is shown in Figure~\ref{fig:muampli}. A scan of the collected light as a function of the muon momentum  was performed inserting 50 $\rm{\mu m}$ thick mylar foils in front of the fibres until the muons are stopped in it. The muon signal is clearly visible for all the various different energy deposits. 

\paragraph{Event selection}
{\bf Positron selection.} For $10^8$ muon stopped per second, a rate of $2\cdot10^6$ event/s per fibre $0.25\times0.25\; \rm{mm^2}$ is expected. Two-fold are the main background sources for positron signal selection: the thermal noise and the muon stops. The first is present since  the positron signal is expected  to be at a  level of few photo-electrons and therefore at the same level of the thermal noise, with a rate of few hundred KHz (in the best case). The second since  the higher muon signal could overlap with the positron one. This muon signal can come from uncorrelated muons hitting the same fibre fired by the positron or from the stopping muon from which the positron originates. Both can be rejected using the Timing Counter detector as an external trigger (MEG trigger) in a time window of 20 ns. All the listed backgrounds are reduced to a contamination level of a few percent. In conclusion the positron selection efficiency  based on an external trigger is expected to be $\geq95\%$. 

{\bf Muon selection.} The muon rate can be measured setting a relatively high amplitude threshold, which removes completely the positron and thermal noise contributions.

\subsubsection{Conclusion} 
In conclusion, preliminary tests have shown that minimum ionizing particles can be detected with 250 $\rm{\mu m}$ scintillating fibres coupled to SiPM. The light collection and the detection efficiency was increased using an Aluminum deposit on the polished free one-end of the fibre. A detection efficiency of approximately $80\%$ was measured. A different reflector ($\rm{TiO_2}$) is in preparation and could improve the light collection and the detection efficiency even more. The positron signal is clearly detected rejecting the thermal noise with an external trigger. No cooling system for the SiPM is needed. The large pulse-height difference between positron and muon allows one to distinguish between the two particles. A prototype is under-construction in order to finalize the project.

%% file: 13_Appendix/RDC.tex
\subsection{Radiative muon decay veto counter : RDC}
The dominant source of $\gamma$-rays in the accidental background events in the MEG analysis
window is the radiative muon decays (RMD). In the present detector, $53\%$ of $\gamma$-rays above
48\,MeV are from RMDs and the fraction is more in the upgraded detector because of the reduced
background $\gamma$-rays thanks to the new positron tracker.
In a radiative decay with a high energy
$\gamma$-ray, a low momentum positron, typically lower than 2\,MeV, is emitted in a high probability.
On the other hand, in Michel decay, the probability of a muon to decay with such a low momentum positron is
low.
In the CrystalBox detector \cite{bolton_1988_prd}, RMD events with low momentum positrons were
clearly detected and used in the physics analysis, which discarded 12.3\% of the candidate
events with a loss in the signal detection efficiency of 0.5\%.
In the MEGA experiment \cite{PhysRevD.65.112002}, internal bremsstrahlung veto counters (IBV) were mounted to
detect low momentum positrons. The original design was to decrease 80\% of background events.
However those counters were not used in the final physics analysis because only 3.5\% of
high energy $\gamma$-rays with a pair positron were detected with IBV hits, and the rate was
independent of the $\gamma$-ray energy.

In the MEG detector, the bending radii of those low momentum positrons are typically
smaller than 4\,cm and 9\,cm at the center and the end of the magnet, respectively.
The radiative decay veto counters (RDC) therefore have to be mounted on the muon
beam axis. The detection of the low momentum positrons can be done using plastic
scintillation counters of
about 250\,$\mu$m thickness. The counter at the upstream side is used also to reduce the
momentum of the muon beam.
By removing or thinning the degrader in the beam transport solenoid, which is
presently 300\,$\mu$m thick Mylar, the total thickness of the material before the target can 
be the same as the present detector.

\begin{figure}[tbc]
\begin{center}
\includegraphics[width=.75\linewidth]{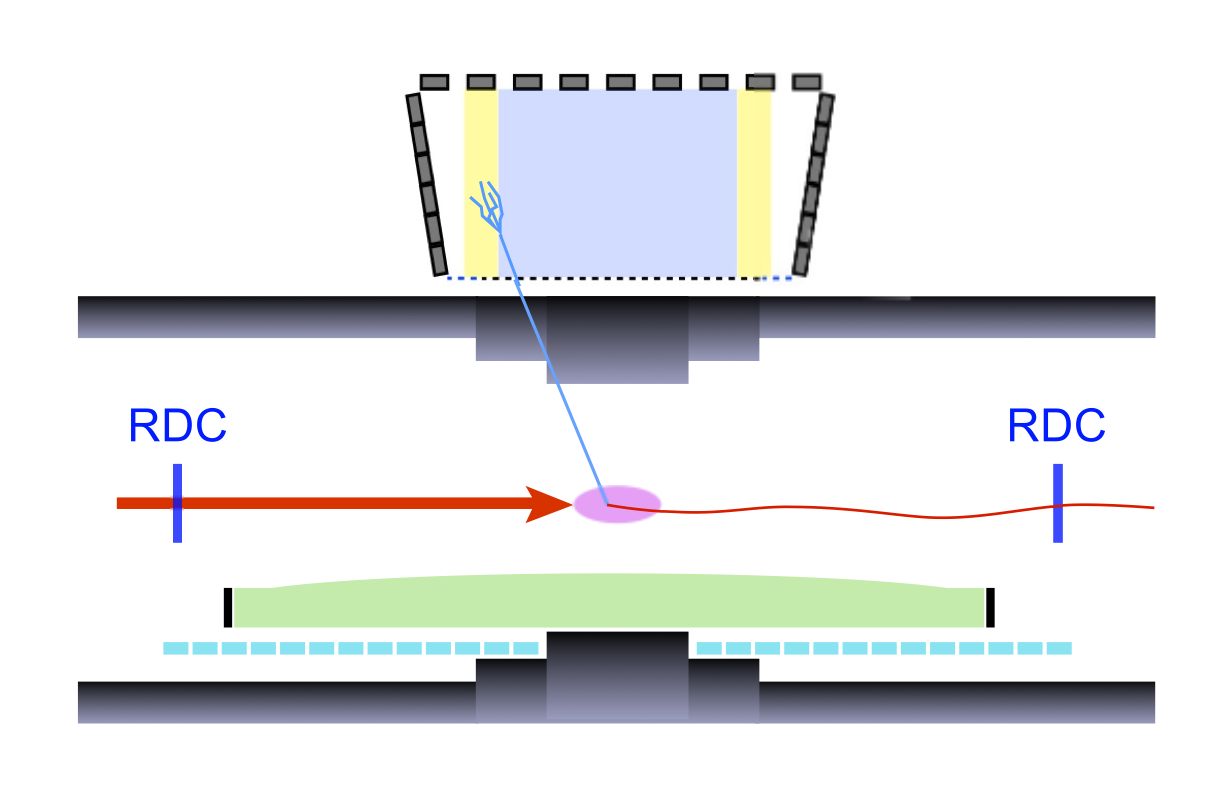}
\caption{\label{fig:rdc_3d}
 A configuration of the upgraded detector with radiative decay counters located at the both
    ends of the spectrometer. In this figure, a high-energy $\gamma$-ray and a low-momentum
    positron from a muon decay on the target are detected by the LXe detector and the downstream RDC,
             respectively.}
\end{center}
\end{figure}

\begin{figure}[tbc]
\begin{center}
\includegraphics[width=.5\linewidth]{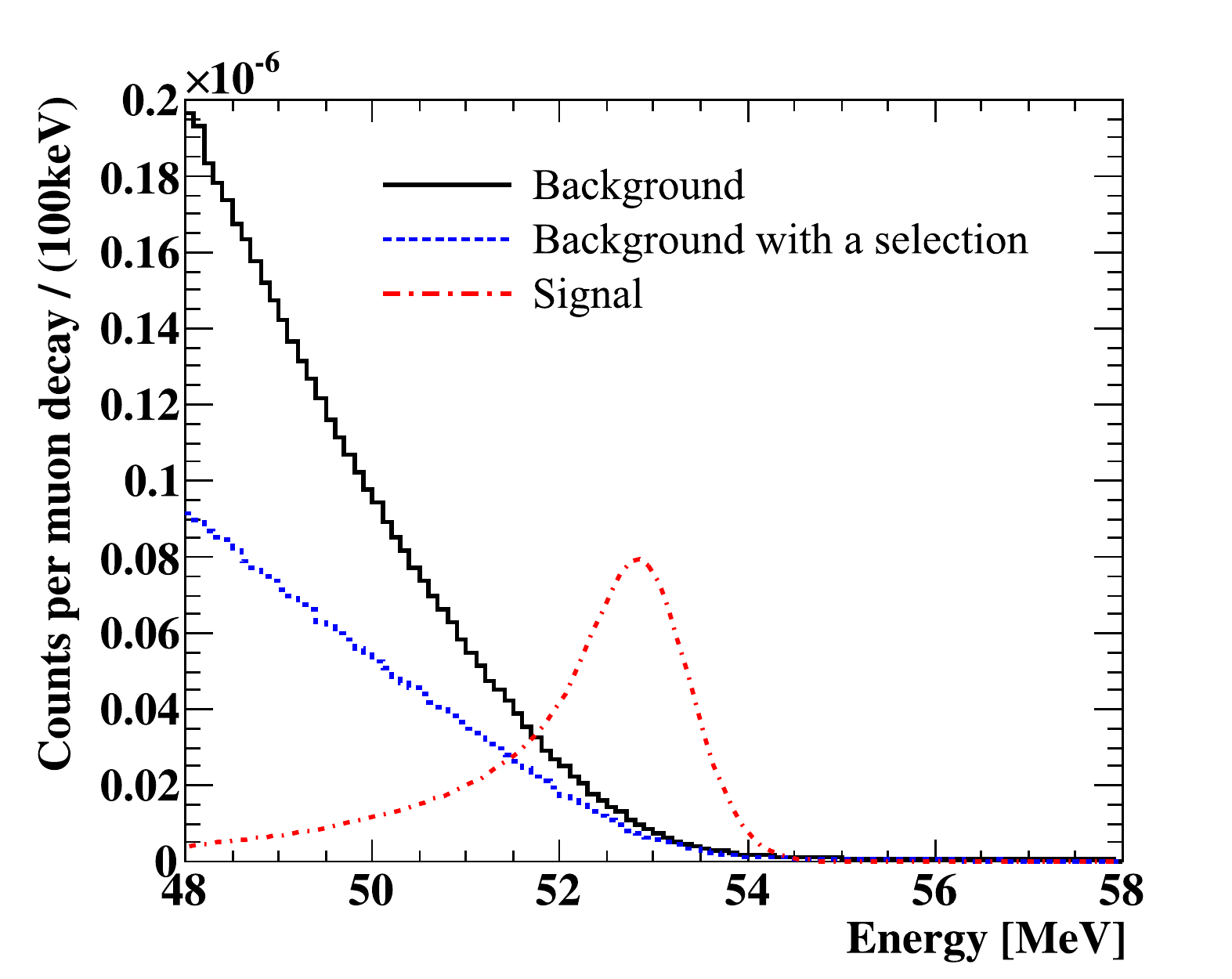}
\caption{\label{fig:rdc_bg} Simulated spectra of $\gamma$-rays detected by the upgraded
   LXe detector.
The dotted blue histogram shows a spectrum after a rejection of RMD events identified by RDC.
   The signal spectrum is arbitrary scaled.}
\end{center}
\end{figure}

A RDC module consists of 192 vertically aligned scintillation fibers and several scintillation plates.
The thickness of the fibers and plates is 250\,$\mu$m.
The fibers are used at the central part to minimize the dead-time due to the high
hit-rate, and the plates are used at the edge parts of the counter to not increase number of
read-out channels too much.
Scintillation photons are
transported through optical fibers. Sixteen optical fibers are bundled and coupled to a
1$\times$1\,mm$^2$ SiPM. In total 28 SiPMs are used for the upstream and the downstream RDCs.

The performance of RDC was studied using MC simulation by inserting thin plastic
scintillators at the both ends of the magnet (Fig.~\ref{fig:rdc_3d}).
The position is far from the center of the magnet and outside of the tracking volume;
therefore neither the significant increase of background $\gamma$-rays nor the reduction of the
signal detection efficiency are expected.
The tagging efficiency of radiative muon decays was evaluated to be $70\%$ for radiative
decays with $\gamma$ energy
higher than 48\,MeV when the coincidence time window between RDC and the LXe detector is
chosen to be 8\,nsec.
Figure~\ref{fig:rdc_bg} shows a spectrum of background $\gamma$-rays detected by the
upgraded LXe detector, and the same after a rejection of RMD events identified by RDC.
The probability for signal events for coinciding accidentally with RDC hits is about 15\%
for the same coincidence time window.
Instead of using RDC for rejecting signal candidate events,
the probability density function of the time difference between a RDC hit and a $\gamma$-ray detected by the LXe detector can
be included in the physics analysis so that signal efficiency is not reduced.



%% file: 13_Appendix/Positron_Tracker_TPC.tex
\subsection{TPC-based tracker option}

\subsubsection{Concept}
\begin{figure}[htb]
\begin{center}
\subfigure[3D view]{
   \includegraphics[height=6cm]
    {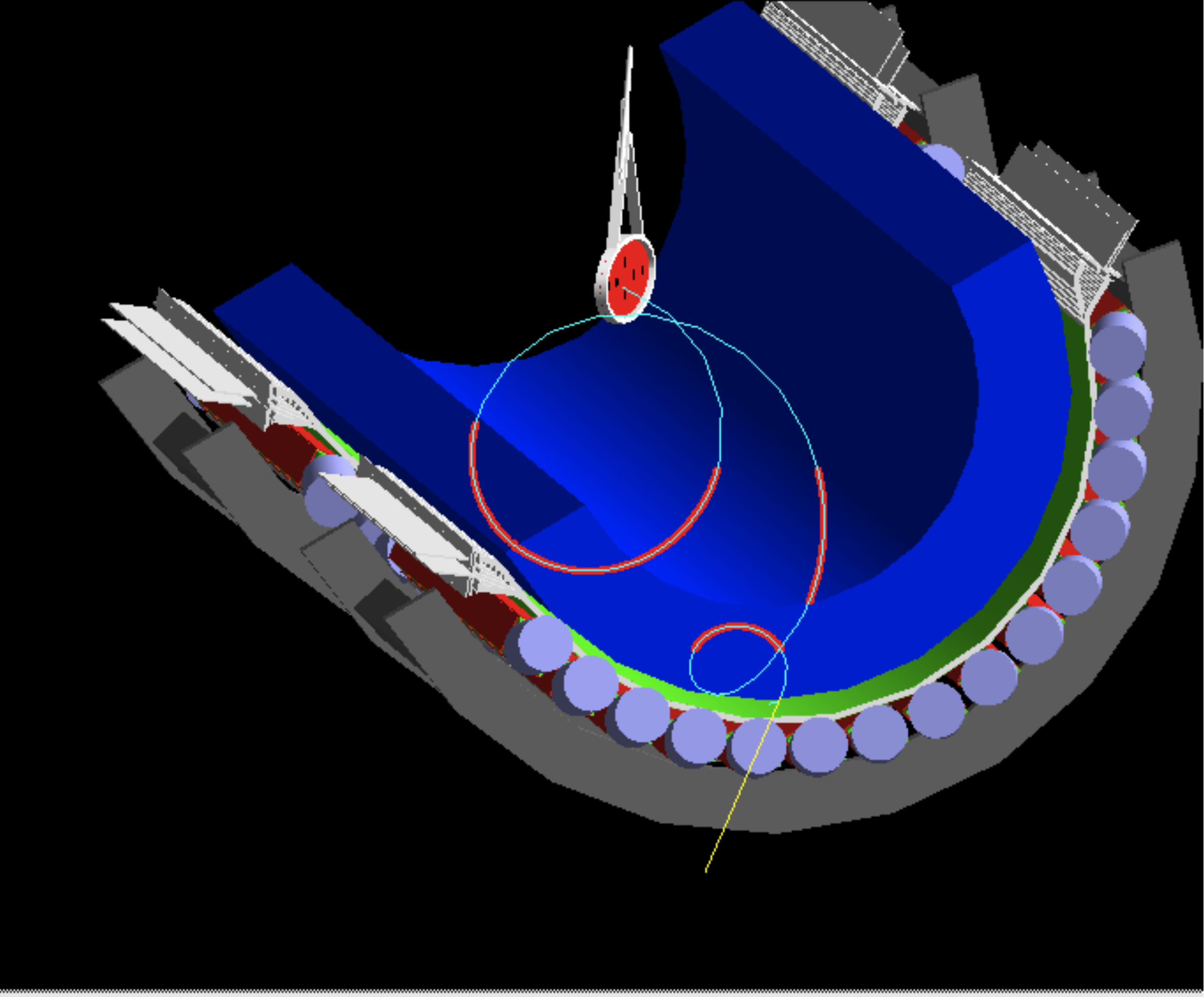}
    }
\subfigure[$x-y$ view]{
   \includegraphics[height=6cm]
    {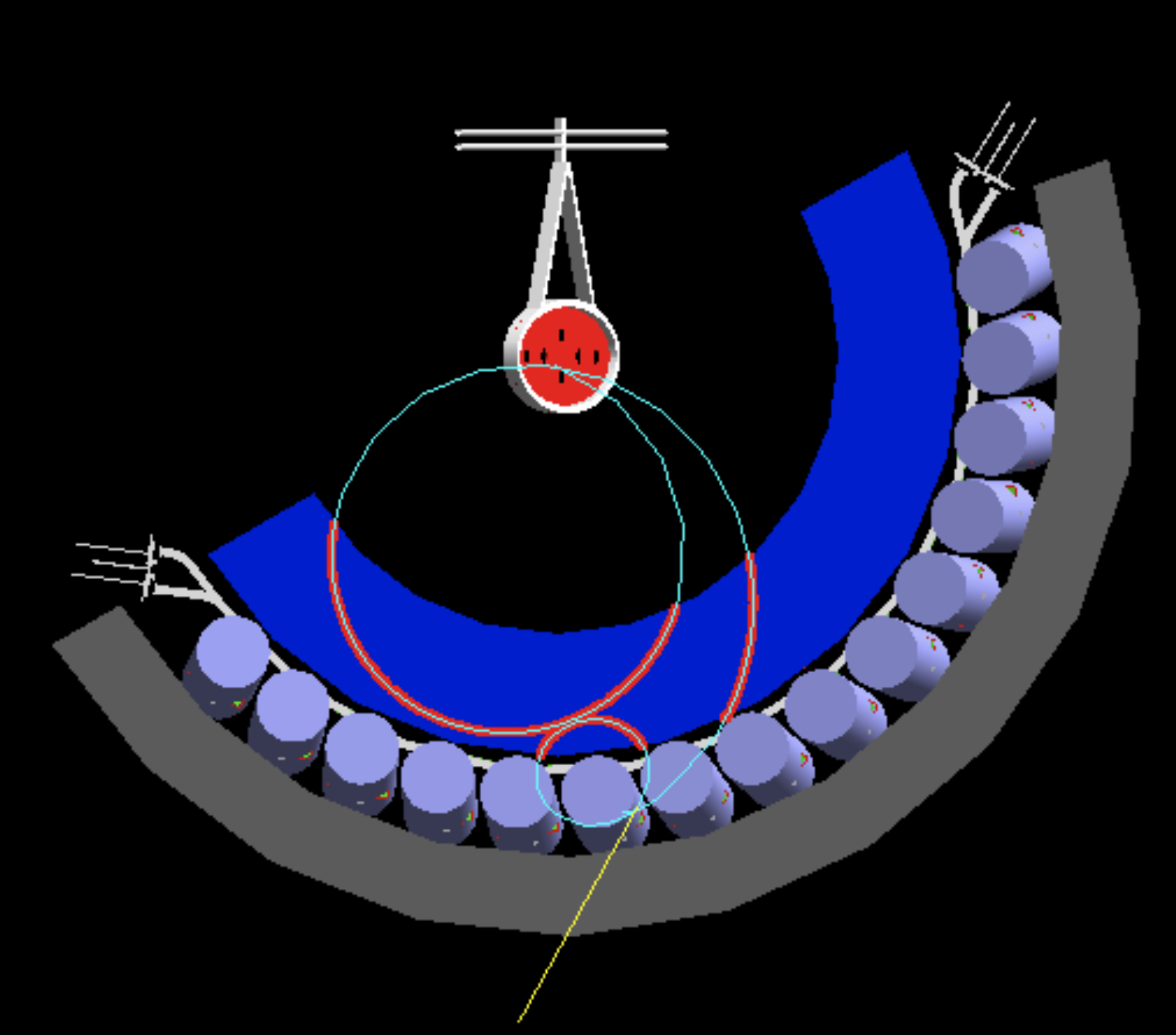}
    }
\subfigure[$z-x$ view]{
   \includegraphics[width=0.8\linewidth]
    {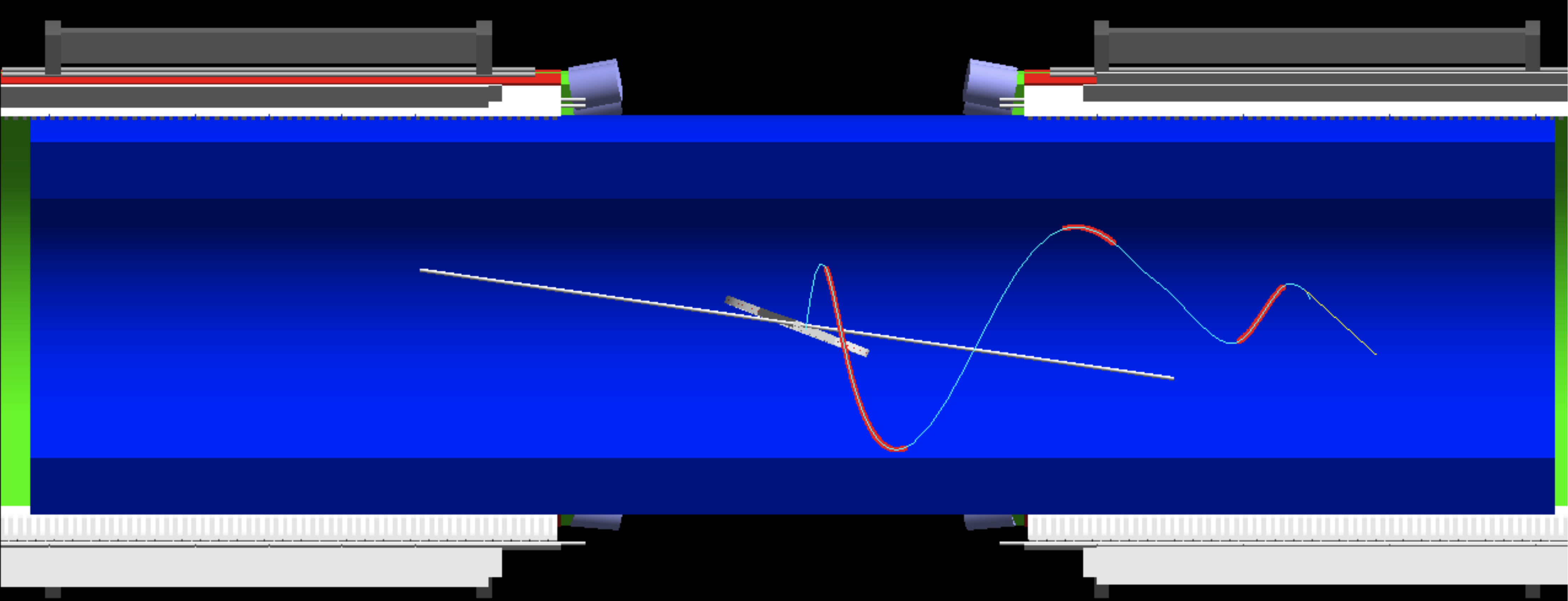}
    }
\caption{\label{fig:eventdisp}
  Event display of TPC-based tracker by Geant4-base simulation
  with three view points.
  Blue trajectory indicates the generated positron
  and red indicates the hits recorded in the TPC fiducial, respectively.
}
\end{center}
\end{figure}

As one of alternatives for the positron tracker,
{a Time-Projection-Chamber(TPC)-based tracker} was considered for the upgrade.

The TPC provides a complete 3-dimensional (3D) pictures
of the ionization deposited in an active gas volume.
It acts somewhat like a bubble chamber.
The TPC's 3D localization makes it extremely useful
in tracking charged particles in a high-track-density environment.
This can improve the MEG positron spectrometer performance
in a significant way.

As described in the previous sections,
we aim to improve the detection efficiency of the positron
spectrometer by reducing the amount of material used in the components.
In the present tracker, the amount of material existing inside the
fiducial volume is substantially small while
there  is significant amount of material in between the drift chamber
and timing counter; this cause deterioration
of the position detection efficiency.
The proposed drift chamber design solves this problem by adopting a
unique gas volume. The TPC design also takes the same approach to
address this.

In addition to this we plan to improve the tracking accuracy by
increasing the number of sampling points along a positron track.
The present tracker system is based on the modularized 
drift chambers, and thus, the number of possible
ionization positions, the so-called {\it hits}, is limited.
Consequently, the reconstructed momentum resolution
and angular resolutions are somehow limited.
With this principle, the TPC-based tracker is also possible
to bring tremendous improvement for that,
because the TPC can provide a complete 3D pictures of charged track
in its fiducial volume.\\

In order to adopt the TPC-based detector idea for the MEG
positron spectrometer,
there is one important issue to be addressed first,
namely, the orientation of time-projection.
The commonly adopted method by many experiments
is projecting the charged trajectory onto the plane
which is perpendicular to the beam axis,
{\it i.e.} projecting along the beam axis.
However, it is difficult to adhere the same idea for the 
MEG positron spectrometer,
because of the limited space to install the front-end electronics.
In addition, very long projection distance is necessary
in case we adopt time projection along the beam axis, approximately 3 m,
which will potentially lead to a technical problem to maintain the
electric field,
and also poor spatial resolution caused by electron diffusion during drift.

In consequence, for the MEG positron spectrometer,
we have to adopt the ``radially-projecting'' TPC
which consists of:
\begin{itemize}
  \item{electrodes which establish an electric field to make electrons drift radially, and }
  \item{electron detection device cylindrically curved to fit the shape of the inner face of the COBRA magnet.}
\end{itemize}
In order to investigate the possibility to realize 
such a challenging detector idea,
intensive simulation work is progressing,
and some prototypes are being planned.
The simulated event display for the TPC-based positron spectrometer
is shown in Fig. \ref{fig:eventdisp} with three different view points.

\subsubsection{Design}
\begin{figure}[htb]
  \begin{center}
    \subfigure[Calculated electric field map]{
     \includegraphics[width=0.48\linewidth]
     {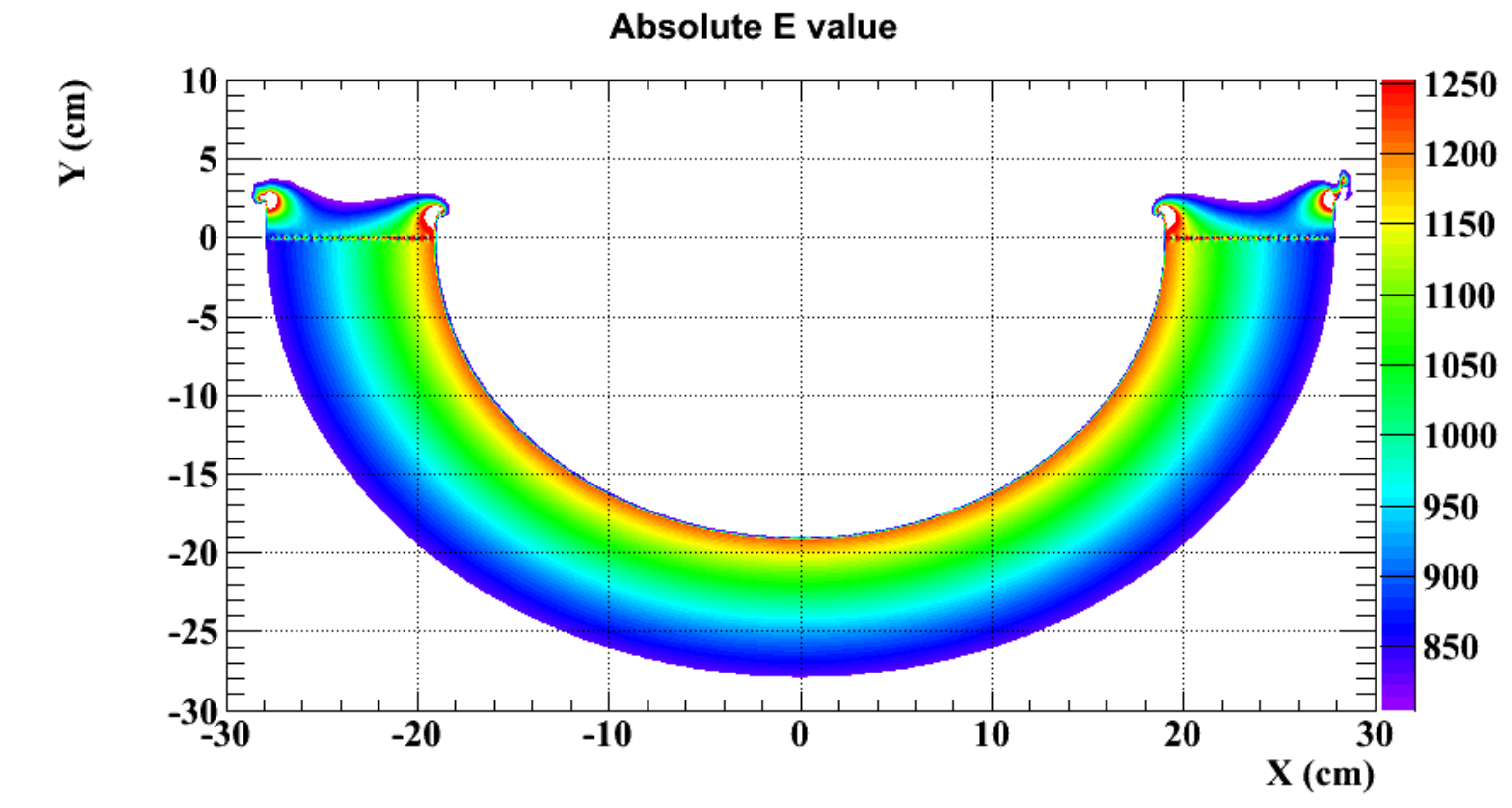}
      }
    \subfigure[Simulated drift electron lines]{
     \includegraphics[width=0.48\linewidth]
     {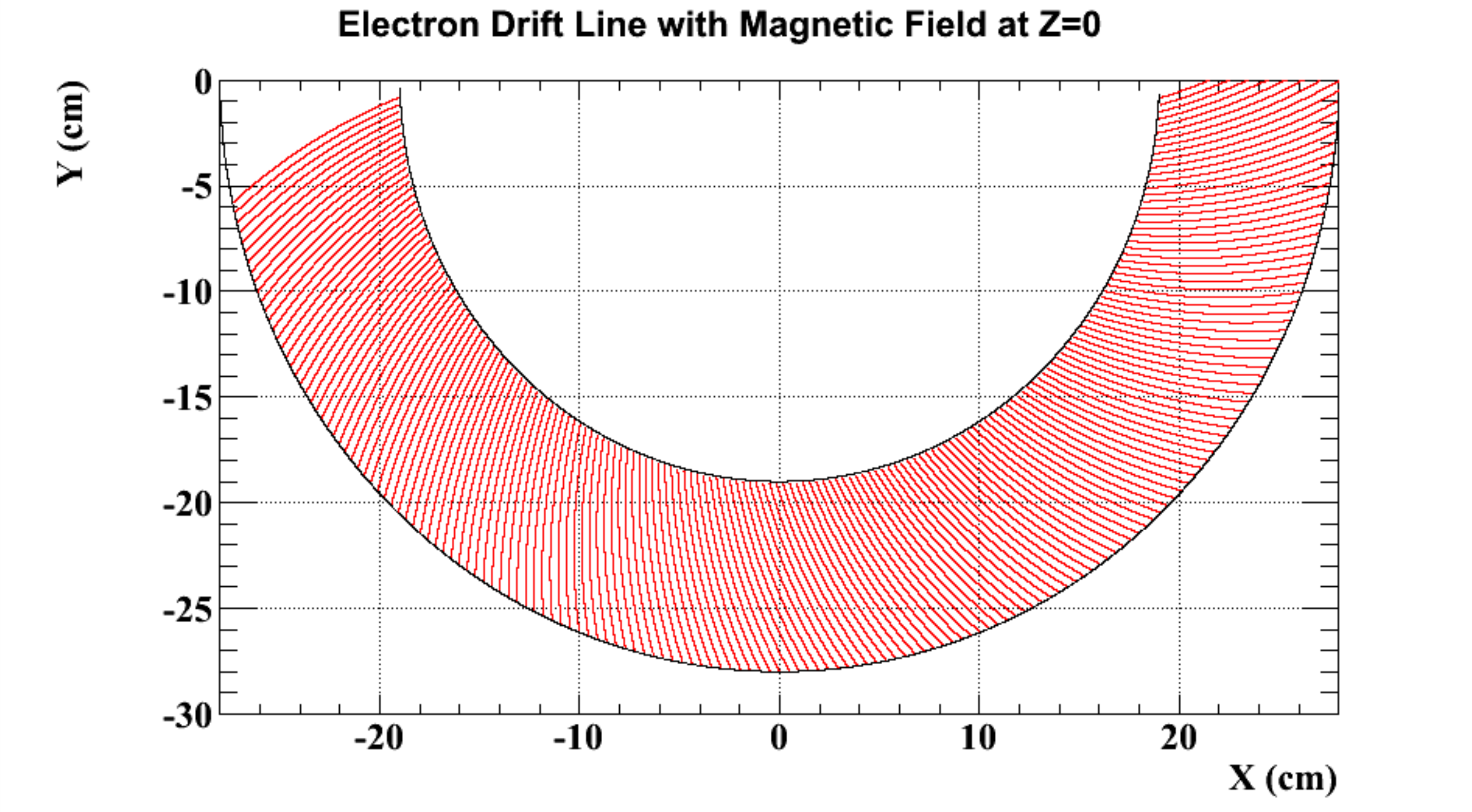}
    }
    \caption{\label{fig:radial_efield}
    Calculated field map and drift-line.
  }
  \end{center}
\end{figure}

At first, in order to have a look at the time-projection
as described in the previous sub-section,
the calculated electric field is implemented in the MC framework.
Here we assume the 9 cm of maximum drift distance and curved electrode
to be matched with the volume inside COBRA magnet.
Fig. \ref{fig:radial_efield} (a) shows the obtained electric field map
which is calculated by the commercial FEA program \cite{ansys}.
In this example, 9 kV is applied on the anode,
and field-cage wires are equipped at 5 mm intervals
at both ends of fiducial volume, $y = 0$,
in order to maintain the uniformity of electric field.
By extracting this calculated electric field map
with the COBRA magnetic field map,
the time-projection trajectory inside the fiducial volume
can be obtained as shown in Fig. \ref{fig:radial_efield} (b).
In this figure, $z = 0$ of coordinate is adopted as an example
to determine a cut view, because COBRA magnetic field
is highly graded.
As shown in this figure, even if we have a graded magnetic field,
each ionization position can be reconstructed by
the combination of $z$-coordinate and the projection time uniquely.\\

Thus, now it is required to consider how to measure the $z$-coordinate
reasonably.
The front-end detector of TPC is normally realized by a 
wire-chamber-base detector or a MPGD\footnote{Micro Pixel Gas Detector}-base
detector.
For our case, the projection plane is cylindrical,
hence the MPGD-base detector, like GEM \cite{Sauli2007}, is suitable.
However, now we are facing the next issue, {\it i.e.}
we have to build a cylindrical GEM detector
to cover the inner surface of COBRA magnet.
So far, three cylindrical GEM detector have been successfully fabricated
in the particle/nuclear experimental field:
(1) NA61 tracker at CERN \cite{Sauli2007}, 
(2) KLOE upgrade detector at LNF \cite{KLOE},
and (3) BoNuS TPC at JLab \cite{BoNuS}.
Thus, in principle, it has been established to build the
cylindrical GEM detector technically,
nevertheless it is very difficult to keep the precise gaps
between each GEM layers over the all cylindrical surface.
\begin{figure}[htb]
  \begin{center}
    \subfigure[Cylindrical GEM for NA61 at CERN]{
     \includegraphics[height=6cm]
     {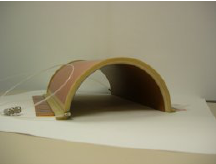}
      }
    \subfigure[Cylindrical GEM for KLOE-2 at LNF]{
     \includegraphics[height=6cm]
     {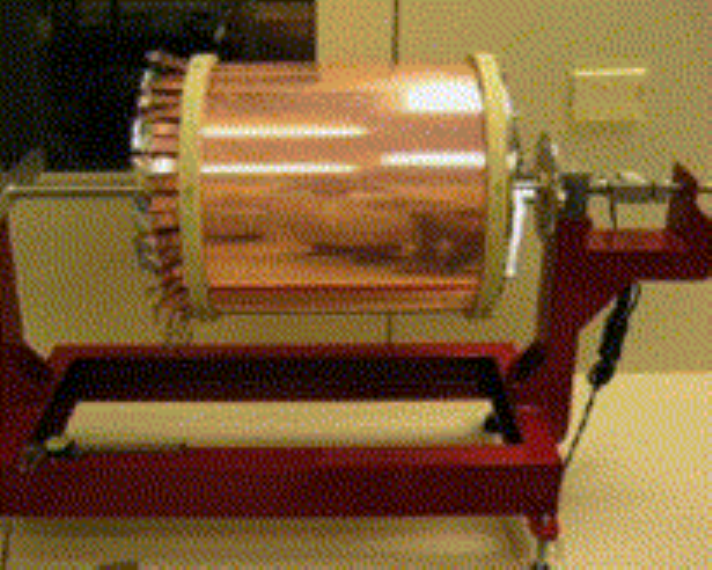}
    }
    \caption{\label{fig:cylindrical_gem}
    Examples of the cylindrical GEM detector.
  }
  \end{center}
\end{figure}
Two examples of the fabricated cylindrical GEM detector are shown 
in Fig. \ref{fig:cylindrical_gem}.
Furthermore, all realized detectors so far are not big size,
$\sim 20$ cm of diameter at maximum,
while it is required to build a $\sim 60$ cm of diameter
cylindrical GEM for the MEG positron spectrometer.\\

\begin{figure}[htb]
  \begin{center}
    \subfigure[$\Delta T$ with scaled E-fields]{
     \includegraphics[height=5cm]
     {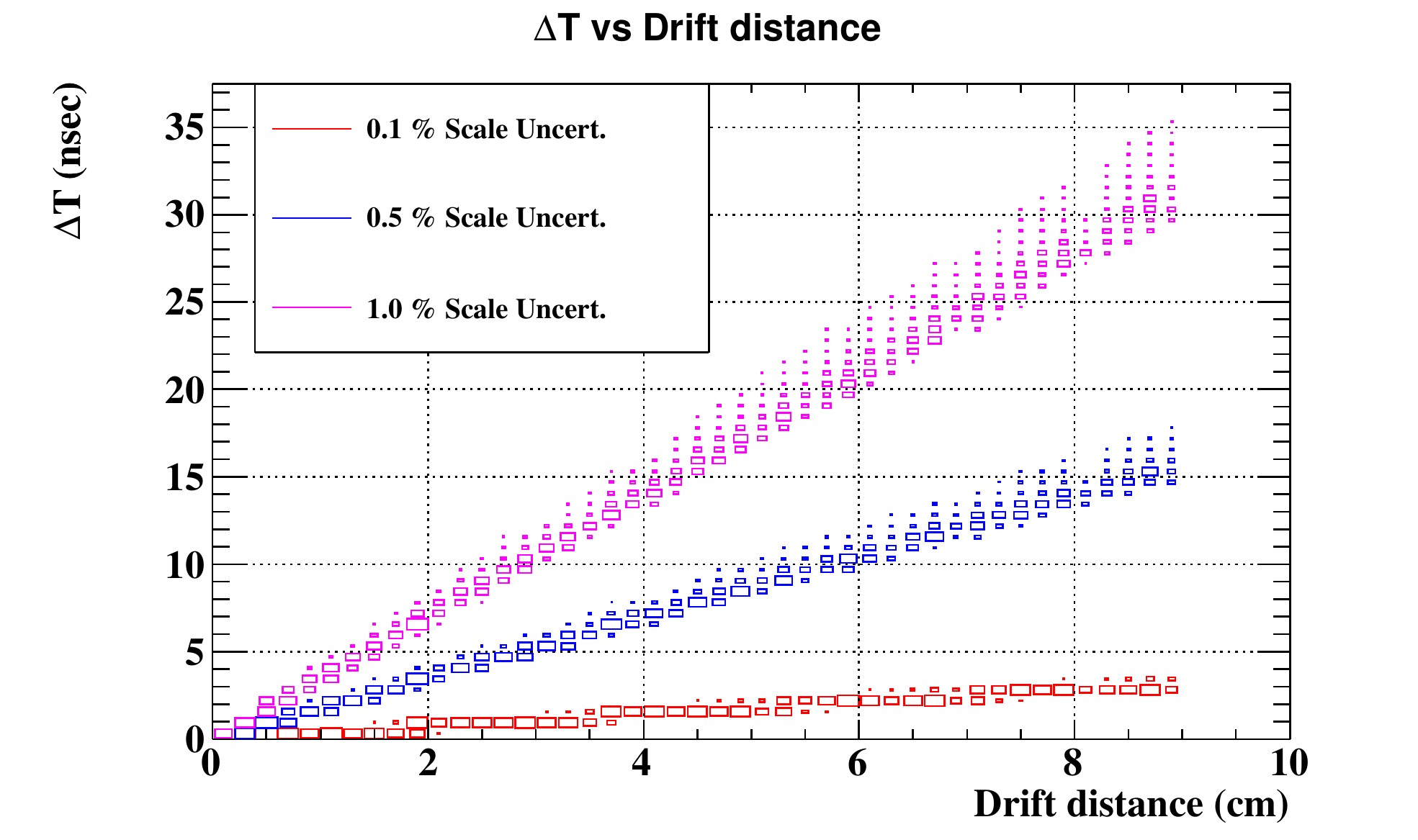}
      }
    \subfigure[$\Delta T$ with randomly perturbed E-fields]{
     \includegraphics[height=5cm]
     {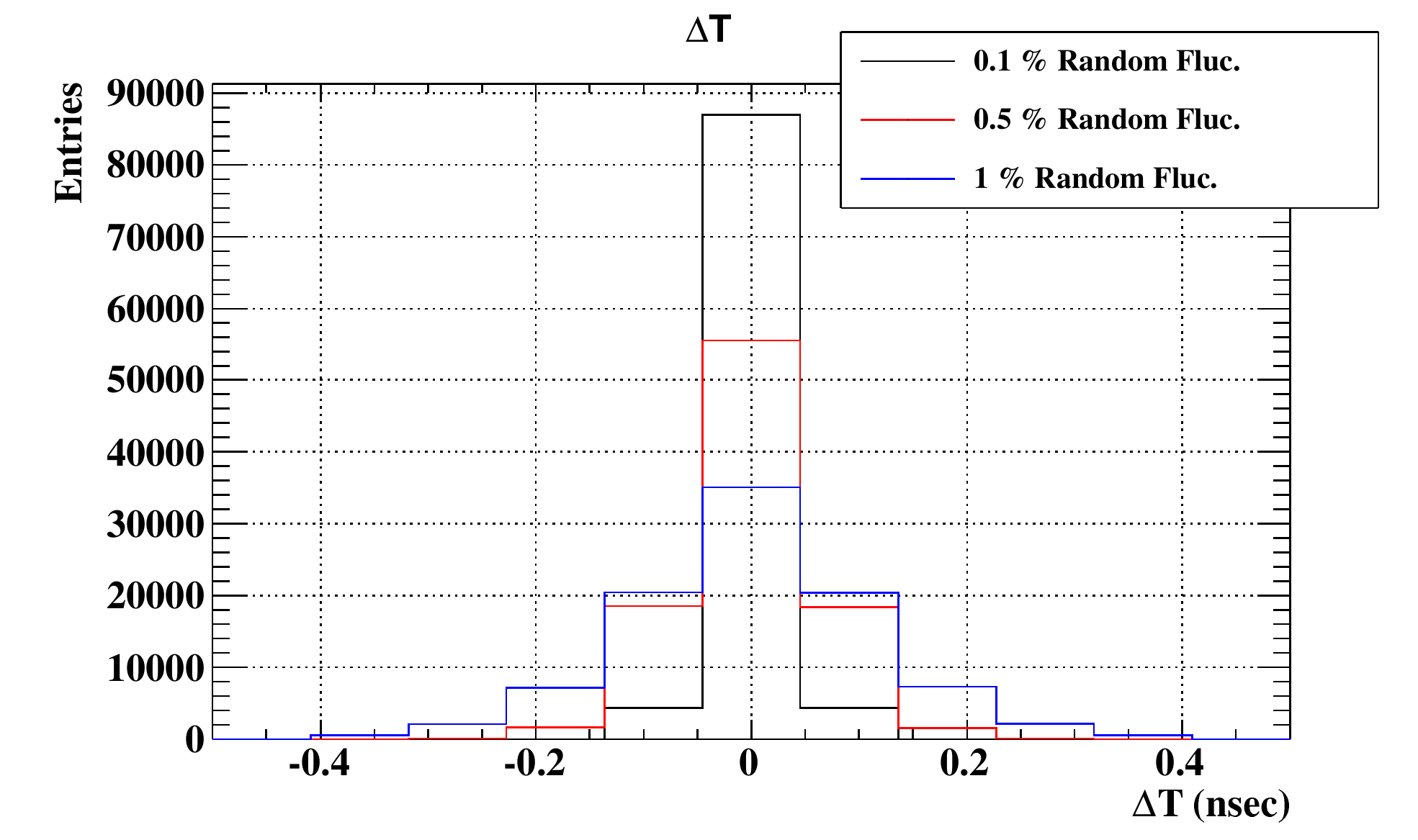}
    }
    \caption{\label{fig:EField_Uncertainty}
    Projection-time differences as a function of electric field uncertainties
  }
  \end{center}
\end{figure}

Front-end detector for the TPC may be realized by 
the cylindrical GEM detector, on the other hand,
we have to carefully consider how to build the electrode 
to maintain the TPC electric field.
Such an electric field should be;
\begin{itemize}
  \item{high field ($\sim 1$ kV/cm), and}
  \item{precisely uniform (or precisely controlled).}
\end{itemize}
Fig. \ref{fig:EField_Uncertainty} shows how electric field uncertainties affect 
the tracking performance in the MC simulation.
Both plots in Fig. \ref{fig:EField_Uncertainty} show the difference of
projected time with some perturbed electric fields.
In Fig. \ref{fig:EField_Uncertainty} (a), the electric fields
are scaled by 0.1, 0.5, and 1 \%, respectively,
while the electric fields are randomly perturbed
by factor of 0.1, 0.5, and 1 \%, respectively,
in Fig. \ref{fig:EField_Uncertainty} (b).
According to these results, a randomly fluctuated
electric field does not have a severe effect on the tracker performance,
while scaled fields lead serious systematic uncertainties.
In consequence, in order to obtain a reasonable spatial resolution
by means of time-projection, 0.2 \% of electric field uniformity
is required at worst.

It is necessary to actualize such a good uniformity on the electric field
by using the ultra-thin electrode foil,
because all the incoming positrons pass through
the inner surface of TPC, namely, the inner electrode,
as clearly shown in Fig. \ref{fig:eventdisp}.
During the development of present MEG drift chamber,
the specially dedicated technique to make a
ultra-thin electrode foil which is 12.5 $\mu$m thickness of polyimide 
with precisely patterned aluminum layer
which is 250 nm of deposition thickness over 1 m length.
This 250 nm of deposition thickness is compromised due to patterning precision,
however, this deposition thickness can be reduced
down to 50 nm level for the TPC detector,
because patterning is not necessary for the TPC electrode.
By assuming 50 nm of aluminum deposition on the electrode foil,
the required uniformity of electric field is achievable,
nevertheless the fabrication method will be a next issue to be discussed.

\subsubsection{Expected Performance}
By implementing the items described in previous sub-sections,
{\it e.g.} radial electric field with assumed precision,
cylindrical GEM as a front-end detector of TPC,
ultra-thin electrode foil, {\it etc.},
the detector performances are estimated by the MC simulation.

In the MC simulation, the following parameters are assumed;
\begin{itemize}
  \item{event generation: same as the present MEG MC}
  \item{gas mixture: He/CO$_{2}$/C$_{2}$H$_{6}$($70\%:20\%:10\%$)}
  \item{diffusion coefficient: 150$\mu$m/$\sqrt{{\rm cm}}$}
  \item{inner electrode: 40 $\mu$m of polyimide with 1 $\mu$m of Cu deposition}
\end{itemize}
The event generation scheme is based on the present MEG MC simulation,
and it would be further tuned, in particular, 
the event rate is varied to simulate the capability of possible high rate
experiment.
The gas mixture is currently assumed to use 
the He/CO$_{2}$/C$_{2}$H$_{6}$ mixture,
however, almost nothing is decided yet about gas mixture,
just a use of helium-base gas mixture is mandated,
because multiple scattering effect should be one of the dominant 
source to deteriorate the tracking resolution for 52.8 MeV/$c$ positron.
In any case, the gas mixture is the most important item to study 
at the next step, not only for the simulation but also for the hardware R\&D.
Some numbers of diffusion coefficient are assumed to calculate 
the diffusion effect for the time-projection, 
and tentative electrode materials are also implemented
even it is very conservative setting.
\begin{figure}[htb]
  \begin{center}
    \subfigure[Sample event display of track finding]{
     \includegraphics[height=6cm]
     {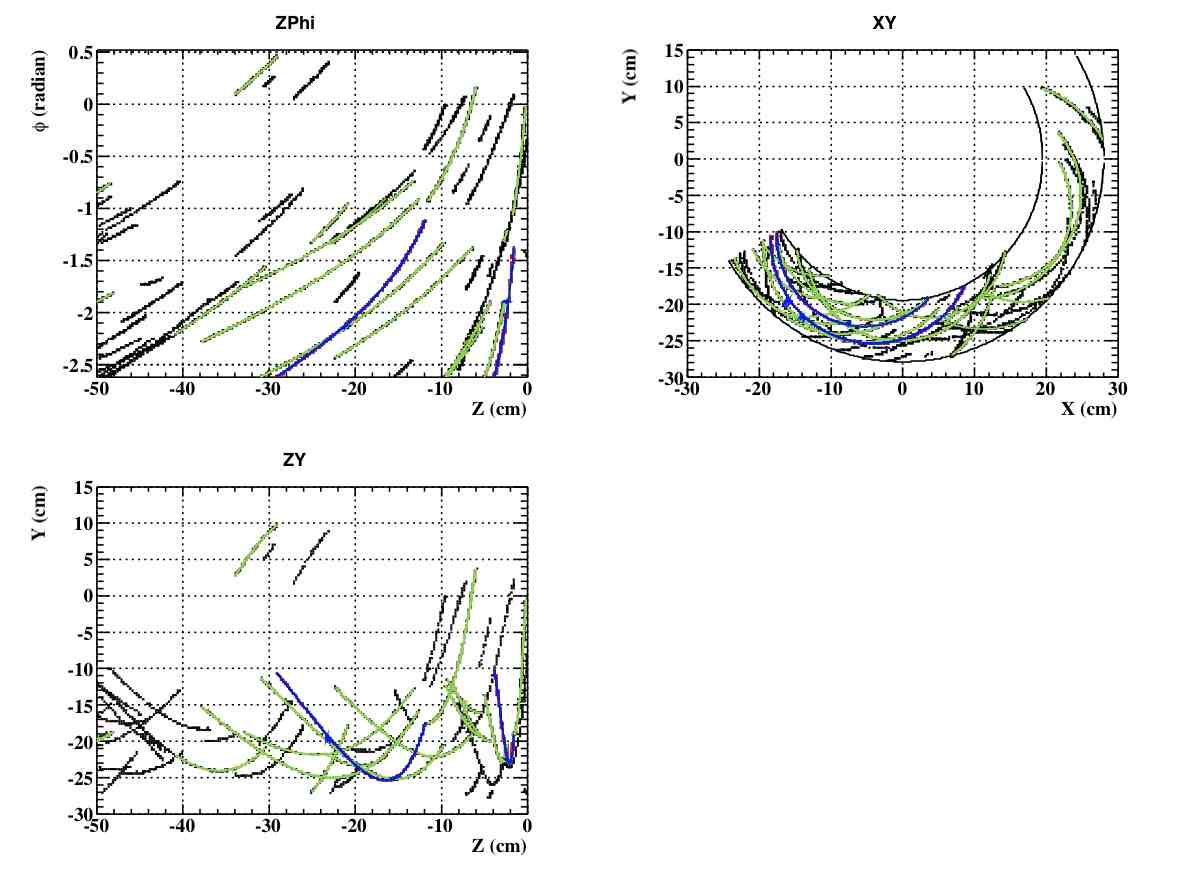}
      }
    \subfigure[Track finding power as a function of beam rate]{
     \includegraphics[height=6cm]
     {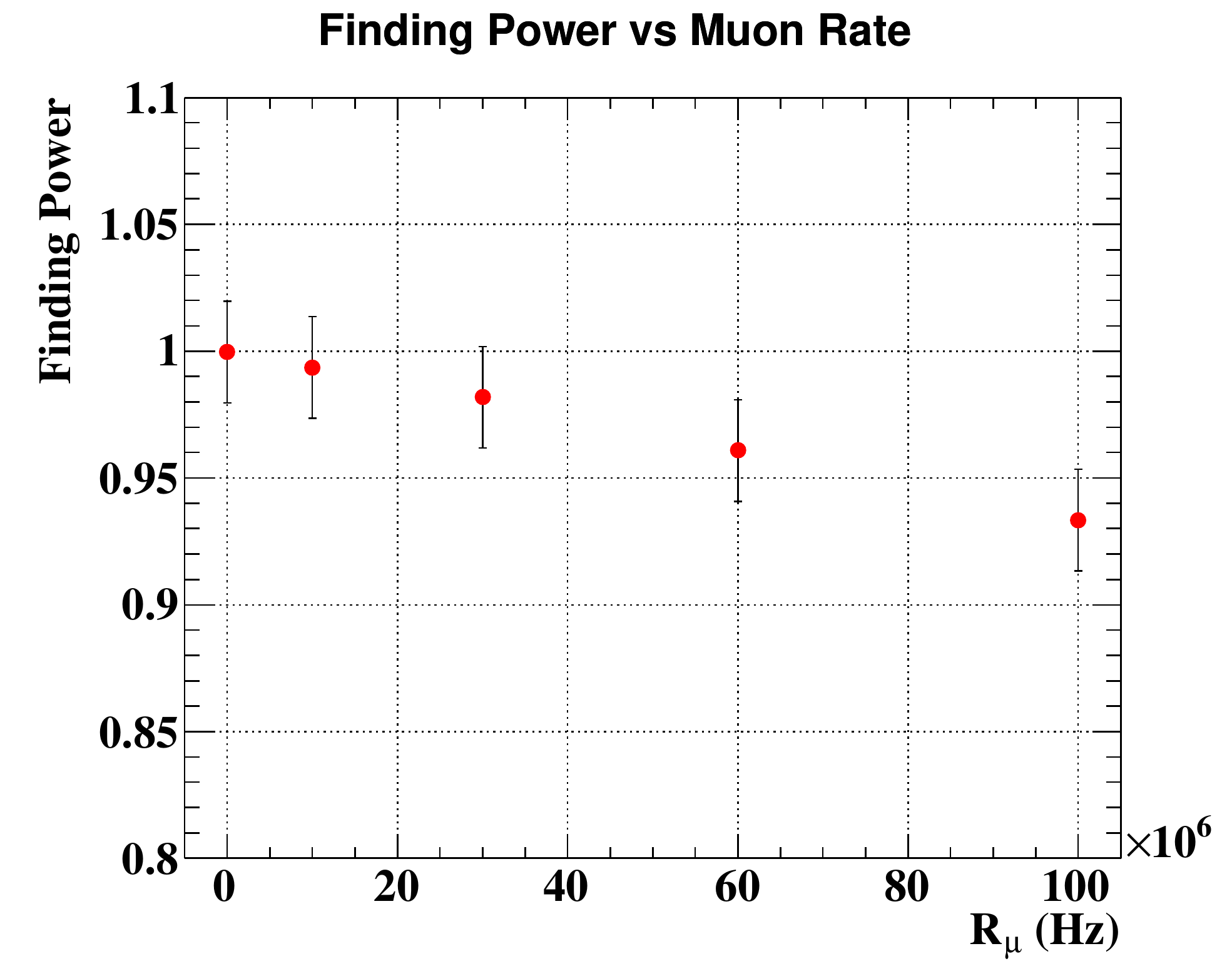}
    }
    \caption{\label{fig:track_finding}
    Track finding simulation
  }
  \end{center}
\end{figure}
The simulated track finding is shown in Fig. \ref{fig:track_finding}
with MC events.
Fig. \ref{fig:track_finding} (a) shows an example of track finding;
blue trajectory indicates the found signal track which fires a trigger
while all others indicate the pile-up events.
Fig. \ref{fig:track_finding} (b) shows the obtained
track-finding power as a function of the beam intensity.
As shown, $\sim 93\%$ of finding power is expected 
at the 1$\times 10^{8} $($\mu$/s) of beam intensity.
So far, we applied just a local clustering method to find the track. 
No fancy algorithms, like a Hough transformation, are applied any more.\\

After the track finding, track fitting is performed which is based on
the Kalman-filter-base track fitter developed for the present
MEG tracking code.
Fig. \ref{fig:TPC_momreso} show the reconstructed
momentum/angular resolutions as a distribution of residuals
between the reconstructed and the true(MC) information.
\begin{figure}[htb]
  \begin{center}
     \includegraphics[width=0.9\linewidth]
     {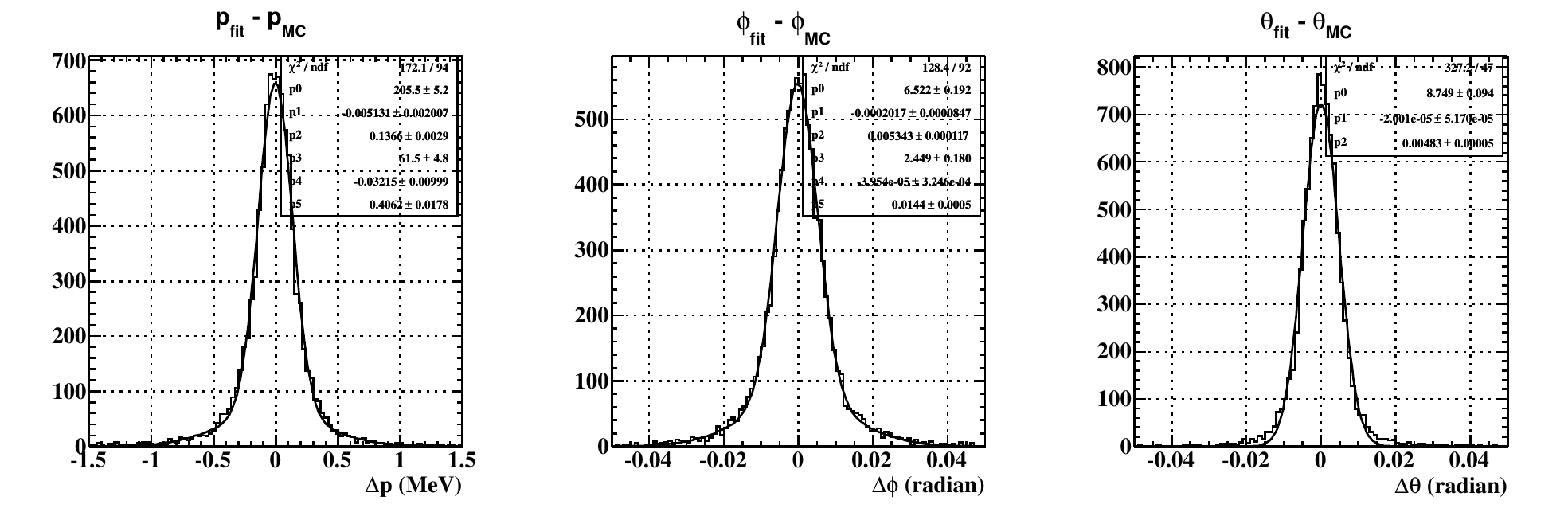}
    \caption{\label{fig:TPC_momreso}
    Example of reconstructed momentum and angular resolutions
  }
  \end{center}
\end{figure}
Of course, the accuracy of reconstruction strongly depends on
the diffusion (gas mixture), intrinsic spatial resolution (pad size), 
{\it etc.}, and hence, the intrinsic spatial resolution dependence
and the diffusion coefficient dependence of the track-reconstruction
accuracy are also studied.
\begin{figure}[htb]
  \begin{center}
    \subfigure[Resolution as a function of diffusion coefficient]{
     \includegraphics[height=6cm]
     {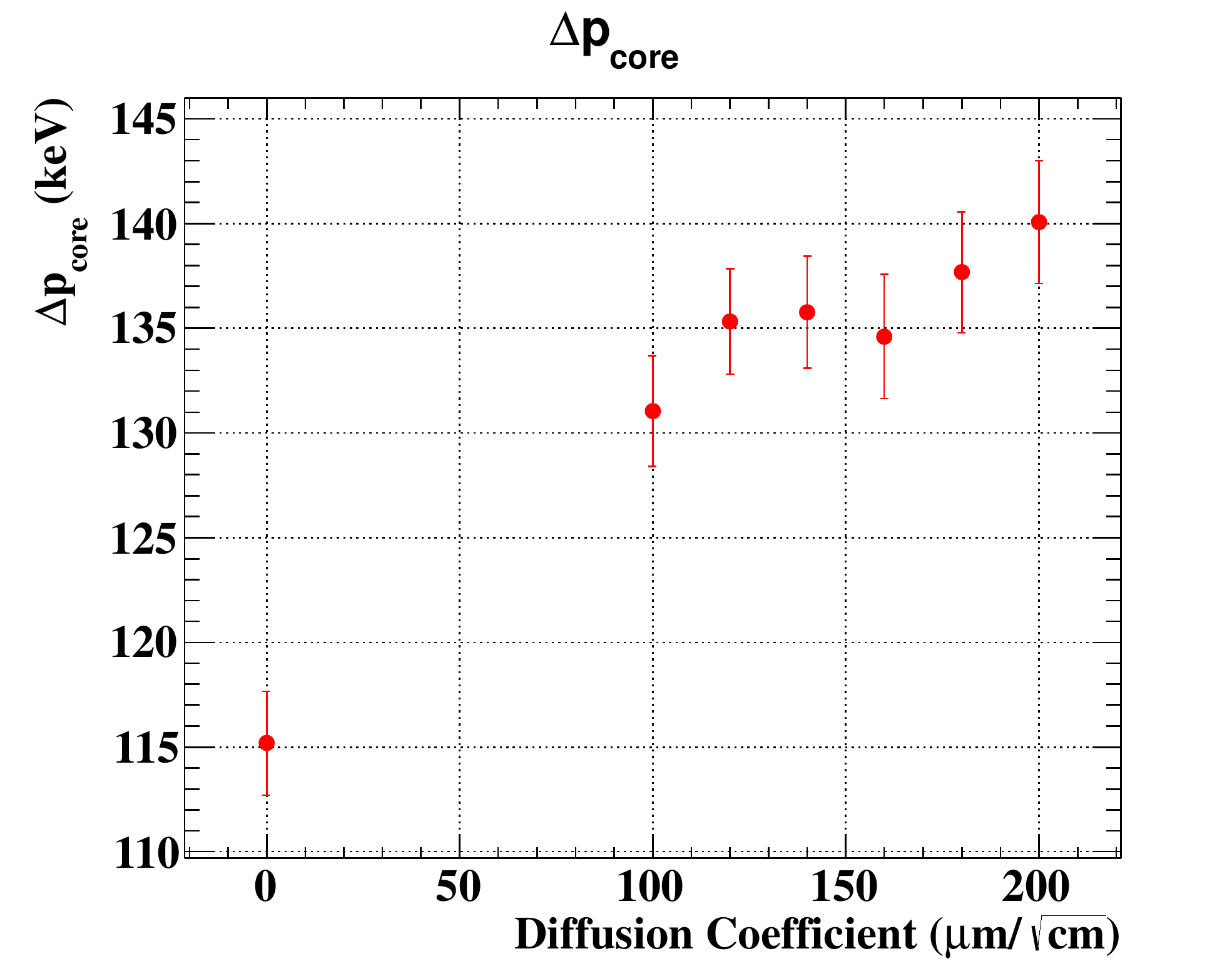}
      }
    \subfigure[Resolution as a function of intrinsic spatial resolution]{
     \includegraphics[height=6cm]
     {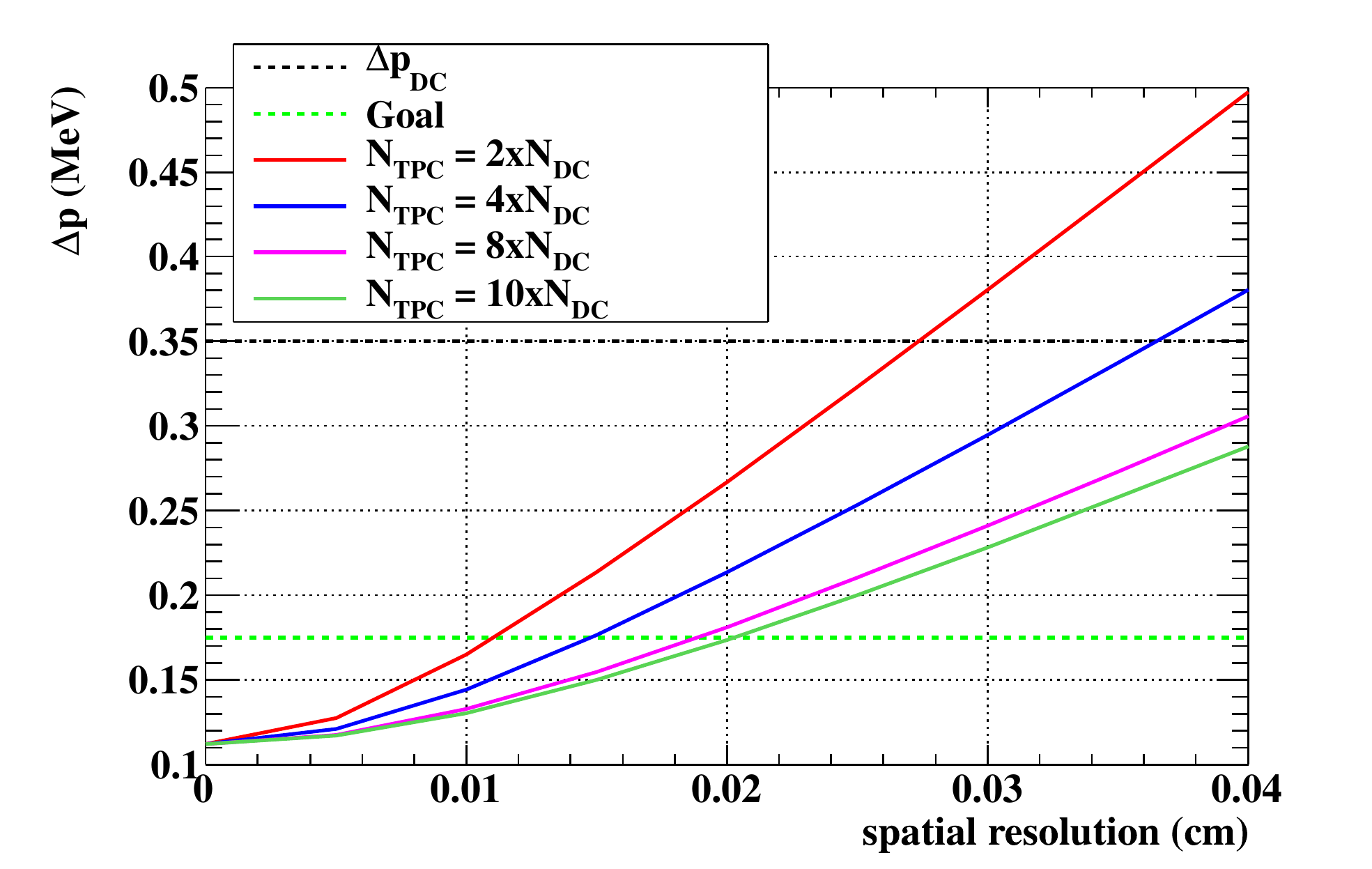}
    }
    \caption{\label{fig:TPC_momreso}
    Reconstructed momentum resolution studies
  }
  \end{center}
\end{figure}
According to the MC simulation studies,
130-160 keV/$c$ of momentum resolutions could be expected
depending on diffusions and spatial resolutions, as shown in 
Fig. \ref{fig:TPC_momreso}.
In addition to the momentum resolution, $4 \div 9$ mrad of
angular resolution and $80 \div 90\%$ of detection efficiency
are also expected.

\subsubsection{Prototyping Plan}
\begin{figure}[htb]
  \begin{center}
     \includegraphics[height=4.5cm]
     {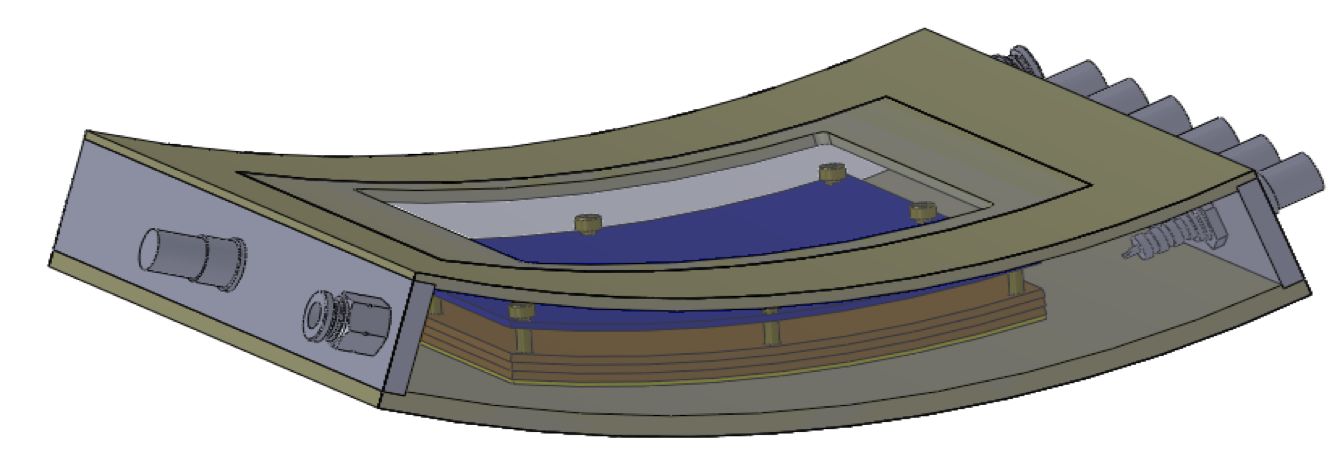}
    \caption{\label{fig:TPC_prototype3d}
    3D view of the first prototype of cylindrical GEM
  }
  \end{center}
\end{figure}

Here some prototyping plans are presented.
First of all, it is certainly required to clarify if 
it is possible to build the cylindrical GEM detector even with 
larger dimension than the previously actualized detector
which should be matched with our COBRA magnet geometry.
Thus we are now building the first prototype
which is shown in Fig. \ref{fig:TPC_prototype3d}.
As shown in this figure, a three layered GEM detector is adopted as 
the first prototype.
The GEM foil has approximately 10$\times$10 cm$^{2}$ of area
and pattern characteristics is optimized for the E16 experiment
at J-PARC, gap is 2 mm and 10$^4$ order of gain in total is expected.
The most important challenge of this prototype is
this three-layered GEM detector is bent with
30 cm of curvature which is corresponding to COBRA radius.
This first prototype currently under construction will be used to
study the following items:
\begin{itemize}
  \item{how to build the cylindrical GEM for COBRA,}
  \item{use of helium-base gas mixture for GEM,}
  \item{fundamental characteristics; {\it e.g.} gas gain, rate dependence, 
        diffusion coefficient, efficiency, intrinsic spatial resolutions,}
  \item{behavior in the magnetic field, in particular, COBRA field.}
\end{itemize}

After this first prototype, we will be able to step into
the second prototype which has much larger dimension.

%% file: 13_Appendix/Positron_Tracker_SVT.tex
%
\subsection{Silicon vertex tracker option: SVT}
\subsubsection{Concept}
Solid state detectors have been widely used for tracking charged particles in high-energy physics. Recently silicon pixel detectors have been successfully implemented in LHC detectors to obtain high precision position resolution in high rate environments.
The advantages to use such detectors are 
\begin{itemize}
	\item good stability of operation under high rate,
	\item high rate tolerance with low occupancy,
	\item high spatial resolution ($O(10~\mathrm{\mu m})$).
\end{itemize}
These features are attractive for our upgrade project as well. However, those devices have been considered to be unsuitable for low energy experiments because of the material budget limiting the performance. To overcome this issue, many R\&D are now under way mainly for future lepton collider experiments such as super B factories and ILC. 
Several devices such as DEPFET \cite{Lutz2001}, MAPS \cite{Turchetta2006}, and SOI \cite{Arai2010} are getting feasible for practical usage. With these new technologies very thin sensors down to several tens micro meter could be available in a few years from now.
Among them, the High-Voltage Monolithic Active Pixel Sensor (HV-MAPS) \cite{Peric2007} which is proposed to be used in the $\mu\to {\rm eee}$ experiment at PSI (Mu3e) \cite{mu3eLoI} is in particular a good candidate, because many of the experimental requirements are shared between $\meg$ and $\mu\to {\rm eee}$ searches.  Here, we propose an option of MEG-upgrade positron tracker using such new silicon device inspired by the Mu3e project.

\begin{figure}[htb]                                                                                                                                                                  
\begin{center}
\includegraphics[width=1\linewidth]{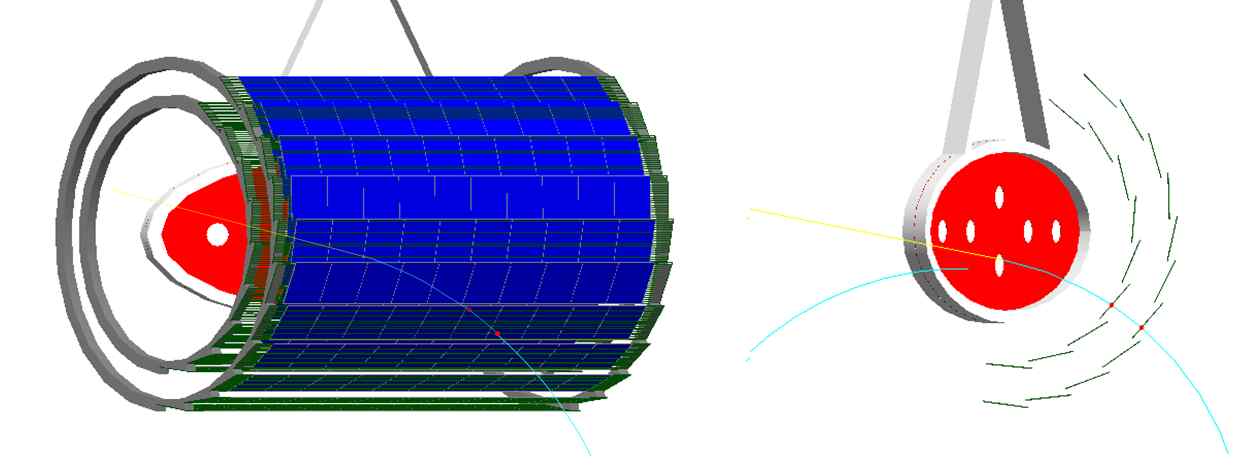}
\caption{\label{fig:svt}Schematic view of double layer SVT.
}
\end{center}
\end{figure}

In application to a $\meg$ search, one has to carefully design the detector to minimize the generation of photons becoming accidental background source. Not only positrons from $\megc$ decays in acceptance region but also all possible particles have to be considered. The dimension of the tracker system is already fixed by the magnet. Because of the size, making a full tracker with silicon sensors is not feasible to fit our limited time scale and budget.
From these limitations, introducing a silicon tracker as a vertex detector is an effective and efficient solution. Fig.~\ref{fig:svt} shows the conceptual design of the silicon vertex tracker (SVT). The silicon sensors are placed only at small radii to form a couple of layers, and cover only the signal positron acceptance, namely the opposite side of the photon detector. This configuration works to minimize the generation of photons pointing to the photon detector as well as to reduce the area to be covered. Because measurements with a couple of layers are not sufficient to reconstruct positron trajectories, a main tracker is necessary outside SVT. Hence, SVT is not an alternative solution of the new tracker, but an optional detector. 
Even with a few measurement points, SVT gives us powerful functions in tracking:
\begin{itemize}
	\item providing precise measurements of muon decay verteces,
	\item providing precise measurements of positron emission angles,
	\item improving momentum measurement by providing precise initial point of tracking.
\end{itemize}
These functions can be obtained by putting the tracker close to the vertex points where the hit rate is extremely high.

One may think such an additional detector is not necessary if we construct a new tracker proposed in previous sections. We think it is important to have a redundancy of functions among detectors to promise the proposed performance. In this sense, SVT perfectly works. It might happen that the expected position resolution of the main tracker cannot be achieved for example because of a bad noise situation (as that happened for current DC system) until the final detector is constructed and the performance is verified. Even in such a case, SVT makes it possible to achieve the proposed resolutions.
However, the greatest role of SVT would be given by the point that introducing SVT makes the design and construction of main tracker simpler and easier. Requirements for the main tracker will be relaxed and its development can be concentrated on the viewpoints of the detection efficiency and high-rate tolerance. For example, the limitation of beam rate for the proposed  drift chamber is given by the hit rate at the inner-most wires. With SVT, we can think of the modified configuration of the new drift chamber where the radius of the inner-most wires is increased or the inner-most wires are removed without a cost of resolution. 
In such a way, we get the room of further improvement of the sensitivity.

\subsubsection{Design}
\begin{figure}[htbc]
\begin{center}
\includegraphics[width=.32\linewidth]{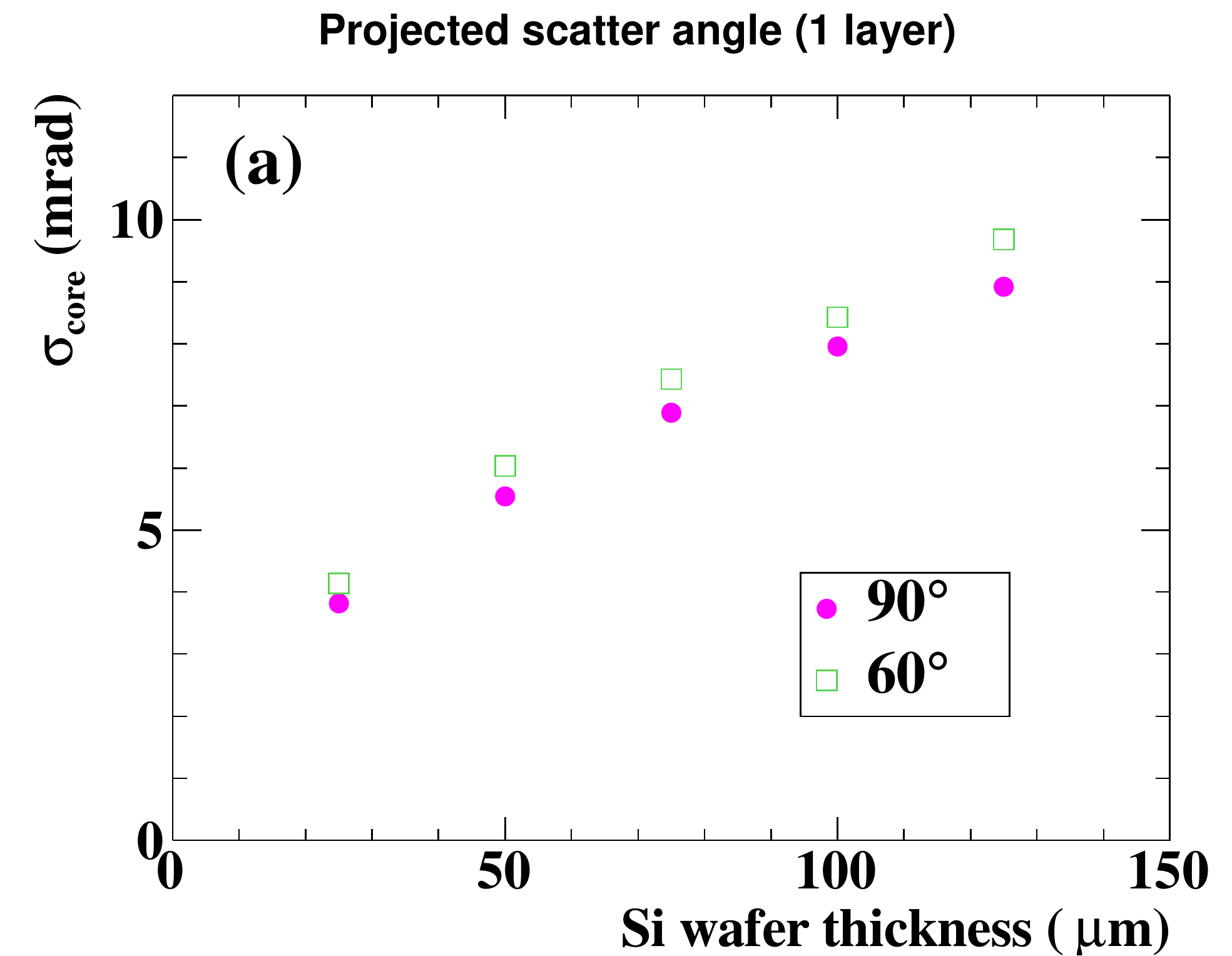}
\includegraphics[width=.32\linewidth]{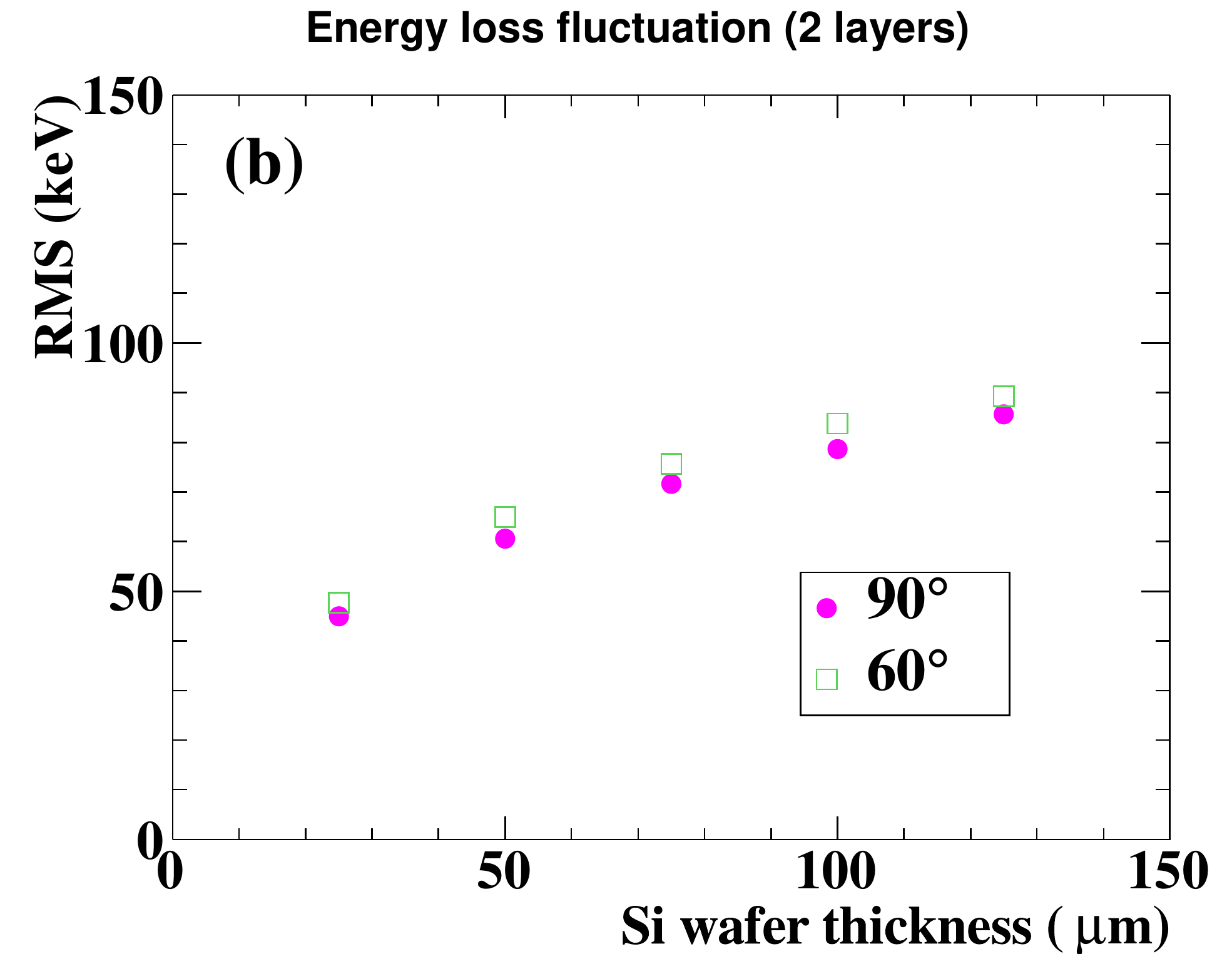}
\includegraphics[width=.32\linewidth]{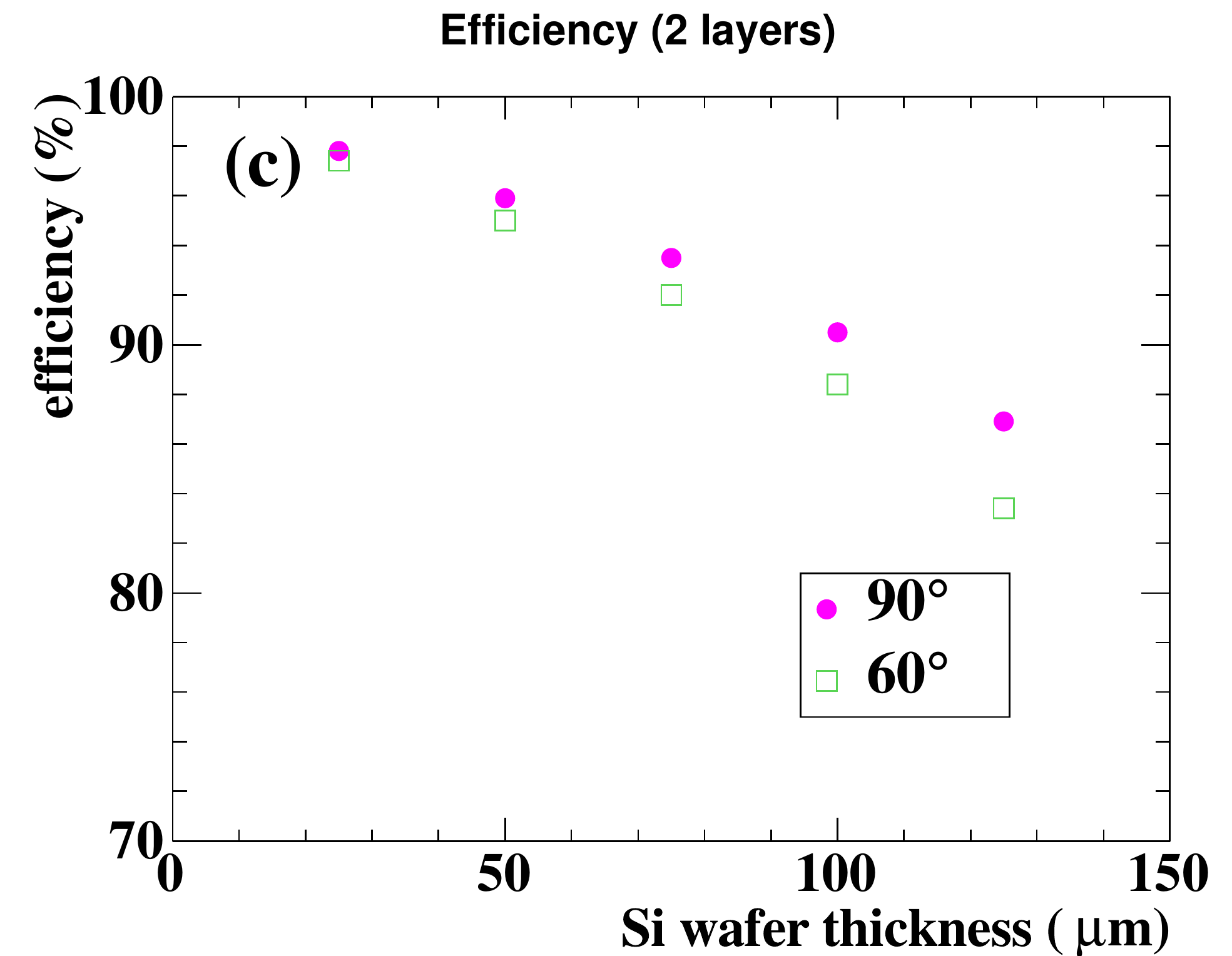}
\caption{\label{fig:svt_limitation} Intrinsic limitation on measurement of 52.8~MeV positrons by the thickness of silicon sensor for two incident angle cases. (a) Limitation on the angle measurement coming from the projected (plane) scatter angle at the first layer. (b) Limitation on the energy measurement coming from the energy loss fluctuation on both layers. (c) Limitation on the detection efficiency. Efficiency here is defined as the fraction of events with the (space) scatter angle by the two layers is less than 50~mrad and the energy loss is less than 300~keV. }
\end{center}
\end{figure}

The most important parameter is the thickness of the silicon sensors. Low momentum positrons at about 50~MeV can be easily scattered by the nuclei of material put on its way. To study the intrinsic limitation from material effect, a simple MC simulation was performed. The results are shown in Fig.~\ref{fig:svt_limitation} for angle- and energy-measurement as well as detection efficiency as a function of the sensor thickness. It is turned out that the sensor of 100~$\mu$m thickness or thicker is unacceptable for our purpose from the angle measurement point of view.
On the other hand, it is also figured out that putting SVT does not affect the energy measurement.
The timing resolution (or the readout speed) gives another requirement for the device when it is used at a high rate. To restrict the number of positrons hitting SVT in an event to less than a few dozens, a resolution of a few hundreds nanosecond is required. From these requirements, HV-MAPS is the first candidate. The advantages of this device are the possibility to be thinned (down to $30~\mu\mathrm{m}$), $100\%$ fill factor, the good S/N, the fast signal collection, high radiation tolerance, the possibility to implement CMOS in-pixel electronics, and a low price. These features as well as the principle of the charged particle detection have been successfully tested \cite{Peric2010,Peric2011}. As the basis of the design, we assume a sensor similar to one proposed in the Mu3e experiment. Numbers assumed in the following study is summarized in Tab.~\ref{tab:SVTParameter}. The digitization and readout circuit will be implemented in the sensor itself by the CMOS technology. Thus, no other components like ASIC are necessary. The digitized signal is readout through flexible print on which the sensors are glued. This flexible print also works as a mechanical support at the barrel part. Since the power consumption of the MAPS-type devices is low enough to be cooled via helium gas, additional cooling system is not necessary.
The material budget is $0.8\times10^{-3}~ \mathrm{X_0}$ per layer.

The momentum resolution in a given magnetic field depends on the number of measurement points, the position resolutions, the length of the measured track segment, and the multiple scattering. Adding SVT measurements effectively works to increase the measurement length, and hence to increase the sagitta. Increasing the number of layers between the first layer and the main tracker does not work for momentum measurement because the effect of multiple scattering dominates the effect of position resolution\footnote{This is opposite to the case for the gaseous detectors where the effect of multiple scattering is smaller than the effect of worse position resolution which can be reduced by increasing the measurement points}. 
Precise measurements of vertex and angle can be achieved with at least two layers. Thus we propose a double-layer configuration for the basic design as schematically shown in Fig.~\ref{fig:svt}.
The radii of layers are one of the important parameters not only for the resolutions but also for the number of readout and mechanical structure. From the resolution point of view, smaller radii are better; the vertex resolution is, in the first order, proportional to the radius of the first layer. The limitation comes from the beam size. With current beam profile, it is not possible to put SVT at radius smaller than $\sim5$~cm.
The interval of layers contributes to the precision of the angle measurements. To minimize the finite size of the pixel, the interval should be larger; an interval of 10~mm corresponds to about 3~mrad contribution, and linearly suppressed as the interval increases.
The actual radial positions will be determined by considering practical mechanical design, while we study with several numbers ranging from 5 to 8~cm.
The area to be covered is defined by the angular acceptance of the whole spectrometer and the beam spot size. Because of the finite beam spot size and slanted target, SVT should cover $\sim 160^\circ$ in $\phi$ and $\sim20$~cm in z direction (about 35\% solid angle). To cover the range, about 200 sensors are necessary, resulting in about 12 million pixels in total.
At a 100~ns readout speed, average number of trajectories making hits on SVT is about three, and the maximum hit rate of a pixel is about 20~Hz (occupancy of $2\times 10^{-6}$) at $1\times 10^8 \mu^+/\mathrm{sec}$.
The layout of the sensors is a usual ladder structure. The slant angle of the sensors from the tangent is optimized for signal positron to pass minimum length. The overlap of sensors is necessary for alignment purpose as well as to compensate possible dead space for the readout.

Other configurations are also under consideration.
A single-layer configuration has a large merit from the material budget viewpoint, and it can be better depending on the performance of the main tracker. 
The single layer SVT and an active target can also be a good solution.
The best configuration will be determined in combination with the main tracker performance. 

\begin{table}[htb]
\caption{\label{tab:SVTParameter}Design parameters of silicon sensors.}
\begin{center}
\begin{tabular}{lr}
\hline \hline
Sensor size         &  $20\times 20~\mathrm{mm}^2$  \\
Sensor thickness &  $50~\mathrm{\mu m}$ \\
Pixel size           & $80\times 80~\mathrm{\mu m}^2$ \\
Readout speed (timestamp) & 100~ns\\
\hline \hline
\end{tabular}
\end{center}
\end{table}

\subsubsection{Expected performance}
\begin{figure}[htb]                                                                                                                                                                  
\begin{center}
\includegraphics[width=.6\linewidth]{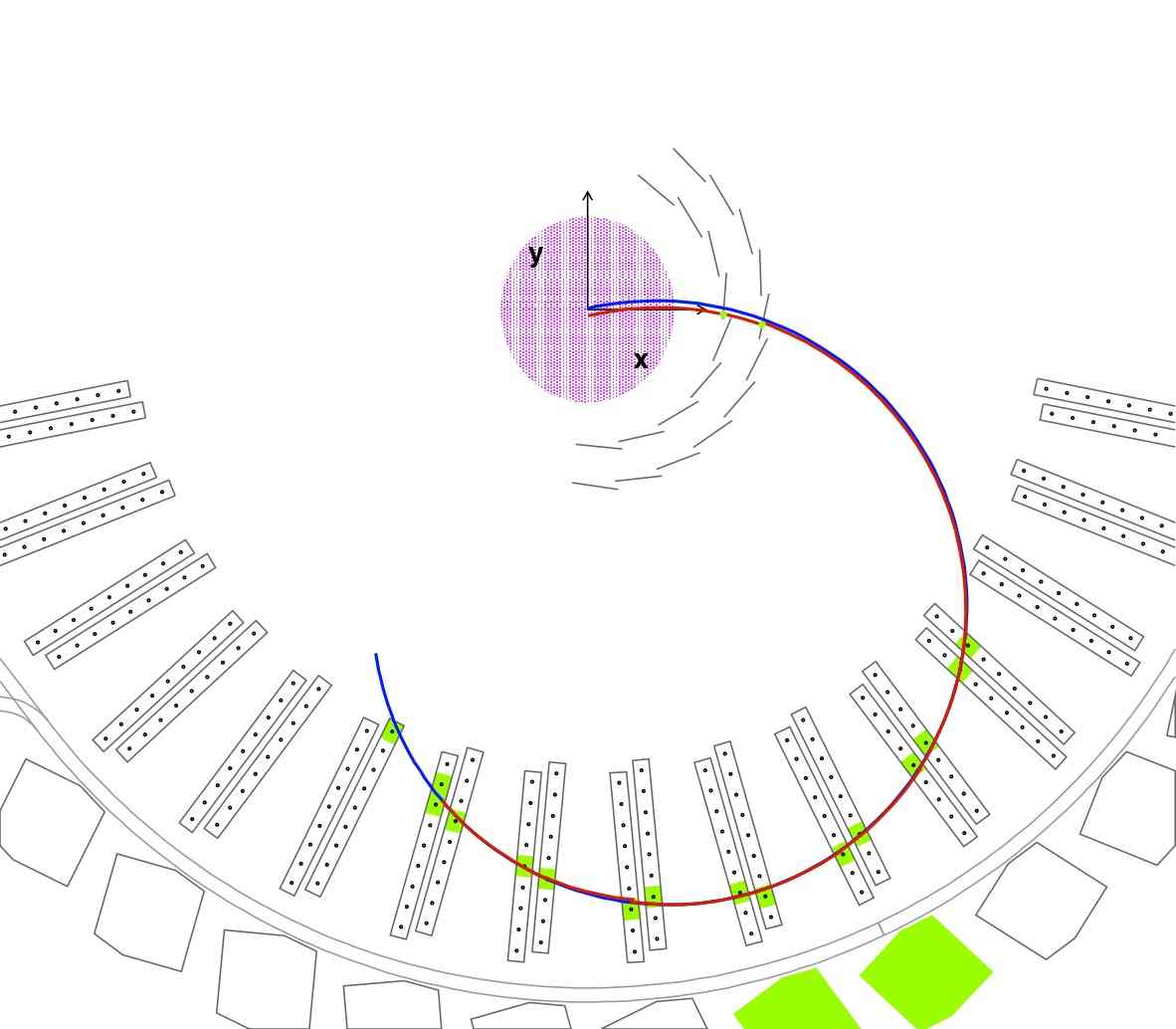}
\caption{\label{fig:svt_rec}Tracking with SVT and current DC for a simulated signal positron. Blue line shows the reconstructed trajectory only using DC measurements, while red one shows that with SVT measurements too.
}
\end{center}
\end{figure}

The expected performance of SVT is being evaluated using MC simulation based on GEANT4.
The functionality of SVT is studied in combination with the current DC system.
Two MC samples of signal positrons, with and without putting SVT, were generated without event mixing, and analyzed with our reconstruction tool.
In the case of with SVT, first the tracking is performed as usual only with DC, and then the interconnection is tried between the extrapolated trajectory and the SVT hits. If the interconnection is successful, the SVT hits are integrated into the Kalman filter and final track fit is performed. 
An example of the track reconstruction is shown in Fig~\ref{fig:svt_rec}.  
The performances of the two cases are evaluated after applying our nominal event selection.
The results are summarized in Tab.~\ref{tab:SVTPerformance}. 
Because of the current reconstruction procedure, adding SVT hits cannot increase the number of reconstructed tracks. Even with such a preliminary reconstruction, the total efficiency after applying the final event selection increases thanks to the improvement of the reconstruction quality.

\begin{table}[htb]
\caption{\label{tab:SVTPerformance}Comparison of the spectrometer (current DC base) performance with and without SVT (double-layer at (5.9, 7.5 cm)). Resolutions are given in sigma.}
\begin{center}
\begin{tabular}{l c c}
\hline \hline
 & DC only & DC \& SVT \\
\hline
Momentum resolution, core component (keV)  &  $233$  & $150$\\
Fraction of the core component (\%) & 83 & 91\\
$\phi$ resolution (mrad) &  $12$ & 7.2 \\
$\theta$ resolution (mrad) & $8.2$ & 7.0 \\
Vertex resolution, Y / Z (mm) & 1.8 / 2.3 & 0.48 / 0.39\\
Efficiency (\%) & 48 & 52\\
\hline \hline
\end{tabular}
\end{center}
\end{table}

\begin{figure}[htbc]
\begin{center}
\includegraphics[width=.45\linewidth]{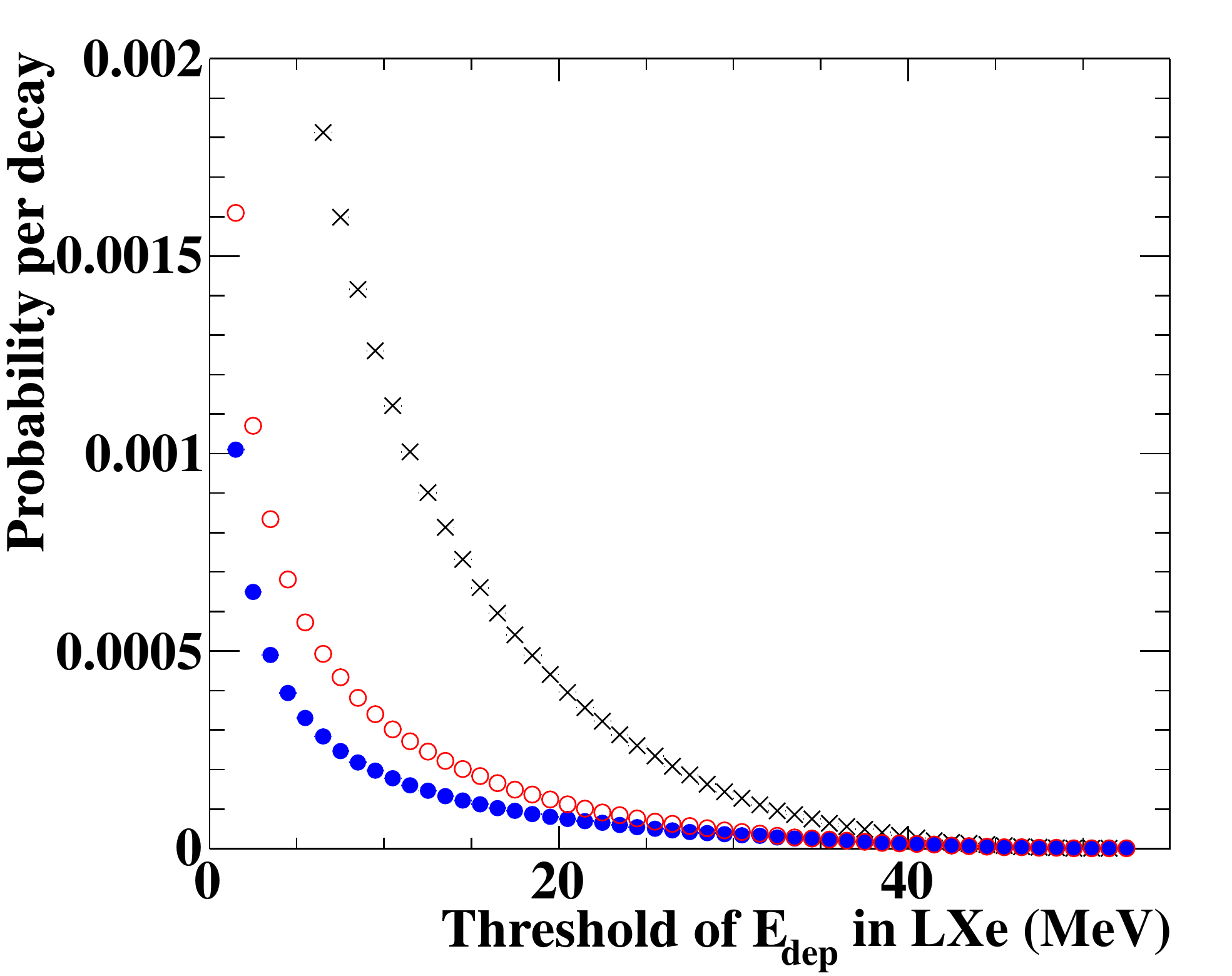}
\includegraphics[width=.45\linewidth]{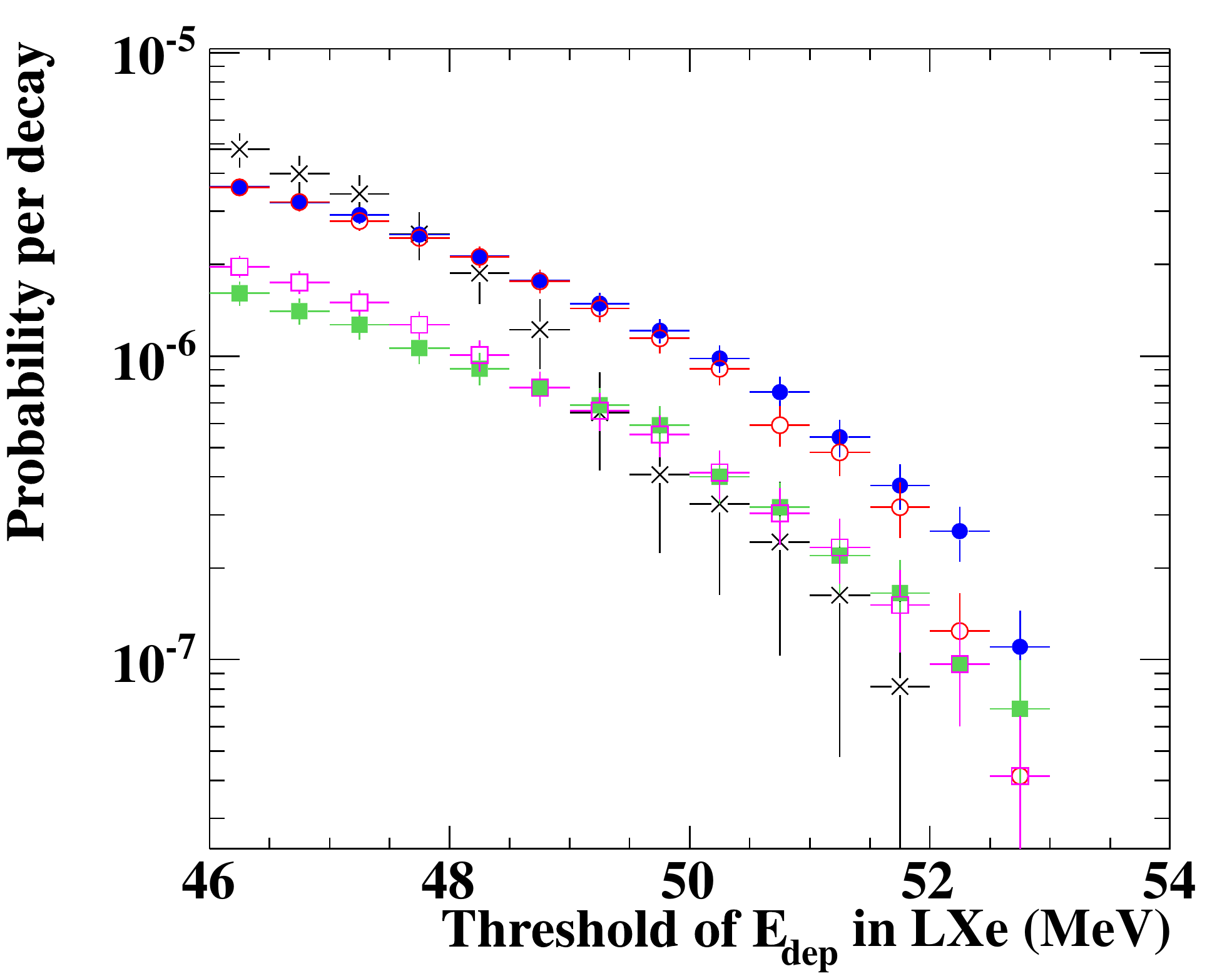}
\caption{\label{fig:svt_photonyield} Integrated photon yield per muon decay. Right figure is the close-up view around the signal region, where the cases of TPC option are also shown. Black cross markers shows photon yield from RMD and the others show those originated from positrons (bremsstrahlung or AIF). Red: DC (current system), blue: DC +SVT, green: TPC, magenta: TPC + SVT configuration, respectively.}
\end{center}
\end{figure}

The effect on the additional photon yield is also evaluated with the MC. The photon yield from Michel positrons (bremsstrahlung or AIF) are shown in Fig.~\ref{fig:svt_photonyield} for current system and that with SVT as a function of the energy deposition in LXe of the photon detector. 
The increase of the yield is clearly seen for low energy part, while at high energy the yield is not increased because of the geometrical configuration.
The main contribution for the high energy photon yield comes from DC, while it will be significantly reduced with  a new tracker in upgrade experiment. Thus, we also investigated the photon yield with TPC tracker.
The results of photon yield which make energy deposit in LXe larger than $0.9\times M_{\mu}/2$ are summarized in Tab.~\ref{tab:SVTPhotonYield}. 
The increment due to SVT for the total photon yield (including one from RMD) can be suppressed lower than $5\%$ even .
Low-energy photons increase the hit rate of the photon detector and can be backgrounds via pileup.
The photon yield originated from positrons increases by $75\%$ for $E_\mathrm{dep} >10~\mathrm{MeV}$, while 
the dominant source at the region is RMD. The total increment due to SVT is about $10\%$.


\begin{table}[htb]
\caption{\label{tab:SVTPhotonYield} Photon yield for energy deposition in LXe larger than $0.9\times M_{\mu}/2$.}
\begin{center}
\begin{tabular}{l c}
\hline \hline
 Detector setup & yield ($10^{-6}$) \\
\hline
RMD & $2.5$\\
AIF, DC (current system) & 2.5\\
AIF, DC + SVT & 2.5 \\
AIF, TPC & 1.1 \\
AIF, TPC + SVT & 1.2\\
\hline \hline
\end{tabular}
\end{center}
\end{table}


%% file: 13_Appendix/ptp.tex
\subsection{R\&D on New Scintillator Material for Timing Counter}
\label{sec:p-Terphenyl}
In the last few years organic scintillators have been investigated 
in order to achieve higher light yield than the usual plastic or liquid scintillators.
It has been shown that very pure organic compound with high quality crystal structure 
can achieve superior light yield performance in combination with fast decay time.
Among a large  number of types of organic compounds only few ones have proven to have practical applications.
These are for example  the stilbene and {\it p}-terphenyl\,\cite{Budakovsky}. 
Both have light yield about 3 times higher than ordinary plastics and  a decay time of few nanoseconds.
It has been found also that the crystal sizes must be larger than the range of the charged particles to be detected, 
otherwise the light output decreases of a large amount. 
In case of particle with small range, the detector can be made of poly-crystal samples. 
Conversely, in our case whose positrons have tracks longer than the typical sizes of scintillators, 
we need to adopt large pieces of single crystals.  
Large single crystals begin to be commercially available: 
single crystals with linear sizes of 10 cm have been recently produced.
Doped {\it p}-terphenyl single crystal is a very appealing option for a fast and high light output pixel of the timing counter. 
Below are listed the measured properties of actual samples from ALKOR company\,\cite{ALKOR}. 
We notice that the data are preliminary  since the investigations are in progress at the INFN of Genova. 
\begin{itemize}
\item Light yield: $2.7 \times 10^4 (2.7 \times 10^4)$
\item Maximum of emission: 390\,nm (420\,nm)
\item Attenuation length (see explanation): $>5$\,cm
\end{itemize}
The emission spectrum has characteristic features with 3 peaks and is  slightly blue-shifted. 
As suggested by previous studies reported in literature this could be caused by two effects: 
first, a reduced self absorption at lower wavelengths due to {\it p}-terphenyl fine powder at the not polished surfaces of the sample and, 
second, to a lower level of dopant, biphenyl butadiene, that act as wavelength red-shifter.
The attenuation length has been obtained with a {\it p}-terphenyl slab with section of $3\times 5 \,\mathrm{mm}^2$ and $50\,\mathrm{mm}$ length. 
The slab has been wrapped with high reflectance foil from 3M and the surfaces are finished at optical level. 
The data have been acquired with two Hamamatsu $1\,\mathrm{mm}^2$ SiPMs at each end and a collimated beam of electrons from a $^{90}$Sr radioactive source. 
Thanks to the high light yield the SiPM pulse amplitude, about  hundred of mV, signal is sent to the digitizer without any loss of signal to noise ratio. 
These features are very appealing for a possible application as active material for the pixels of the timing counter. 
Further assessments are necessary and a proper evaluation of cost to benefit ratio.

%% file: 13_Appendix/LXePMT.tex

\subsection{Development of New Photomultiplier Tube (PMT) for LXe Detector}
\label{sec:LXe PMT}
A small and square-shaped PMT can also be a candidate for the replacement of the current PMT. 
The biggest advantage is that it is a well-proven technology and has been long working quite well in the MEG experiment.  
The drawbacks compared to SiPM are its larger thickness and thick insensitive edge of the pressure-proof package.

Two types of PMTs are under development in collaboration with Hamamatsu Photonics.

\begin{itemize}
\item 1-inch square-shape PMT

The 1-inch square-shape PMT is a smaller version of the current PMT as shown in Fig.\,\ref{fig:1inch square-shape PMT}.
The quantum efficiency (QE) is expected to improve from 15\% to 20-30\% for LXe scintillation photon.
The prototypes are being tested in the test facilities at KEK and Pisa.

\begin{figure}[htb]                                                                                                                                                                  
\begin{center}
\includegraphics[width=6cm]{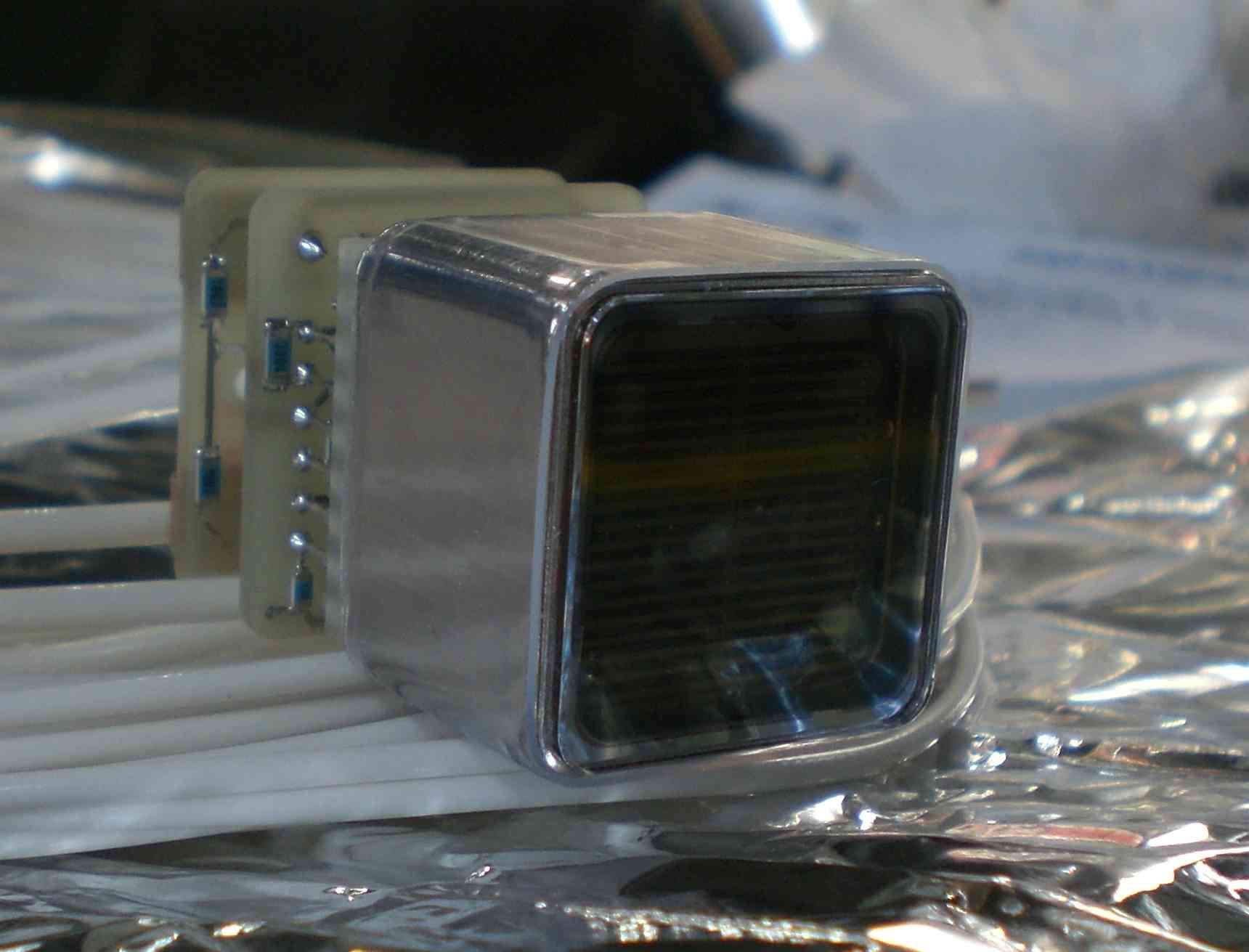}
\caption{\label{fig:1inch square-shape PMT}
   1-inch square-shape PMT developed for the MEG experiment.
}
\end{center}
\end{figure}

\item 2-inch flat panel multi-anode PMT 

The left picture in Fig.\,\ref{fig:2-inch flat panel PMT} shows a sketch of the prototype of
2-inch flat panel multi-anode PMT developed for LXe use.
It is based on Hamamatsu H8500 64-channels multi-anode PMT.
It has the same dynode structure as the current PMT (metal-channel) enabling its compact design,
fast response (TTS of 0.4\,ns) and high gain ($1.5\times 10^6$ at -1100\,V). 
The dimension is $52\times 52\times 27.4\,\mathrm{mm}$ which is not smaller than the current PMT,
but it is position sensitive with $8\times 8$ pixels with a size of $5.8\times 5.8\,\mathrm{mm}^2$ each.
The high voltage is common to all the channels and the channel-to-channel gain variation is $\pm 20$\%.  
A careful calibration would, therefore, be required. 
The same bleeder circuit as for H8500 can be used with Zener diode protection for high rate operation. 

A pressure-proof package design is in progress as shown in Fig.\,\ref{fig:2-inch flat panel PMT} 
and the first working prototype is expected to arrive in 2013.

\begin{figure}[htb]                                                                                                                                                                  
\begin{center}
\includegraphics[width=9cm]{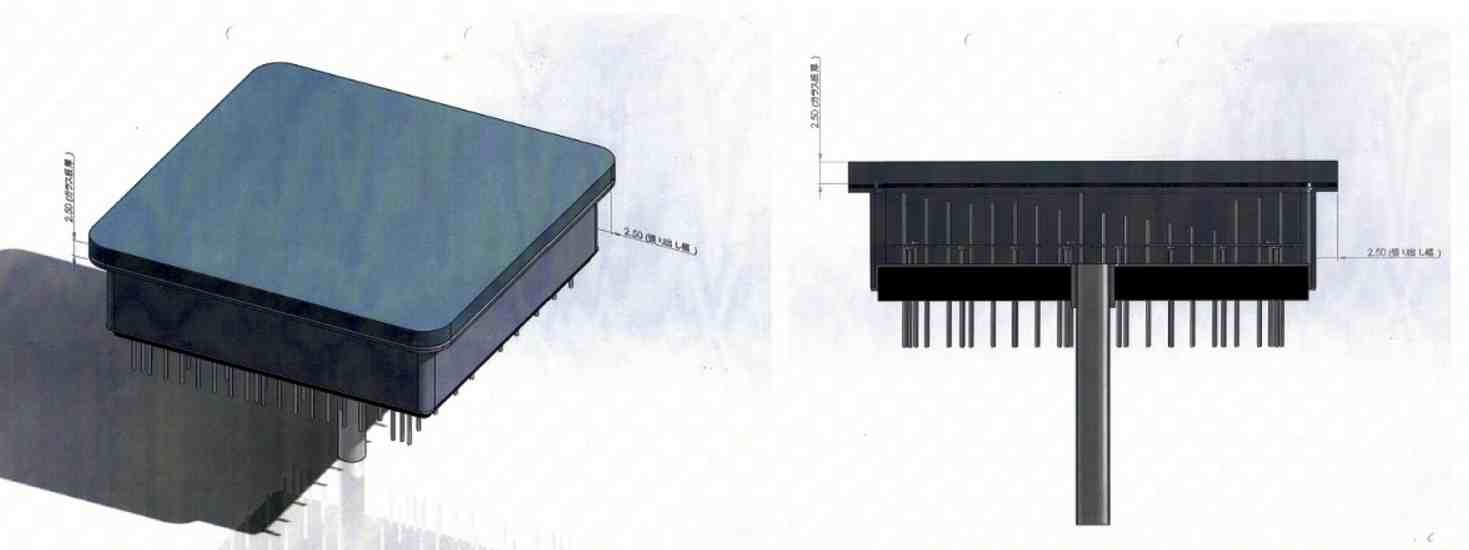}
\includegraphics[width=6cm]{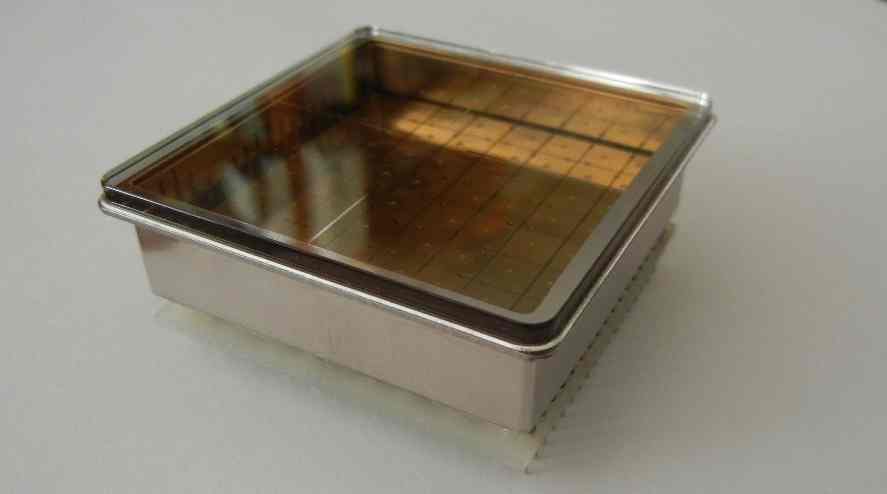}
\caption{\label{fig:2-inch flat panel PMT}
   (Left) Design of the prototype of 2-inch flat panel multi-anode PMT developed for LXe in collaboration with Hamamatsu Photonics 
   and (right) a prototype of the pressure-proof package with a quartz window where a VUV-sensitive photocathode is deposited, 
   but without a dynode structure.
}
\end{center}
\end{figure}

\end{itemize}